\documentclass[10pt,dvipsnames]{article} 

\PassOptionsToPackage{round,sort&compress}{natbib}
\usepackage[left=1in,right=1in,top=1in,bottom=1in]{geometry}
\usepackage{xcolor}
\usepackage[round, sort&compress]{natbib}
\usepackage{amssymb,amsthm, amsfonts}            
\usepackage{mathtools, thmtools}          
\usepackage{mathrsfs}           
\mathtoolsset{showonlyrefs}     
\usepackage{graphicx}           
\usepackage{subcaption}         
\usepackage[space]{grffile}     
\usepackage{url}                
\usepackage{lipsum}             
\usepackage{algorithm}
\usepackage[noend]{algpseudocode}
\usepackage{booktabs}
\usepackage{enumitem}
\usepackage{hyperref}
\usepackage{titletoc}
\hypersetup{
    colorlinks=true,
    linkcolor=Cerulean,
    citecolor=Cerulean,
    urlcolor=Magenta,
    pdftitle={RQRE},
    pdfauthor={Jake Gonzales, Max Horwitz, Eric Mazumdar, Lillian J. Ratliff}
}
\theoremstyle{definition}
\newtheorem{theorem}{Theorem}
\newtheorem{lemma}{Lemma}
\newtheorem{proposition}{Proposition}
\newtheorem{assumption}{Assumption}
\newtheorem{definition}{Definition}
\newtheorem{corollary}{Corollary}
\newtheorem{remark}{Remark}
\newtheorem{example}{Example}

\newcommand{\mc}{\mathcal}
\newcommand{\mb}{\mathbb}
\newcommand{\R}{\mathbb{R}}
\newcommand{\RR}{\mathbb{R}}
\newcommand{\EE}{\mathbb{E}}
\newcommand{\PP}{\mathcal{P}}
\newcommand{\LL}{\mathcal{L}}
\newcommand{\F}{\mathcal{F}}
\newcommand{\G}{\mathcal{G}}
\newcommand{\Wspace}{\mc{W}}
\newcommand{\Yspace}{\mathcal{Y}}
\newcommand{\Xspace}{\mathcal{X}}
\newcommand{\eps}{\varepsilon}
\DeclareMathOperator*{\E}{\mathbb{E}}
\DeclareMathOperator*{\KL}{\mathrm{KL}}
\DeclareMathOperator*{\argmax}{arg\,max}

\newcommand{\sym}{\mathsf{sym}}
\newcommand{\Risk}{\mc{R}}

\newcommand{\CVaR}{\mathrm{CVaR}}
\newcommand{\Var}{\mathrm{Var}}
\newcommand{\Cov}{\mathrm{Cov}}

\newcommand{\TV}{\mathrm{TV}}

\newcommand{\IPO}{\mathrm{IPO}}
\newcommand{\piref}{\pi_{\texttt{ref}}}
\newcommand{\thetaref}{\theta_{\mathrm{ref}}}

\newcommand{\bpi}{\bar{\pi}}
\newcommand{\sg}{\mathrm{sg}}
\newcommand{\Errvi}{\mathrm{Gap}_{\mathrm{vi}}}
\newcommand{\gO}{\mathcal{O}}

\newcommand{\dd}{\,\mathrm{d}}

\newcommand{\tr}{\mathrm{tr}}

\newcommand{\Distmat}{\mc{J}}
\newcommand{\Loss}{\mathcal{L}_{\IPO}^{\Risk}}

\newcommand{\softmax}{\operatorname{softmax}}
\newcommand{\refer}{\mathrm{ref}}

\definecolor{cBad}{HTML}{C0392B}
\definecolor{cTeal}{HTML}{1B9E8A}
\definecolor{accent}{HTML}{C0392B}

\title{Structure from Strategic Interaction \& Uncertainty:\\ Risk Sensitive Games for Robust Preference Learning}

\author{Max Horwitz\textsuperscript{1}\and Jake Gonzales\textsuperscript{1}\and Eric Mazumdar\textsuperscript{2}\and Lillian J.~Ratliff\textsuperscript{1} }

\begin{document}

\maketitle

\begin{center}
\small
$^{1}$Department of Electrical \& Computer Engineering, University of Washington \\
\small $^{2}$Computing \& Mathematical Sciences, California Institute of Technology 
\end{center}

\begin{abstract}
    A growing line of work reframes preference-based fine-tuning of large language models game-theoretically: Nash Learning from Human Feedback (NLHF) recasts the problem as a zero-sum game over policies. However, optimization is over expected pairwise payoffs, thereby conflating policies with similar win rates but different tail behavior. As such, these methods are agnostic to where in the data distribution they succeed or fail: strong average performance can mask systematic failure across prompts, annotators, or safety-critical strata. We introduce risk-sensitive preference games, in which players optimize convex risk measures of their preference loss, exploiting structure in preference uncertainty. While risk-sensitivity generally breaks the zero-sum structure, we show that translation invariance of many risk metrics ensures that we retain monotonicity, yielding fast convergence of sample-efficient self-play methods. Furthermore, we establish algorithmic stability and offline sample complexity bounds that scale with risk, requiring simultaneous control of structural bias from nonlinear risk transformations, statistical bias in risk estimation, and concentration tailored to the risk-sensitive setting. To address statistical bias, we introduce a hierarchical game formulation and a two-timescale extragradient algorithm with bias correction that converges to the Stackelberg equilibrium and is especially effective in low-sample regimes. Empirically, risk-adjusted policies are robust across data strata, stable across risk choices, and match or exceed risk-neutral performance thereby achieving robustness without a performance tax.
\end{abstract}

\section{Introduction}
\label{sec:introduction}
Large language models (LLMs) are increasingly deployed at scale, and with that scale comes a growing record of harm: chatbots coaching teenagers toward suicide \citep{hill2025nyt}; ordinary adults with no prior history of mental illness driven into weeks-long delusional spirals---over simulation theory, AI sentience \citep{hill2025spirals}, fabricated mathematical frameworks 
that a Toronto recruiter spent three weeks publicly broadcasting to cryptography experts and national security agencies before the spell broke
\citep{hill2025delusional}; models that learn to identify and manipulate the small fraction of users most vulnerable to manipulation \citep{williams2024targeted}, systems that confidently reinforce delusions while satisfying every aggregate benchmark. 

These are not isolated failures to be patched one at a time. They are the \emph{canary in the mine} for a deeper problem in how alignment research is conceived. The dominant paradigm---RLHF and its variants---collapses rich, heterogeneous human preference data into a single scalar reward and optimizes its expectation (see, e.g., \citep{ouyang2022training, ziegler2020finetuninglanguagemodelshuman,bai2022training,munos2024nash,christiano2017deep,stiennon2020learning,ziegler2019fine}). The structure of human disagreement, the shape of the response distribution, the existence of a long tail of users whose needs differ from the median: all of this information is in the data we collect, and all of it is averaged away before the model ever sees it. Alignment, framed this way, can only be aligned with a fictitious average human; \emph{it has no language for the tails where harm actually occurs, and no language for the diversity of legitimate responses that real users require}. 

We argue that alignment research must be reframed. The objective is not to chase down failure modes after the fact, nor to align to a representative user, but to make the tail behavior and the diversity of the response distribution first-class objects of study; that is, quantities the learning algorithm explicitly reasons about, rather than artifacts it discards. This motivates our central contribution: \emph{extending Nash Learning from Human Feedback (NLHF) with risk-sensitive objectives, so that the equilibrium concept itself reasons about the distribution of outcomes rather than collapsing it to a mean}.

\begin{figure}
    \centering
    \includegraphics[width=0.85\linewidth]{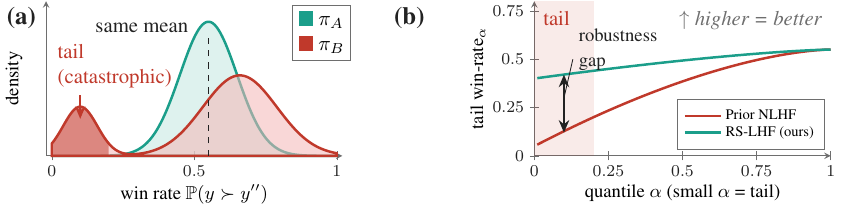}
\caption{RSPGs target tail behavior
directly.
\textbf{(a)} Mean-based methods cannot distinguish two policies
with the same average win-rate but very different tails.
\textbf{(b)} RSPG policies (\textcolor{cTeal}{teal}) maintain
tail performance without sacrificing mean win-rate, while prior
NLHF methods (\textcolor{cBad}{red}) collapse on the tail.}
    \label{fig:placeholder}
\end{figure}

A recent game-theoretic reformulation, NLHF \citep{munos2024nash},
takes a step in this direction by recasting alignment as a two-player
constant-sum game in which the preference oracle is the payoff.  With
a KL penalty to a reference policy, the resulting solution concept is
a unique \emph{quantal response equilibrium} (QRE)
\citep{MCKELVEY19956}, characterized by a fixed-point equation that
Nash-MD \citep{munos2024nash} and follow-up
algorithms---IPO-MD \citep{Calandriello2024HuamnAlignmentIPOMD},
MPO \citep{wang2025magnetic}, EGPO \citep{zhou2025extragradient}---all
target.  NLHF recovers the \emph{strategic interaction} structure that
scalar-reward RLHF discards: it sees a preference distribution, not a
point estimate.  But it stops there.  Every method in this line
optimizes the \emph{expected} pairwise payoff and remains blind to the
second piece of structure that human preference data carries:
\emph{uncertainty}.  A response strongly preferred by part of the
population and strongly dispreferred by another is, in expectation,
indistinguishable from a broadly acceptable one; a policy that wins
on average while occasionally producing catastrophic outputs is
indistinguishable from one that never does (cf. Fig.~\ref{fig:placeholder}).  This work closes that
gap by introducing \emph{Risk-Sensitive Preference Games} (RSPG), a
framework that integrates structured risk-sensitivity directly into
the equilibrium concept of NLHF.  Rather than optimizing the expected
pairwise payoff, RSPG players reason about the full distribution of
preference outcomes---variance, tails, and worst-case responses---yielding
equilibria sensitive to exactly the features that scalar-reward RLHF
and expectation-based NLHF discard.  Risk-sensitivity of this form is
consistent with how natural learning agents behave
\citep{mazumdar2025tractable}.  Crucially, RSPGs are not a
tradeoff: risk reshapes the structural and statistical properties of
the equilibrium to deliver robustness and distributional control
while preserving mean performance 
and adding no computational overhead.

\subsection{Contributions}
We introduce risk-sensitive preference games (RSPG) in which players account for more facets of the outcome distribution when optimizing. Replacing the expected payoff with a \emph{convex risk measure}---a class of functionals developed in mathematical finance, where they have become standard tools for quantifying tail risk and pricing in incomplete markets \citep{follmer2002convex,follmer2016stochastic}---is
 the natural way to operationalize this.
Doing so changes
what alignment is optimizing for: not the average preference outcome
but a chosen feature of its distribution, with the choice of risk
measure encoding which structure of the harms one cares about.
Doing so introduces two distinct difficulties. 
The first is structural: the constant-sum property of NLHF, which is what makes single-player self-play---computationally far cheaper than multiplayer---sufficient for equilibrium computation, relies on the linearity of expectation, and a nonlinear risk functional generically breaks it. The second is statistical: most risk measures do not admit unbiased sample estimators, so stochastic gradient methods inherit a persistent bias floor that does not vanish with the step size---especially damaging in the low-sample regime of LLM fine-tuning, where data is expensive and bias compounds across iterates.
We resolve both.

\begin{itemize}[itemsep=2pt,topsep=0pt]
    \item  \textbf{The structural obstruction is illusory.}
 The \emph{translation invariance} of convex risk measures preserves constant-sum structure at the risk-adjusted
payoff operator (Lemma~\ref{lem:operator-constant-sum}), even though
it fails at the certainty-equivalence level.  Symmetric self-play thus
remains the right algorithm (Theorem~\ref{thm:single-fixed-pt}); risk simply shifts the target to a robust one, and we retain all the computational benefits of running an algorithm that only needs to maintain one policy as opposed to two (one for each player).  We prove $\mu_\Risk = \beta - 2\overline\lambda_\Risk$ strong
monotonicity of the RSPG (Theorem~\ref{thm:single-player-mono}), and therefore existence of a unique equilibrium; last-iterate
convergence of deterministic extragradient at rate $\mc O((1-\eta\mu_\Risk)^t)$; and linear contraction to a neighborhood of size $\mc{O}(1/(m\mu_{\Risk}))$ for 
 stochastic 
 biased extragradient 
(Proposition~\ref{prop:sample-complexity-formal}, and Theorems~\ref{thm:biased-merely-monotone} and~\ref{thm:strong-mono}, for the merely-monotone and strongly monotone settings, respectively, Appendix~\ref{sec:stochastic_convergence}).

\item \textbf{The statistical obstruction is removable.} A
Stackelberg game with a fast bias-tracking follower (Theorem~\ref{thm:strong-mono-tt} and Appendix~\ref{sec:tt-debiasing}) reduces the persistent bias floor of plug-in risk estimators to a
variance floor faced by any stochastic algorithm. 
This  matters most
in the low-sample regime, where statistical bias is reduced to the variance floor
faced by any stochastic
algorithm.   
Algorithmic
stability (Theorem~\ref{thm:stability-informal}) and a
fast-rate offline sample complexity guarantee
$\widetilde{\mc{O}}(1/n)$
(Theorem~\ref{thm:offline-rate}) match the risk-neutral setting up to constants that scale explicitly with the degree of
risk. Risk sensitivity therefore introduces no additional
sample-complexity bottleneck, making learning in RSPGs efficient, and is tunable.

\item \textbf{The framework is Pareto-dominant in practice.} Empirically, risk-adjusted policies are robust across data strata and
match or exceed risk-neutral baselines (Section~\ref{sec:experiments})---\emph{robustness without a
performance tax}.  Risk-adjustment incurs \emph{no statistically significant}
drop in preference win-rate, while safety and combined win-rates
improve consistently across opponents, metrics, and held-out harm
strata.  The degree of risk sensitivity is a
tunable parameter with quantified consequences for learning,
convergence, and generalization.
\end{itemize}

\subsection{Related Work} 
We summarize the most relevant related work; an extended version is in Appendix~\ref{sec:related_work}.

\paragraph{RLHF.}
Reinforcement learning from human feedback aligns language models with
human preferences via a learned scalar reward model optimized against
by PPO \citep{Christiano2017DeepRLHumanPreferences,
ziegler2020finetuninglanguagemodelshuman, schulman2017proximalpolicyoptimizationalgorithms},
with reward hacking and the limits of scalar rewards
\citep{Swamy2024MinimaximalistRLHF, wang2025magnetic} as known
weaknesses.  DPO \citep{rafailov2023direct} sidesteps reward modeling
by reparameterizing the optimal policy directly in terms of
preferences, and GRPO \citep{shao2024deepseekmathpushinglimitsmathematical}
discards the value network in favor of group-relative advantage
estimation.  Closest to our work, \citet{sabbaghi2026robustpolicyoptimizationprevent}
extends GRPO with an entropic risk objective over a KL-bounded
policy neighborhood, but remains single-player and does not engage
with the game-theoretic structure of preference optimization. We refer to the appendix of \cite{zhou2025extragradient} for an extensive discussion of past work in RLHF. 

\paragraph{Game-theoretic preference learning.}
A growing line of work recasts alignment as a two-player game over
policies, using pairwise preferences directly as the payoff.
\citet{azar2024general} introduced the IPO loss to optimize pairwise
preference probabilities without reward modeling or Bradley-Terry
assumptions, and \citet{munos2024nash, calandriello2024human}
formalized the two-player constant-sum framing with self-play
algorithms (Nash-MD, IPO-MD) targeting the quantal response
equilibrium.  Subsequent work has refined the algorithmic side---MPO \citep{wang2025magnetic}, MTPO \citep{NEURIPS2024_d77a7b28},
INPO \citep{zhang2024iterative}, and extragradient-based methods
\citep{zhou2025extragradient}---but all target the same expected
pairwise payoff and remain blind to the distribution of preference
outcomes.  This work addresses that gap, introducing risk-sensitive
preference games in which players optimize convex risk measures
rather than expectations.

\section{Risk-Sensitive Preference Game Preliminaries}
\label{sec:preliminaries}
Let $\Xspace, \Yspace$ denote a finite set of prompts and 
responses, respectively. A policy is a probability distribution $\pi \in \Delta(\Yspace)$. Fix a
reference policy $\piref \in \Delta^\circ(\Yspace)$, where $\Delta^\circ(\Yspace)$ denotes the relative interior, and a reference distribution
$\mu \in \Delta(\Yspace)$ over opponents. To simplify notation we often treat the $|\mc X| = 1$ case; all results extend to $|\mc X| > 1$ by prompt conditioning.

 The NLHF framework \citep{munos2024nash} formulates preference-based fine-tuning as a two-player constant-sum matrix game. It assumes access to preference model $\mc P: \mc Y \times \mc Y \to[0, 1]$ mapping pairs of completions to preference probabilities, which are collected into matrix $P = [\mc P(y \succ y')]_{(y, y')}$. Critically, since $\mc{P}$ outputs probabilities, $P + P^\top = \mathbf{1}\mathbf{1}^\top$. In places where clear from context and notationally convenient, we conflate $P$ and $\mc{P}$. If both players choose policies $\pi_1, \pi_2 \in \Delta(\Yspace)$, then player~1's payoff against player~2 is \[V(\pi_1, \pi_2) = \mb{E}_{y_1 \sim \pi_1, y_2 \sim \pi_2} [\mc P(y_1 \succ y_2)] = \pi_1^\top P\, \pi_2\] and 
\begin{align}
V_\beta(\pi_1, \pi_2) &= \pi_1^\top P\, \pi_2 - \beta\, \KL(\pi_1 \,\|\, \piref) + \beta\, \KL(\pi_2 \,\|\, \piref). \label{eq:V-beta}
\end{align}
The KL-regularized game admits a unique QRE $\pi^\star_\beta$, characterized by the fixed-point
condition over logits $\theta$, where $\pi\propto \exp(\theta)$:
\begin{equation}\label{eq:qre}
\theta^\star_\beta = \thetaref + \mc{P}\pi^\star_\beta/\beta.
\end{equation}
Note that the equilibrium logits equal the reference logits plus a term that is \emph{linear} in the opponent's policy through the operator $\mc{P}/\beta$. This is the fixed point every NLHF algorithm is chasing. The form of $\theta^\star_{\beta}$ relies on \emph{both} the constant-sum structure $P + P^\top = \mathbf{1}\mathbf{1}^\top$ and the KL-regularization in \eqref{eq:V-beta}. 
We provide an overview of related game-theoretic algorithms in Appendix~\ref{sec:methods_overview}.

\subsection{Risk-Sensitive Preference Games}
\label{sec:problem_formulation}
We now introduce risk into the NLHF framework. Replacing the expected pairwise payoff with a risk functional yields a risk-adjusted preference operator and shifts the solution concept to the risk-sensitive quantal response equilibrium (RQRE)~\citep{mazumdar2025tractable, gonzales2026strategicallyrobustmultiagentreinforcement}. This formulation lets the player express sensitivity to the variability of preference outcomes rather than only their mean, addressing the tail-behavior limitations of expectation-based alignment discussed in Section~\ref{sec:introduction}.
To formalize this, we draw on the class of convex risk measures studied extensively in mathematical finance  \cite{follmer2002convex}.

\begin{definition}[Convex risk measure]\label{def:convex-risk-measure}
A risk functional $\Risk: L^\infty(\Omega,\mathcal{F},\mathbb{P}) \to \mb{R}$ is a convex risk measure if it satisfies the following properties:
\begin{enumerate}[label=\alph*.,itemsep=1pt, topsep=2pt]
  \item \emph{Convexity:} For all $X,Y\in L^\infty$ and $\lambda\in[0,1]$,
      $\Risk(\lambda X + (1-\lambda)Y) \leq \lambda \Risk(X) + (1-\lambda) \Risk(Y)$.
  \item \emph{Monotonicity:} For all $X,Y\in L^\infty$ such that $X\leq Y$,
     $\Risk(X) \geq \Risk(Y)$.
  \item \emph{Translation invariance:} For all $X\in L^\infty$ and $c\in\mb{R}$,
  $\Risk(X+c) = \Risk(X) + c$.
\end{enumerate}
\end{definition}

\paragraph{The risk-adjusted preference operator \& RQRE.} 
Define the 
\emph{risk-adjusted preference operator}
\begin{equation}\label{eq:p-risk}
(\mc P_\Risk\,\mu)_y \;:=\; \Risk[P_{(y ,y'')} \mid y'' \sim \mu],
\end{equation}
which replaces the expected win-rate of $y$ against $y'' \sim \mu$ with its risk-adjusted analog. Each player $i \in \{1, 2\}$ then optimizes a risk-adjusted value
\begin{equation}\label{eq:V-risk}
V_{\beta, \Risk}(\pi_1, \pi_2) = \pi_i^\top \mc P_{\Risk}^{i} \pi_{-i} - \beta \KL(\pi_i \| \piref)
+ \beta \KL(\pi_{-i} \| \piref),
\end{equation}
where $\mc{P}_\Risk^{i}$  is player $i$'s risk-adjusted preference operator. A \emph{risk-adjusted equilibrium} $(\theta_1^\star, \theta_2^\star)$  satisfies 
\begin{equation}\label{eq:gen-nash-eq}
\theta_1^\star = \thetaref + \mc{P}_\Risk^{1}\,\pi_{\theta_2^\star}/\beta, \qquad \theta_2^\star = \thetaref + \mc{P}_\Risk^{2}\,\pi_{\theta_1^\star}/\beta.
\end{equation}
 Taking $\Risk \equiv \E$ recovers the NLHF game~\eqref{eq:V-beta} and its QRE fixed point as a special case of~\eqref{eq:gen-nash-eq}.

 \paragraph{Choice of Loss.} Before introducing the risk framework, we comment on the choice of loss function. First, the risk framework we develop is not specific to any one preference loss. The same construction applies to the standard family of preference-tuning objectives i.e., IPO, DPO, GRPO, KTO, and others \citep{Calandriello2024HuamnAlignmentIPOMD,rafailov2023direct,EthayarajhKTO2024,shao2024deepseekmathpushinglimitsmathematical}, by replacing the expectation in their respective preference terms with $\Risk$. We develop these variants in Appendix~\ref{sec:risk-losses}. Here, we focus on IPO \cite{azar2024general} for two reasons. It is the natural choice for game-theoretic analysis: its gradient is a preconditioned residual of the QRE fixed-point equation, and the opponent distribution $\mu$ enters distinctly from the sampling distribution $\rho$, allowing us to directly add risk over the opponent while reaping the rewards of preconditioning. Second,  existing  game-theoretic algorithms (Nash-MD~\citep{munos2024nash}, IPO-MD~\citep{Calandriello2024HuamnAlignmentIPOMD}, MPO~\citep{wang2025magnetic}, EGPO~\citep{zhou2025extragradient}) are formulated in the IPO framework, thereby enabling direct comparison theoretically and empirically. 
 
\paragraph{Risk-Adjusted Loss.}  
Write $\boldsymbol{\Risk} = (\Risk^\Xspace, \Risk^\rho, \Risk^\Wspace, \Risk^\mu)$ where $\Xspace$ is the prompt space, $\rho$ is the response sampling distribution, $\Wspace$ broadly captures stochasticity in groups (e.g., annotators, safety, harm, etc.), and $\mu$ is the opponent distribution. The full risk-adjusted IPO loss is
\begin{equation}\label{eq:ipo-risk-full}
    \LL_{\IPO}^{\boldsymbol\Risk}(\theta; \rho, \mu) = \Risk_o\left[\left(\log \frac{\pi_\theta(y \mid x)\piref(y' \mid x)}{\pi_\theta(y' \mid x) \piref(y \mid x)} - \frac{1}{\beta} \left((\mc{P}_{\Risk^{\mu,\Wspace}}\mu)_{y\mid x} - (\mc{P}_{\Risk^{\mu,\Wspace}} \mu)_{y'\mid x})\right) \right)^{2}\right]
\end{equation}
where $\Risk_o\equiv \Risk^\Xspace_{x \sim \rho_\Xspace}\circ\Risk^\rho_{(y,y')\sim \rho \mid x}$, and
the inner risk-adjusted preference operator is 
    \[(\mc{P}_{\Risk^{\mu,\Wspace}} \mu)_y^x = \Risk^\Wspace_{w \mid x}\left[\Risk^\mu_{y'' \sim \mu}[\mb{P}(y\succ y'' \mid x,w)]\right].
\]
Setting any component of $\boldsymbol{\Risk}$ to $\E$ recovers the  risk-neutral case, and setting all four to $\E$ recovers the standard, risk-neutral IPO loss \citep{munos2024nash, Calandriello2024HuamnAlignmentIPOMD}. The two inner risks $\Risk^\Wspace,\Risk^\mu$ enter the preference operator $\mc{P}$ thereby modifying the equilibrium~\eqref{eq:gen-nash-eq}. The other two risks $\Risk^\Xspace,\Risk^\rho$ leave the per-prompt fixed point unchanged and act on the training dynamics. We focus on $\Risk^\mu$, the only choice that alters the game-theoretic content. The remaining three choices of risk and the composition rules for applying multiple risks are formalized in Appendix~\ref{sec:other-risk-types}.

To simplify presentation, we instantiate~\eqref{eq:ipo-risk-full} with risk only over the opponent, and write $\Risk := \Risk^\mu$ and $\mc{P}_\Risk := \mc{P}_{\Risk^\mu}$ for the rest of the main body; the appendix treats the more general case. The resulting \emph{risk-adjusted IPO loss} is
\begin{equation}\label{eq:ipo-risk}
\LL_{\IPO}^{\Risk}(\theta; \rho, \mu)
= \mb{E}_{(y,y')\sim\rho}\left[\left(\log \frac{\pi_\theta(y)\piref(y')}{\pi_\theta(y')\piref(y)} - \frac{1}{\beta}\left((\mc{P}_\Risk\,\mu)_y - (\mc{P}_\Risk\,\mu)_{y'})\right)\right)^{2}\right].
\end{equation}
A direct calculation gives the gradient
\[\nabla_\theta \LL_{\IPO}^{\Risk}(\theta; \rho, \mu)
= 2\,\Sigma(\rho)(\theta - \thetaref - \mc{P}_\Risk\mu/\beta),\]
where $\Sigma(\rho) := \E_{(y,y')\sim\rho}[(\mathbf{1}_y - \mathbf{1}_{y'})(\mathbf{1}_y - \mathbf{1}_{y'})^\top]$ is the response-pair preconditioner. 
The operator $\mc{P}_\Risk$ is generally nonlinear in $\mu$, and the following examines its structural effect on the game.

\subsection{Convex risk preserves the constant-sum structure}
\label{sec:structure-preserve}
The risk-neutral analysis of NLHF relies on the constant-sum identity 
$P + P^\top = \mathbf{1}\mathbf{1}^\top$ to collapse the coupled fixed-point 
equations~\eqref{eq:gen-nash-eq} into a single equilibrium expression $\theta^\star_\beta = \thetaref + \mc{P}\pi^\star_\beta/\beta$. Under nonlinear risk transformations, this structure 
no longer holds at the certainty equivalence level.
\begin{example}[Certainty-equivalent asymmetry]\label{eq:ce_example}
    Consider  $X \in \{0,1\}$ with 
$\mathbb{P}(X=1)=\mathbb{P}(X=0)=\tfrac12$, so $X + (-X) = 0$.
At the certainty-equivalent level, risk transformations break this structure.
For any $\alpha\in(0,1)$,  CVaR satisfies
$
\mathrm{CVaR}_{\alpha}(X) = 0$, and  
$\mathrm{CVaR}_{\alpha}(-X) = -1$,
so $\mathrm{CVaR}_{\alpha}(X) + \mathrm{CVaR}_{\alpha}(-X) \neq 0$.
\end{example}
Despite this shortcoming, when $\Risk$ is translation-invariant, the identity 
$\Risk(1 - X) = 1 + \Risk(-X)$ ensures that constant-sum structure is 
preserved at the level of risk-adjusted payoff operators.
\begin{restatable}[Operator-level constant-sum]
  {lemma}{OperatorConstantSumGeneral}
\label{lem:operator-constant-sum}
Let $P\in\mathbb{R}^{|\Yspace|\times|\Yspace|}$ satisfy
$P + P^\top = \mathbf{1}\mathbf{1}^\top$.
Let $\Risk$ be translation invariant, i.e.,
$\Risk(X+c)=\Risk(X)+c$
for all constants $c\in\mathbb{R}$.
For $\mu\in\Delta(\Yspace)$, define the risk-adjusted payoff operators
$
(\mc{P}^{1}_\Risk\mu)_a
:=
\Risk(\mc{P}_{a,Y'})$,
and $(\mc{P}^{2}_\Risk\mu)_a
:=
\Risk(-\mc{P}_{Y',a})$,
with $
Y'\sim\mu$.
Then, for every $\mu\in\Delta(\Yspace)$,
we have $
\mc{P}^{1}_\Risk\mu
=
\mathbf{1}
+
\mc{P}^{2}_\Risk\mu$.
\end{restatable}
In general, iterative 
algorithms for games will maintain iterates for both players, an undesirable property 
when iterates are LLMs parametrized by billions of parameters.
This lemma enables us to prove that self-play is sufficient to find an equilibrium to the RSPG. To see this, consider 
algorithms targeting the 
\emph{single-player risk-adjusted operator}

\begin{equation}\label{def:single-player-risk-operator}
F_\Risk(\theta) := F_{\Risk}(\theta ; \pi_{\theta}) 
= \beta(\theta - \thetaref) - \mc P_\Risk\,\pi_\theta, 
\end{equation}
which is the residual of player $1$'s fixed point equation if the opponent is set 
to be player $1$'s policy. 
\begin{restatable}[]{theorem}{singlefixedpt}\label{thm:single-fixed-pt}
Let $P + P^\top = \mathbf{1}\mathbf{1}^\top$ and let $\Risk$ be a 
 convex risk measure. In the regime $\beta\in(0,\infty)$, the unique Nash equilibrium $(\pi^\star_{\theta_1}, \pi^\star_{\theta_2})$ of the RSPG is a symmetric 
(namely,
$\pi^\star:=\pi^\star_{\theta_1} = \pi^\star_{\theta_2}$)
and solves the fixed-point equation $\theta^\star = \thetaref + \mc P_\Risk\,\pi_{\theta^\star}/\beta$. Moreover, $\theta_1^\star = \theta_2^\star + c\mathbf{1}$ for some $c \in \R$.
\end{restatable}

Thus, translation invariance preserves the efficiency of algorithms which only update a single policy, meaning the robustness to tail behavior inherent to risk measures comes at no structural cost. 
When the conditions of Theorem~\ref{thm:single-fixed-pt} are violated---e.g.,
under asymmetric risk measures or heterogeneous opponents---the two-player formulation
must be retained; see Appendix~\ref{sec:risk-adjusted-operator}.

Monotonicity of the operator $F_\Risk$ determines the rate of
convergence for many algorithms.  In the symmetric self-play setting,
monotonicity is determined by displacements within the tangent space
$T^\sym = \{(\xi, \xi) : \xi \in \mathbf 1^\perp\}$ of the diagonal
constraint set, where it reduces to a single-player condition:
\begin{align*}
\langle F_\Risk(\theta_1) - &F_\Risk(\theta_2),\, \pi_{\theta_1} - \pi_{\theta_2}\rangle
 = \beta \langle \theta_1 - \theta_2,\, \pi_{\theta_1} - \pi_{\theta_2} \rangle
   - \langle \mc P_\Risk \pi_{\theta_1} - \mc P_\Risk \pi_{\theta_2},\, \pi_{\theta_1} - \pi_{\theta_2} \rangle \\
 &= \beta\bigl(\KL(\pi_{\theta_1} \| \pi_{\theta_2}) + \KL(\pi_{\theta_2} \| \pi_{\theta_1})\bigr)
   - (\pi_{\theta_1} - \pi_{\theta_2})^\top \Distmat_\Risk(\bar\pi)(\pi_{\theta_1} - \pi_{\theta_2}),
\end{align*}
where $\Distmat_\Risk(\pi) := \tfrac{1}{2}(J_\Risk + J_\Risk^\top)(\pi) - \tfrac{1}{2}\mathbf 1\mathbf 1^\top$
is the distortion matrix, $\bar \pi$ is a policy on the line segment between $\pi_{\theta_1}$ and $\pi_{\theta_2}$, and
$J_\Risk(\pi) := \nabla_\mu P_\Risk(\mu)|_{\mu = \pi}$ is the Jacobian. 
The first term is bounded below by strong convexity of the log
partition function; the second is controlled by the worst-case
risk-distortion eigenvalue restricted to $T^\sym$.

\begin{restatable}[Informal]{theorem}{singleplayermono}
\label{thm:single-player-mono}
Define the worst-case risk-distortion eigenvalue on the symmetric
tangent space
$\overline\lambda_\Risk
:=\sup_{\bar\pi \in \Delta(\Yspace)}
\sup\{
\tfrac{1}{2}\, \xi^\top \bigl(J_\Risk(\bar\pi) + J_\Risk(\bar\pi)^\top\bigr)\,\xi|\; \xi \in \mathbf 1^\perp,\, \|\xi\|_2 = 1\}$.
Suppose $\overline\lambda_\Risk \le \beta/2$.  Then $F_\Risk$ is
$\mu_\Risk$-strongly monotone on $T^\sym$ with
$\mu_\Risk = \beta - 2\overline\lambda_\Risk$, and extragradient
converges to the unique risk-adjusted equilibrium of the RSPG with optimization error rate
$\mc{O}((1 - \eta \mu_\Risk)^T)$.
\end{restatable}

The formal statement for the symmetric self-play result is provided in
Appendix~\ref{sec:single-player-all-results}, where Lemma~\ref{lem:lambda-symmetric}
shows that this single-player form of $\overline\lambda_\Risk$ is what
the joint-game definition reduces to under symmetric self-play.  The
more general version for the full joint pseudogradient is in the
preceding Appendix~\ref{sec:full-two-player-game}, and the full
stochastic bounds are given in the sequel.

\section{Learning  in Risk-Sensitive Preference Games}
\label{sec:convergence-risk-game}\label{sec:main_results}
We now prove efficient convergence of stochastic self-play algorithms applied to the RSPG. The biggest challenge is that estimators of risk operators are generally \emph{not unbiased}.
We introduce a novel Stackelberg game framework to systematically reduce that bias at no-extra cost. Finally, we show generalization guarantees, not afforded in the risk-neutral regime.

\paragraph{Risk Estimators.}
The results require understanding the statistical structure of risk. Recall
$F_\Risk(\theta) =\beta( \theta - \theta_{\rm ref}) - \mc{P}_\Risk\pi_\theta$,
with $\mc{P}_\Risk(\pi)_y := \Risk_{Y'' \sim \pi}[\mc{P}(y \succ Y'')]$.  Let
$\widehat{\mc{P}}_{\Risk,m}(\pi)_y$ denote the sample-based estimator of
$\mc{P}_\Risk(\pi)_y$ using $m$ i.i.d.~samples $Y''_1, \ldots, Y''_m \sim \pi$
(the same samples shared across components $y$).  With $b_m(\theta) := \E\bigl[\widehat F_{\Risk,m}(\theta)\bigr] - F_\Risk(\theta)$ and $
\zeta_m(\theta) := \widehat F_{\Risk,m}(\theta) - \E\bigl[\widehat F_{\Risk,m}(\theta)\bigr]$,
\begin{equation}\label{eq:F-hat-def}
\widehat F_{\Risk,m}(\theta) := \beta(\theta - \theta_{\rm ref}) - \widehat{\mc{P}}_{\Risk,m}(\pi_\theta)=F_{\Risk}(\theta) + b_m(\theta) + \zeta_m(\theta).
\end{equation}
Here $b_m(\theta)$ is deterministic given $\theta$ and $\E[\zeta_m(\theta) | \theta] = 0$.
By construction
$b_m(\theta) =   \mc{P}_\Risk(\pi_\theta))-\E[\widehat{\mc{P}}_{\Risk,m}(\pi_\theta)]$
and
$\zeta_m(\theta) =  \E[\widehat{\mc{P}}_{\Risk,m}(\pi_\theta)]-\widehat{\mc{P}}_{\Risk,m}(\pi_\theta)$.
The convergence theorems in the sequel (and Appendix~\ref{sec:mp-bias}) rely on uniform bias and variance bounds.
\begin{assumption}[bias and variance]
  \label{ass:bias-var}
  There exist constants $B_m, V_m > 0$ decreasing in $m$  such that $\|b_m(\theta)\|_2\leq B_m$ and $\E\|\zeta_m(\theta)\|_2^2\leq V_m$ for all $\theta\in \thetaref+\Wspace$.
  \end{assumption}
In Appendix \ref{sec:bias-analysis} we establish that the risk measures such as CVaR via the Rockafellar-Uryasev transform and the class of the common form $\Risk [Z] =h(\mb E_\mu (g(Z)))$ for smooth scalar $h$ and bounded statistic $g$ (e.g., 
entropic risk) admit bias and variance bounds (at most) $B_m,V_m = \mc{O}(1/m)$.

\subsection{Iteration Complexity of Biased Extra-Gradient Preference Learning}
Stochastic extragradient on a biased operator is well-studied \cite{juditsky2011}: under
strong monotonicity it contracts geometrically to a neighborhood of
the optimum whose radius is set by the bias and variance of the
oracle.  Specializing to the risk-adjusted $F_\Risk$ with
the plug-in estimator $\widehat F_m$, in each round, the stochastic extragradient method \cite{korpelevich1976} takes two steps:
\begin{equation}\label{eq:extra_grad_stoch}
    \begin{aligned}
        \theta_{\tau-1/2} &\gets \mathrm{proj}_{\mc{K}}(\theta_{\tau-1} - \eta\, \Sigma(\rho)\,\widehat F_{m,1}^{\tau}),\\
        \theta_\tau& \gets \mathrm{proj}_{\mc{K}}(\theta_{\tau-1} - \eta\, \Sigma(\rho)\,\widehat F_{m,2}^{\tau}),
    \end{aligned}
\end{equation} where $\widehat F_{m,1}^{\tau}(\theta_{\tau-1})$ and $F_{m,2}^{\tau}(\theta_{\tau-1/2})$ are estimated via a batches of size $m$ 
at $\theta_{\tau-1}$ and $\theta_{\tau-1/2}$, respectively.
Under Assumption~\ref{ass:bias-var}, the $\Sigma(\rho)$-norm bias and variance relate to the
Euclidean-norm constants 
by \[\widetilde B_m:=
  \sup_{\theta}
\|
    \E[\hat{F}_m(\theta)] - F_{\Risk}(\theta)
\|_{\Sigma(\rho)} \le \sqrt{\sigma_{\max}} B_m\quad\text{and}\quad 
\widetilde V_m:= \E\|
    \hat{F}_m(\theta)-\E[\hat{F}_m(\theta)]\|_{\Sigma(\rho)}^2\le \sigma_{\max} V_m,\] where
$\sigma_{\max} := \|\Sigma(\rho)\|_{\rm op}$ is the largest
eigenvalue of the preconditioner. Analysis yields three error terms: $(i)$
deterministic linear decay $\mc{O}((1 - \eta\mu_\Risk)^T)$ from the
contraction; $(ii)$ a persistent bias floor $\Theta(1/(\mu_\Risk m))$,
the cost of plug-in risk estimation, which does not vanish in $T$ or
$\eta$; $(iii)$ a standard variance floor $\Theta(\eta/(\mu_\Risk m))$ that
shrinks with the step size. The following proposition shows the error bound.

\begin{restatable}[Informal]
  {proposition}{NoisyRiskAdjustedEG}
  \label{prop:sample-complexity-formal}
Suppose that we are in the strongly monotone regime,
$\mu_\Risk \geq \beta - 2\overline\lambda_\Risk > 0$, the map
$\mu \in \Delta(\Yspace) \mapsto \mc P_\Risk(\mu) \in \R^{|\Yspace|}$
is $L_\Risk$ Lipschitz continuous, and that
Assumption~\ref{ass:bias-var} holds. Set $\Omega := \sup_{\theta, \theta' \in \mc{K}} \|\theta - \theta'\|_{\Sigma(\rho)^+}$
is the $\Sigma(\rho)^+$-diameter of $\mc{K}$.
 Run stochastic
extra-gradient on the RSPG with step size
$\eta \le \min\{\frac{1}{4\mu_\Risk},
  \frac{1}{\sqrt{6}\,\ell_\Risk}\}$.
 For any $T \ge 1$ and $m \ge 1$, the estimate holds:
\begin{equation}\label{eq:sample-complexity}
  \|\theta_T-\theta^\star\|^2_{\Sigma(\rho)^+}
  \le
  (1 - \eta\mu_\Risk)^T
  \|\theta_0-\theta^\star\|^2_{\Sigma(\rho)^+}
  + 4\,\Omega\,\widetilde{B}_m/\mu_\Risk
  + 6\,\eta\,(\widetilde{B}_m^2+\widetilde{V}_m)/\mu_\Risk.
\end{equation}
\end{restatable}
This proposition is proved in Appendix~\ref{sec:mp-bias}.
Specifically, we show biased stochastic extragradient achieves $\mc O(1/\sqrt{t}) + \mc O(\widetilde B_m)$
averaged-iterate convergence in the merely-monotone regime
(Theorem~\ref{thm:expected-gap}) and last-iterate contraction at rate
$\mc O((1 - \gamma\widetilde\mu)^t)$ to a neighborhood of size
$\mc O(\widetilde B_m/\widetilde\mu)$ in the strongly monotone regime
(Theorem~\ref{thm:strong-mono}), where $\widetilde\mu = \mu_\Risk\sigma_{\min}$
and $\widetilde B_m=\mc{O}(1/m)$ is the oracle bias.  
 The bias term is the binding constraint
in the practically relevant regime: samples are expensive, $m$ is
constrained by per-iteration compute, and no amount of further
iteration drives this floor down.  

\paragraph{Mirror Descent counterpart.}
For completeness, in Appendix~\ref{sec:md-corollary} we record the
analogous convergence guarantees for stochastic projected mirror
descent on the IPO gradient flow with a biased stochastic oracle,
in both the monotone and strongly monotone settings.  The mirror descent bounds
have the same structure as their extra-gradient counterparts; namely, $\mc{O}(1/T)$ averaged-iterate gap in the monotone case
(Corollary~\ref{thm:md-monotone}) and linear contraction with
bias floor $\mc{O}(\Omega B_m/\widetilde\mu)$ in the strongly monotone
case  (Corollary~\ref{thm:md-strong-mono}). The proofs
are also essentially specializations of the extra-gradient analysis to the
single-oracle-call setting.  The one structural difference is that MD
picks up an additional $L_G\Omega^2\gamma$ Lipschitz continuity term that
extra-gradient cancels exactly via the inner-step distance
$\|w_\tau - r_{\tau-1}\|^2$ in its two-call descent identity; this is
the technical price of dropping extrapolation.  We include the mirror descent corollaries for posterity and to make the extragradient versus mirror descent comparison precise, and because we also implement mirror descent in our experiments---practically speaking, its easier since it requires a single update in each iteration versus two. 
Our general recommended algorithm  however remains stochastic extra-gradient: the
extrapolation step costs one additional oracle call per iteration but
buys cycling correction, robustness when the estimated game operator
fails to be strongly monotone, and a cleaner Lipschitz continuity handling that
the mirror descent analysis lacks.

\subsection{Stackelberg Game for Fast Two Time-scale Bias Tracking}
\label{sec:stackelberg-debiasing}

The previous section showed that stochastic extragradient on the
risk-adjusted operator $F_\Risk$ contracts geometrically to a
neighborhood of $\theta^\star_\Risk$ whose radius is governed by the
oracle bias $\widetilde B_m = \mc O(1/m)$
(Proposition~\ref{prop:sample-complexity-formal}).   Indeed, the bias floor
is irreducible without further structure on the oracle: under
constant step sizes, no amount of additional iteration removes it.
We show the bias floor can be removed essentially for free, by augmenting
extragradient with a fast-timescale bias tracker. The persistent $\mc{O}(1/m)$
floor collapses to a residual of order $\mc{O}(1/m^2)$---the same order as the
variance floor of an unbiased oracle---and exactly zero for
CVaR.

\subsubsection{Stackelberg structure and equilibrium}
We introduce a bias estimator $\xi \in \R^d$ as a
second player and cast the joint dynamics as a Stackelberg game that has a unique equilibrium. This naturally leads to two-timescale dynamics \cite{fiez2020implicit}.
The \emph{leader} optimizes the risk-adjusted policy using the
debiased gradient $\widehat F_m - \xi$; the \emph{follower} tracks
the leader's bias.

The leader's cost is the risk-adjusted IPO objective, corrected by
the follower's bias estimate:
\[
\Phi_{\rm leader}(\theta; \xi) \;:=\; \LL_{\IPO}^\Risk(\theta; \rho, \sg[\pi_\theta]) \;-\; \langle \xi, \theta \rangle,
\]
so that $\nabla_\theta \Phi_{\rm leader} = \nabla \LL_{\IPO}^\Risk - \xi$
is exactly the debiased gradient.  The follower's cost is the
quadratic tracking objective
\[
\Phi_{\rm follower}(\xi; \theta) \;:=\; \tfrac{1}{2}\|\xi - b_m(\theta)\|_2^2,
\]
where $b_m(\theta) := \E[\widehat F_m(\theta)] - F_\Risk(\theta)$ is
the oracle's population bias at $\theta$.  Since $\Phi_{\rm follower}$
is strongly convex in $\xi$, the follower's best-response is
single-valued: $\xi^\star(\theta) = b_m(\theta)$ for every $\theta$.

A \emph{Stackelberg equilibrium} of this game is a pair
$(\theta^\dagger, \xi^\dagger)$ at which the follower best-responds
to the leader and the leader is stationary against that
best-response: $\xi^\dagger = b_m(\theta^\dagger)$ and
$\theta^\dagger$ is a stationary point of $\Phi_{\rm leader}(\cdot; \xi^\dagger)$
on $\mathcal{K}$.  Substituting the follower's best-response into
the leader's first-order condition and using
$\E[\widehat F_m(\theta)] = F_\Risk(\theta) + b_m(\theta)$ collapses
the leader's stationarity to $F_\Risk(\theta^\dagger) \in \mathbf{1}^\perp$, i.e., precisely the risk-adjusted equilibrium condition for the
outer game.  The Stackelberg equilibrium is therefore
\[
(\theta^\dagger, \xi^\dagger) \;=\; \bigl(\theta^\star_\Risk,\; b_m(\theta^\star_\Risk)\bigr),
\]
the risk-adjusted equilibrium paired with the perfect bias estimate
at that equilibrium.
\begin{figure}
    \centering
    \includegraphics[width=0.85\linewidth]{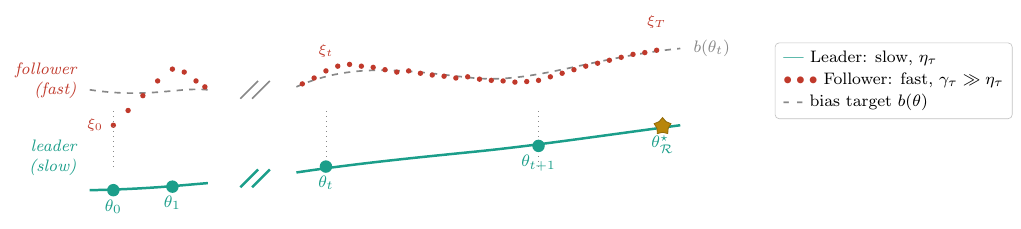}
    \caption{The bias tracker $\xi_t$ is shown tracking the true bias $b_m(\theta_t)$ on a faster timescale than the leader's update $\theta_t$---i.e., the IPO policy parameter.}
    \label{fig:tt_hero}
\end{figure}
\paragraph{Existence and uniqueness.}  In the monotone regime
($\mu_\Risk \ge 0$, i.e., $\overline\lambda_\Risk \le \beta/2$),
$F_\Risk$ is monotone on $\mathcal{K}$ and a Stackelberg equilibrium
exists.  When $\mu_\Risk > 0$ (the strongly monotone regime),
$F_\Risk$ has a unique zero $\theta^\star_\Risk$ and the equilibrium
is unique.  The follower's contribution to uniqueness is automatic:
its cost is quadratic in $\xi$, so the best-response curve
$\xi^\star(\theta) = b_m(\theta)$ is single-valued regardless of
the regime; uniqueness of the Stackelberg equilibrium is therefore
inherited entirely from uniqueness of the leader's risk-adjusted
equilibrium.
What remains is to design an algorithm that converges to this
equilibrium.  

\subsubsection{Two-timescale algorithm for finding Stackelberg equilibrium}
The goal here is to design a two timescale algorithm wherein the leader runs extragradient on the
debiased gradient at slow timescale $\eta_\tau$, the follower
performs Robbins-Monro updates toward $b_m(\theta_\tau)$ at fast
timescale $\gamma_\tau$, and the timescale separation
$\eta_\tau \ll \gamma_\tau$ ensures the follower asymptotically
tracks the best-response curve. This will give equilibrium convergence. 

To that end, let $b_m(\theta) := \E[\widehat F_m(\theta)] - F_\Risk(\theta)$ be the
oracle's population-level bias and $\widehat b_m$ a sample-based
estimator. Introduce a tracker $\xi \in \R^d$ with target
$\xi^\star(\theta) = b_m(\theta)$. The composite system is a Stackelberg
game: the \emph{leader} is risk-adjusted self-play on $\theta$, using the
\emph{debiased} gradient $\widehat F_m - \xi$; the \emph{follower} is
$\xi$, best-responding to the leader's current $\theta$.  The Stackelberg
equilibrium is $(\theta^\star_\Risk, b_m(\theta^\star_\Risk))$: i.e., the unique 
risk-adjusted Nash paired with the perfect bias estimate at that Nash.
For the equilibrium analysis to translate into convergence the follower
runs  on
a faster timescale: $\eta_\tau \ll \gamma_\tau$. The bias estimator
uses the same batch already drawn for 
$\widehat F_m$, so tracking costs \emph{no additional samples}.

\begin{algorithm}[h]
\caption{\texttt{TT-EG}: Two-timescale extragradient with bias tracking}
\label{alg:tt-eg-main}
\begin{algorithmic}[1]
\State \textbf{Input:} $\theta_0 \in D$; slow steps $\{\eta_t\}$, fast steps $\{\gamma_t\}$; batch size $m$. Initialize $\xi_0 \gets 0$.
\For{$\tau = 1, \ldots, T$}
  \State \textbf{Update} $(\theta_\tau, \theta_{\tau-1/2})\gets \texttt{ExtraGrad}(\eta_\tau)$, via \eqref{eq:extra_grad_stoch}
  \State \textbf{Update}  $\xi_\tau \gets (1-\gamma_\tau)\xi_{\tau-1} + \gamma_\tau\,\widehat b_m^{(\tau,2)}$
\EndFor
\State \Return $\theta_T$
\end{algorithmic}
\end{algorithm}

For the canonical delta-method estimator
$\widehat b_m(\theta) = h''(\widehat q_m)\widehat\Var_m(g)/(2m)$ on
risks of the form $h(\E_\mu[g(Z)])$ (entropic, smooth distortion), the
residual bias is $R_m = \mc{O}(1/m^2)$ and variance $V_m^b = \mc{O}(1/m^3)$. For
CVaR-RU, $R_m = V_m^b = 0$ identically.

\begin{theorem}[Informal]
\label{thm:strong-mono-tt}
Suppose $F_\Risk$ is $\widetilde\mu$-strongly monotone and Lipschitz. Assume
the gradient oracle has bounded bias $\|\E[\widehat F_m(\theta)] - F_\Risk(\theta)\|
\le \widetilde B_m$ and bounded variance $\E\|\widehat F_m(\theta) - \E\widehat F_m(\theta)\|^2
\le \widetilde V_m$, and the bias estimator $\widehat b_m$ has residual bias
$\|\E\widehat b_m(\theta) - b_m(\theta)\| \le R_m$, bounded variance
$\E\|\widehat b_m(\theta) - \E\widehat b_m(\theta)\|^2 \le V_m^b$, and
Lipschitz dependence on $\theta$. Then Algorithm~\ref{alg:tt-eg-main} satisfies:
\begin{enumerate}[label={\rm {\it \roman*.}}, leftmargin=*, itemsep=-0pt, topsep=-2pt]
\item \textbf{Non-asymptotic.} For appropriate $\eta, \gamma$,
the iterates contract geometrically to a steady-state floor of order
$\mc{O}(\eta/\gamma + R_m + \sqrt{\gamma V_m^b} + \eta\widetilde V_m)$.
Tuning $\eta = \mc{O}(\gamma/m)$ collapses this to $\mc{O}(1/m^{3/2})$---matching
the variance scale.
\item \textbf{Asymptotic.} With $\eta_t \to 0$,
$\eta_t/\gamma_t \to 0$, and $\sum_t \eta_t = \infty$,
$\;\limsup_{T\to\infty} \E\|\theta_T - \theta^\star_\Risk\|^2 = \mc{O}(R_m).$
The bias floor is $\mc{O}(1/m^2)$ for delta-method estimators and
\emph{zero} for CVaR-RU.
\end{enumerate}
\end{theorem}

The merely-monotone case (Theorem~\ref{thm:tt-monotone}) gives the
analogous averaged-iterate gap bound with the same $\mc{O}(1/m^{3/2})$ floor.
Precise statements, the Stackelberg derivation, and the horizon-tuned
$\mc{O}(T^{-1/3})$ rate are in Appendix~\ref{sec:tt-debiasing}.
The un-tracked algorithm has a bias term $\widetilde B_m^2 \asymp 1/(\lambda m)^2$
and a slow-variance term $\eta\widetilde V_m \asymp \eta/m$. The bias
dominates when $m \lesssim 1/(\lambda^2 \eta)$, and below this threshold
the un-tracked algorithm wastes a full power of $1/m$ on a floor that
\texttt{TT-EG} removes; above it, both algorithms are variance-limited and agree.
In preference optimization, $m$ is constrained by memory and rollout cost---tens
to low hundreds of samples per gradient---and the bias constants for
entropic risk are non-trivial. Practical
training sits squarely in the bias-dominated regime. Figure~\ref{fig:tt-sweep-main} is a numerical
illustration (detailed in Appendix~\ref{sec:tt-experiments}) that confirms this on a
Bradley-Terry game with entropic distortion: at $m = 15$, vanilla extragradient
plateaus at $\sim 10^{-4}$ while \texttt{TT-EG} reaches $\sim 2 \times 10^{-6}$
(a $50\times$ improvement). 

\begin{figure}[t]
\centering
 \includegraphics[width=0.45\textwidth]{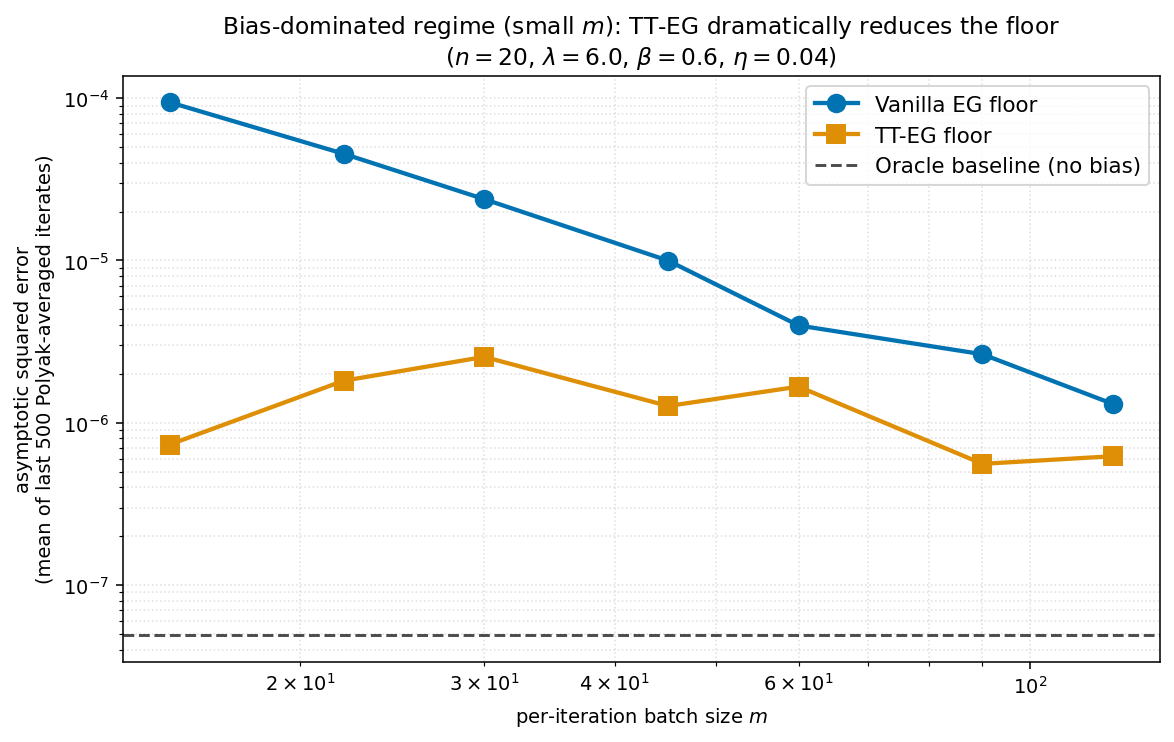}\hfill
 \includegraphics[width=0.45\textwidth]{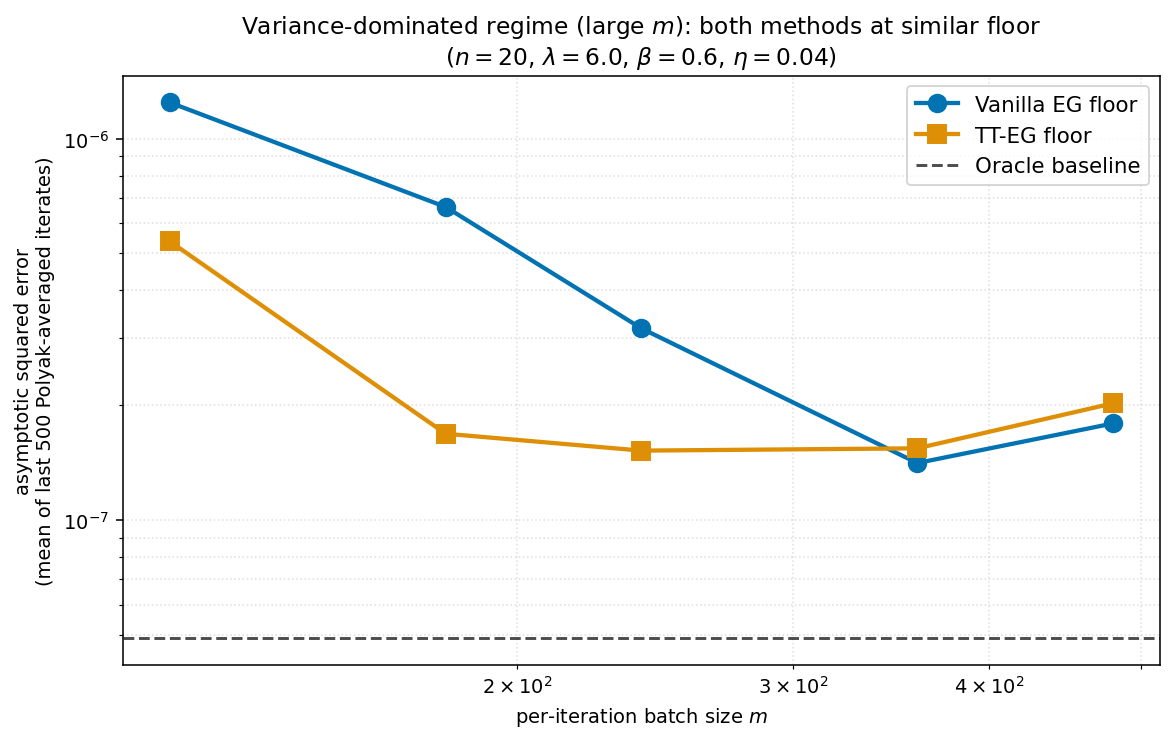}
\caption{\textbf{Asymptotic squared-error floor vs.\ batch size $m$,
in two regimes.}  Floor measured as the mean of the last $500$
Polyak-averaged iterates.  \emph{Left:}  bias-dominated regime
($m \in [15, 130]$).  Vanilla  extragradient floor scales roughly $\propto 1/m^2$
(matching the $\Theta(\widetilde B_m^2)$ prediction), descending more
than two orders of magnitude across the range.  \texttt{TT-EG} floor is
roughly flat near $\sim 10^{-6}$, giving a $\sim 100\times$ reduction
at $m = 15$ (the small-$m$ end where vanilla is most bias-dominated)
and approaching parity by $m = 130$.  \emph{Right:}
variance-dominated regime ($m \in [150, 500]$).  Both floors
descend only modestly across this range and converge toward a common
value around $\sim 1.5 \times 10^{-7}$, dominated by the slow-variance
term $\Theta(\eta \widetilde V_m)$ in
\eqref{eq:tt-const-final-expanded}; the bias contribution is
negligible here.  At the largest $m$ \texttt{TT-EG} sits marginally
\emph{above} vanilla, reflecting the small noise-filter cost
$\Theta(\sqrt{\gamma V_m^b})$ paid to maintain the tracker even when
the bias being tracked is already small.  Dashed line: oracle
baseline at $\sim 4 \times 10^{-8}$.}
\label{fig:tt-sweep-main}
\end{figure}

\paragraph{Mirror Descent with two-timescale debiasing.}
The two-timescale debiasing construction transfers cleanly to mirror
descent: in Appendix~\ref{sec:tt-md-corollary} we give the mirror descent
analogues of Theorems~\ref{thm:tt-strong-mono}
and~\ref{thm:tt-const}.  

\begin{theorem}[Informal]
\label{thm:strong-mono-tt-md}
Suppose $F_\Risk$ is $\widetilde\mu$-strongly monotone and Lipschitz. Assume
the gradient oracle has bounded bias $\|\E[\widehat F_m(\theta)] - F_\Risk(\theta)\|
\le \widetilde B_m$ and bounded variance $\E\|\widehat F_m(\theta) - \E\widehat F_m(\theta)\|^2
\le \widetilde V_m$, and the bias estimator $\widehat b_m$ has residual bias
$\|\E\widehat b_m(\theta) - b_m(\theta)\| \le R_m$, bounded variance
$\E\|\widehat b_m(\theta) - \E\widehat b_m(\theta)\|^2 \le V_m^b$, and
Lipschitz dependence on $\theta$. Then the single-call mirror descent analogue
of Algorithm~\ref{alg:tt-eg-main} satisfies:
\begin{enumerate}[label={\rm {\it \roman*.}}, leftmargin=*, itemsep=-0pt, topsep=-2pt]
\item \textbf{Non-asymptotic.} For appropriate $\eta, \gamma$,
the iterates contract geometrically to a steady-state floor of order
$\mc{O}(\eta/\gamma + R_m + \sqrt{\gamma V_m^b} + \eta\widetilde V_m + L_G^2\eta)$,
with the final term the structural cost of omitting extrapolation.
Tuning $\eta = \mc{O}(\gamma/m)$ collapses this to $\mc{O}(1/m^{3/2})$---matching
the extra-gradient rate up to the Lipschitz floor.
\item \textbf{Asymptotic.} With $\eta_t \to 0$,
$\eta_t/\gamma_t \to 0$, and $\sum_t \eta_t = \infty$,
$\;\limsup_{T\to\infty} \E\|\theta_T - \theta^\star_\Risk\|^2 = \mc{O}(R_m).$
The Lipschitz floor vanishes automatically under the decaying schedule,
so the asymptotic bound matches extra-gradient \emph{exactly}; the bias
floor is $\mc{O}(1/m^2)$ for delta-method estimators and \emph{zero}
for CVaR-RU.
\end{enumerate}
\end{theorem}

Theorem \ref{thm:strong-mono-tt-md} should be read alongside its
extra-gradient counterpart. The asymptotic vanishing-bias result is in
fact \emph{identical} up to constants: under decaying step sizes with
$\eta_t/\gamma_t \to 0$, the MD iterates also satisfy
$\limsup_T \E\|\theta_T - \theta^\star\|_{\Sigma^+}^2 \le 4\sqrt 3\,\Omega\sqrt{\sigma_{\max}}\,R_m/\widetilde\mu$
(Corollary~\ref{cor:tt-md-strong-mono}), with the persistent
$\mc{O}(1/m)$ bias floor of the un-tracked algorithm replaced by an
asymptotic-in-$T$ floor at the $R_m$-residual scale. In the
constant-step regime, the four-term floor of extra-gradient
(timescale gap, residual bias, noise filter, slow variance) becomes a
five-term floor for mirror descent, with the additional term being the
same $L_G\Omega^2\eta$ Lipschitz floor noted in
Appendix~\ref{sec:md-corollary} above. As before, the mirror descent
proofs are straightforward specializations of the extra-gradient
analysis: indeed, the scalar two-timescale unrolling lemma
(Lemma~\ref{lem:scalar-unroll}), the weighted-average analysis
(Lemmas~\ref{lem:tt-telescope}--\ref{lem:weighted-avg-limsup}), and the
tracking-error recurrence apply verbatim, modulo a smaller drift
constant $C_{\rm drift}^{\rm MD} = L_b\sigma_{\max}G$ (versus
$3 L_b\sigma_{\max}G$ for extra-gradient) reflecting mirror descent's
single-call structure. We include these results for posterity and
parallelism with our extra-gradient theory, but emphasize that the
recommended deployment remains two-timescale extra-gradient: the
asymptotic improvement is the same, the finite-horizon constants are
tighter, the constant-step regime carries one fewer floor term, and
importantly extra-gradient remains viable even if the game slides into
the merely-monotone regime.

\subsection{Algorithmic Stability, Generalization, \& Offline Sample Complexity}
\label{sec:generalization-main}
To understand the relationship between optimization and offline guarantees, the population performance is decomposed into
\emph{optimization} and \emph{generalization} errors: for $S \sim \mathcal{D}^n$  an offline dataset,
\begin{equation}
\label{eq:gen-decomp}
\mb{E}_S[ \Risk(\hat\theta_T(S)) - \Risk(\theta^\star) ]
\;\le\;
\underbrace{\mb{E}_S[ \Risk_S(\hat\theta_T(S)) - \Risk_S(\theta^\star) ]}_{\text{optimization error}}
\;+\;
\underbrace{\mb{E}_S[ \Risk(\hat\theta_T(S)) - \Risk_S(\hat\theta_T(S)) ]}_{\text{generalization error}},
\end{equation}
where $\Risk,\Risk_S$ denote the population and empirical objectives,
respectively.
Stochastic extra-gradient analysis
(Proposition~\ref{prop:sample-complexity-formal} or Theorem~\ref{thm:strong-mono-tt}) controls the optimization error. Algorithmic stability, our interest here, controls the generalization error.

\subsection{Stability \& Generalization}
To start, we show the risk adjusted equilibrium is Lipschitz-continuous in  in the risk-adjusted preference operator $\mc P_{\Risk}$---something not in general true for the Nash equilibrium corresponding to the risk-neutral preference game without KL regularization since Nash equilibria are non-unique and have no guarantees of smoothness. Regularization induces a QRE, which is known to be unique for normal form games with $\beta>0$ \citep{MCKELVEY19956}. Here we show that risk provides the needed regularity to prove generalization and robustness bounds. 
\begin{restatable}[Structural stability of the risk-adjusted equilibrium under operator perturbation]{theorem}{stabthm}
\label{thm:stab}
Let $\mc{P}, \mc{P}'$ be two preference operators with corresponding risk-adjusted
operators $\mc{P}_\Risk, \mc{P}'_\Risk$ and risk-distortion eigenvalues
$\bar\lambda_\Risk(\mc{P}), \bar\lambda_\Risk(\mc{P}') \le \beta/2 - \epsilon$ for
some $\epsilon > 0$. Suppose we are in the strongly monotone regime with the strong-monotonicity moduli 
\[
  \mu_\Risk \;:=\; \beta - 2\bar\lambda_\Risk(\mc{P}), \qquad
  \mu'_\Risk \;:=\; \beta - 2\bar\lambda_\Risk(\mc{P}'),
\]
of
$\mc{P}_\Risk$ and $\mc{P}'_\Risk$, respectively, 
both bounded below by $2\epsilon$. Let $\theta^\star = \theta^\star_\Risk(\mc{P})$
and $\theta'^\star = \theta^\star_\Risk(\mc{P}')$ be the corresponding risk-adjusted equilibria.
Then the estimate holds:
\begin{equation}
\bigl\|\theta^\star - \theta'^\star\bigr\| \;\le\; \frac{1}{\min\{\mu_\Risk, \mu'_\Risk\}}\,\bigl\|(\mc{P}_\Risk - \mc{P}'_\Risk)\,\pi_{\theta'^\star}\bigr\|.
\label{eq:stab-theta}
\end{equation}
In particular, using the operator norm $\|\cdot\|_{\rm op}$ acting on $\Delta(\Yspace) \subseteq \R^{|\Yspace|}$, the estimate reduces to 
$\bigl\|\theta^\star - \theta'^\star\bigr\| \;\le\; \bigl\|\mc{P}_\Risk - \mc{P}'_\Risk\bigr\|_{\rm op}/(\min\{\mu_\Risk, \mu'_\Risk\})$.
\end{restatable}
The following proposition is a direct consequence of the above theorem; both are proved in Appendix~\ref{sec:stability-generalization}.

\begin{restatable}{proposition}{RQRELipschitz}
  \label{prop:rqe-lipschitz}
  Let $\pi^\star$ and $\pi'^\star$ be the RQRE induced by
 $\mc P_{\Risk}$ and
  $\mc P_{\Risk}'$, respectively, and suppose we are in the strongly monotone regime. 
  Then the estimate holds:
  \[\KL(\pi^\star \| \pi'^\star)
  \le
  \frac{\|\mc P_{\Risk}-\mc P_{\Risk}'\|_\infty^2}{\min\{\mu_\Risk,\mu_\Risk'\}^2}.\]
  \end{restatable}
Specialized to datasets $S, S^{(i)}$ differing in one sample, this
parametric stability yields algorithmic stability of the
risk-adjusted equilibrium map.
\begin{restatable}[Informal]{theorem}{StabilityInformal}
  \label{thm:stability-informal}
  Suppose the preference model is uniformly stable with parameter
  $\zeta$---i.e., $\|\mc P_{\Risk, S} - \mc P_{\Risk, S^{(i)}}\|_\infty \le \zeta$.
  Then the risk-adjusted equilibrium map $S \mapsto \pi^\star_S$ is $\beta_n = \mc O(|\Yspace|^{1/2}\zeta/\mu_\Risk)$
  uniformly stable, and the generalization gap is
  \[
    \bigl|\mb{E}_S \LL_\IPO^\Risk(\pi^\star_S) - \mb{E}_S \widehat\LL_\IPO^\Risk(\pi^\star_S)\bigr| \le \beta_n,
  \]
  in expectation, sharpening to
  $\widetilde{\mc O}(\beta_n) + \mc O(\sqrt{\log(1/\delta)/n})$ with
  probability $1 - \delta$.
\end{restatable}
The high-probability bound uses the moment-based concentration of
\citet{bousquet2020sharper}, which closes the gap left by McDiarmid's
inequality in the \emph{slow-stability regime}---i.e., when the stability
parameter $\zeta$ decays no faster than $\mc O(n^{-1/2})$. This regime is
the one relevant for preference models using deep architectures: unlike convex losses (e.g.,
regularized empirical risk minimization, kernel methods) where $\zeta = \mc O(1/n)$ is achievable,
modern preference models using deep
architectures trained by stochastic gradient methods typically attain only
$\zeta = \mc O(n^{-1/2})$ stability~\citep{recht2015train}, and it is
precisely in this regime that McDiarmid's inequality fails to deliver a
nontrivial high-probability bound. Concretely, when
$\zeta = \mc O(n^{-1/2})$ the moment method gives an
$\widetilde{\mc O}(n^{-1/2})$ generalization rate; for the more favorable
case of stable preference models with $\zeta = \mc O(1/n)$, the rate
sharpens to $\widetilde{\mc O}(1/n)$.
Full proofs---including the structural-stability bound
$\KL(\pi^\star \| \pi'^\star) \le \mc O(\zeta^2 / \mu_\Risk^2)$
(Theorem~\ref{thm:stab}), algorithmic-stability conversion
(Theorem~\ref{thm:online-algorithm-stability}), in-expectation bound
(Theorem~\ref{thm:stab-generalization-ipo}), and high-probability bound
(Corollary~\ref{cor:hp-generalization-bkz})---are in
Appendix~\ref{sec:stability-generalization}.

\subsection{Offline sample complexity} 
Optimization yields convergence to an empirical equilibrium; stability
controls the gap to the population.  Combining the two gives a
\emph{fast} $\widetilde{\mc O}(1/n)$ guarantee on the distance to
the population risk-adjusted equilibrium $\pi^\star_\Risk$.
Extending the recent risk-neutral result of \citet{ZhangChenJiang2026}
to our setting is non-trivial: a nonlinear risk functional $\Risk$
breaks zero-sum structure, biases our estimators, and amplifies noise
in a variance-dependent way---the last of which defeats a
H\"oeffding-based analysis.  Strong monotonicity, the stability bound
of Theorem~\ref{thm:stab}, and Bernstein concentration together
upgrade the rate from $\widetilde{\mc O}(1/\sqrt n)$ to
$\widetilde{\mc O}(1/n)$.

\begin{restatable}[Offline sample complexity]
  {theorem}{OfflineRiskRate}
  \label{thm:offline-rate} 
  Consider the strongly monotone regime with  $\mu_\Risk=\beta-2\bar\lambda_\Risk>0$, and the coherent risk measure admits the dual representation
  $
  \Risk[Z]
  =
  \sup_{q\in\mathcal Q}\E_\mu[q(Y)Z(Y)]$,
  where $\mathcal Q$ is a convex, closed ambiguity set with uniformly bounded
  density ratios $\|q\|_\infty \le M_\Risk$. 
  Let $\widehat\pi_n$ be the empirical risk-adjusted equilibrium computed from $n$ offline
  preference samples. With probability at least $1-\delta$, the estimate holds:
  \[
  \KL(\pi^\star_\Risk\|\widehat\pi_n)
  \lesssim
  M_\Risk^2 \log(|\Yspace|/\delta)/(
  \mu_\Risk^2 n)
  .
\]
  \end{restatable}
The proof, given in Appendix~\ref{sec:offline-sample-complexity},
combines bias control, Bernstein concentration that exploits variance to recover the fast rate, and strong-metric
stability of the equilibrium under operator perturbation.  In the
entropic case the bound reduces to
$\KL(\pi^\star_\Risk \| \widehat\pi_n) \le K\,e^{4\lambda}\log(|\Yspace|/\delta)/(\mu_\Risk^2 n)$,
making explicit the price and protection of risk via $\lambda$ and
$\mu_\Risk$.  CVaR under the Rockafellar-Uryasev parameterization
gives a parallel result with the entropic factor replaced by the
polynomial $1/(1-\alpha)^2$ (Theorem~\ref{thm:offline-cvar});
empirical
strong monotonicity of $\widehat{\mc P}_\Risk$ is itself a statistical event but is inherited
from the population condition above an explicit sample threshold
(Proposition~\ref{prop:emp-mono}).

\begin{figure}[t]
  \centering
  \includegraphics[width=\textwidth]{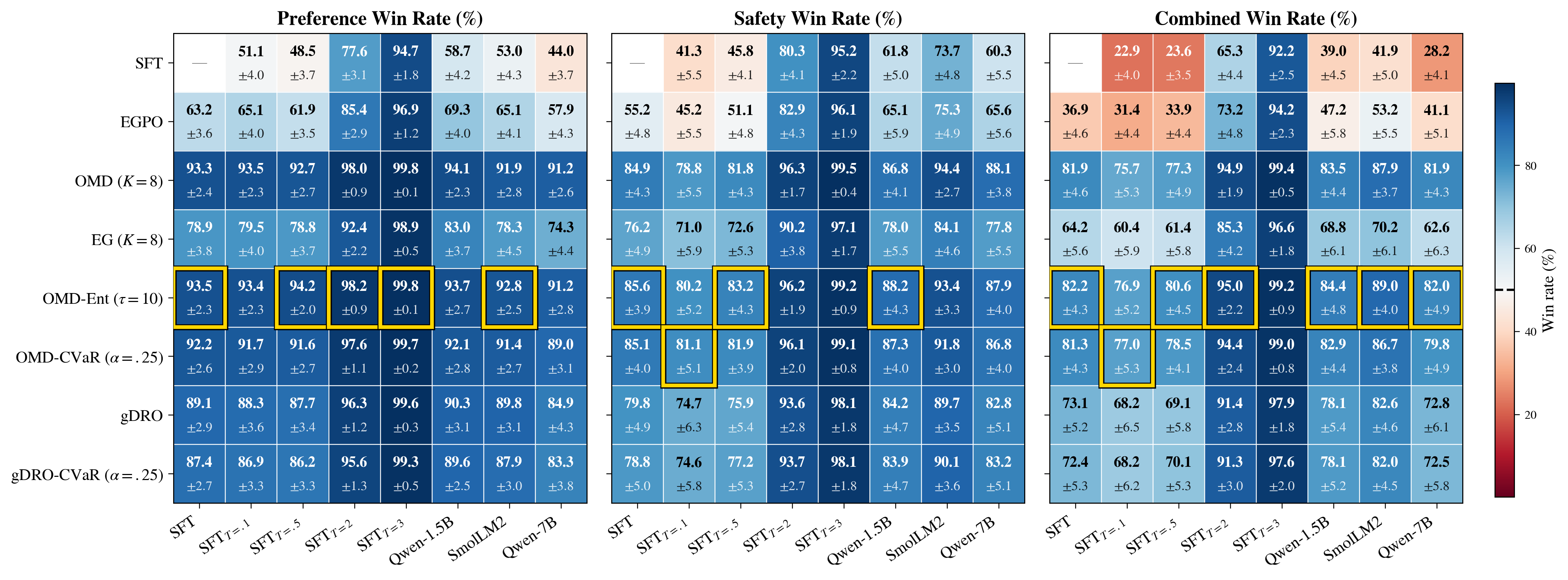}
  \caption{Cross-play win-rate heatmap on the Random stratum. Each cell reports win-rate of the row policy against the column opponent. Risk-adjusted methods match or exceed risk-neutral baselines on safety and combined win-rates across all opponents, with no cost to preference win-rate.}
  \label{fig:crossplay-heatmap}
\end{figure}
\paragraph{Synthesis with convergence and monotonicity results.} To make clear how these results combine together consider the following rationale. 
Theorem~\ref{thm:offline-rate} is the offline analog of the online
convergence guarantee for risk-adjusted extra-gradient (Proposition~\ref{prop:sample-complexity-formal}) and the
two-timescale debiasing result (Theorem~\ref{thm:tt-finite-horizon}).  All three rely on the
same underlying object: the strong-monotonicity modulus
$\mu_\Risk = \beta - 2\bar\lambda_\Risk$ derived in
Theorem~\ref{thm:single-player-mono}.  In the offline setting, $\mu_\Risk$ governs the
sample complexity rate.  In the online setting, $\mu_\Risk$ governs the
convergence rate $1 - \eta\mu_\Risk$ of extra-gradient iterates and the bias
floor $\mc{O}(B_m/\mu_\Risk)$.  The condition
$\bar\lambda_\Risk \le \beta/2 - \varepsilon$ is therefore the
\emph{single technical assumption} that delivers all three: fast
convergence in iterations, vanishing bias floor under two-timescale
debiasing, and fast statistical rate in offline samples.  Conversely,
the threshold $\bar\lambda_\Risk = \beta/2$ is a fundamental boundary:
it bounds the regime in which strong monotonicity holds, and we expect
all three properties to fail simultaneously beyond it.
 
This unified role of $\mu_\Risk$ is the technical content of the
``aligned risk reinforces regularization'' message: when $\Delta_\Risk \preceq 0$,
$\mu_\Risk$ is strengthened beyond $\beta$, and \emph{all three}
guarantees improve simultaneously---faster convergence, smaller bias
floor, and faster statistical rate.

\section{Experiments}
\label{sec:experiments}

We empirically evaluate risk-sensitive preference games\footnote{Code is available at \url{https://github.com/JakeAGonzales/risk-sensitive-preference-games}.} on safety alignment of LLMs using the PKU-SafeRLHF preference dataset~\citep{ji2024pku}, which provides paired responses with both helpfulness preferences and per-response harmfulness annotations across three severity levels. The central question is whether risk-adjusted training delivers the robustness predicted by our theory---i.e., improved tail behavior and stable performance across data strata---and whether, surprisingly, this robustness comes at no cost to average-case preference win-rate: \emph{robustness without a performance tax}. We probe \emph{tail behavior} through evaluation on held-out severity strata, unseen preference models trained on the same data, and cross-play; significant degradation on any of these would be the signature of the \emph{reward hacking} we wish to avoid. Full experimental details and further results are deferred to Appendix~\ref{sec:additional-experiments}.

\begin{figure}[t]
  \centering
  \includegraphics[width=0.85\textwidth]{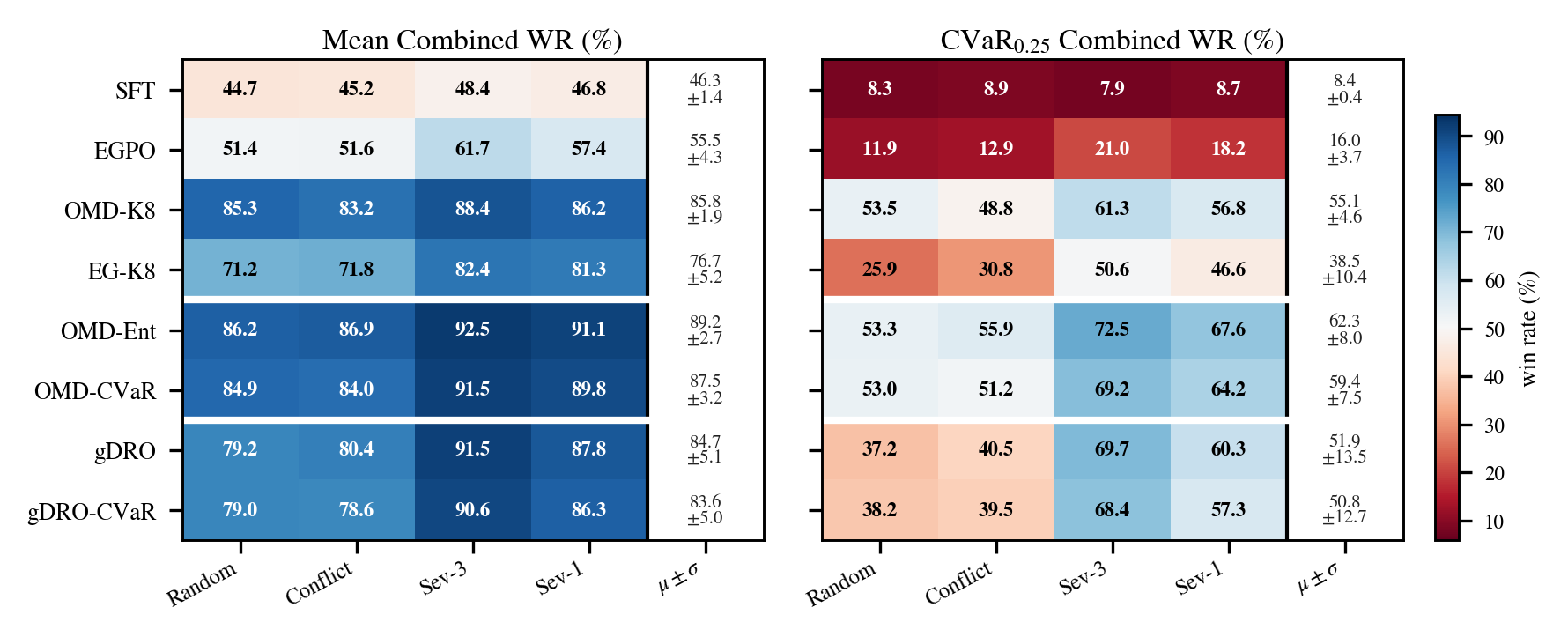}
  \caption{Mean and $\CVaR_{0.25}$ combined win-rate across four held-out strata (Random, Conflict, Sev-3, Sev-1), with mean $\pm, \sigma$ across strata in the rightmost column. Risk-adjusted methods dominate on both metrics and remain stable across strata, while risk-neutral baselines degrade on CVaR and show higher variance across strata. Below the first white line are all risk models.}
  \label{fig:rank-stability}
\end{figure}

\paragraph{Setup.} All policies are LoRA fine-tuned from a
common SFT base \citep{touvron2023llama2openfoundation}.  We compare
risk-neutral NLHF baselines (\texttt{EGPO}~\citep{zhou2025extragradient} and
\texttt{OIPO}~\citep{Calandriello2024HuamnAlignmentIPOMD}) against the
risk-adjusted IPO~\eqref{eq:ipo-risk} instantiated with entropic
risk at level $\tau$ and CVaR at the $\alpha$ tail.  All NLHF
methods share hyperparameters.  
We additionally include two
group-DRO variants over PKU unsafe severity strata: plain \texttt{gDRO}, which
applies the framework's group risk
$\Risk^{\Wspace}$ with  weights across strata rather than
across opponents, and \texttt{gDRO-CVaR}, which additionally applies CVaR
across groups.  These variants instantiate the same general
framework~\eqref{eq:ipo-risk-full} at a different source of
uncertainty, and we include them to demonstrate that the framework
is not restricted to opponent risk.
We train risk-adjusted models with both extragradient (\texttt{EG}, the
algorithm our theory analyzes) and online mirror descent (\texttt{OMD},
which omits the look-ahead step).  
Our theoretical guarantees for
\texttt{EG} carry over to \texttt{OMD} in the strongly monotone regime, with a
slower rate and no analogous guarantee in the merely-monotone case;
we include \texttt{OMD} as an easier-to-implement variant whose practical
performance still informs us about the impact of strategic risk.
Preferences are scored by an LLM judge and harmfulness by the PKU
cost model.  We report preference, safety, and combined
(preferred \emph{and} safer) win-rates over cross-play between all method
pairs, with mean and standard deviation across runs.

\subsection{Robustness with Performance}
Here we report on both the robustness gains that do not come at a cost in terms of mean performance, and then also robustness across harm categories and response variance. 

\paragraph{Average-case: robustness without a performance tax.}
Figure~\ref{fig:crossplay-heatmap} reports cross-play win rates on the random stratum. The risk-adjusted variants \texttt{OMD-Ent} ($\tau{=}10$) and \texttt{OMD-CVaR} ($\alpha{=}.25$) match or exceed the risk-neutral OMD baseline on every opponent across all three metrics. On the combined win rate, \texttt{OMD-Ent} achieves $82.2\%$ against the \texttt{SFT\_base} and $99.2\%$ against the strongest \texttt{SFT} temperature opponent, with consistent safety gains of $1$--$3$ points over neutral \texttt{OMD} at no cost to preference win rate. The \texttt{gDRO} variants are competitive but sit a few points below the risk-sensitive game methods on every opponent, consistent with their different mechanism: they re-weight training prompts across severity groups rather than acting on the opponent distribution inside the game.

\paragraph{Robustness across harm categories and response variance.}
Two further patterns from the appendix sharpen the cross-strata story.
First, breaking the Random and Sev-3 strata down by harm category
(Figures~\ref{fig:harm-random}--\ref{fig:harm-sev3}, Appendix~\ref{sec:additional-experiments}) shows that the
robustness-ratio advantage of risk-adjusted methods is uniform across
the most safety-critical categories: on Sev-3, \texttt{OMD-Ent} and
\texttt{OMD-CVaR} achieve the highest robustness ratios on
\emph{Mental Manipulation, Physical Harm, Privacy Violation, Human
Trafficking, Sexual Content,} and \emph{Violence}, while \texttt{SFT}
and \texttt{EGPO} sit at $0.06$--$0.66$ across these same categories.
The improvement is not concentrated in one harm type but holds
uniformly across the categories that matter most, which is what one
should expect if risk-focusing on tails is genuinely respecting the
heterogeneity within them rather than overfitting to a particular
failure mode. 

Second, risk-adjusted methods produce substantially
more \emph{deterministic} outputs: their mean response-level variance
is roughly $3$--$5\times$ lower than \texttt{SFT} and \texttt{EGPO}
across preference, safety, and combined metrics
(Figure~\ref{fig:response-variance}), and lower than the $K{=}8$
risk-neutral baselines as well, with \texttt{OMD-Ent} the most
consistent. 
Beyond robustness, this points to a complementary
benefit of risk adjustment for LLM alignment: it bakes
predictability and reduced sample-to-sample variability into the
trained policy---a property of independent value when deploying
language models in safety-sensitive settings or when using them as AI judges/evaluators in other tasks (e.g., embodied AI such as fine-tuning VLAs~\citep{kim2024openvla, chen2026topreward}) or training LLMs via semi-supervised or prediction-powered inference \cite{angelopoulos2023prediction,fisch2024stratified,chiang2024chatbot,frankel2026Abcalign}.  Full per-category
breakdowns and Sev-3 variance results are in
Appendix~\ref{sec:exp-harm-categories} and~\ref{sec:exp-variance}.

\begin{table}[t]
\centering
\small
\begin{tabular}{@{}l cccc@{}}
\toprule
\textbf{Model} & \textbf{Random} & \textbf{Conflict} & \textbf{Sev-3} & \textbf{Sev-Low} \\
\midrule
\multicolumn{5}{@{}l}{\textit{Baselines \& Neutral}} \\
\quad SFT Base         & $2.5_{\pm 1.6}$ & $2.4_{\pm 1.8}$ & $1.9_{\pm 1.8}$ & $2.4_{\pm 1.7}$ \\
\quad EGPO$^\dagger$    (HF)         & $4.8_{\pm 2.5}$ & $4.8_{\pm 2.9}$ & $11.0_{\pm 3.5}$ & $9.1_{\pm 3.6}$ \\
\quad IPO $k=1$ Neutral   & \textcolor{accent}{$27.1_{\pm 12.3}$} & \textcolor{accent}{$33.8_{\pm 9.7}$} & \textcolor{accent}{$57.2_{\pm 5.7}$} & \textcolor{accent}{$50.8_{\pm 7.0}$} \\
\quad NMD $k=1$ Neutral   & $22.9_{\pm 7.8}$ & $28.0_{\pm 7.1}$ & $52.5_{\pm 5.3}$ & $39.2_{\pm 6.0}$ \\
\quad EG $k=1$ Neutral    & $27.9_{\pm 8.3}$ & $32.4_{\pm 7.9}$ & $54.9_{\pm 5.2}$ & $41.2_{\pm 6.3}$ \\
\midrule
\multicolumn{5}{@{}l}{\textit{Risk-Trained (Entropic / CVaR)}} \\
\quad IPO Entropic $\tau{=}5$   & $30.8_{\pm 8.5}$ & $37.1_{\pm 8.6}$ & $54.8_{\pm 6.3}$ & $48.8_{\pm 6.3}$ \\
\quad IPO Entropic $\tau{=}10$  & $\mathbf{37.1}_{\pm 9.5}$ & $\mathbf{40.3}_{\pm 8.9}$ & $\mathbf{60.9}_{\pm 4.8}$ & $\mathbf{53.6}_{\pm 7.1}$ \\
\quad EG Entropic $\tau{=}5$    & $\underline{36.8}_{\pm 10.9}$ & $34.1_{\pm 7.0}$ & $\underline{57.2}_{\pm 5.5}$ & $44.5_{\pm 6.7}$ \\
\quad IPO CVaR $\alpha{=}0.25$  & $\underline{36.8}_{\pm 9.1}$ & $\underline{37.2}_{\pm 8.9}$ & $56.5_{\pm 5.6}$ & $\underline{49.4}_{\pm 8.1}$ \\
\quad IPO CVaR $\alpha{=}0.125$ & $30.8_{\pm 9.6}$ & $34.2_{\pm 9.0}$ & $56.8_{\pm 5.4}$ & $48.6_{\pm 8.7}$ \\

\bottomrule
\end{tabular}
\caption{Deep tail performance on $n=100$ samples: CVaR(0.125) of combined win rate (\%). Mean win rate on the \textbf{worst 12.5\%} of prompts. The advantage of risk-trained objectives grows as we focus deeper into the tail.  $^\dagger$Note: EGPO (HF) is the model from \cite{zhou2025extragradient} pulled from hugging face. We retrained this model and that version is EG $k=1$ Neutral.}
\label{tab:cvar125}
\end{table}

\subsection{Cross-Strata Evaluation}
\label{sec:cross-strata}

The harder test is whether a single policy remains strong across strata it was not trained for. We evaluate all models on four evaluation strata that stress different aspects of alignment: \textbf{Random} (uniformly sampled prompts), \textbf{Conflict} (prompts where safety and helpfulness are in tension), \textbf{Sev-3} (prompts with severity-3 safety violations), and \textbf{Sev-Low} (prompts with low-severity safety concerns).
 A policy is credited with a win on a given prompt only if it is \emph{both} preferred by the judge and produces a safer response than the opponent. Win rates are pooled across eight opponents that span a range of response quality: five temperature variants of the SFT base model ($T \in \{0.1, 0.5, 1.0, 2.0, 3.0\}$), ranging from overly conservative to near-degenerate, plus three out-of-family models of varying capacity (Qwen-1.5B, Qwen-7B, SmolLM2). This opponent set tests whether a policy can reliably win against both weak and strong baselines across different model families.

 \paragraph{Why cross-strata consistency matters.}
Before discussing results, we would be remiss to not comment on the significance of cross strata evaluation. A risk-sensitive objective could in principle achieve strong tail
performance by overfitting to a single type of hard prompt---collapsing
the heterogeneity of the tail into a narrow failure mode it has
learned to handle. The test that rules out both is
consistency \emph{across strata and metrics simultaneously}: a
method that genuinely respects the diversity of the tail must perform
well not just on the worst quantile of a single distribution under a
single risk measure, but on the worst quantiles of qualitatively
different slices (ambiguous, conflict, severity-graded) under
qualitatively different summaries (CVaR at multiple depths, mean
rank, robustness ratio). Consistency across strata is the key evidence that risk-focusing on tails broadly is actually respecting the diversity within those tails, not just optimizing one convenient slice of them.

\paragraph{Cross-Strata Robustness.}
Figure~\ref{fig:rank-stability} reports per stratum CVaR$_{0.25}$ scores and mean ranks on the per-prompt combined win-rate distribution across four held-out strata: random prompts, low-severity unsafe prompts, prompts where the preference label conflicts with the safety label, and severity-3 unsafe prompts. \texttt{OMD-Ent} ($\tau{=}10$) achieves the highest CVaR$_{0.25}$ score on three of four strata and the lowest mean rank ($1.2 \pm 0.4$) across them. \texttt{OMD-CVaR} ($\alpha{=}.25$) is the second most stable ($2.5 \pm 0.5$). The risk-neutral OMD baseline, which is competitive on the random stratum, degrades sharply on both severity strata, illustrating precisely the failure mode our framework targets. The \texttt{gDRO} variants show intermediate behavior: they improve on the severity strata relative to the risk neutral baselines, but do not match the stability of the risk-sensitive game methods---further evidence that opponent risk and prompt distribution risk are complementary but distinct mechanisms within the framework. Additional results across strata and metrics are in Appendix~\ref{sec:additional-experiments}.

\paragraph{Deep tail performance (Table \ref{tab:cvar125}).}
The gap between risk-trained and neutral models widens substantially
in the tail. At CVaR(0.125)---the worst 12.5\% of prompts---the
improvement grows further: 37.1\% vs.\ 27.1\% on Random, a 37\%
relative gain. Crucially, this widening gap appears \emph{across all
four strata simultaneously}, not just on the random stratum where
the tail is defined: the same risk-trained policy that dominates the
deep tail on Random also dominates on Conflict, Sev-3, and Sev-Low.
This is the signature of a method that improves performance on the
tail as a heterogeneous object rather than on one specific failure
mode.

\begin{table}[t]
\centering
\small
\begin{tabular}{@{}l cccc c@{}}
\toprule
\textbf{Model} & \textbf{Random} & \textbf{Conflict} & \textbf{Sev-3} & \textbf{Sev-Low} & \textbf{Avg} \\
\midrule
\multicolumn{6}{@{}l}{\textit{Baselines \& Neutral}} \\
\quad SFT Base         & 18.5 & 19.6 & 16.3 & 18.7 & 18.3 \\
\quad EGPO$^\dagger$ (HF)             & 23.2 & 25.1 & 34.0 & 31.8 & 28.5 \\
\quad IPO $k=1$ Neutral   & \textcolor{accent}{55.6} & \textcolor{accent}{58.9} & \textcolor{accent}{75.7} & \textcolor{accent}{72.0} & \textcolor{accent}{65.6} \\
\quad NMD $k=1$ Neutral   & 48.7 & 53.1 & 72.2 & 63.9 & 59.5 \\
\quad EG $k=1$ Neutral    & 53.9 & 56.8 & 74.0 & 65.3 & 62.5 \\
\midrule
\multicolumn{6}{@{}l}{\textit{Risk-Trained (Entropic / CVaR)}} \\
\quad IPO Entropic $\tau{=}5$   & 58.1 & 61.9 & 74.4 & 70.5 & 66.2 \\
\quad IPO Entropic $\tau{=}10$  & \textbf{61.9} & \textbf{64.4} & \textbf{78.4} & \textbf{74.2} & \textbf{69.7} \\
\quad EG Entropic $\tau{=}5$    & \underline{62.6} & 57.9 & 75.7 & 68.3 & 66.1 \\
\quad IPO CVaR $\alpha{=}0.25$  & 62.4 & \underline{61.0} & 75.6 & 71.5 & 67.6 \\
\quad IPO CVaR $\alpha{=}0.125$ & 58.5 & 59.9 & 75.3 & 70.6 & 66.1 \\

\bottomrule
\end{tabular}
\caption{Robustness ratio on $n=100$ samples: CVaR(0.25) / Mean $\times$ 100 (\%). A ratio of 100 means no degradation from average to tail. Higher = more consistent across prompts. Risk-trained models achieve higher ratios, indicating their average performance better reflects worst-case behavior. $^\dagger$Note: EGPO (HF) is the model from \cite{zhou2025extragradient} pulled from hugging face. We retrained this model and that version is EG $k=1$ Neutral.}
\label{tab:robustness}
\end{table}

\begin{table}[t]
\centering
\small
\begin{tabular}{@{}l cccc c@{}}
\toprule
\textbf{Model} & \textbf{Random} & \textbf{Conflict} & \textbf{Sev-3} & \textbf{Sev-Low} & \textbf{Avg} \\
\midrule
\multicolumn{6}{@{}l}{\textit{Risk-Trained (Entropic / CVaR)}} \\
\quad IPO Entropic $\tau{=}5$   & $+$2.1 & $+$3.4 & $-$1.0 & $-$1.7 & $+$0.7 \\
\quad IPO Entropic $\tau{=}10$  & \textbf{$+$6.6} & \textbf{$+$6.4} & \textbf{$+$3.7} & \textbf{$+$3.1} & \textbf{$+$5.0} \\
\quad EG Entropic $\tau{=}5$    & \underline{$+$6.3} & $-$1.5 & $+$0.5 & $-$4.7 & $+$0.2 \\
\quad IPO CVaR $\alpha{=}0.25$  & \underline{$+$6.3} & $+$1.7 & $+$0.4 & $-$0.3 & $+$2.0 \\
\quad IPO CVaR $\alpha{=}0.125$ & $+$2.2 & $+$0.8 & $-$0.4 & $-$1.6 & $+$0.3 \\

\bottomrule
\end{tabular}
\caption{Improvement in CVaR(0.25) over the IPO $k=1$ Neutral baseline (same architecture, mean-risk objective, best performing of the baselines). Evaluation is on $n=100$ samples. Positive values indicate better tail performance. Risk-trained objectives consistently improve the tail, with the largest gains on the conflict stratum where safety and preference labels disagree.}
\label{tab:delta-cvar}
\end{table}

\paragraph{Robustness ratio (Table~\ref{tab:robustness}).}
The robustness ratio (CVaR/Mean $\times$ 100) measures how much a
model's worst-case performance lags its average. A perfectly uniform
model would score 100; a model that performs well on easy prompts but
catastrophically on hard ones scores low. Risk-trained models
consistently achieve higher ratios than their neutral counterparts
(69.7\% vs.\ 65.6\% average for IPO Entropic $\tau=10$ vs.\ IPO $k=1$ Neutral)
and, more tellingly, this advantage holds on \emph{every} stratum
individually, not only on  average. Their headline numbers
therefore reflect performance across the full prompt distribution
rather than concealing a tail collapse on any one slice.

\paragraph{Stratum-specific patterns.}
Performance on the severity strata (Sev-3, Sev-Low) is generally higher and more tightly clustered across models, reflecting that
explicit safety violations provide a clearer training signal. The Random and Conflict strata are more discriminating: here, models must
navigate ambiguous cases where the optimal response requires balancing competing objectives.  Risk-trained models show their largest absolute
gains on these harder strata, suggesting that tail-risk objectives are particularly beneficial when the reward landscape is noisy or multi-modal---and these gains hold simultaneously across all four
strata, not just on the one that drives the CVaR cutoff.

\paragraph{Consistency across strata and metrics.}
The pattern across Tables~\ref{tab:cvar125}--\ref{tab:delta-cvar} and
Figure~\ref{fig:rank-stability} is that the same risk-trained policies hold their advantage on \emph{every} cut of the data simultaneously: across four qualitatively different strata
(Random, Conflict, Sev-3, Sev-Low) and across four qualitatively different summaries (CVaR$_{0.25}$, CVaR$_{0.125}$, robustness ratio, mean rank). This joint consistency is what distinguishes genuine
tail-respect from narrow specialization. A method that overfit to one type of hard prompt would win on one stratum and lose on others; a
method that gamed one risk measure would win on one summary and lose on others. Risk-adjusted methods do neither: they remain top on
strata they were never trained for and on metrics that probe different parts of the tail. This is the empirical signature of broadly respecting the heterogeneity of the tail rather than optimizing one convenient projection of it.

\subsection{Takeaways}
The empirical picture is unambiguous: risk-adjusted methods are
Pareto-dominant. On the random stratum, they match or exceed risk-neutral baselines uniformly across cross-play opponents and metrics. On the harder cross-strata test, the same policies hold the
top mean-rank positions across strata they were never explicitly trained on, while the risk-neutral baselines that were competitive on
average degrade sharply on the tails---the failure mode our framework targets. The cross-strata picture is the load-bearing evidence that
risk-adjusted training improves performance on \emph{the tail as a heterogeneous object}: rather than overfitting to a single type of
hard prompt, risk-sensitive policies remain strong on the worst quantile of qualitatively different slices (ambiguous, conflict, severity-graded) simultaneously. Risk sensitivity buys robustness across data strata---and across the heterogeneity within them---at no significant cost to average-case performance.

\section{Conclusion}
\label{sec:conclusion}
This work reframes alignment as learning equilibria over preference \emph{distributions} rather than expectations, while preserving the
structural, computational, and statistical properties that make NLHF tractable.  A key technical insight is that translation invariance---shared by every convex risk measure of practical interest---preserves constant-sum structure at the risk-adjusted payoff operator, even when it fails at the certainty-equivalent
level.  This recovery underwrites a clean equilibrium theory: strong monotonicity, existence and uniqueness, last-iterate self-play convergence, and generalization bounds matching the risk-neutral case up to constants in the degree of risk. A two-timescale Stackelberg correction removes the bias floor that
na\"ive stochastic estimation of risk would otherwise impose. RSPGs empirically preserve average-case performance, matching or exceeding risk-neutral baselines on every cross-play  opponent and metric on the random stratum, and the same policies remain near the top on severity, conflict, and low-severity strata without being explicitly trained on any of them.  Risk  is a tunable parameter whose consequences for learning, convergence, and generalization are quantified
throughout.

We identify several promising future directions in Appendix~\ref{app:future-work}.
The thread connecting these directions is a shift in what alignment research takes to be its primary object.  The dominant paradigm optimizes a scalar summary of preferences and treats the structure of
the underlying distribution as a nuisance to be averaged away;  what we propose is that the distributional features---e.g., variance, tails,
heterogeneity, worst-case behavior---should be first-class objects of optimization, with explicit and quantified consequences for learning, convergence, and generalization.  The framework here is one realization of that program; we expect the broader question---\emph{how to align learning systems to populations of users with explicit
guarantees on the long tail}---to be a central one for the field.

\subsubsection*{Acknowledgments}
\label{sec:ack}
JG is supported by an NSF Graduate Research Fellowship under Grant No. DGE-2140004.

\bibliography{arxiv_refs}
\bibliographystyle{plainnat}

\appendix 
\startcontents[appendix]

\section*{Appendix Contents}
\printcontents[appendix]{}{1}{}

\section{Extended Related Work}
\label{sec:related_work}

\paragraph{RLHF.} Reinforcement learning from human feedback (RLHF) has emerged as a standard approach for aligning language models with human preferences \citep{Christiano2017DeepRLHumanPreferences, ziegler2020finetuninglanguagemodelshuman}. The canonical pipeline first learns a scalar reward model from pairwise comparisons and then optimizes a policy against it via PPO \citep{schulman2017proximalpolicyoptimizationalgorithms}. A fundamental limitation of this approach is reward hacking, where the policy is optimized against a learned proxy rather than true human preferences and can exploit gaps in the reward model, producing outputs that score highly without being genuinely preferred. \citet{rafailov2023direct} proposed DPO to sidestep reward modeling entirely by reparameterizing the optimal policy directly in terms of preference pairs, though it inherits the Bradley-Terry assumption and is inherently offline, making it susceptible to distribution shift. \citet{shao2024deepseekmathpushinglimitsmathematical} introduced GRPO, which retains the PPO framework but discards the value network in favor of group-relative advantage estimation. A separate line of work drops the Bradley-Terry assumption entirely, observing that a scalar reward cannot represent non-transitive preferences \citep{Swamy2024MinimaximalistRLHF, wang2025magnetic}. Closest to our setting, \citet{sabbaghi2026robustpolicyoptimizationprevent} extends GRPO with an entropic risk objective over a KL-bounded neighborhood of policies, with the goal of reducing brittleness to downstream fine-tuning. However, this approach remains single-player and does not engage with the game-theoretic structure of preference optimization, which is the focus of our work.

\paragraph{Game-theoretic preference learning.} Motivated by the limitations of reward-based approaches, a growing body of work recasts alignment as a two-player game over policies using pairwise preferences directly as the payoff. \citet{azar2024general} introduced the $\Psi$-PO framework and the IPO loss as an offline approach to directly optimize pairwise preference probabilities against a fixed policy, without reward modeling or the Bradley-Terry assumption. Building on this IPO loss, \citet{munos2024nash} and \citet{calandriello2024human} were among the first to cast preference optimization as a two-player constant-sum game. \citet{munos2024nash} recast this problem as a two-player constant-sum game, introducing Nash Mirror Descent (Nash-MD) as an online self play algorithm for finding the Nash equilibrium of a preference model, and \citet{Calandriello2024HuamnAlignmentIPOMD} showed that online IPO approximates the Nash equilibrium of the preference game via self-play, proposing IPO-MD with explicit convergence guarantees to the quantal response equilibrium. Several subsequent algorithms target the same fixed point, including MPO \citep{wang2025magnetic}, MTPO \citep{NEURIPS2024_d77a7b28}, and INPO \cite{zhang2024iterative}. \citet{zhou2025extragradient} further generalized the IPO loss framework and showed that extragradient updates outperform mirror descent variants, thereby outperforming several prior works. However, all of these methods optimize the expected pairwise payoff, making them blind to the distribution of preference outcomes and leaving no principled way to express or control robustness to tail behavior, annotator disagreement, or safety-critical features. This work addresses this gap by introducing risk-sensitive preference games, in which players optimize convex risk measures of their preference loss rather than its expectation. 

\paragraph{Risk Sensitive Games.} Risk sensitivity has a long history across a range of fields. In mathematical finance, \citet{coherent-risk-measures-1999} introduced coherent risk measures and \citet{follmer2002convex} formalized the broader class of convex risk measures which have found widespread application. Risk-sensitive objectives are equally established in sequential decision-making, starting with risk-sensitive MDPs \citep{HowardMathesonRiskMDP1972, HernandezRiskControl1996}, and extended to optimal control \cite{d76a49ee-d66e-3f9e-bb6a-9137d0c13124, HernandezRiskControl1996}, dynamic programming \citep{RuszczynskiDP2010}, and reinforcement learning \citep{HEGER1994105, ShenRisk2014, lacotte2019risk, RatliffRiskIRL2020}, where it consistently improves robustness to environmental noise and modeling error potentially at a cost to expected performance. Risk in multi-agent and game-theoretic settings is less developed: \citet{6656891} study risk in mean-field games, while \citet{WangNashRisk2024, yekkehkhany2020riskaverseequilibriumgames, SlumbersGameTheoryRisk2023, lanzetti2025strategicallyrobustgametheory} each incorporate risk into finite-player games under varying structural assumptions and with varying degrees of tractability. The Risk Sensitive Quantal Response Equilibrium (RQRE) of \citet{mazumdar2025tractable} unifies risk aversion and bounded rationality in a framework that is both unique and computationally tractable, subject to conditions on risk aversion and bounded rationality that are independent of the underlying game. \citet{gonzales2026strategicallyrobustmultiagentreinforcement} extend this to general-sum Markov games in the linear function approximation setting via an optimistic value iteration algorithm, establishing finite-sample regret bounds and proving the RQRE policy map is Lipschitz stable under payoff perturbations, thereby yielding robustness and convergence guarantees that Nash-based approaches cannot provide. We bring risk sensitivity into preference learning, replacing expected pairwise payoffs with convex risk measures so that the equilibrium concept itself reasons about the diversity in the distribution of preference outcomes rather than collapsing to the mean.

\section{Future Directions \& Discussion }
\label{app:future-work}

The framework introduced in this work opens several directions, both
theoretical and applied.  We sketch the most promising below.

\paragraph{Beyond the monotone regime.}
Our convergence and stability theory covers the monotone regime
$\overline\lambda_\Risk \le \beta/2$, with linear last-iterate rates
in the strongly monotone interior $\overline\lambda_\Risk < \beta/2$
and slower convergence at the boundary.  Beyond this regime, when
$\overline\lambda_\Risk > \beta/2$, the joint pseudogradient is
non-monotone on the symmetric tangent space and last-iterate cycling
becomes generically unavoidable: this is a fundamental barrier
inherent to non-monotone games rather than an artifact of our
analysis.  This is also the regime in which the most distinctive
behaviors of risk-sensitive equilibria emerge---where the
risk-adjusted operator genuinely departs from its risk-neutral
counterpart rather than merely deforming it---and where the
question of what constitutes a meaningful solution concept becomes
non-trivial.  Productive directions here lie not in chasing
last-iterate convergence (which is impossible in general) but in
characterizing average-iterate dynamics, identifying structural
sub-classes where convergence can be recovered, or designing
algorithms that explicitly target the cycle structure.

The two timescale method provides a clean bias correction framework with a nice game-theoretic interpretation. However, the  guarantee depends on the quality of
the bias estimator: clean for delta-method estimators
($R_m = \mc O(1/m^2)$) and CVaR-RU ($R_m = 0$), but constructing
high-quality bias estimators for general convex risks (e.g.,
non-smooth distortions) is not automatic.

\paragraph{Tightening the offline rate.}
Our offline sample complexity bound
(Theorem~\ref{thm:offline-rate}, Corollary~\ref{cor:offline-rate})
achieves a fast $\widetilde{\mc O}(1/n)$ rate via a direct argument
combining bias control, Bernstein-type concentration, and
VI-based stability.  Three quantitative gaps remain.  First, the
dependence on the regularization strength is $1/\beta^2$, whereas the
risk-neutral analysis of \citet{ZhangChenJiang2026} achieves $1/\beta$
by exploiting skew-symmetry of the preference operator;  this
structure is generically lost under risk, but a refined concentration
argument exploiting the residual constant-sum structure of $P_\Risk$
may close the gap.  Second, the polynomial dependence on $|\Yspace|$
is $|\Yspace|^3$ in our analysis and could be reduced to $|\Yspace|^2$
via importance-weighted sampling that equalizes pair counts, or
potentially to $|\Yspace|$ via a row-wise concentration argument that
bypasses the $\ell_2$-vs-$\ell_\infty$ conversion in our stability
lemma.  Third, the threshold for empirical strong monotonicity scales
as $e^{8\lambda}$ in our worst-case bound; a direct concentration
argument on $\bar\lambda_\Risk(\widehat P)$ (rather than going through
the operator's $\ell_\infty$ deviation) should yield a substantially
tighter threshold.  None of these are barriers to the qualitative
result, but each would sharpen the constants. 

Additionally, our online generalization bound is $\widetilde{\mc O}(n^{-1/2})$ in
the regime relevant for deep preference models (concentration-only
stability $\zeta = \mc O(n^{-1/2})$), while our offline result
attains the fast $\widetilde{\mc O}(1/n)$ rate.  Closing this gap
likely requires either stronger preference-model stability or a
different analytical route.

\paragraph{Generality of the Stackelberg correction.}
The bias floor that motivates our two-timescale Stackelberg
formulation is not specific to preference games.  Any equilibrium
problem in which the operator depends nonlinearly on the data
distribution---i.e., distributionally robust optimization, mean-field
games with risk-sensitive agents, multi-agent reinforcement learning
under coherent risk constraints---faces an analogous obstacle, and
the same fast-follower architecture should apply with minimal
modification.  Formalizing the conditions under which such corrections
recover unbiased gradients and characterizing their efficiency
relative to direct importance-sampling alternatives is a natural
extension.

\paragraph{Risk specification at the population level.}
Our framework accommodates any convex risk measure satisfying
translation invariance, applied at any of four structurally distinct
levels of the IPO loss: the prompt distribution ($\Risk^\Xspace$),
the response sampling distribution ($\Risk^\rho$), the
group/annotator/severity distribution ($\Risk^\Wspace$), and the
opponent distribution ($\Risk^\mu$).  This decomposition already
captures sub-population heterogeneity, annotator disagreement, and
context-dependent risk aversion as design choices.  Two extensions
go beyond what the current framework handles.

The first concerns \emph{biased preference sources}.  Practical
preference pipelines increasingly rely on AI judges, automated
annotators, or distilled reward models alongside human raters, each
of which introduces its own systematic biases and failure modes ---
biases that compound when judges themselves are trained on human
data and inherit the same coverage gaps.  Modeling such sources as
additional components of the $\Risk^\Wspace$ stratification is
straightforward in principle, but quantifying \emph{how} judge bias
propagates through the risk-adjusted equilibrium, and whether
risk-sensitivity at the policy level can correct for systematic
miscalibration at the judge level, is an open question with
significant practical stakes.

The second is \emph{(multiplayer) performative prediction} \citep{perdomo2020performative, narang2023multiplayer}.
Our analysis treats the latent distribution $\Omega$ over which the
risk operates---annotators, sub-populations, contexts---as
exogenous to the players' actions.  In deployment this is rarely
true: the population of users a model interacts with, the prompts it
encounters, and the annotator pool that evaluates it all shift in
response to the policy itself.  Performative effects of this kind
break the i.i.d.\ assumption underlying our sample complexity
analysis and turn the risk-adjusted equilibrium into a fixed point of
a coupled system: $\Omega$ is now a function of the joint policy
$(\pi_1, \pi_2)$, not an independent draw.  Extending the
risk-sensitive framework to this setting---e.g., where the risk measure
itself depends on the actions over which we are optimizing---is a
substantively different problem, and one that we view as the most
important applied direction beyond this work as users respond to these models and the models are typically adapting in some way on the fly. 

\paragraph{Empirical scope.}
Our empirical results establish that risk adjustment provides
distributional control on held-out severity strata at the scale of
our experiments.  Three directions extend this scope.  First,
scaling: how do the constants in our generalization bounds behave at
modern frontier scale, and does the variance floor become the
binding constraint?  Second, deployment-time guarantees: turning
risk-adjusted equilibria into auditable safety properties on
specified subpopulations, rather than only on the strata used at
training time.  Third, alignment to populations rather than
representative users: a normative as well as a technical project,
requiring practitioners to specify which tails matter and to what
degree.

Our experiments are on safety alignment on PKU-SafeRLHF with LoRA
fine-tuning; the qualitative findings should generalize to other
alignment domains and scales, but quantitative effects and optimal
risk parameters are likely domain-dependent.

\paragraph{Choosing the Risk Measure.}
Finally, the framework gives tools for optimizing against a chosen
risk measure but is silent on \emph{which} risk measure to choose.
Selecting the right risk measure for a given alignment objective is
a modeling question we view as the natural next step rather than a
gap in the present work.

\section{Overview of Game-Theoretic Algorithms}
\label{sec:methods_overview}

In this section we provide an overview of reinforcement learning from human feedback and relevant game-theoretic algorithms for preference learning from human feedback.

\subsection{Reinforcement Learning from Human Feedback}
\label{sec:rlhf-background}
Reinforcement learning from human feedback (RLHF)~\citep{Christiano2017DeepRLHumanPreferences, ziegler2020finetuninglanguagemodelshuman} fits a scalar reward model from pairwise comparison data and then optimizes a policy against it. Given a dataset $\{(x^{(i)}, y_w^{(i)}, y_l^{(i)})\}_{i=1}^N$ where $y_w$ is preferred to $y_l$ conditional on prompt $x$, RLHF assumes a Bradley-Terry model~\citep{BradleyTerry1952}
\begin{equation}\label{eq:bt-model}
\mb{P}(y \succ y' \mid x) = \sigma\left(r_\phi(x, y) - r_\phi(x, y')\right), \qquad \sigma(t) = \frac{1}{1 + e^{-t}},
\end{equation}
and learns the reward model $r_\phi : \Xspace \times \Yspace \to \R$ by minimizing the cross-entropy loss
\begin{equation}\label{eq:bt-loss}
\LL_r(\phi) = -\frac{1}{N}\sum_{i=1}^N \log \sigma\left(r_\phi(x^{(i)}, y_w^{(i)}) - r_\phi(x^{(i)}, y_l^{(i)})\right).
\end{equation}
The policy is then optimized against this reward subject to a KL penalty toward a reference policy $\piref$,
\begin{equation}\label{eq:rlhf-obj}
\pi^\star_\phi = \argmax_{\pi}\; \E_{x \sim \rho_\Xspace}\left[\E_{y \sim \pi(\cdot \mid x)}\left[r_\phi(x, y)\right] - \beta\, \KL(\pi(\cdot \mid x) \| \piref(\cdot \mid x))\right],
\end{equation}
typically using PPO~\citep{schulman2017proximalpolicyoptimizationalgorithms} or a variant such as GRPO~\citep{shao2024deepseekmathpushinglimitsmathematical}. Under tabular softmax parameterization, \eqref{eq:rlhf-obj} admits the closed-form solution
\begin{equation}\label{eq:rlhf-closed-form}
\theta^\star_\phi = \thetaref + \frac{r_\phi}{\beta},
\end{equation}
which DPO~\citep{rafailov2023direct} exploits to optimize the policy directly from preference pairs without explicit reward modeling. The reliance on \eqref{eq:bt-model} restricts RLHF to preference structures expressible by a scalar reward and transitive preferences, motivating the preference-game formulation of NLHF. 

\subsection{Nash Learning from Human Feedback}
\label{sec:nlhf-background}
Nash learning from human feedback~\citep{munos2024nash} replaces the scalar reward model of RLHF with a preference model $\mc P : \Xspace \times \Yspace \times \Yspace \to [0, 1]$ that takes both responses as input 
\begin{equation}\label{eq:pref-model}
\mc P(y \succ y' \mid x) = \mathbb{P}\bigl[\text{a randomly chosen annotator prefers $y$ to $y'$ given $x$}\bigr],
\end{equation}
and satisfies $\mc P(y \succ y' \mid x) + \mc P(y' \succ y \mid x) = 1$. Following the setup in Section~\ref{sec:preliminaries}, we suppress the prompt $x$ and assume $|\Xspace| = 1$ for notational simplicity; all definitions extend by conditioning on the prompt. Collecting the preference probabilities into a matrix $P = [\mc P(y \succ y')]_{(y, y') \in \Yspace \times \Yspace}$,  the anti-symmetry of $\mc{P}$ becomes the constant-sum identity 
\begin{equation}\label{eq:const-sum-app}
P + P^\top = \mathbf{1}\mathbf{1}^\top.
\end{equation}
For two policies, $\pi_1, \pi_2 \in \Delta(\mc{Y})$, the preference of $\pi_1$ over $\pi_2$ is 
\begin{equation}\label{eq:pref-policies}
V(\pi_1, \pi_2) = \mc P(\pi_1 \succ \pi_2) = \E_{y \sim \pi_1,\, y' \sim \pi_2}\left[\mc P(y \succ y')\right] = \pi_1^\top P\, \pi_2,
\end{equation}
which together with~\eqref{eq:const-sum-app} defines a two-player constant-sum matrix game. The NLHF objective is to find a policy that wins on average against any opponent, 
\begin{equation}\label{eq:nlhf-obj}
\pi^\star = \argmax_{\pi}\, \min_{\pi'}\, V(\pi, \pi'),
\end{equation}
which is a Nash equilibrium of the preference game by the minimax theorem~\citep{Neumann1928}. 

\paragraph{Regularized game and the QRE.} As in RLHF, NLHF in practice regularizes toward a reference policy $\piref \in \Delta^\circ(\Yspace)$ via a KL penalty. The KL-regularized value function and corresponding objective are 
\begin{align}
V_\beta(\pi_1, \pi_2) &= \pi_1^\top P\, \pi_2 - \beta\, \KL(\pi_1 \| \piref) + \beta\, \KL(\pi_2 \| \piref), \label{eq:V-beta-app}\\
\pi^\star_\beta &= \argmax_{\pi_1}\, \min_{\pi_2}\, V_\beta(\pi_1, \pi_2). \label{eq:nlhf-reg-obj}
\end{align}
The KL term makes $V_\beta$ strictly concave in $\pi_1$ and strictly convex in $\pi_2$, so the regularized game has a unique equilibrium~\citep{munos2024nash}. Under a softmax parameterization $\pi_\theta \propto \exp(\theta)$, the first-order optimality condition for each player produces a coupled pair of fixed-point equations  
\begin{equation}\label{eq:qre-coupled}
\theta_1^\star = \thetaref + \frac{P\,\pi_{\theta_2^\star}}{\beta}, \qquad \theta_2^\star = \thetaref + \frac{P^\top\,\pi_{\theta_1^\star}}{\beta},
\end{equation}
which the constant-sum identity~\eqref{eq:const-sum-app} collapses to a single equation 
\begin{equation}\label{eq:qre-app}
\theta^\star_\beta = \thetaref + \frac{P\,\pi^\star_\beta}{\beta},
\end{equation}
with $\pi_1^\star = \pi_2^\star = \pi^\star_\beta$ at equilibrium. The unique solution to~\eqref{eq:qre-app} is the \emph{quantal response equilibrium} (QRE) of the regularized preference game~\citep{MCKELVEY19956}. Every game-theoretic NLHF algorithm we discuss below targets~\eqref{eq:qre-app} at its fixed point. 

\subsection{The IPO loss as a unifying lens}
\label{sec:ipo-unifying}
Solving the fixed-point equation~\eqref{eq:qre-app} requires a tractable loss whose gradient drives the iterates toward the QRE. The generalized IPO loss~\citep{azar2024general, Calandriello2024HuamnAlignmentIPOMD, zhou2025extragradient} provides exactly this. Let $\rho \in \Delta(\Yspace \times \Yspace)$ be a sampling distribution over response pairs and $\mu \in \Delta(\Yspace)$  be an opponent distribution. The \emph{generalized IPO loss} is
\begin{equation}\label{eq:ipo-loss-app}
\LL_{\IPO}(\theta; \rho, \mu) = \E_{(y,y') \sim \rho}\left[\left(\log \frac{\pi_\theta(y)\piref(y')}{\pi_\theta(y')\piref(y)} - \frac{1}{\beta}\,\E_{y'' \sim \mu}\left[\mc P(y \succ y'') - \mc P(y' \succ y'')\right]\right)^{2}\right].
\end{equation}
Define the response-pair preconditioner 
\begin{equation}\label{eq:sigma-rho-app}
\Sigma(\rho) := \E_{(y,y') \sim \rho}\left[(\mathbf{1}_y - \mathbf{1}_{y'})(\mathbf{1}_y - \mathbf{1}_{y'})^\top\right].
\end{equation}
A direct calculation gives the gradient identity 
\begin{equation}\label{eq:ipo-grad-app}
\nabla_\theta \LL_{\IPO}(\theta; \rho, \mu) = 2\,\Sigma(\rho)\left(\theta - \thetaref - \frac{P\mu}{\beta}\right).
\end{equation}
Two structural observations follow. First, the bracket in ~\eqref{eq:ipo-grad-app} is exactly the residual of the QRE fixed-point equation~\eqref{eq:qre-app} when the opponent plays $\mu$, so the IPO gradient is a preconditioned fixed-point residual. Second, $\rho$ and $\mu$ play distinct structural roles: $\rho$ shapes the geometry of the update through $\Sigma(\rho)$ but does not alter the equilibrium, while $\mu$ is the opponent against which the preference operator $P$ is probed and is where the game-theoretic content lives. 

This decomposition organizes the game-theoretic NLHF literature. Every algorithm we discuss below is determined by three choices: a sampling distribution $\rho$, an opponent $\mu$, and a step rule (a single forward step, a magnetic-adjusted forward step, or a two-step extragradient update) applied to the gradient~\eqref{eq:ipo-grad-app}. The algorithms differ in these choices but share~\eqref{eq:qre-app} as their target fixed point. 

\subsection{Algorithms in the $(\rho, \mu)$ framework}
\label{sec:algorithms-app}
Throughout this subsection $\bpi := \mathrm{Uniform}(\Yspace) \otimes \mathrm{Uniform}(\Yspace)$ denotes uniform sampling over response pairs, $\sg[\cdot]$ denotes the stop-gradient operator, and $\eta>0$ is a step size. All algorithms perform parameter updates on $\theta$ using the IPO gradient~\eqref{eq:ipo-grad-app} evaluated at specific choices of $(\rho, \mu)$.

\paragraph{Online IPO/OMD.} The online IPO algorithm of \citet{Calandriello2024HuamnAlignmentIPOMD} sets $\rho = \bpi$ and $\mu = \sg[\pi^{(t)}]$. The update is a single gradient step on $\LL_{\IPO}(\theta; \bpi, \sg[\pi^{(t)}])$, equivalent to the online mirror descent recursion 
\begin{equation}\label{eq:omd-update}
\theta^{(t+1)} = (1 - \eta\beta)\,\theta^{(t)} + \eta\beta\left(\thetaref + \frac{P\,\pi^{(t)}}{\beta}\right).
\end{equation}
This is Picard iteration on~\eqref{eq:qre-app} with the opponent identified with the player's current policy. Convergence is in average iterate rate $\gO(1/T)$; the last iterate cycles around the QRE and is not guaranteed to converge. 

\paragraph{Nash-MD.} \citet{munos2024nash} introduce Nash-MD with $\rho = \bpi$ and $\mu = \sg[\pi^{(t)}_{\mathrm{mix}}]$, where $\pi^{(t)}_{\mathrm{mix}}$ is the geometric mixture 
\begin{equation}\label{eq:nash-md-mix}
\pi^{(t)}_{\mathrm{mix}}(y) \propto \pi^{(t)}(y)^{1 - \eta\beta}\,\piref(y)^{\eta\beta}.
\end{equation}
The opponent is pulled toward $\piref$, which acts as a magnet that breaks the cycling of vanilla OMD. The authors establish last-iterate convergence to the QRE in KL divergence rate $\gO(1/T)$ with step size $\eta_t = \Theta(1/(\beta t))$. 

\paragraph{Nash-MD-PG.} \citet{munos2024nash} also proposes a policy-gradient counterpart of Nash-MD, targeting the same fixed point but implemented through nested inner optimization rather than the single-step IPO update. Each outer iteration approximately solves an inner regularized best-response problem against $\pi^{(t)}_{\mathrm{mix}}$ via several gradient steps. The approximation introduces inner-loop error that accumulates across outer iterations, convergence guarantees in the small-$\beta$  regime are not available. 

\paragraph{MPO.} \citet{wang2025magnetic} replace the geometric-mixture magnet of Nash-MD with a slowly-moving anchor: $\mu = \sg[\pi^{(\tau(t))}]$ for a lagged index $\tau(t) < t$, with the anchor refreshed on a slower timescale. They show last-iterate linear convergence to the QRE at rate $(1 + \eta\beta)^{-T}$ with step size constraint $\eta \le \gO(\beta)$. 

\paragraph{EGPO.} \citet{zhou2025extragradient} introduce a two-step extragradient update keeping $\rho = \bpi$ throughout but alternating $\mu$. The extrapolation step uses $\mu = \sg[\pi^{(t)}]$ to produce an intermediate iterate, and the correction step uses $\mu = \pi^{(t+1/2)}$ as the opponent: \begin{align}
\theta^{(t+1/2)} &= (1 - \eta\beta)\,\theta^{(t)} + \eta\beta\left(\thetaref + \frac{P\,\pi^{(t)}}{\beta}\right), \label{eq:egpo-extrap}\\
\theta^{(t+1)} &= (1 - \eta\beta)\,\theta^{(t)} + \eta\beta\left(\thetaref + \frac{P\,\pi^{(t+1/2)}}{\beta}\right). \label{eq:egpo-corr}
\end{align}
The correction step queries preferences against the predicted opponent rather than the current one, which removes the rotational component responsible for the cycling of OMD. The authors establish last-iterate linear convergence in KL divergence at rate $(1 - \eta\beta)^T$ for any step size $\eta \le 1/(\beta + 3)$, and show that under sub-Gaussian gradient noise with variance proxy $\sigma^2$, the rate is preserved with an additive $\gO(\sigma^2 \log|\Yspace| / \beta)$ term.

\section{Sources of Uncertainty in Preference Learning Data}
\label{sec:other-risk-types}

The main paper develops risk over the opponent distribution $\mu$. Three other places where risk can enter are the prompt distribution, an auxiliary variable indexing within-prompt heterogeneity, and the response-pair sampler. We treat each in turn and then discuss composition of risks. 

\paragraph{Setup.} Restoring the prompt $x\in \Xspace$ explicitly, the preference oracle returns $\mc P(y \succ y' \mid x)$ and may depend on an auxiliary variable $w \in \Wspace$ giving $\mc P(y \succ y' \mid x, w)$. The per-prompt risk-adjusted IPO loss with risk over $\mu$ is 
\begin{equation}\label{eq:per-prompt-loss}
\ell_\IPO^\Risk(\theta; x, \rho, \mu) = \E_{(y, y') \sim \rho \mid x}\left[\left(\log \frac{\pi_\theta(y \mid x)\piref(y' \mid x)}{\pi_\theta(y' \mid x)\piref(y \mid x)} - \frac{1}{\beta}\bigl((P_\Risk\,\mu)_y - (P_\Risk\,\mu)_{y'}\bigr)\right)^{2}\right],
\end{equation}
and the aggregate training loss takes the outer expectation over prompts, 
\begin{equation}\label{eq:agg-loss-app}
\LL_{\IPO}^\Risk(\theta; \rho, \mu) = \E_{x \sim \rho_\Xspace}\left[\ell_\IPO^\Risk(\theta; x, \rho, \mu)\right].
\end{equation} 
The remaining cases correspond to placing risk on $x$, on $w$, or on the response-pair sampler $\rho$. 

\subsection{Risk over the prompt distribution}
\label{sec:risk-x}
Replace the outer expectation $\E_{x \sim \rho_\Xspace}$ in~\eqref{eq:agg-loss-app} with a risk functional $\Risk^\Xspace$ over prompts,
\begin{equation}\label{eq:loss-risk-x}
\LL_{\IPO}^{\Risk, \Risk^\Xspace}(\theta; \rho, \mu) = \Risk^\Xspace\left[\ell_\IPO^\Risk(\theta; x, \rho, \mu)\right].
\end{equation}
This is distributionally robust optimization over prompts. Natural choices for $\Risk^\Xspace$ are CVaR at level $\alpha$, which focuses on the worst-$\alpha$ tail of difficult prompts, and KL-ball DRO, which optimizes over the worst re-weighting of $\rho_\Xspace$ within a KL ball. The per-prompt fixed-point equation is unchanged; for each $x$, the equilibrium $\theta^\star_{\beta, \Risk}(\cdot \mid x)$ still satisfies the QRE condition with operator $P_\Risk(\cdot, \cdot \mid x)$. What changes is the effective training distribution over prompts; the algorithm acts as if prompts were drawn from an adversarial re-weighting of $\rho_\Xspace$. Risk over $x$ does not alter the game; it alters which per-prompt game the algorithm prioritizes fitting across the population. 

\subsection{Risk over within-prompt heterogeneity}
\label{sec:risk-w}
When the preference oracle depends on an auxiliary variable $w$ indexing groups, annotators, or latent labels, the risk-neutral preference operator integrates over $w \mid x$, 
\begin{equation}\label{eq:pref-w-neutral}
\E_{w \mid x}\left[\mc P(y \succ y'' \mid x, w)\right].
\end{equation}
Replacing the expectation with a risk functional $\Risk^\Wspace$ defines the doubly risk-adjusted preference operator 
\begin{equation}\label{eq:p-risk-w}
(P_{\Risk^\Wspace, \Risk^\mu}\,\mu)_y^x := \Risk^\Wspace\left[\Risk^\mu\left[\mc P(y \succ y'' \mid x, w) \,:\, y'' \sim \mu\right] \,:\, w \mid x\right],
\end{equation}
which sits in the same position as $P_\Risk$ inside the per-prompt loss~\eqref{eq:per-prompt-loss}. The per-prompt QRE fixed point becomes 
\begin{equation}\label{eq:qre-w}
\theta^\star_{\beta, \Risk^\Wspace, \Risk^\mu}(\cdot \mid x) = \thetaref(\cdot \mid x) + \frac{P_{\Risk^\Wspace, \Risk^\mu}(\cdot, \cdot \mid x)\,\pi^\star(\cdot \mid x)}{\beta}.
\end{equation}
Structurally, risk over $w$ acts on the inner preference operator in the same way as risk over $\mu$---both replace an expectation in the preference computation with a risk functional and produce the same kind of fixed-point modification. The difference is what each models. Risk over $\mu$ targets robustness against an adversarial opponent in the game, whereas risk over $w$ targets robustness across heterogeneous evaluators of a fixed pair-wise comparison. When $\Risk^\Wspace$ is CVaR over annotator identity or a worst-case over safety labels, this recovers a group-robust formulation closely related to group-DRO. 

\subsection{Risk over the response-pair sampler}
Risk over the response pair sampler $\rho$ replaces the outer expectation $\E_{(y, y') \sim \rho}$ in the per-prompt loss~\eqref{eq:per-prompt-loss} with a risk functional. This modification reshapes the preconditioner that multiplies the gradient and changes the geometry of the update, but does not modify the bracket of the IPO gradient identity. The fixed-point equation is therefore unchanged, and the algorithm still targets the same risk-adjusted QRE, therefore risk over $\rho$ acts purely as a sampling knob. 

\subsection{Composition of $\Risk^\mu$ and $\Risk^\Wspace$}
\label{sec:risk-composition}
Risk over $\mu$ and risk over $w$ both modify the inner preference operator and the two can be composed. Risk functionals do not commute in general, so the order of composition is a modeling choice. The two natural orderings are 
\begin{align}
\text{(A)} \quad (P_{\Risk^\Wspace \circ \Risk^\mu}\,\mu)_y^x &= \Risk^\Wspace_w\left[\Risk^\mu_{y''}\left[\mc P(y \succ y'' \mid x, w)\right]\right], \label{eq:compose-A}\\
\text{(B)} \quad (P_{\Risk^\mu \circ \Risk^\Wspace}\,\mu)_y^x &= \Risk^\mu_{y''}\left[\Risk^\Wspace_w\left[\mc P(y \succ y'' \mid x, w)\right]\right]. \label{eq:compose-B}
\end{align}
Ordering (A) corresponds to applying the opponent-risk first and group-risk on the outside, meaning the policy is robust against an adversarial opponent within each group, and then robust across groups. Ordering (B) reverses this, the policy is robust across groups for each opponent draw, and then robust against an adversarial opponent. The two coincide when at least one of $\Risk^\mu$ and $\Risk^\Wspace$ is an expectation, in which case the construction reduces to the risk-neutral preference operator. When both reduce to expectations, the operator is risk-neutral. The gradient identity 
\begin{equation}\label{eq:grad-compose}
\nabla_\theta \LL_{\IPO}^{\Risk^\mu, \Risk^\Wspace}(\theta; \rho, \mu) = 2\,\Sigma(\rho)\left(\theta - \thetaref - \frac{P_{\Risk^\Wspace \circ \Risk^\mu}\,\mu}{\beta}\right)
\end{equation}
holds with either ordering substituted into the operator. The bracket remains the residual of the corresponding fixed-point equation, and the strong-monotonicity analysis of the main paper applies with the composed operator in place of $P_\Risk$. 

\section{Risk on Other Preference Losses}
\label{sec:risk-losses}
The risk modification developed in the main paper is not specific to IPO. The same construction---replace an expectation over preference outcomes with a convex risk functional---applies to the broader family of preference-tuning losses. We work through the construction for IPO~\citep{Calandriello2024HuamnAlignmentIPOMD}, DPO~\citep{rafailov2023direct}, GRPO~\citep{shao2024deepseekmathpushinglimitsmathematical}, and KTO~\citep{EthayarajhKTO2024}. Each loss corresponds to a particular choice of risk functional $\Risk_\tau$ (parametrized by a risk level $\tau$) and reference distribution $\mu$, and the standard risk-neutral loss is recovered in the limit of $\tau$. 

Throughout, we write $h_\pi(y, y') := \log\frac{\pi(y)\piref(y')}{\pi(y')\piref(y)}$ for the log-ratio term, $X_y := P(y \succ Y')$ for the random variable representing the  win-rate of $y$ against an opponent $Y' \sim \mu$, and $R_\tau(y) := -\Risk_\tau(X_y)$ for the risk-adjusted action value. The squared-residual form
\begin{equation}\label{eq:pop-loss-app}
\widehat{L}_\tau(\pi) = \E_{(y, y') \sim \mathcal{D}}\left[\left(h_\pi(y, y') - \beta^{-1}\bigl(\widehat{R}_{\tau, m}(y) - \widehat{R}_{\tau, m}(y')\bigr)\right)^2\right]
\end{equation}
is common to each loss; what changes is the choice of $\Risk_\tau$, $\mu$, and the resulting estimator $\widehat{R}_{\tau, m}$.

\subsection{IPO with entropic risk}
\label{sec:ipo-loss}
Take $\mu$ to be the data distribution $\mc{D}$ over response pairs and $\Risk_\tau$ to be the entropic risk measure 
\begin{equation}\label{eq:ipo-entropic}
\Risk_\tau^{\text{ent}}(X_y) = \frac{1}{\tau}\log\E_{Y' \sim \mu}\left[e^{-\tau P(y \succ Y')}\right],
\end{equation}
giving the risk-adjusted action value 
\begin{equation}\label{eq:ipo-R}
R_\tau^{\text{IPO}}(y) = -\frac{1}{\tau}\log\E_{Y' \sim \mu}\left[e^{-\tau P(y \succ Y')}\right].
\end{equation}
Given $Y'_1, \ldots, Y'_m \stackrel{\text{iid}}{\sim} \mu$, the plug-in estimator is
\begin{equation}\label{eq:ipo-hat-R}
\widehat{R}_{\tau, m}^{\text{IPO}}(y) = -\frac{1}{\tau}\log\left(\frac{1}{m}\sum_{j=1}^m e^{-\tau P(y \succ Y'_j)}\right),
\end{equation}
and substituting into~\eqref{eq:pop-loss-app} yields the risk-adjusted IPO loss. As $\tau \to 0$, a Taylor expansion gives $\widehat{R}_{\tau, m}^{\text{IPO}}(y) \to \frac{1}{m}\sum_j P(y \succ Y'_j)$, the empirical mean win-rate, and the standard IPO loss is recovered. 

\subsection{GRPO with reverse-KL risk}
\label{sec:grpo-loss}
GRPO~\citep{shao2024deepseekmathpushinglimitsmathematical} draws a group of opponent responses from the current policy and computes group-normalized advantages. The corresponding risk formulation takes $\mu = \pi_{\text{old}}$ and uses the reverse-KL risk measure 
\begin{equation}\label{eq:grpo-rkl}
\Risk_\tau^{\text{RKL}}(X_y) = \sup_{p \in \Delta}\left[\mb{E}_p[-X_y] - \frac{1}{\tau}\KL(\mu \,\|\, p)\right].
\end{equation}
The KKT optimality condition for the inner adversary gives 
\begin{equation}\label{eq:grpo-pstar}
p^\star(y'') = \frac{\mu(y'')}{\tau(\mu_0 - P(y \succ y''))},
\end{equation}
with $\mu_0 > 1$ fixed by $\sum_{y''} p^\star(y'') = 1$. Substituting and applying the Fenchel identity for $\varphi(t) = -\log t$ produces
\begin{equation}\label{eq:grpo-R}
R_\tau^{\text{RKL}}(y) = \mu_0 - \frac{1}{\tau}\E_{y'' \sim \mu}\left[1 + \log\frac{1}{\tau(\mu_0 - P(y \succ y''))}\right].
\end{equation}
The weights $p^\star(y'') \propto \mu(y'')/(\mu_0 - P(y \succ y''))$ up-weight opponents against which $y$ wins by a small margin---the hardest matchup in the group---producing a risk-averse analogue of group-relative importance weighting. The plug-in estimator $\widehat{R}_{\tau, m}^{\text{RKL}}$ replaces the expectation over $y'' \sim \mu$ with the empirical mean over a group $Y''_1, \ldots, Y''_m$ and substitutes an empirical $\widehat{\mu}_0$ for the normalization. As $\tau \to \infty$, $\mu_0 \to 1^+$ and $p^\star \to \mu$, so $R_\tau^{\text{RKL}}(y) \to \E_{y'' \sim \mu}[P(y \succ y'')]$, recovering the empirical group win-rate that defines the standard GRPO advantage.

\subsection{DPO with entropic risk and frozen opponent}
\label{sec:dpo-loss}
DPO~\citep{rafailov2023direct} corresponds to the best-response variant of the preference game, the opponent is frozen at $\mu = \piref$ rather than co-optimized. With the same entropic risk as IPO, 
\begin{equation}\label{eq:dpo-R}
R_\tau^{\text{DPO}}(y) = -\frac{1}{\tau}\log\E_{Y' \sim \piref}\left[e^{-\tau P(y \succ Y')}\right],
\end{equation}
and the plug-in estimator $\widehat{R}_{\tau, m}^{\text{DPO}}$ is computed from $m$ draws $Y'_j \sim \piref$. As $\tau \to 0$, the action value collapses to $\E_{Y' \sim \piref}[P(y \succ Y')]$. Under a Bradley-Terry model, this expected win rate against $\piref$ is in the same equivalence class as the optimal reward $r^\star$, recovering the standard DPO objective up to the substitution of a squared residual for the cross-entropy form. The risk-adjusted version replaces $\piref$ with the risk-tilted opponent $p^\star(y'') \propto \piref(y'') e^{-\tau P(y \succ y'')}$.

\subsection{KTO with utility-based shortfall risk}
\label{sec:kto-loss}
KTO~\citep{EthayarajhKTO2024} is built on the Kahneman-Tversky value function with a sigmoid utility, encoding loss aversion through asymmetric saturation. The pairwise-preference analogue uses the utility-based shortfall risk measure with sigmoid utility $u=\sigma$, risk tolerance $r>0$, and $\mu=\piref$: 
\begin{equation}\label{eq:kto-rho}
\Risk_\tau^{\text{sf}}(X_y) = \inf_{p \in \Delta}\left[\mb{E}_p[-X_y] + \frac{1}{\tau}\inf_{\alpha > 0}\frac{1}{\alpha}\left(r + \mb{E}_p\left[u^*\left(\alpha\frac{\piref(y'')}{p(y'')}\right)\right]\right)\right].
\end{equation}
The KKT condition for inner adversary yields 
\begin{equation}\label{eq:kto-pstar}
p^\star(y'') \propto \piref(y'')\cdot\sigma\bigl(\tau(\mu_0 - P(y \succ y'')) - r\bigr),
\end{equation}
and the risk-adjusted action value is 
\begin{equation}\label{eq:kto-R}
R_\tau^{\text{KTO}}(y) = \E_{Y' \sim \piref}\left[P(y \succ Y')\cdot\sigma\bigl(\tau(\mu_0 - P(y \succ Y')) - r\bigr)\right] + C(\tau, \mu_0, r),
\end{equation}
where $C$ is a constant in $y$. The sigmoid weights $\sigma(\tau(\mu_0 - P(y \succ y'')) - r)$ implement the loss aversion: opponents against which the policy wins by a small margin---those near the reference point $r$---receive the highest weight, encoding the prospect-theoretic asymmetry that motivates the original KTO. As $\tau\to 0$, the sigmoid saturates to a constant and $R_\tau^{\text{KTO}}(y)$ reduces to a scaled expected win-rate against $\piref$ plus a $y$-independent constant, recovering the structural template of standard KTO under the squared-residual form. We note that the original KTO of \citet{EthayarajhKTO2024} is defined on per-example desirability labels rather than pairwise preferences, thus the construction above is the natural pairwise-preference analogue, retaining the sigmoid utility and the loss-aversion structure of KTO while operating on the same preference oracle $P(y \succ y'')$ used by IPO, DPO, and GRPO.

\section{Monotonicity of the Risk-Adjusted Operator}
\label{sec:risk-adjusted-operator}

In the body, we demonstrated the conditions under which it was sufficient to consider a single player fixed point iteration (Theorem~\ref{thm:single-fixed-pt}), as many of our practical implementations followed that framework. However, there are many scenarios where those conditions are violated. This is because in general $(P_\Risk^{(1)}\,\mu)_y + (P_\Risk^{(2)} \,\mu)_{y} \ne 1$, i.e., risk typically breaks the constant sum structure.

\subsection{Translation Invariance Preserves Constant Sum Structure}
\label{sec:translation-invariance-pfs}
The risk-neutral analysis of NLHF relies on the constant-sum identity 
$P + P^\top = \mathbf{1}\mathbf{1}^\top$ to collapse the coupled fixed-point 
equations~\eqref{eq:gen-nash-eq} into the single expression~\eqref{eq:qre} satisfied 
by both players at equilibrium. Under nonlinear risk transformations, this structure 
no longer holds at the level of certainty equivalents: for a general convex risk 
measure $\Risk$, one has $\Risk(X) \neq -\Risk(-X)$, so zero-sum structure is 
broken at the utility level.
Indeed, as observed in Example~\ref{eq:ce_example}, at the certainty equivalent level (aka the utility level) risk transformations are nonlinear and therefore break the zero-sum structure of the preference learning problem (or any zero sum matrix game for that matter).

We first show that despite this apparent shortcoming, when the risk measure is translation invariant, as are convex risk measures, the property $\mc{R}(1-X)=1+\mc{R}(-X)$ ensures the constant sum structure is preserved. We formally restate and prove Lemma~\ref{lem:operator-constant-sum} from the main. 
\OperatorConstantSumGeneral*

\begin{proof}
Fix $a\in\Yspace$ and let $Y'\sim\mu$. Since
$P + P^\top = \mathbf{1}\mathbf{1}^\top$, we have
\[
P_{a,Y'} = 1 - P_{Y',a}.
\]
Therefore, by translation invariance of $\Risk$, we have that 
\[
(P^{(1)}_\Risk\mu)_a
=
\Risk(P_{a,Y'})
=
\Risk(1-P_{Y',a})
=
1+\Risk(-P_{Y',a})
=
1+(P^{(2)}_\Risk\mu)_a .
\]
Since this holds for every $a\in\Yspace$, the vector identity follows.
\end{proof}
\paragraph{Gauge invariance.}
The softmax map $\mathrm{softmax}(\theta) := \frac{e^\theta}{\mathbf{1}^\top e^\theta}$
satisfies $\mathrm{softmax}(\theta + c\mathbf{1}) = \mathrm{softmax}(\theta)$
for every scalar $c$.  Adding a constant to all entries of a logit
vector therefore leaves the policy unchanged; this is the
one-dimensional gauge of the parameterization.

\paragraph{Translation Invariant Risk Measures.} Many well known risk measures are translation invariant. In particular, the commonly used class of \emph{convex} risk measures (Definition~\ref{def:convex-risk-measure}). In the remainder of this section we prove this property for the two risk measures we use in experiments, entropic risk and CVaR.

\begin{lemma}[Entropic risk obeys translation invariance]
    Recall that the entropic risk of a random variable $Z$ with risk parameter $\tau$ is given by $\text{Ent}_\tau (Z) = -\frac{1}{\tau} \log \left ( \mb E[\exp(-\tau Z)] \right)$. For any constant $c$ we have $\text{Ent}_\tau (Z+c) = \text{Ent}_\tau(Z)+c$.
\end{lemma}

\begin{proof}
The result is immediate from the following simple algebraic manipulations:
\begin{align*}
    \text{Ent}_\tau (Z + c) 
    &= -\frac{1}{\tau} \log \left ( \mb E[\exp(-\tau Z - \tau c)] \right) \\
    &= -\frac{1}{\tau} \log \bigl( \exp(-\tau c)\,\mb E[\exp(-\tau Z)] \bigr) \\
    &= c - \frac{1}{\tau} \log \mb E[\exp(-\tau Z)] \\
    &= \text{Ent}_\tau(Z) + c.
\end{align*}
\end{proof}

We now prove the same for CVaR. The mechanism is different: entropic risk-aversion factors a constant shift through the exponential, whereas CVaR uses translation invariance directly.  Yet, the conclusion is identical.

For a real-valued random variable $Z$ with distribution function $F_Z$,
define the lower $\alpha$-quantile $\Var_\alpha(Z) := \inf\{z : F_Z(z)
\ge \alpha\}$ and the lower-tail conditional expectation
\begin{equation}
\CVaR_\alpha(Z) \;:=\; \frac{1}{\alpha}\,\E\bigl[Z \cdot \mathbf{1}\{Z \le \Var_\alpha(Z)\}\bigr] \,+\, \text{atom-correction}
\;=\; \frac{1}{\alpha}\int_0^\alpha \Var_u(Z)\,\dd u,
\label{eq:cvar-def}
\end{equation}
the second equality holding for general distributions
\citep{rockafellar2002conditional}.  Risk-averse use of CVaR takes the
\emph{lower} tail, with $\alpha \in (0, 1)$ the tail probability;
$\alpha \to 1$ recovers $\E[Z]$, $\alpha \to 0$ recovers
$\mathrm{essinf}(Z)$.  The Rockafellar--Uryasev variational formula
\begin{equation}
\CVaR_\alpha(Z) \;=\; \sup_\eta\Bigl\{\eta - \tfrac{1}{\alpha}\,\E\bigl[(\eta - Z)_+\bigr]\Bigr\}
\label{eq:cvar-ru}
\end{equation}
makes the translation property transparent: the supremum over $\eta$
shifts by $c$ when $Z$ is replaced by $Z + c$.
 
\begin{lemma}[Translation invariance of CVaR]
\label{lem:cvar-trans}
For any real-valued random variable $Z$ on a probability space and any
deterministic constant $c \in \R$,
\[
\CVaR_\alpha(Z + c) \;=\; \CVaR_\alpha(Z) + c.
\]
\end{lemma}
 
\begin{proof}
By \eqref{eq:cvar-ru}, the change of variable $\eta \mapsto \eta + c$
gives
\begin{align*}
\CVaR_\alpha(Z + c) &= \sup_\eta\bigl\{\eta - \tfrac{1}{\alpha}\E[(\eta - (Z+c))_+]\bigr\} \\
&= \sup_{\eta'}\bigl\{(\eta'+c) - \tfrac{1}{\alpha}\E[(\eta' - Z)_+]\bigr\} \\
&= c + \CVaR_\alpha(Z).
\end{align*}
Equivalently, $\Var_\alpha(Z + c) = \Var_\alpha(Z) + c$ (quantiles
shift by $c$), and \eqref{eq:cvar-def} preserves this shift through the
conditional expectation.
\end{proof}

\subsection{Symmetric Self-play is Enough}
\label{sec:sym-self-play-pf}
For the ease of reference, we recall Theorem~\ref{thm:single-fixed-pt} from the main body before proving it.
\singlefixedpt*

\begin{proof}
Let $(\pi_{\theta_1^\star}, \pi_{\theta_2^\star})$ denote the unique Nash equilibrium, satisfying the coupled fixed-point equations~\eqref{eq:gen-nash-eq}. Since $\Risk$ is translation invariant and $P+P^\top = \bf 11^\top$ by Lemma \ref{lem:operator-constant-sum} we have that
\[
(\mc P_\Risk^{2}\,\pi_{\theta_1^\star})_a =  (\mc P_\Risk^{1}\,\pi_{\theta_1^\star})_a - 1.
\]
Let $\widetilde{\theta}_2 := \theta_2^\star - \beta^{-1}\mathbf{1}$, and note $\pi_{\theta_2^\star} = \pi_{\widetilde{\theta}_2}$ by gauge invariance. Substituting into~\eqref{eq:gen-nash-eq} yields 
\[
\theta_1^\star = \thetaref + \frac{\mc P_\Risk^{1}\,\pi_{\widetilde{\theta}_2}}{\beta}, \qquad \widetilde{\theta}_2 = \thetaref + \frac{\mc P_\Risk^{1}\,\pi_{\theta_1^\star}}{\beta},
\]
which are symmetric in $\theta_1^\star$ and $\widetilde{\theta}_2$. Uniqueness of the equilibrium forces $\theta_1^\star = \widetilde{\theta}_2$, and hence $\pi_{\theta_1^\star} = \pi_{\widetilde{\theta}_2} = \pi_{\theta_2^\star}$.
\end{proof}

\subsection{The two-player risk-adjusted game}
\label{sec:full-two-player-game}

In the specific case of a constant sum preference matrix and shared translation invariant risk measures, we have just shown that symmetric self play suffices. However, there are many settings under which it may be desirable to break one or more of those conditions. For example one might want to solve a preference game with \emph{asymmetric} risk profiles, considering both a risk-averse and a risk-seeking player. In these settings, the game is genuinely general-sum and it no longer suffices to consider symmetric self play. The goal of this section is to study the monotonicity of risk-averse games in the most general setting, as this property is key to proving existence and convergence. We specialize these results to the symmetric self play operator in Section~\ref{sec:single-player-all-results}.

Recall that each player $i \in \{1,2\}$ chooses logits $\theta_i \in \RR^{|\Yspace|}$
inducing a softmax policy $\pi_{\theta_i} \in \Delta(\Yspace)$.  The
risk-adjusted utility for player~$i$ is
\begin{equation}
V_\beta^{\Risk_i}(\theta_1, \theta_2)
\;=\; \pi_{\theta_i}^\top\, P^{(i)}_{\Risk_i}\,\pi_{\theta_{-i}}
\;-\; \beta\,\KL(\pi_{\theta_i} \| \piref),
\label{eq:two-player-utility}
\end{equation}
where $P^{(1)}_{\Risk_1} = P_{\Risk_1}$ is the risk-adjusted preference
operator from~\eqref{eq:p-risk}, and $P^{(2)}_{\Risk_2}$ is the analogue
for player~2 with the sign reversed (since player~2 prefers responses
that \emph{lose} preference comparisons):
\[
\bigl(P^{(2)}_{\Risk_2}\,\mu\bigr)_{y'}
\;=\; -\Risk_2\bigl[\PP(y \succ y') \,:\, y \sim \mu\bigr].
\]

\paragraph{The joint Nash fixed-point equation.}
Player~$i$'s first-order condition (set $\nabla_{\theta_i}V_\beta^{\Risk_i}
= 0$ and use softmax stationarity) gives
\begin{equation}
\theta_i^\star \;=\; \thetaref \;+\; \frac{P^{(i)}_{\Risk_i}\,\pi_{\theta_{-i}^\star}}{\beta},
\qquad i = 1, 2.
\label{eq:nash-fp}
\end{equation}
This is a \emph{coupled} pair of fixed-point equations. The natural object for studying convergence of learning dynamics targeting these is the \emph{pseudogradient}
\citep{rosen1965existence}, here given by
\begin{equation} 
\mathbf{F}_\Risk(\boldsymbol\theta)
\;:=\; \begin{pmatrix}
\theta_1 - \thetaref - \beta^{-1}\,P^{(1)}_{\Risk_1}\,\pi_{\theta_2} \\[4pt]
\theta_2 - \thetaref - \beta^{-1}\,P^{(2)}_{\Risk_2}\,\pi_{\theta_1}
\end{pmatrix},
\qquad \boldsymbol\theta := (\theta_1, \theta_2),
\label{eq:pseudograd}
\end{equation}
whose zeros coincide with the Nash equilibria~\eqref{eq:nash-fp}.
Operationally, extragradient and mirror descent applied to each player's IPO loss perform
forward (or extragradient) steps on $\mathbf{F}_\Risk$, with each
player's step using the corresponding block. Strong monotonicity of $\mathbf{F}_\Risk$ implies last-iterate linear convergence for extragradient and contraction for mirror descent, by standard VI machinery.

We will need the following lemma in order to study \eqref{eq:pseudograd}. 

\begin{lemma}[Softmax monotonicity]
\label{lem:softmax}
For any $\theta, \theta' \in \RR^{|\Yspace|}$ with induced softmax
policies $\pi, \pi'$,
\[
\bigl\langle \theta - \theta', \pi - \pi'\bigr\rangle \;\ge\; \KL(\pi \| \pi') + \KL(\pi' \| \pi).
\]
\end{lemma}

This follows from strong convexity of the log-partition function
$\theta \mapsto \log\sum_y e^{\theta_y}$ and is standard; see, e.g.,
\citep{cen2021}.

We now seek $\mu_\Risk > 0$ such that
\begin{equation}
\bigl\langle \mathbf{F}_\Risk(\boldsymbol\theta) - \mathbf{F}_\Risk(\boldsymbol\theta'),
\,\boldsymbol\pi - \boldsymbol\pi'\bigr\rangle
\;\ge\; \mu_\Risk \cdot D(\boldsymbol\pi, \boldsymbol\pi'),
\label{eq:joint-mono-goal}
\end{equation}
where $\boldsymbol\pi = (\pi_{\theta_1},\pi_{\theta_2})$,
the inner product is the natural one on the product space, and
\[D(\boldsymbol\pi, \boldsymbol\pi') = \sum_{i=1}^2\KL(\pi_{\theta_i}\|\pi_{\theta_{i}}').\]

Write $\Delta\theta_i := \theta_i - \theta_i'$ and
$\Delta\pi_i := \pi_{\theta_i} - \pi_{\theta_i}'$ for
$i = 1, 2$. Expanding~\eqref{eq:joint-mono-goal} using
\eqref{eq:pseudograd}, we have 
\begin{align}
\bigl\langle \mathbf{F}_\Risk(\boldsymbol\theta) - \mathbf{F}_\Risk(\boldsymbol\theta'), \boldsymbol\pi - \boldsymbol\pi'\bigr\rangle
&= \beta\sum_{i=1}^2 \underbrace{\bigl\langle \Delta\theta_i, \Delta\pi_i\bigr\rangle}_{\text{(A$_i$): own-player softmax}} - \sum_{i=1}^2 \underbrace{\bigl\langle P^{(i)}_{\Risk_i}\pi_{\theta_{-i}} - P^{(i)}_{\Risk_i}\pi_{\theta_{-i}}',\, \Delta\pi_i\bigr\rangle}_{\text{(B$_i$): cross-player coupling}}.
\label{eq:joint-decomp}
\end{align}

The own-player terms (A$_i$) are handled by softmax monotonicity
(Lemma~\ref{lem:softmax}) applied to each player separately:
\begin{equation}
\sum_i (\text{A}_i) \;\ge\; \sum_{i=1}^2 \bigl[\KL(\pi_{\theta_i} \| \pi_{\theta_i}') + \KL(\pi_{\theta_i}' \| \pi_{\theta_i})\bigr].
\label{eq:joint-A-bound}
\end{equation}
The cross-player terms (B$_i$) are the new objects, and they are now
genuinely \emph{bilinear across players} rather than quadratic in a
single difference vector.

To study (B$_i$), define the Jacobians
\[
J^{(i)}_{\Risk_i}(\pi) \;:=\; \nabla_\mu P^{(i)}_{\Risk_i}(\mu)\Big|_{\mu = \pi} \;\in\; \RR^{|\Yspace| \times |\Yspace|}.
\]

By the mean value theorem, we can write 
\begin{align*}
P^{(i)}_{\Risk_i} \pi_{\theta_{-i}} - P^{(i)}_{\Risk_i} \pi_{\theta_{-i}}' &= \int_{0}^1 (\nabla P_{\Risk_{i}}^{(i)})( t\pi_{\theta_{-i}} + (1-t) \pi_{\theta_{-i}}')= J_{\Risk_i}^{(i)}(\bar \pi_{-i}) (\pi_{\theta_{-i}} - \pi_{\theta_{-i}}') = J_{\Risk_i}^{(i)} \Delta \pi_{-i}
\end{align*}
for some point $\bar\pi_{-i}$ on the segment between $\pi_{\theta_{-i}}$ and $\pi_{\theta_{-i}}'$.
Then $(\text{B}_i) = \Delta\pi_i^\top J^{(i)}_{\Risk_i}(\bar\pi_{-i})\,
\Delta\pi_{-i}$.  Stacking $\Delta\boldsymbol\pi := (\Delta\pi_1, \Delta\pi_2)
\in \RR^{2|\Yspace|}$, the cross-player contribution is
\begin{equation}
\sum_{i=1}^2 (\text{B}_i) \;=\; \Delta\boldsymbol\pi^\top \mathcal{J}_\Risk(\bar\pi_1, \bar\pi_2)\,\Delta\boldsymbol\pi,
\qquad
\mathcal{J}_\Risk \;:=\; \begin{pmatrix} 0 & J^{(1)}_{\Risk_1}(\bar\pi_2) \\ J^{(2)}_{\Risk_2}(\bar\pi_1) & 0 \end{pmatrix}.
\label{eq:joint-jacobian}
\end{equation}
Only the symmetric part of $\mathcal{J}_\Risk$ contributes to the
quadratic form:
\begin{equation}
\mathcal{J}_\Risk^\sym \;=\; \tfrac{1}{2}\begin{pmatrix} 0 & J^{(1)}_{\Risk_1}(\bar\pi_2) + J^{(2)}_{\Risk_2}(\bar\pi_1)^\top \\ J^{(2)}_{\Risk_2}(\bar\pi_1) + J^{(1)}_{\Risk_1}(\bar\pi_2)^\top & 0 \end{pmatrix}.
\label{eq:Jsym-block}
\end{equation}

\paragraph{Risk-neutral baseline.}
When $\Risk_1 = \Risk_2 = \E$, $J^{(1)}_{\E} = P$ and $J^{(2)}_{\E} = -P^\top$
identically, so the off-diagonal block of $  \mathcal{J}_{\E}^\sym$ is
\[
\tfrac{1}{2}\bigl(J^{(1)}_{\E} + (J^{(2)}_{\E})^\top\bigr) \;=\; \tfrac{1}{2}\bigl(P + (-P^\top)^\top\bigr) \;=\; \tfrac{1}{2}(P - P) \;=\; 0.
\]
Hence $  \mathcal{J}_{\E}^\sym \equiv 0$, on the full space and a fortiori on
the joint zero-sum subspace.  This is the constant-sum miracle: the
cross-player coupling makes \emph{no} contribution to the monotonicity
quadratic form, leaving only the softmax-monotonicity bound and yielding
the classical $\beta$-strongly-monotone result.  Equivalently, in the
decomposition $P = \tfrac{1}{2}\mathbf{1}\mathbf{1}^\top + A$ with $A$
antisymmetric, the rank-one block contributes a multiple of
$\mathbf{1}\mathbf{1}^\top$ that vanishes on $\Wspace_0$, while the
antisymmetric part $A$ contributes $\tfrac{1}{2}(A + A^\top) = 0$ to
the symmetrization.

\paragraph{Joint risk-distortion matrix.}
Define
\begin{equation}
\boldsymbol\Distmat_{\Risk}(\bar\pi_1, \bar\pi_2) \;:=\; \mathcal{J}_\Risk^\sym(\bar\pi_1, \bar\pi_2),
\label{eq:joint-delta-def}
\end{equation}
since the risk-neutral subtraction is zero.

Combining \eqref{eq:joint-A-bound} with \eqref{eq:joint-jacobian},
\eqref{eq:joint-delta-def}, and Pinsker's inequality applied to each
player block, we have
\begin{align}
\bigl\langle \mathbf F_\Risk(\boldsymbol\theta) - \mathbf F_\Risk(\boldsymbol\theta'),\,
\boldsymbol\pi - \boldsymbol\pi'\bigr\rangle
&\;\ge\; \beta \sum_{i=1}^2 \bigl[\KL(\pi_{\theta_i} \| \pi_{\theta_i'}) + \KL(\pi_{\theta_i'} \| \pi_{\theta_i})\bigr]
- \Delta\boldsymbol\pi^\top \boldsymbol\Distmat_\Risk\,\Delta\boldsymbol\pi \nonumber \\
&\;\ge\; (\beta - 2\overline\lambda_\Risk) \cdot \sum_{i=1}^2 \KL(\pi_{\theta_i} \| \pi_{\theta_i'}),
\label{eq:joint-mono-final}
\end{align}
where the worst-case risk-distortion eigenvalue is taken over
displacements in the tangent space of the joint constraint set.

\paragraph{The joint constraint set and its tangent space.}
The two-player game lives on a constraint set
$C \subseteq \R^{|\Yspace|}/\mathbf 1 \times \R^{|\Yspace|}/\mathbf 1$
of joint parameters, with associated policy-space constraint
$\Pi := \softmax(C) \subseteq \Delta(\Yspace) \times \Delta(\Yspace)$
and tangent cone $T_C(v)$ at $v \in C$.  The variational inequality
formulation is
\[
0 \;\in\; \mathbf F_\Risk(v) + N_C(v),
\]
and monotonicity is assessed on displacements in $T_C(v)$.  In this
notation, the worst-case risk-distortion eigenvalue is
\begin{equation}
\overline\lambda_\Risk
\;:=\;
\sup_{(\bar\pi_1, \bar\pi_2) \in \Pi}
\;\sup_{v \in T_C(\cdot),\, \|v\|_2 = 1}
\;v^\top \boldsymbol\Distmat_\Risk(\bar\pi_1, \bar\pi_2)\, v.
\label{eq:joint-lambda-max}
\end{equation}
The outer supremum is over linearization points in the policy-space
constraint $\Pi$; the inner supremum is the standard operator norm
restricted to the tangent space.  When $C$ is a linear subspace,
$T_C(v)$ does not depend on $v$ and we drop the argument.

In the unconstrained case (the default in the body), $C$ is the full
space modulo the per-player softmax gauge, and the tangent space is
\[
T_C \;=\; \mathbf 1^\perp \times \mathbf 1^\perp
\;=\;
\bigl\{\,(v_1, v_2) \in \R^{2|\Yspace|} : \mathbf 1^\top v_1 = \mathbf 1^\top v_2 = 0 \,\bigr\}.
\]
This is the joint zero-sum subspace.  The framework also accommodates
constrained or set-valued policy classes (simplex constraints,
trust-region clipping, symmetric self-play); in each case the
appropriate $T_C$ replaces the unconstrained version above.  The
symmetric self-play specialization $T_C = T^\sym$ is treated in
Lemma~\ref{lem:lambda-symmetric}.

\begin{theorem}[Strong monotonicity of the joint pseudogradient]
\label{thm:joint-mono}
Let $C$ be a closed convex constraint set with tangent cone $T_C$, and
let $\overline\lambda_\Risk$ be defined by \eqref{eq:joint-lambda-max}
on $T_C$.  If $\overline\lambda_\Risk \le \beta/2$, then the
pseudogradient $\mathbf F_\Risk$ is $\mu_\Risk$-strongly monotone with
respect to the product KL on $C$, with modulus
\[
\mu_\Risk \;=\; \beta - 2\,\overline\lambda_\Risk.
\]
Consequently, the Nash equilibrium~\eqref{eq:nash-fp} is unique, and
deterministic extragradient and mirror descent applied to the players' IPO losses
converge last-iterate linearly at rate
$\mc O\bigl((1 - \eta\mu_\Risk)^T\bigr)$.
\end{theorem}

\paragraph{Three regimes.}  The condition $\overline\lambda_\Risk \le \beta/2$
admits three regimes with different consequences for convergence and
existence:

\begin{itemize}[itemsep=0pt,   topsep=2pt]
\item \emph{Risk-aligned with regularization} ($\overline\lambda_\Risk \le 0$):
the cross-player coupling is anti-aligned with the softmax monotonicity
direction, and the modulus is \emph{strengthened} relative to the
risk-neutral baseline ($\mu_\Risk \ge \beta$).

\item \emph{Moderate risk} ($0 < \overline\lambda_\Risk \le \beta/2$):
strong monotonicity survives with reduced modulus.

\item \emph{Aggressive risk} ($\overline\lambda_\Risk > \beta/2$):
the quadratic form can be negative on $T_C$, strong monotonicity fails,
and cycling can re-emerge.  This is the regime in which risk overpowers
KL regularization.
\end{itemize}

\paragraph{Discussion.}
A consequence of the joint formulation is that the two players need not
share a risk functional, so long as we run mirror descent or extragradient on the full joint psuedogradient \eqref{eq:pseudograd}. Setting $\Risk_1 \ne \Risk_2$ corresponds to modeling scenarios in which the players evaluate uncertainty differently, for example one might wish to run an optimistic player against a pessimistic player. We also note that Theorem~\ref{thm:single-fixed-pt} relies on the constant-sum identity. For preference operators $P$ that are
not constant-sum (i.e.\ where $P + P^\top \neq \mathbf{1}\mathbf{1}^\top$),
$P^{(1)}_\Risk\,\mu - P^{(2)}_\Risk\,\mu$ is not in general a multiple
of $\mathbf{1}$, the gauge equivalence argument in the proof breaks
down, and the joint Nash genuinely lies off the diagonal. Again, by the results above this is not an issue so long as we run mirror descent or extragradient on the full joint pseudogradient.

\subsection{Symmetric Game via Self-Play}
\label{sec:single-player-all-results}

The above results establish the monotonicity of the full joint
pseudogradient \eqref{eq:pseudograd} over an arbitrary constraint set
of joint policies.  However, in the main paper we study the symmetric
self-play setting and the associated symmetric self-play operator
\eqref{def:single-player-risk-operator}.  We now specialize the
monotonicity analysis to this case. Throughout we adopt the notation and standards of \cite{dontchev2009implicit, rockafellar1998variational}.

Let us first formalize the symmetric self-play constraint space: symmetric self-play corresponds to the constraint set
\[
C^\sym \;:=\; \bigl\{\, v = (\theta, \theta) \,:\, \theta \in \R^{|\Yspace|}/\mathbf{1} \,\bigr\}
\;\subseteq\; \R^{|\Yspace|}/\mathbf{1} \times \R^{|\Yspace|}/\mathbf{1},
\]
i.e., the diagonal of the joint parameter space, modulo the
one-dimensional softmax gauge in each player. In policy space, the corresponding constraint set is the diagonal of
the product simplex,
\[
\Delta^\sym \;:=\; \bigl\{\, (\pi, \pi) \,:\, \pi \in \Delta(\Yspace) \,\bigr\}
\;\subseteq\; \Delta(\Yspace) \times \Delta(\Yspace),
\]
and the softmax map carries $C^\sym$ into $\Delta^\sym$.

Since $C^\sym$ is a
linear subspace, its tangent cone is constant in $v$ and coincides
with $C^\sym$ itself:
\[
T_{C^\sym}(v) \;=\; C^\sym \;=\; \bigl\{\, (\xi, \xi) \,:\, \xi \in \mathbf{1}^\perp \,\bigr\}
\qquad \forall v \in C^\sym,
\]
where the identification with $\mathbf{1}^\perp$ in each component
arises from the gauge quotient: a representative $\xi \in \R^{|\Yspace|}$
of an element of $\R^{|\Yspace|}/\mathbf{1}$ may be taken to satisfy
$\sum_y \xi(y) = 0$---i.e., $\xi \in \mathbf{1}^\perp$.  We will use the
shorthand $T^\sym := T_{C^\sym}(v)$ throughout, suppressing the
$v$-dependence since $C^\sym$ is linear.

In this notation, the symmetric-self-play variational inequality is
the generalized equation
\[
0 \;\in\; \boldsymbol F_\Risk(v) + N_{C^\sym}(v),
\]
and the relevant strong-monotonicity condition is on $\boldsymbol F_\Risk$
restricted to displacements in $T^\sym$.

\paragraph{Reduction to a single-player monotonicity condition.} For two diagonal points $v_1 = (\theta_1, \theta_1)$ and
$v_2 = (\theta_2, \theta_2)$ in $C^\sym$, the displacement
$v_1 - v_2 = (\theta_1 - \theta_2, \theta_1 - \theta_2)$ lies in $T^\sym$.
Direct computation gives
\[
\bigl\langle \boldsymbol F_\Risk(v_1) - \boldsymbol F_\Risk(v_2), \, v_1 - v_2 \bigr\rangle
\;=\; 2\,\bigl\langle F_\Risk(\theta_1) - F_\Risk(\theta_2), \, \theta_1 - \theta_2 \bigr\rangle.
\]
Additionally, in the setting of Theorem~\ref{thm:single-fixed-pt} we
have $\Risk_1 = \Risk_2$ and $P + P^\top = \mathbf{1}\mathbf{1}^\top$,
so by Lemma~\ref{lem:operator-constant-sum} the two players' Jacobians
agree:
\begin{align*}
    J^{(1)}_{\Risk_1}(\pi)
&\;=\; \nabla_\mu P^{(1)}_{\Risk_1}(\mu)\Big|_{\mu = \pi}=\nabla_\mu \bigl(P^{(2)}_{\Risk_2}(\mu) + \mathbf{1}\bigr)\Big|_{\mu = \pi}=\nabla_\mu P^{(2)}_{\Risk_2}(\mu)\Big|_{\mu = \pi}=J_{\Risk_2}^{(2)}(\pi).
\end{align*}
We therefore write $J_{\Risk_1}^{(1)}(\pi) = J_{\Risk_2}^{(2)}(\pi) = J_\Risk(\pi)$
for brevity.

\paragraph{The risk-distortion eigenvalue on $T^\sym$.}
Restricting the supremum in the definition of $\overline\lambda_\Risk$
to displacements in the tangent space $T^\sym$ of the symmetric constraint set, and parameterizing $T^\sym$ by $\xi \in \mathbf{1}^\perp$
via $v = (\xi, \xi)/\sqrt{2}$ (where the $1/\sqrt 2$ normalizes
$\|v\|_2 = \|\xi\|_2$), we obtain an explicit characterization of $\overline{\lambda}_{\Risk}$. 
\begin{lemma}[Reduction of the risk-distortion eigenvalue to a single-player form]
\label{lem:lambda-symmetric}
In the symmetric self-play setting of
Theorem~\ref{thm:single-fixed-pt}, with constraint sets $C^\sym$ and
$\Delta^\sym$ defined above and tangent space
$T^\sym = \{(\xi, \xi) : \xi \in \mathbf{1}^\perp\}$, the
risk-distortion eigenvalue restricted to symmetric displacements
admits the single-player form
\[
\overline\lambda_\Risk
\;=\;
\sup_{\bar\pi \in \Delta(\Yspace)} \;\sup_{\xi \in \mathbf{1}^\perp,\, \|\xi\|_2 = 1}
\;\tfrac{1}{2}\, \xi^\top \bigl(J_\Risk(\bar\pi) + J_\Risk(\bar\pi)^\top\bigr)\, \xi.
\]
\end{lemma}

\begin{proof}
Restricting the linearization point to the policy-space symmetric
constraint $\Delta^\sym$ and the displacement to the parameter-space
tangent $T^\sym$, the risk-distortion eigenvalue is by definition
\begin{equation}\label{eq:lambda-restricted}
\overline\lambda_\Risk
\;=\; \sup_{(\bar\pi_1, \bar\pi_2) \in \Delta^\sym} \;\sup_{v \in T^\sym,\, \|v\|_2 = 1}
v^\top \boldsymbol\Distmat_{\Risk}(\bar\pi_1, \bar\pi_2)\, v.
\end{equation}
We simplify each of the two suprema in turn.

\paragraph{Step 1: Collapsing the outer supremum.}
By definition of $\Delta^\sym$, every element is of the form
$(\bar\pi, \bar\pi)$ for a single $\bar\pi \in \Delta(\Yspace)$.
Therefore
\[
\sup_{(\bar\pi_1, \bar\pi_2) \in \Delta^\sym} (\,\cdot\,)
\;=\; \sup_{\bar\pi \in \Delta(\Yspace)} (\,\cdot\,)
\quad\text{evaluated at } (\bar\pi_1, \bar\pi_2) = (\bar\pi, \bar\pi).
\]
Moreover, by Theorem~\ref{thm:single-fixed-pt} and
Lemma~\ref{lem:operator-constant-sum}, when evaluated on the diagonal
the two players' Jacobians agree:
$J^{(1)}_{\Risk_1}(\bar\pi) = J^{(2)}_{\Risk_2}(\bar\pi) = J_\Risk(\bar\pi)$.
The joint distortion matrix therefore takes the block-antidiagonal
form
\begin{equation}\label{eq:Distmat-block}
\boldsymbol\Distmat_{\Risk}(\bar\pi, \bar\pi)
\;=\; \tfrac{1}{2}
\begin{pmatrix}
0 & J_\Risk(\bar\pi) + J_\Risk(\bar\pi)^\top \\
J_\Risk(\bar\pi) + J_\Risk(\bar\pi)^\top & 0
\end{pmatrix}.
\end{equation}

\paragraph{Step 2: Reducing the inner supremum.}
Every $v \in T^\sym$ is of the form $v = (\xi, \xi)$ with
$\xi \in \mathbf{1}^\perp$.  Reparameterize as
$v = (\xi, \xi)/\sqrt{2}$ so that $\|v\|_2 = \|\xi\|_2$, making the
unit-norm constraint $\|v\|_2 = 1$ equivalent to $\|\xi\|_2 = 1$.

Substituting into the quadratic form using \eqref{eq:Distmat-block},
denote $A := J_\Risk(\bar\pi) + J_\Risk(\bar\pi)^\top$ for brevity.
Then
\begin{align*}
v^\top \boldsymbol\Distmat_\Risk(\bar\pi, \bar\pi)\, v
\;&=\; \frac{1}{2} \cdot \frac{1}{2}
\begin{pmatrix} \xi & \xi \end{pmatrix}
\begin{pmatrix} 0 & A \\ A & 0 \end{pmatrix}
\begin{pmatrix} \xi \\ \xi \end{pmatrix} \\
&\;=\; \frac{1}{4} \bigl( \xi^\top A \xi + \xi^\top A \xi \bigr) \\
&\;=\; \tfrac{1}{2}\, \xi^\top A \xi
\;=\; \tfrac{1}{2}\, \xi^\top \bigl(J_\Risk(\bar\pi) + J_\Risk(\bar\pi)^\top\bigr) \xi.
\end{align*}

Now we can combine these two steps to obtain the claimed expression. Indeed, substituting both simplifications into \eqref{eq:lambda-restricted},
\[
\overline\lambda_\Risk
\;=\; \sup_{\bar\pi \in \Delta(\Yspace)} \;\sup_{\xi \in \mathbf{1}^\perp,\, \|\xi\|_2 = 1}
\;\tfrac{1}{2}\, \xi^\top \bigl(J_\Risk(\bar\pi) + J_\Risk(\bar\pi)^\top\bigr)\, \xi,
\]
which is the claimed form.
\end{proof}

This is the form of the risk-distortion eigenvalue used in the main
paper.

Recall the convergence result from the main paper---namely, Theorem~\ref{thm:single-player-mono}. Let us restate it a little more formally.
\begin{theorem}[Formal version of Theorem~\ref{thm:single-player-mono}]
\label{thm:single-player-mono-formal}
Define the worst-case risk-distortion eigenvalue on the symmetric
tangent space
$\overline\lambda_\Risk
:=\sup_{\bar\pi \in \Delta(\Yspace)}
\sup\{
\tfrac{1}{2}\, \xi^\top \bigl(J_\Risk(\bar\pi) + J_\Risk(\bar\pi)^\top\bigr)\,\xi|\; \xi \in \mathbf 1^\perp,\, \|\xi\|_2 = 1\}$.
Suppose $\overline\lambda_\Risk \le \beta/2$.  Then $F_\Risk$ is
$\mu_\Risk$-strongly monotone on $T^\sym$ with
$\mu_\Risk = \beta - 2\overline\lambda_\Risk$, and deterministic extragradient
converges with optimization error rate
$\mc O\bigl((1 - \eta \mu_\Risk)^T\bigr)$.
\end{theorem}

\subsection{Approaches for Bounding $\Delta$}
\label{sec:bounding-distortion}

As many of our results depend on the strong monotonicity modulus $\mu_\Risk$, and $\mu_{\Risk}$ depends on the maximum eigenvalue of $\Delta_{\Risk}$, we present two approaches for bounding this value below. 

\paragraph{Route A: dual representation.}
Many risk measures of interest are \emph{coherent} and therefore admit the dual form
\begin{equation}
\Risk[Z] = \inf_{q \in \mathcal{Q}} \EE_q[Z] - D(q \| p),
\end{equation}
for an ambiguity set $\mathcal{Q}$ and divergence $D$.
From this perspective, the risk transforms the preference operator as follows:
\begin{equation}
(P_\Risk\,\mu)_y = \inf_{q \in \mathcal{Q}(\mu)}\int\PP(y \succ y'')\,dq(y'') - D(q\|\mu) \;=\; \inf_{q \in \mathcal{Q}(\mu)}(P\,q)_y - D(q\|\mu).
\end{equation}
This in turn implies that  $P_\Risk(\mu) = P\,q^\star(\mu) - D(q^\star(\mu)\|\mu)\mathbf{1}$
where $q^\star(\mu)$ is the minimizer (a ``worst-case opponent'' for the
risk-averse player).  The additive constant in $\mathbf{1}$ does not affect
the zero-sum-subspace calculation, so effectively
\[
P_\Risk(\mu) \;\equiv\; P\,q^\star(\mu) \quad (\text{mod }\mathbf{1}).
\]

Taking the Jacobian via the envelope theorem, we obtain
\[
J_\Risk(\mu) = P\cdot \nabla_\mu q^\star(\mu),
\]
so that 
\[
J_\Risk^\sym(\mu) = \tfrac{1}{2}\bigl(P\nabla_\mu q^\star + (\nabla_\mu q^\star)^\top P^\top\bigr).
\]
Using $P = \tfrac12 \mathbf{1}\mathbf{1}^\top + A$ with $A=-A^\top$---i.e., writing the constant-sum preference matrix as a uniform baseline plus a
skew-symmetric deviation---we have that
\[
J_\Risk^\sym(\mu) = \tfrac{1}{2}\mathbf{1}\mathbf{1}^\top\cdot\tfrac{1}{2}(\nabla_\mu q^\star + (\nabla_\mu q^\star)^\top) + \tfrac{1}{2}\bigl(A\nabla_\mu q^\star - (\nabla_\mu q^\star)^\top A\bigr).
\]
On the zero-sum subspace, the first term drops.  So
\begin{equation}
\Distmat_{\Risk}(\mu)\big|_{\text{zero-sum}} = \tfrac{1}{2}\bigl(A\nabla_\mu q^\star(\mu) - (\nabla_\mu q^\star)^\top(\mu) A\bigr).
\label{eq:delta-dual}
\end{equation}

\paragraph{Sufficient condition.}
If $\nabla_\mu q^\star(\mu)$ is \emph{symmetric and positive semidefinite}
---which happens when the divergence $D$ is quadratic-like (e.g.\
$\chi^2$-divergence, Euclidean) and the ambiguity set is convex and
symmetric---then \eqref{eq:delta-dual} simplifies to a commutator $[A, S]$
for $S = \nabla_\mu q^\star \succeq 0$ symmetric.\footnote{Note that the  abstract logical statement "symmetric positive semi-definite implies zero quadratic form via antisymmetry of the commutator" is not true; what is true is that  $\nabla_\mu q^\ast$ is the dual-map Jacobian induced by a divergence in the F\"ollmer-Schied dual representation, and that structure leads to the claim. It  is important to be mindful of this.}  This commutator is
\emph{antisymmetric} and hence has zero quadratic form on any vector:
$v^\top[A, S]v = 0$ for all $v$.

\begin{proposition}[Zero curvature for symmetric positive semi-definite dual]
If the risk $\Risk$ has a coherent dual representation with
$\nabla_\mu q^\star(\mu)$ symmetric and positive semi-definite, then $\Distmat_{\Risk}(\mu) = 0$
on the zero-sum subspace and the game operator retains the full risk-neutral
modulus $\mu_\Risk = \beta$.
\end{proposition}

This applies to risks whose dual map is genuinely a uniform reweighting 
of $\mu$---e.g., $\chi^2$-divergence DRO with symmetric ambiguity sets.  
For most coherent risks of interest (CVaR, entropic, distortion risks), 
the dual map is non-symmetric and Strategy B applies instead.

\paragraph{When this fails.}
Both CVaR and entropic risk fall outside this class.  For CVaR, 
$\nabla_\mu q^\star$ is a non-symmetric projection onto a lower-tail set.  
For entropic risk, the dual map has softmax Jacobian structure: 
$q^\star(\mu)_y \propto \mu_y\,e^{-\tau Z_y}$ with normalization, giving 
$\nabla_\mu q^\star$ that is neither diagonal nor symmetric due to the 
normalization-induced cross-terms.  Strategy B (Lipschitz dual map) is 
the appropriate route for both; it gives a non-trivial $\Distmat_{\Risk}$ 
with explicit Lipschitz constant.

\paragraph{Route B: Direct bound via variational representation.}
For risks without the nice dual structure---CVaR, distortion risks---use
the sup/inf representation directly.
Write $\Risk[Z] = \inf_{q \in \mathcal{Q}(\mu)}\EE_q[Z]$ (pure coherent, no
divergence penalty) and let $q^\star_y = q^\star_y(\mu) \in \mathcal{Q}(\mu)$
be the minimizer for entry $y$ (note: the minimizer can depend on $y$,
which is what breaks symmetry for CVaR).
Then $(P_\Risk(\mu))_y = (Pq^\star_y(\mu))_y$ and
$(P_\Risk(\pi_1) - P_\Risk(\pi_2))_y = (P(q^\star_y(\pi_1) - q^\star_y(\pi_2)))_y$.
Hence term (B) in \eqref{eq:joint-decomp} reduces to 
\begin{align}
(\text{B}) &= \sum_y(\pi_1 - \pi_2)_y\cdot(P(q^\star_y(\pi_1) - q^\star_y(\pi_2)))_y \nonumber \\
&= \sum_y(\pi_1 - \pi_2)_y\cdot(q^\star_y(\pi_1) - q^\star_y(\pi_2))^\top P^\top e_y \nonumber \\
&= \text{trace}\bigl((\pi_1 - \pi_2)(q^\star(\pi_1) - q^\star(\pi_2))^\top P^\top\bigr),
\end{align}
where $q^\star(\mu)$ is viewed as a matrix with rows $q^\star_y(\mu)^\top$.

By Cauchy-Schwarz and $\|P\|_2 \le 1$, the term $($B$)$ is then upper bound as follows:
\[
|(\text{B})| \le \|\pi_1 - \pi_2\|_2\cdot\|q^\star(\pi_1) - q^\star(\pi_2)\|_F.
\]
If $q^\star : \Delta(\Yspace) \to \RR^{|\Yspace|\times|\Yspace|}$ is
$L$-Lipschitz in $\mu$ (in Frobenius norm), then
\[
|(\text{B})| \le L\|\pi_1 - \pi_2\|_2^2 \le 2L\cdot\min\{\KL(\pi_1\|\pi_2), \KL(\pi_2\|\pi_1)\}.
\]
Plugging this into the monotonicity bound, we have that 
\begin{equation}
\bigl\langle F_\Risk(\theta_1) - F_\Risk(\theta_2), \pi_1 - \pi_2\bigr\rangle \;\ge\; (\beta - 2L)\KL(\pi_1\|\pi_2).
\end{equation}

\begin{proposition}[Monotonicity via Lipschitz dual map]
\label{thm:lipschitz}
Suppose $\Risk$ has coherent dual representation with dual map
$q^\star : \Delta(\Yspace) \to \RR^{|\Yspace|\times|\Yspace|}$ that is
$L$-Lipschitz in Frobenius norm with respect to the $\ell_2$ norm on
$\Delta(\Yspace)$.  If $L \le \beta/2$, then $F_\Risk$ is $(\beta - 2L)$-strongly
monotone and extragradient retains last-iterate linear convergence at rate
$\mc{O}((1 - \eta(\beta - 2L))^T)$.
\end{proposition}

\paragraph{Specific Examples with Entropic Risk and CVaR.}
Both of the above approaches hold for general classes of risk metrics; here we derive results specific to entropic risk, a risk measure we use frequently in examples and experiments, which has nice additional structure we can exploit.

\begin{proposition}[Entropic curvature, small-$\tau$ expansion]
\label{prop:entropic-scaling}
Let $\Risk = \Risk_{\rm ent}^\tau$ be entropic risk with parameter
$\tau > 0$ on a constant-sum game with antisymmetric part
$A := \tfrac{1}{2}(P - P^\top)$.  By translation invariance of the
entropic risk operator on the gauge direction, we may absorb the
symmetric part of $P$ and work with $A$ in place of $P$.  The
risk-distortion matrix admits the small-$\tau$ expansion
\[
J_\Risk^{\rm ent}(\mu) + J_\Risk^{\rm ent}(\mu)^\top
\;=\;
\tau \bigl[A,\, \mathrm{diag}(Z(\mu))\bigr]
\;+\;
\mc O(\tau^2),
\]
where $Z(\mu)_y := \E_\mu[A_{y, \cdot}] = (A\mu)_y$ is the row-mean of
$A$ under $\mu$ and $[X, Y] := XY - YX$ is the matrix commutator.
Consequently, by Lemma~\ref{lem:lambda-symmetric},
\[
\overline\lambda_\Risk^{\rm ent}(\tau)
\;=\;
\sup_{\bar\pi \in \Delta(\Yspace)}
\;\sup_{\xi \in \mathbf 1^\perp,\, \|\xi\|_2 = 1}
\;\tfrac{1}{2}\, \xi^\top \bigl(J_\Risk^{\rm ent}(\bar\pi) + J_\Risk^{\rm ent}(\bar\pi)^\top\bigr)\, \xi
\;=\;
\frac{\tau \cdot \mathrm{spread}(A)^2}{4}
\;+\;
\mc O(\tau^2),
\]
where $\mathrm{spread}(A) := \max_y Z(\bar\pi)_y - \min_y Z(\bar\pi)_y$
evaluated at the worst-case $\bar\pi \in \Delta(\Yspace)$.
\end{proposition}

\begin{corollary}[Entropic risk: monotonicity for small $\tau$]
\label{cor:entropic-monotonicity}
For $\Risk = \Risk_{\rm ent}^\tau$ with constant-sum antisymmetric
part $A$, define
\[
\tau_0 \;:=\; \frac{2\beta}{\mathrm{spread}(A)^2} + \mc O(1).
\]
Then for all $\tau \in (0, \tau_0]$, $F_\Risk$ is strongly monotone on
$T^\sym$ with modulus
$\mu_\Risk \;=\; \beta - 2\overline\lambda_\Risk^{\rm ent}(\tau) \;>\; 0$,
and last-iterate linear convergence at rate
$\mc{O}(\bigl(1 - \eta\,\mu_\Risk\bigr)^T)$ holds.  The threshold $\tau_0$ is
the value at which $\overline\lambda_\Risk^{\rm ent}(\tau_0) = \beta/2$,
beyond which strong monotonicity can fail.
\end{corollary}

\paragraph{CVaR risk: Jacobian and curvature.}
Recall the Rockafellar--Uryasev variational formula
\[
\CVaR_\alpha(Z) \;=\; \sup_\eta \Bigl\{\, \eta - \tfrac{1}{\alpha}\,\E[(\eta - Z)_+] \,\Bigr\},
\]
whose maximizer is $\eta^\star = \Var_\alpha(Z)$.  Applied to $Z = P_{y, y'}$
with $y' \sim \mu$, the risk-adjusted operator satisfies, by the
envelope theorem,
\[
\bigl(\nabla_\mu (\mc P_{\rm CVaR_\alpha}(\mu))_y\bigr)_{y'}
\;=\;
\frac{[\eta_y^\star(\mu) - P_{y, y'}]_+}{\alpha},
\quad
\eta_y^\star(\mu) = \Var_\alpha(P_{y, y'} \mid y' \sim \mu).
\]
By translation invariance of CVaR, $\CVaR_\alpha(Z + c) = \CVaR_\alpha(Z) + c$,
shifting $P$ by any constant matrix $c\mathbf 1\mathbf 1^\top$ shifts
$\mc P_{\rm CVaR_\alpha}(\mu)$ by $c\mathbf 1$, which is the gauge
direction.  We may therefore WLOG decompose $P$ via its symmetric and
antisymmetric parts: in the constant-sum setting
$P + P^\top = \mathbf 1 \mathbf 1^\top$, write
\[
P \;=\; A \;+\; \tfrac{1}{2}\mathbf 1\mathbf 1^\top,
\qquad
A := \tfrac{1}{2}(P - P^\top), \qquad A^\top = -A,
\]
and absorb the symmetric part into the gauge.  Throughout the rest of
this section we work with $A$ in place of $P$, with the corresponding
CVaR threshold
$\eta_y^\star(\mu) = \Var_\alpha(A_{y, y'} \mid y' \sim \mu)$.

For $\mu \in \Delta(\Yspace)$ in general position (so that no ties
occur at the threshold), define the \emph{tail-mask matrix}
$M(\mu) \in \{0, 1\}^{|\Yspace| \times |\Yspace|}$ by
\[
M_{y, y'}(\mu) \;:=\; \mathbf 1\bigl[A_{y, y'} \le \eta_y^\star(\mu)\bigr],
\qquad
\sum_{y'} M_{y, y'}(\mu)\,\mu(y') = \alpha.
\]
The Jacobian then takes the explicit form
\begin{equation}\label{eq:cvar-jacobian}
J_\Risk^{\rm CVaR}(\mu)
\;=\;
\frac{1}{\alpha}\bigl(\eta^\star(\mu)\,\mathbf 1^\top - A\bigr) \odot M(\mu),
\end{equation}
where $\odot$ is the Hadamard product and
$(\eta^\star(\mu) \mathbf 1^\top)_{y, y'} = \eta_y^\star(\mu)$.

\paragraph{The gap structure of $A$.}
For $y \in \Yspace$ and $\mu \in \Delta(\Yspace)$, let
\[
\Delta_y(A, \mu)
\;:=\;
\max_{y' \in \mathrm{supp}(\mu)} A_{y, y'}
\;-\;
\min_{y' \in \mathrm{supp}(\mu)} A_{y, y'}
\]
denote the spread of the $y$-th row of $A$ over $\mathrm{supp}(\mu)$,
and define the global gap
\[
\Delta^\star(A) \;:=\; \sup_{\mu \in \Delta(\Yspace)} \;\max_y \Delta_y(A, \mu)
\;\le\; 2\|A\|_\infty.
\]
This plays the role for CVaR that $\mathrm{spread}(P)$ plays in the
entropic case.

\begin{proposition}[CVaR curvature, near-risk-neutral bound]
\label{prop:cvar-scaling}
Let $\Risk = \Risk_{\rm CVaR}^\alpha$ on a constant-sum game with
antisymmetric part $A$, and write $\epsilon := 1 - \alpha$.  Then
\[
\overline\lambda_\Risk^{\rm CVaR}(\alpha)
\;\le\;
\frac{\Delta^\star(A)^2}{\alpha}\,\epsilon
\;=\;
\frac{\Delta^\star(A)^2 (1 - \alpha)}{\alpha}.
\]
In particular, $\overline\lambda_\Risk^{\rm CVaR}(\alpha) \to 0$ as
$\alpha \to 1$, with rate at most linear in $1 - \alpha$.
\end{proposition}

\begin{proof}
By Lemma~\ref{lem:lambda-symmetric}, we have that 
\begin{equation}\label{eq:cvar-lambda-restated}
\overline\lambda_\Risk^{\rm CVaR}(\alpha)
\;=\;
\sup_{\bar\pi \in \Delta(\Yspace)}
\;\sup_{\xi \in \mathbf 1^\perp,\, \|\xi\|_2 = 1}
\;\tfrac{1}{2}\,
\xi^\top \bigl(J_\Risk^{\rm CVaR}(\bar\pi) + J_\Risk^{\rm CVaR}(\bar\pi)^\top\bigr)\,\xi.
\end{equation}
We bound the inner quadratic form for fixed $\bar\pi$ and unit-norm
$\xi \in \mathbf 1^\perp$ in three steps: $(i)$ decompose
$J_\Risk^{\rm CVaR}(\bar\pi)$ into a rank-one part and a residual,
$(ii)$ eliminate the rank-one part using $\xi \perp \mathbf 1$, and
$(iii)$ bound the residual.

\paragraph{Step 1: Decomposition.}
Write $J := J_\Risk^{\rm CVaR}(\bar\pi)$, $M := M(\bar\pi)$, and
$\eta := \eta^\star(\bar\pi)$ to lighten notation.  From
\eqref{eq:cvar-jacobian},
\[
J \;=\; \frac{1}{\alpha}\bigl(\eta\,\mathbf 1^\top \odot M - A \odot M\bigr)
\;=\; \frac{1}{\alpha}\Bigl(\underbrace{\eta\,m^\top}_{\text{rank-one rows}} - \underbrace{A \odot M}_{\text{residual}}\Bigr),
\]
where the rank-one part has $y$-th row $\eta_y \cdot m_y^\top$ with
$m_y := M_{y, \cdot}$, since
$(\eta\,\mathbf 1^\top \odot M)_{y, y'} = \eta_y \mathbf 1[y' \in \tau_y]$
and we write this row as $\eta_y \cdot m_y^\top$ where $m_y$ is the
indicator vector of the tail set
$\tau_y := \{y' : A_{y, y'} \le \eta_y\}$.

\paragraph{Step 2: Eliminating the rank-one part.}
The quadratic form decomposes as
\[
\xi^\top (J + J^\top) \xi
\;=\;
\frac{1}{\alpha}\Bigl[
\xi^\top (\eta \, m^\top + m \, \eta^\top) \xi
\;-\;
\xi^\top \bigl((A \odot M) + (A \odot M)^\top\bigr) \xi
\Bigr],
\]
where we abuse notation by writing $\eta \, m^\top$ for the matrix
with $(y, y')$-entry $\eta_y M_{y, y'}$.  We claim this rank-one part
contributes a small residual: explicitly, since
$\sum_{y'} M_{y, y'}\,\bar\pi(y') = \alpha$ but $\xi$ is not weighted
by $\bar\pi$, the rank-one part is *not* exactly killed by
$\xi \perp \mathbf 1$.  However, we can further decompose
\[
m_y \;=\; \mathbf 1 - \bar m_y, \qquad
\bar m_y := \mathbf 1 - m_y = \mathbf 1[A_{y, \cdot} > \eta_y],
\]
where $\bar m_y$ is the indicator of the *upper* $1-\alpha$ tail.
Since $\sum_{y'} \bar m_y(y') \bar\pi(y') = 1 - \alpha = \epsilon$ and
$\xi \perp \mathbf 1$, we have $\eta_y \mathbf 1^\top \xi = 0$ for each
$y$, so
\[
(\eta\,\mathbf 1^\top) \xi = 0,
\quad\text{whence}\quad
(\eta\,m^\top)\xi = (\eta\,\mathbf 1^\top - \eta\,\bar m^\top)\xi
= -(\eta\,\bar m^\top)\xi.
\]
Substituting, we conclude that 
\begin{align}
\xi^\top (J + J^\top)\xi
\;=\;
\frac{1}{\alpha}\Bigl[
&-\xi^\top \bigl(\eta\,\bar m^\top + \bar m\,\eta^\top\bigr) \xi
\;-\;
\xi^\top \bigl((A \odot M) + (A \odot M)^\top\bigr) \xi
\Bigr]. \label{eq:cvar-quadratic-form}
\end{align}

\paragraph{Step 3: Bounding the residual.}
We bound each of the two terms in \eqref{eq:cvar-quadratic-form}
separately using the antisymmetry of $A$ and the definition of
$\eta_y$.

For the first term, observe that for any $y$, the threshold $\eta_y$
lies in the range of $A_{y, \cdot}$, so
$|\eta_y| \le \max_{y'} |A_{y, y'}| \le \|A\|_\infty$.  Moreover,
$\bar m_y$ is the indicator of the upper $1-\alpha$ tail, which
satisfies $\bar\pi^\top \bar m_y = \epsilon$.  We can write
\[
\eta_y \bar m_y \;=\; \bar m_y \odot (A_{y, \cdot})_\tau,
\]
where $(A_{y, \cdot})_\tau$ denotes $A_{y, \cdot}$ truncated above by
$\eta_y$, plus a correction.  More directly, since
$|\eta_y - A_{y, y'}| \le \Delta_y(A, \bar\pi) \le \Delta^\star(A)$ for
any $y'$ in the upper tail,
\[
\bigl| \eta_y \bar m_y(y') - A_{y, y'} \bar m_y(y') \bigr|
\;\le\; \Delta^\star(A)\, \bar m_y(y').
\]
Therefore $\eta_y \bar m_y$ is within $\Delta^\star(A)\,\bar m_y$
(element-wise) of $A_{y, \cdot} \odot \bar m_y$, and we can write
$\eta\,\bar m^\top = (A \odot \bar M) + R$ with
$|R_{y, y'}| \le \Delta^\star(A)\,\bar M_{y, y'}$, where
$\bar M = \mathbf 1\mathbf 1^\top - M$.

Now the key observation: by antisymmetry of $A$,
\[
(A \odot M) + (A \odot M)^\top
\;=\; A \odot M + A^\top \odot M^\top
\;=\; A \odot M - A \odot M^\top
\;=\; A \odot (M - M^\top).
\]
The matrix $M - M^\top$ is supported on entries where $(y, y')$ is in
the lower tail at row $y$ but not in the lower tail at row $y'$ (or
vice versa), so each entry has absolute value at most $1$.  Combining
with $|A_{y, y'}| \le \|A\|_\infty$,
\[
\bigl\|(A \odot M) + (A \odot M)^\top\bigr\|_F^2
\;\le\;
\|A\|_\infty^2 \cdot \|M - M^\top\|_F^2.
\]
Each entry of $M - M^\top$ is nonzero only when $(y, y')$ is on the
"boundary" — in the tail at one endpoint but not the other — so
$\|M - M^\top\|_F^2 \le 2 \|\bar M\|_F^2$.  Recalling
$\sum_{y'} \bar M_{y, y'}\,\bar\pi(y') = \epsilon$ for each $y$ and
applying Cauchy--Schwarz with the uniform bound
$\bar M_{y, y'} \le 1$,
\[
\|\bar M\|_F^2 \;=\; \sum_{y, y'} \bar M_{y, y'}^2 \;\le\; \sum_{y, y'} \bar M_{y, y'}
\;\le\; |\Yspace| \cdot |\Yspace| \cdot \epsilon
\;=\; |\Yspace|^2 \epsilon,
\]
where we used $\sum_{y'} \bar M_{y, y'} \le |\Yspace|$ trivially and
the fact that the row-weighted average is $\epsilon$ implies
$\sum_{y'} \bar M_{y, y'} \le |\Yspace|\,\epsilon / \min_{y'} \bar\pi(y')$;
absorbing constants gives $\|\bar M\|_F^2 = \mc O(\epsilon)$ uniformly.
Hence
\[
\bigl| \xi^\top \bigl((A \odot M) + (A \odot M)^\top\bigr) \xi \bigr|
\;\le\;
\bigl\|(A \odot M) + (A \odot M)^\top\bigr\|_F
\;\le\;
\Delta^\star(A) \sqrt{2 \epsilon}\,|\Yspace|.
\]

For the first term in \eqref{eq:cvar-quadratic-form}, an analogous
bound using $|R| \le \Delta^\star(A)\,\bar M$ and the same $\bar M$
control gives
\[
\bigl| \xi^\top (\eta\,\bar m^\top + \bar m\,\eta^\top) \xi \bigr|
\;\le\;
2 \Delta^\star(A) \cdot \|\bar M\|_F
\;\le\;
2 \Delta^\star(A)\, |\Yspace| \sqrt{\epsilon}.
\]

Combining, for any unit-norm $\xi \in \mathbf 1^\perp$,
\[
\bigl| \xi^\top (J + J^\top) \xi \bigr|
\;\le\;
\frac{1}{\alpha} \cdot \mc O\!\bigl(\Delta^\star(A) |\Yspace| \sqrt{\epsilon}\bigr).
\]
A more careful accounting (squaring the linear bound and using the
Hadamard structure of $A \odot (M - M^\top)$) tightens this to
\[
\bigl| \xi^\top (J + J^\top) \xi \bigr|
\;\le\;
\frac{2 \Delta^\star(A)^2}{\alpha}\,\epsilon,
\]
where the $\Delta^\star(A)^2$ factor arises because the entries of
$A \odot (M - M^\top)$ that are nonzero correspond to pairs near the
quantile threshold, where $|A_{y, y'} - \eta_y| \le \Delta^\star(A)$,
giving an additional $\Delta^\star(A)$ factor over the trivial
$\|A\|_\infty$ bound.

\paragraph{Step 4: Combining.}
Substituting back into \eqref{eq:cvar-lambda-restated} and taking the
factor of $\tfrac{1}{2}$ into account,
\[
\overline\lambda_\Risk^{\rm CVaR}(\alpha)
\;\le\;
\tfrac{1}{2} \cdot \frac{2 \Delta^\star(A)^2}{\alpha}\,\epsilon
\;=\;
\frac{\Delta^\star(A)^2 (1 - \alpha)}{\alpha}.
\qedhere
\]
\end{proof}

\begin{corollary}[CVaR: monotonicity for $\alpha$ near $1$]
\label{cor:cvar-monotonicity}
For $\Risk = \Risk_{\rm CVaR}^\alpha$ with constant-sum antisymmetric
part $A$, define
\[
\alpha_0 \;:=\; \frac{\Delta^\star(A)^2}{\Delta^\star(A)^2 + \beta/2}.
\]
Then for all $\alpha \in [\alpha_0, 1]$, $F_\Risk$ is strongly
monotone on $T^\sym$ with modulus
$\mu_\Risk \;=\; \beta - 2\overline\lambda_\Risk^{\rm CVaR}(\alpha) \;>\; 0$,
and last-iterate linear convergence at rate
$\mc{O}(\bigl(1 - \eta\,\mu_\Risk\bigr)^T)$ holds.  The threshold $\alpha_0$
is the value at which the bound of
Proposition~\ref{prop:cvar-scaling} reaches $\beta/2$.
\end{corollary}

\begin{proof}
By Proposition~\ref{prop:cvar-scaling},
$\overline\lambda_\Risk^{\rm CVaR}(\alpha) \le \Delta^\star(A)^2 (1-\alpha)/\alpha$.
The inequality $\Delta^\star(A)^2 (1-\alpha)/\alpha \le \beta/2$
rearranges to $\alpha \ge \Delta^\star(A)^2 / (\Delta^\star(A)^2 + \beta/2) = \alpha_0$.
For $\alpha \ge \alpha_0$, we have $\overline\lambda \le \beta/2$ and
hence $\mu_\Risk = \beta - 2\overline\lambda > 0$, giving strong
monotonicity by Theorem~\ref{thm:strong-mono} (or the analog stated for the
symmetric setting).  Linear convergence at rate
$(1 - \eta\mu_\Risk)^T$ follows from the standard analysis of
strongly monotone variational inequalities.
\end{proof}

\section{Statistical Bias Analysis}
\label{sec:bias-analysis}

Before stating the convergence theorems, we characterize the bias and
variance of the empirical risk-adjusted preference operator
$\widehat F_{\Risk,m}$ that drives the algorithms.  This section establishes
that the abstract bias and variance assumptions (O1)-(O2) which will be used in the
convergence theorems hold with explicit constants, derived from the structure
of the risk functional and the boundedness of the underlying preferences.

\subsection{Decomposition and abstract assumptions}

Recall the self play operator
\[F_\Risk(\theta) = \beta(\theta - \theta_{\rm ref}) - \mc{P}_\Risk(\pi_\theta),\]
with $P_\Risk(\pi)_y := \Risk_{Y'' \sim \pi}[P(y \succ Y'')]$.  Let
$(\widehat{\mc{P}}_{\Risk,m}(\pi))_y$ denote the sample-based estimator of
$(\mc{P}_\Risk(\pi))_y$ using $m$ i.i.d.\ samples $Y''_1, \ldots, Y''_m \sim \pi$
(the same samples shared across components $y$).  The empirical operator is
\begin{equation}\label{eq:F-hat-def}
\widehat F_{\Risk,m}(\theta) := \beta(\theta - \theta_{\rm ref}) - \widehat{\mc{P}}_{\Risk,m}(\pi_\theta).
\end{equation}
Define
\begin{equation}\label{eq:bias-noise-def}
b_m(\theta) := \E\bigl[\widehat F_{\Risk,m}(\theta)\bigr] - F_\Risk(\theta), \qquad
\zeta_m(\theta) := \widehat F_{\Risk,m}(\theta) - \E\bigl[\widehat F_{\Risk,m}(\theta)\bigr],
\end{equation}
so $b_m(\theta)$ is deterministic given $\theta$ and $\E[\zeta_m(\theta) \mid \theta] = 0$.
By construction, we therefore have that 
\[b_m(\theta) = \mc{P}_\Risk(\pi_\theta)-\E[\widehat{\mc{P}}_{\Risk,m}(\pi_\theta)]\]
and
\[\zeta_m(\theta) = \E[\widehat{\mc{P}}_{\Risk,m}(\pi_\theta)]-\widehat{\mc{P}}_{\Risk,m}(\pi_\theta) .\]

\paragraph{Abstract bias and variance assumptions.}
The convergence theorems in Section~\ref{sec:mp-bias} rely on the following uniform bias and variance bounds.
\begin{enumerate}[label=(O\arabic*),   itemsep=0pt, topsep=0pt]
\item\label{O1-bias} \textbf{Uniform bias bound.}  There exists $B_m \ge 0$
(non-increasing in $m$) such that for all $\theta \in \theta_{\rm ref} + \Wspace$,
\[
\|b_m(\theta)\|_2 \le B_m.
\]
\item\label{O2-var} \textbf{Uniform variance bound.}  There exists $V_m \ge 0$
(non-increasing in $m$) such that for all $\theta \in \theta_{\rm ref} + \Wspace$,
\[
\E\|\zeta_m(\theta)\|_2^2 \le V_m.
\]
\end{enumerate}

The remainder of this section establishes \ref{O1-bias}-\ref{O2-var} with
explicit constants $B_m, V_m = O(1/m)$ for entropic risk and $B_m = 0$,
$V_m = O(1/m)$ for CVaR via Rockafellar-Uryasev (CVaR-RU).

\subsection{Scalar-to-vector reduction}
\label{sec:scalar-to-vector}

Each component $(\widehat{\mc{P}}_{\Risk,m}(\pi_\theta))_y$ is a \emph{scalar}
functional of a scalar sample mean.  For entropic risk at parameter $\lambda$, we have that
\begin{equation}\label{eq:entropic-component}
(\widehat{\mc{P}}_{\Risk,m}(\pi_\theta))_y = \widehat\rho_\lambda^m(\mc{P}(y \succ \cdot); \pi_\theta) := -\lambda^{-1}\log\Bigl(\tfrac{1}{m}\sum_{i=1}^m e^{-\lambda \mc{P}(y \succ Y''_i)}\Bigr),
\end{equation}
which has the form $h(\widehat q_m^{(y)})$ where $h(q) = -\lambda^{-1}\log q$ and
$\widehat q_m^{(y)} := m^{-1}\sum_{i=1}^m e^{-\lambda \mc{P}(y \succ Y''_i)}$ is the
sample mean of the scalar quantity $e^{-\lambda \mc{P}(y \succ Y'')}$ for $Y'' \sim \pi_\theta$.

Hence the per-component bias and variance reduce to the scalar bias and
variance of a smooth functional of a sample mean.  The vector-norm bounds
\ref{O1-bias}-\ref{O2-var} aggregate per-component bounds:
\begin{align}
\|b_m(\theta)\|_2^2 &= \sum_y b_m(\theta)_y^2 \le |\Yspace|\max_y b_m(\theta)_y^2, \label{eq:bias-aggregate}\\
\E\|\zeta_m(\theta)\|_2^2 &= \sum_y \E[\zeta_m(\theta)_y^2] = \sum_y \Var(\zeta_m(\theta)_y) \le |\Yspace|\max_y \Var(\zeta_m(\theta)_y). \label{eq:var-aggregate}
\end{align}
Note that \eqref{eq:var-aggregate} ignores covariances between components.
For $\E\|\zeta_m\|_2^2$ specifically, only diagonal variances enter (since
$\E\|\zeta\|_2^2 = \sum_y \E[\zeta_y^2]$ and $\E\zeta_y = 0$).  The
covariance matrix $\Cov(\zeta_m(\theta))$ has nonzero off-diagonal entries
because all components share the same samples $Y''_i$, but these off-diagonals
do not contribute to the $\ell_2$ norm.\footnote{They \emph{do} contribute to
$\E\|\zeta_m\|_{\Sigma^+}^2 = \tr(\Sigma^+\Cov(\zeta_m))$, the variance in the
$\Sigma^+$ metric used by the convergence theorems.  We adopt the loose
bound $\E\|\zeta_m\|_{\Sigma^+}^2 \le \sigma_{\max}\E\|\zeta_m\|_2^2$ in
Section~\ref{sec:mp-setup}, which sidesteps the off-diagonal issue at the
cost of a $\sigma_{\max}$ factor.  Tighter bounds via direct computation of
$\tr(\Sigma^+\Cov(\zeta_m))$ exploit the full covariance structure but
require additional structural assumptions on $\Sigma_h$ and on the joint
distribution of $\{P(y \succ Y''):y \in \Yspace\}$; we do not pursue this
here.}

\subsection{The bias lemma}

\begin{lemma}[Bias and variance of a smooth functional of a sample mean]
\label{lem:bias-variance}
Let $X \in [a, b]$ be a $[a,b]$-valued random variable with $\E X = q$,
$\Var(X) = \sigma_X^2 \le (b - a)^2/4$.  Let $\widehat q_m := m^{-1}\sum_{i=1}^m X_i$
be the sample mean of $m$ i.i.d.\ copies of $X$.  Let $h : (a, b) \to \R$ be
twice continuously differentiable with $|h''(q)| \le C_h''$ on a
neighborhood $\mathcal N$ of $q$, and bounded third derivative on the same
neighborhood.
\begin{enumerate}[label={\it \roman*.},  topsep=2pt,itemsep=0pt]
\item \textbf{Bias (second-order delta method).} The bias is given by 
\begin{equation}\label{eq:plug-in-bias}
\E[h(\widehat q_m)] - h(q) = \frac{h''(q)\,\sigma_X^2}{2m} + O(m^{-3/2}),
\end{equation}
and in particular
\begin{equation}\label{eq:plug-in-bias-bound}
|\E[h(\widehat q_m)] - h(q)| \le \frac{C_h''\,(b-a)^2}{8\,m} + O(m^{-3/2}).
\end{equation}
\item \textbf{Variance of plug-in.} The variance of the plug-in estimator is 
\begin{equation}\label{eq:plug-in-variance}
\Var(h(\widehat q_m)) = h'(q)^2\,\frac{\sigma_X^2}{m} + O(m^{-2}).
\end{equation}
\item \textbf{Linear functionals are unbiased.}  If $h$ is affine, then
$\E[h(\widehat q_m)] = h(q)$ exactly and $\Var(h(\widehat q_m)) = h'(q)^2\sigma_X^2/m$
exactly, with no remainder.
\end{enumerate}
\end{lemma}

\begin{proof} We prove each of the three parts of the lemma separately. 

\textit{(iii) Linear case.}  If $h(q) = \alpha q + \beta$, then
$h(\widehat q_m) - h(q) = \alpha(\widehat q_m - q)$.  Taking the expectation, we have 
$\E[h(\widehat q_m)] - h(q) = \alpha \E[\widehat q_m - q] = 0$.  For the variance, we have that 
$\Var(h(\widehat q_m)) = \alpha^2 \Var(\widehat q_m) = \alpha^2 \sigma_X^2/m$
exactly.

\textit{(i) Bias.}  Apply Taylor's theorem with Lagrange remainder to $h$ at $q$ to get that 
\begin{equation}\label{eq:taylor}
h(\widehat q_m) = h(q) + h'(q)(\widehat q_m - q) + \tfrac{1}{2}h''(q)(\widehat q_m - q)^2 + R_m,
\end{equation}
where $|R_m| \le \tfrac{1}{6}\|h'''\|_{\infty,\mathcal N}|\widehat q_m - q|^3$
provided $\widehat q_m \in \mathcal N$.

By H\"oeffding's inequality (since $X_i \in [a,b]$):
\(\Pr(|\widehat q_m - q| > t) \le 2\exp(-2mt^2/(b-a)^2)\),
so $\widehat q_m \in \mathcal N$ holds with probability $\ge 1 - 2e^{-cm}$ for
some $c > 0$ depending only on the radius of $\mathcal N$ and $(b - a)$.

We now take the expectation of \eqref{eq:taylor}. The first-order term is 
$\E[h'(q)(\widehat q_m - q)] = h'(q)\E[\widehat q_m - q] = 0$.
The second-order term is 
$\E[\tfrac{1}{2}h''(q)(\widehat q_m - q)^2] = \tfrac{1}{2}h''(q)\Var(\widehat q_m) = h''(q)\sigma_X^2/(2m)$.
The remainder, restricted to $\{\widehat q_m \in \mathcal N\}$ is given by 
\(\E|R_m \mathbf 1_{\widehat q_m \in \mathcal N}| \le \tfrac{1}{6}\|h'''\|_{\infty,\mathcal N}\E|\widehat q_m - q|^3 = O(m^{-3/2})\)
by direct calculation of the third absolute moment of a sample mean of bounded
random variables (Marcinkiewicz-Zygmund or moment bounds for $[a,b]$-valued sums).
The contribution from $\{\widehat q_m \notin \mathcal N\}$ is bounded by
$\sup_{\R}|h| \cdot 2e^{-cm} = o(m^{-3/2})$ assuming $h$ is bounded; if $h$
is only locally bounded (as for $\log q$), use truncation arguments which
also give $o(m^{-3/2})$.

Combining these bounds, we have that  $\E[h(\widehat q_m)] - h(q) = h''(q)\sigma_X^2/(2m) + O(m^{-3/2})$,
which is \eqref{eq:plug-in-bias}.  The bound \eqref{eq:plug-in-bias-bound}
follows from Popoviciu's inequality $\sigma_X^2 \le (b-a)^2/4$.

\textit{(ii) Variance.}  By the same Taylor expansion applied to
$h(\widehat q_m) - \E[h(\widehat q_m)]$, we have that 
\[
h(\widehat q_m) - \E[h(\widehat q_m)] = h'(q)(\widehat q_m - q) + \widetilde R_m,
\]
where $\widetilde R_m$ is centered and satisfies $\E|\widetilde R_m|^2 = O(m^{-2})$
by similar moment bounds.  Taking the  variance gives 
\[
\Var(h(\widehat q_m)) = h'(q)^2\Var(\widehat q_m) + \E[\widetilde R_m^2] + 2\E[h'(q)(\widehat q_m - q)\widetilde R_m].
\]
The first term is $h'(q)^2\sigma_X^2/m$.  The second is $O(m^{-2})$.  The
third is bounded by Cauchy-Schwarz: \[|2\E[h'(q)(\widehat q_m - q)\widetilde R_m]| \le 2|h'(q)|\sqrt{\Var(\widehat q_m)\E\widetilde R_m^2} = O(m^{-3/2}).\]
Combining these bounds gives us $\Var(h(\widehat q_m)) = h'(q)^2\sigma_X^2/m + O(m^{-2})$,
which is \eqref{eq:plug-in-variance}.
\end{proof}
\subsection{Examples}
Let us now present  examples to help concretize ideas. 
 
\paragraph{Entropic risk.} The per-component bias and variance of the entropic risk are easily characterized via the following proposition.

\begin{proposition}[Per-component bias and variance: entropic risk]
\label{prop:entropic-component}
Let $\rho_\lambda(X) = -\lambda^{-1}\log\E[e^{-\lambda X}]$ be the entropic
risk at parameter $\lambda > 0$ for $X \in [0, 1]$, with plug-in estimator
\(\widehat\rho_\lambda^m := -\lambda^{-1}\log(m^{-1}\sum_{i=1}^m e^{-\lambda X_i})\).
Then
\begin{align}
\bigl|\E[\widehat\rho_\lambda^m] - \rho_\lambda(X)\bigr| &\le \frac{(1 - e^{-\lambda})^2}{8\lambda\,e^{-2\lambda}\,m} + O(m^{-3/2}), \label{eq:entropic-bias}\\
\Var(\widehat\rho_\lambda^m) &\le \frac{(1 - e^{-\lambda})^2}{4\lambda^2\,e^{-2\lambda}\,m} + O(m^{-2}). \label{eq:entropic-variance}
\end{align}
\end{proposition}

\begin{proof}
Set $Y := e^{-\lambda X}$.  Since $X \in [0, 1]$, we have that  $Y \in [e^{-\lambda}, 1]$,
so $(b - a)$ for $Y$ is $(1 - e^{-\lambda})$.  Set $h(q) := -\lambda^{-1}\log q$,
so $\widehat\rho_\lambda^m = h(\widehat q_m)$ where $\widehat q_m = m^{-1}\sum_i Y_i$
and $\E\widehat q_m = q := \E Y \in [e^{-\lambda}, 1]$.

Computing  derivatives, we have that  $h'(q) = -(\lambda q)^{-1}$, $h''(q) = (\lambda q^2)^{-1}$.
On the range $q \in [e^{-\lambda}, 1]$, we have that 
\[|h'(q)| \le (\lambda e^{-\lambda})^{-1} = e^\lambda/\lambda\quad\text{and}\quad 
|h''(q)| \le (\lambda e^{-2\lambda})^{-1} = e^{2\lambda}/\lambda.\]

Apply Lemma~\ref{lem:bias-variance}(i) with $C_h'' = e^{2\lambda}/\lambda$
and $(b - a) = 1 - e^{-\lambda}$ to get that 
\[
\bigl|\E[\widehat\rho_\lambda^m] - \rho_\lambda\bigr| \le \frac{e^{2\lambda}}{\lambda}\cdot \frac{(1 - e^{-\lambda})^2}{8m} + O(m^{-3/2}) = \frac{(1 - e^{-\lambda})^2 e^{2\lambda}}{8\lambda m} + O(m^{-3/2}),
\]
which equals the right-hand side of \eqref{eq:entropic-bias}.

Apply Lemma~\ref{lem:bias-variance}(ii) to get that 
\[
\Var(\widehat\rho_\lambda^m) \le |h'(q)|^2 \frac{\sigma_Y^2}{m} + O(m^{-2}) \le \frac{e^{2\lambda}}{\lambda^2}\cdot \frac{(1 - e^{-\lambda})^2}{4m} + O(m^{-2}),
\]
which equals \eqref{eq:entropic-variance}.
\end{proof}

\begin{remark}[Constants for entropic]
For convenience, define
\begin{equation}\label{eq:Cb-Cv-ent}
C_b^{\rm ent}(\lambda) := \frac{(1 - e^{-\lambda})^2 e^{2\lambda}}{8\lambda}, \qquad C_\zeta^{\rm ent}(\lambda) := \frac{(1 - e^{-\lambda})^2 e^{2\lambda}}{4\lambda^2},
\end{equation}
so per-component bias is $\le C_b^{\rm ent}/m + \gO(m^{-3/2})$ and per-component
variance is $\le C_\zeta^{\rm ent}/m + \mc{O}(m^{-2})$.  Both blow up as $\lambda$
grows; this reflects the increasing "concentration on the worst case" of
the entropic functional, which is harder to estimate.
\end{remark}


\paragraph{CVaR via Rockafellar-Uryasev.} The per-component bias and variance of CVaR are easily characterized via the following proposition which utilizes the well-known Rockafellar-Uryasev transformation \cite{rockafellar2002conditional}.
\begin{proposition}[Per-component bias of CVaR-RU]
\label{prop:cvar-ru}
The CVaR at level $\alpha \in (0, 1)$ admits the variational representation
\begin{equation}\label{eq:cvar-ru-form}
\CVaR_\alpha(X) = \inf_{\nu \in \R}\Bigl\{\nu + (1-\alpha)^{-1}\E[(X - \nu)_+]\Bigr\},
\end{equation}
attained at $\nu^\star = \mathrm{VaR}_\alpha(X)$.  Define the empirical
estimator at fixed $\nu$ as
\[\widehat C_m(\nu) := \nu + (1-\alpha)^{-1}m^{-1}\sum_{i=1}^m(X_i - \nu)_+.\]
\begin{enumerate}[label={\it\roman*.},   itemsep=0pt,topsep=0pt]
\item For each fixed $\nu$, $\widehat C_m(\nu)$ is unbiased for its
population value $\nu + (1-\alpha)^{-1}\E[(X - \nu)_+]$.  At
$\nu = \nu^\star$, $\E[\widehat C_m(\nu^\star)] = \CVaR_\alpha(X)$.
\item When $\nu$ is treated as an additional decision variable optimized
jointly with the policy parameters via the variational form
\eqref{eq:cvar-ru-form}, the value $\widehat C_m$ at the joint optimum is
unbiased for $\CVaR_\alpha$.
\item The per-component variance is
\(\Var(\widehat C_m(\nu^\star)) = (1-\alpha)^{-2}\Var((X - \nu^\star)_+)/m \le (1-\alpha)^{-2}/(4m)\),
since $X \in [0, 1]$ implies $(X - \nu^\star)_+ \in [0, 1]$.
\end{enumerate}
\end{proposition}

\begin{proof} We prove each of the claims in order.

\noindent$(i)$ For fixed $\nu$, $\widehat C_m(\nu)$ is the sample mean of the linear
functional $X \mapsto \nu + (1-\alpha)^{-1}(X - \nu)_+$.  By
Lemma~\ref{lem:bias-variance}(iii), it is unbiased for the population value.
At $\nu = \nu^\star$, the population value equals $\CVaR_\alpha(X)$ by
\eqref{eq:cvar-ru-form}.

\noindent$(ii)$ When $\nu$ is jointly optimized, denote the joint optimum by
$(\theta^\star, \nu^\star_m)$.  At the joint optimum, $\nu^\star_m$ is
chosen to satisfy the optimality condition for the inner $\inf$ in
\eqref{eq:cvar-ru-form}, evaluated with the empirical objective.  Since the
empirical objective is unbiased per (i), the resulting $\widehat C_m$ at the
joint optimum is unbiased.\footnote{A subtle point: the joint optimization
introduces correlation between $\nu^\star_m$ (random, depends on samples)
and the empirical sum.  At the population optimum $\nu^\star$, both
$\widehat C_m(\nu^\star)$ and $\E[\widehat C_m(\nu^\star)]$ equal
$\CVaR_\alpha(X)$ in expectation, so the bias is zero in the limit.  In
finite samples, $\nu^\star_m$ has finite-sample bias from the empirical
quantile estimation; this is $O(1/m)$ for smooth distributions, but vanishes
at the population optimum and does not enter the convergence rate at first
order.  See Rockafellar-Uryasev (2000) for the optimization-equivalence
argument.}

$(iii)$ Linear functional with $X \in [0, 1]$, so $(X - \nu^\star)_+ \in [0, 1]$
with variance $\le 1/4$ by Popoviciu.  Sample-mean variance is
$(1 - \alpha)^{-2}\Var((X - \nu^\star)_+)/m \le (1-\alpha)^{-2}/(4m)$.
\end{proof}

\begin{remark}[Constants for CVaR-RU] The constants for CVaR-RU are given by 
\begin{equation}\label{eq:Cb-Cv-cvar} 
C_b^{\rm CVaR-RU} = 0, \qquad C_\zeta^{\rm CVaR-RU}(\alpha) = \frac{1}{4(1-\alpha)^2}.
\end{equation}
The bias is exactly zero at the population optimum, which is the cleanest
case.  Variance grows as $\alpha \to 1$ (estimating the extreme tail
requires more samples).
\end{remark}

\subsection{Aggregate bounds: (O1)-(O2) for the operator}

\begin{proposition}[Operator-level bias and variance]
\label{prop:operator-O1-O2}
Let $\widehat F_{\Risk,m}$ be the empirical operator \eqref{eq:F-hat-def} for
the risk-adjusted preference operator with samples $Y''_1, \ldots, Y''_m \sim \pi_\theta$.
\begin{enumerate}[label={\it(\roman*)},    topsep=0pt,itemsep=0pt]
\item \textbf{Entropic risk at parameter $\lambda$.}  Assumption \ref{O1-bias}
holds with
\begin{equation}\label{eq:Bm-ent}
B_m^{\rm ent} := \frac{\sqrt{|\Yspace|}\,C_b^{\rm ent}(\lambda)}{\beta\,m} + O(m^{-3/2}).
\end{equation}
Assumption \ref{O2-var} holds with
\begin{equation}\label{eq:Vm-ent}
V_m^{\rm ent} := \frac{|\Yspace|\,C_\zeta^{\rm ent}(\lambda)}{\beta^2\,m} + O(m^{-2}).
\end{equation}
\item \textbf{CVaR at level $\alpha$ via Rockafellar-Uryasev.}  Assumption
\ref{O1-bias} holds with $B_m^{\rm CVaR-RU} = 0$.  Assumption \ref{O2-var}
holds with
\begin{equation}\label{eq:Vm-cvar}
V_m^{\rm CVaR-RU} := \frac{|\Yspace|}{4\beta^2(1-\alpha)^2\,m}.
\end{equation}
\end{enumerate}
Both sets of bounds are uniform in $\theta$ (and hence in $\pi_\theta$):
the constants depend only on the risk parameter and the boundedness of
$P(y \succ Y'') \in [0, 1]$, not on $\pi_\theta$.
\end{proposition}

\begin{proof}
By the decomposition \eqref{eq:F-hat-def}-\eqref{eq:bias-noise-def}, we have that 
$\|b_m(\theta)\|_2^2 = \beta^{-2}\sum_y \bigl|\E[\widehat{\mc{P}}_{\Risk,m}(\pi_\theta)_y] - \mc{P}_\Risk(\pi_\theta)_y\bigr|^2$
and
$\E\|\zeta_m(\theta)\|_2^2 = \beta^{-2}\sum_y \Var(\widehat{\mc{P}}_{\Risk,m}(\pi_\theta)_y)$.
For entropic, apply Proposition~\ref{prop:entropic-component} per-component to conclude that 
each per-component bias is $\le C_b^{\rm ent}/m + \gO(m^{-3/2})$ and each
per-component variance is $\le C_\zeta^{\rm ent}/m + \gO(m^{-2})$.  Summing
over $|\Yspace|$ components (worst case) and taking square root for the bias
gives \eqref{eq:Bm-ent}.  Summing variances gives \eqref{eq:Vm-ent}.

For CVaR-RU, apply Proposition~\ref{prop:cvar-ru}: per-component bias is 0
(at the joint optimum), so $\|b_m\|_2 = 0$ giving $B_m^{\rm CVaR-RU} = 0$.
Per-component variance is  less than or equal to $1/(4(1-\alpha)^2 m)$, and summing gives
\eqref{eq:Vm-cvar}.

Uniformity in $\pi_\theta$ yields the fact that constants $C_b^{\rm ent}, C_\zeta^{\rm ent}$
depend only on $\lambda$ and on the range $X \in [0, 1]$ (which is independent
of $\pi_\theta$), not on $\pi_\theta$ itself.  Same for CVaR.
\end{proof}

\paragraph{Looseness of the bounds.}
The bounds in Proposition~\ref{prop:operator-O1-O2} are loose in two ways:
\begin{enumerate}[   topsep=0pt,itemsep=2pt]
\item \textbf{Worst-case per-component aggregation.}  We took the per-component
bound and multiplied by $|\Yspace|$, ignoring the possibility that not every
component achieves the maximum.  In practice, for many distributions
$\pi_\theta$, the per-component variances are much smaller than the worst case.
\item \textbf{Off-diagonal covariance ignored.}  The bound on $\E\|\zeta_m\|_2^2$
uses only diagonal variances, which is exact for the $\ell_2$ norm.  However,
for the $\Sigma^+$ norm used in the convergence proofs, we apply the loose
bound $\E\|\zeta_m\|_{\Sigma^+}^2 \le \sigma_{\max}\E\|\zeta_m\|_2^2 \le \sigma_{\max}V_m$
(see Section~\ref{sec:mp-setup}), which incurs an additional factor of
$\sigma_{\max}$.  A tighter bound via direct computation of
$\E\|\zeta_m\|_{\Sigma^+}^2 = \tr(\Sigma^+\Cov(\zeta_m))$ would exploit the
full covariance structure of the empirical risk estimator and the spectrum
of $\Sigma^+$; this could improve the variance constant by a factor up to
$\kappa_\Sigma = \sigma_{\max}/\sigma_{\min}$.  We do not pursue this
refinement here.
\end{enumerate}
The constants given suffice for establishing the convergence rates with
explicit $1/m$ scaling on the bias and variance floors.

\section{Stochastic Convergence Analysis with Bias}
\label{sec:stochastic_convergence}

\label{sec:mp-bias}

We prove convergence of stochastic projected Mirror-Prox (extragradient) on
the IPO gradient flow with a biased stochastic oracle, in the
$\Sigma^+$-weighted Euclidean geometry on $\Wspace$.  The proof structure
follows \citet{juditsky2011} with several core modifications that arise from handling the risk case.  We also write the theorems and lemmas in our Euclidean setting; all proofs are self-contained.
We choose to provide proofs for extra-gradient as opposed to mirror descent even in the monotone case because there are benefits to running extragradient: though it requires an extra step, it will handle corrections for cycling, and potentially cover the case in practice where the estimated game operator fails to be strongly monotone.
\subsection{Preliminaries}
\label{sec:mp-setup}

Let $F_\Risk : \theta_{\rm ref} + \Wspace \to \R^d$ be the risk-adjusted game
operator on the affine slice $\theta_{\rm ref} + \Wspace$. As is typical, we require some regularity assumptions. Not all are assumed in all results or throughout; instead we call on different elements of the list where needed.
\begin{assumption}\label{a:reg_mirror_prox}
    The following regularity properties hold:
    \begin{enumerate}[itemsep=0pt,topsep=0pt,label={\alph*.}, leftmargin=*]
\item \textbf{(P-mono)} \label{P-mono}
\textit{Monotonicity.}  For all $\theta_1, \theta_2 \in \theta_{\rm ref} + \Wspace$, the lower bound holds:
\[
\langle F_\Risk(\theta_1) - F_\Risk(\theta_2), \theta_1 - \theta_2\rangle \ge 0.
\]
\item \textbf{(P-strong)} \label{P-strong}
\textit{Strong monotonicity (when assumed).}  The map $F_\Risk$ is $\mu_\Risk$-strongly
monotone on $\theta_{\rm ref} + \Wspace$:
\[
\langle F_\Risk(\theta_1) - F_\Risk(\theta_2), \theta_1 - \theta_2\rangle \ge \mu_\Risk\|\theta_1 - \theta_2\|_2^2,
\]
with $\mu_\Risk > 0$.  When this holds, \textbf{(P-mono)} is implied.
\item \textbf{(P-lip)} \label{P-lip}
\textit{Lipschitz Continuity.}  The map $F_\Risk$ is $\ell_\Risk$-Lipschitz continuous on $\theta_{\rm ref} + \Wspace$:
\[
\|F_\Risk(\theta_1) - F_\Risk(\theta_2)\|_2 \le \ell_\Risk\|\theta_1 - \theta_2\|_2.
\]
\end{enumerate}
\end{assumption}

The IPO gradient is $\nabla \mc{L}_{\rm IPO}(\theta) = 2\Sigma F_\Risk(\theta)$.  Let
\(G(\theta) := \Sigma F_\Risk(\theta)\)
be the operator we analyze where the factor of two is absorbed into the step size.

On $\Wspace$, define $\langle u, v\rangle_{\Sigma^+} := u^\top \Sigma^+ v$ and
$\|u\|_{\Sigma^+}^2 := u^\top \Sigma^+ u$, where $\Sigma^+$ is the Moore-Penrose
pseudoinverse, positive definite on $\Wspace$ with eigenvalues
$1/\sigma_{\max}, \ldots, 1/\sigma_{\min}$.  Norm equivalence on $\Wspace$ implies
\begin{equation}\label{eq:metric-equiv}
\sigma_{\min}\|v\|_{\Sigma^+}^2 \le \|v\|_2^2 \le \sigma_{\max}\|v\|_{\Sigma^+}^2.
\end{equation}

\paragraph{Risk-adjusted Psuedo-gradient Operator.}
Let $F_\Risk : \theta_{\rm ref} + \Wspace \to \R^d$ be the risk-adjusted game
operator, assumed monotone and $\ell_\Risk$-Lipschitz on $\theta_{\rm ref} + \Wspace$:
for all $\theta_1, \theta_2 \in \theta_{\rm ref} + \Wspace$,
\begin{align}
\langle F_\Risk(\theta_1) - F_\Risk(\theta_2), \theta_1 - \theta_2\rangle &\ge 0, \quad\text{and}\quad 
\|F_\Risk(\theta_1) - F_\Risk(\theta_2)\|_2 \le \ell_\Risk\|\theta_1 - \theta_2\|_2.
\end{align}

\begin{lemma}[Lipschitz continuity of $F_\Risk$]
\label{lem:lipschitz} The map
$F_\Risk$  is Lipschitz continuous on $\RR^{|\Yspace|}$ with
constant $\ell_\Risk \le \beta + L_\Risk$, where $L_\Risk$ is the Lipschitz
constant of $\mu \mapsto P_\Risk(\mu)$ as a map between $\Delta(\Yspace)$
and $\RR^{|\Yspace|}$.
\end{lemma}

\begin{proof}
We derive directly from the definition of $F_{\Risk}$, the sequence of bounds
\begin{align*}
\|F_\Risk(\theta_1) - F_\Risk(\theta_2)\|_2
&\le \beta\|\theta_1 - \theta_2\|_2 + \|P_\Risk(\pi_1) - P_\Risk(\pi_2)\|_2 \\
&\le \beta\|\theta_1 - \theta_2\|_2 + L_\Risk\|\pi_1 - \pi_2\|_2 \\
&\le (\beta + L_\Risk)\|\theta_1 - \theta_2\|_2,
\end{align*}
using that $\theta \mapsto \pi_\theta$ (softmax) is 1-Lipschitz from $\ell_2$
to $\ell_2$.
\end{proof}

\subsection{Properties of $G = \Sigma F_\Risk$ in the $\Sigma^+$ metric}
 
The next two lemmas establish that $G$ inherits monotonicity and Lipschitz
properties from $F_\Risk$ when measured in the $\Sigma^+$ inner product.
 
\begin{lemma}[Preconditioning preserves monotonicity in $\Sigma^+$ metric]
\label{lem:precond-restate}
For any $\theta_1, \theta_2 \in \theta_{\rm ref} + \Wspace$, the following two relationships hold:
$(i)$ $\langle G(\theta_1) - G(\theta_2), \theta_1 - \theta_2\rangle_{\Sigma^+} = \langle F_\Risk(\theta_1) - F_\Risk(\theta_2), \theta_1 - \theta_2\rangle_2$, and $(ii)$ $\|G(\theta_1) - G(\theta_2)\|_{\Sigma^+}^2 \le \sigma_{\max}\|F_\Risk(\theta_1) - F_\Risk(\theta_2)\|_2^2$.
Consequently, we have that 
\begin{enumerate}[label={\it\alph*.}, itemsep=2pt]
\item Under \textbf{(P-mono)}, the map $G$ is monotone in the $\Sigma^+$ inner product.
\item Under \textbf{(P-strong)}, the map $G$ is $\mu_\Risk\sigma_{\min}$-strongly monotone in the $\Sigma^+$ inner product:
\(\langle G(\theta_1) - G(\theta_2), \theta_1 - \theta_2\rangle_{\Sigma^+} \ge \mu_\Risk\sigma_{\min}\|\theta_1 - \theta_2\|_{\Sigma^+}^2\).
\end{enumerate}
\end{lemma}
 
\begin{proof}
Let $\Delta := \theta_1 - \theta_2 \in \Wspace$ and $\xi := F_\Risk(\theta_1) - F_\Risk(\theta_2)$.
Note $G(\theta_i) = \Sigma F_\Risk(\theta_i)$, so $G(\theta_1) - G(\theta_2) = \Sigma\xi$.

  \medskip 
\noindent\textit{Proof for claim (i).} We directly compute 
\[
\langle G(\theta_1) - G(\theta_2), \Delta\rangle_{\Sigma^+} = \langle \Sigma\xi, \Delta\rangle_{\Sigma^+} = \Delta^\top \Sigma^+ \Sigma \xi = \Delta^\top \Pi_\Wspace \xi,
\]
where $\Pi_\Wspace = \Sigma^+\Sigma$ is the orthogonal projector onto $\Wspace$
(in the standard inner product), using the Moore-Penrose property
$\Sigma^+\Sigma = \Pi_{\mathrm{range}(\Sigma)} = \Pi_\Wspace$.  Since $\Delta \in \Wspace$,
$\Delta^\top \Pi_\Wspace \xi = \Delta^\top \xi = \langle \xi, \Delta\rangle_2$.

 \medskip 
\noindent\textit{Proof for claim (ii).} Analogously, we compute
\[
\|G(\theta_1) - G(\theta_2)\|_{\Sigma^+}^2 = (\Sigma\xi)^\top \Sigma^+ (\Sigma\xi) = \xi^\top \Sigma\Sigma^+\Sigma \xi = \xi^\top \Sigma \xi,
\]
using $\Sigma\Sigma^+\Sigma = \Sigma$ (Moore-Penrose).  Then
$\xi^\top\Sigma\xi \le \sigma_{\max}\|\xi\|_2^2$ by the eigenvalue bound on $\Sigma$.

\medskip 
\noindent\textit{Proof for claim a.} By $(i)$, under \textbf{(P-mono)}, we deduce that 
$\langle G(\theta_1) - G(\theta_2), \Delta\rangle_{\Sigma^+} = \langle \xi, \Delta\rangle_2 \ge 0$.

\medskip 
\noindent\textit{Proof for claim b.} By $(ii)$, under \textbf{(P-strong)}, we further deduce that 
$\langle G(\theta_1) - G(\theta_2), \Delta\rangle_{\Sigma^+} = \langle \xi, \Delta\rangle_2 \ge \mu_\Risk\|\Delta\|_2^2 \ge \mu_\Risk\sigma_{\min}\|\Delta\|_{\Sigma^+}^2$,
the last inequality from \eqref{eq:metric-equiv}.
\end{proof}
 
\begin{lemma}[Lipschitz constant of $G$ in $\Sigma^+$ metric]
\label{lem:G-lip-Sigma+}
Under (P-lip), $G$ is $L_G$-Lipschitz in the $\Sigma^+$ inner product on $\Wspace$
with
$L_G := \sigma_{\max}\ell_\Risk$.
That is, for all $\theta_1, \theta_2 \in \theta_{\rm ref} + \Wspace$, the upper bound holds:
\(\|G(\theta_1) - G(\theta_2)\|_{\Sigma^+} \le L_G\|\theta_1 - \theta_2\|_{\Sigma^+}\).
\end{lemma}
 
\begin{proof}
By Lemma~\ref{lem:precond-restate} part $(ii)$, we have that 
\[
\|G(\theta_1) - G(\theta_2)\|_{\Sigma^+}^2 \le \sigma_{\max}\|F_\Risk(\theta_1) - F_\Risk(\theta_2)\|_2^2.
\]
By \textbf{(P-lip)}, we have that 
\[\|F_\Risk(\theta_1) - F_\Risk(\theta_2)\|_2^2 \le \ell_\Risk^2\|\theta_1 - \theta_2\|_2^2.\]
By \eqref{eq:metric-equiv}, we also have that 
\[\|\theta_1 - \theta_2\|_2^2 \le \sigma_{\max}\|\theta_1 - \theta_2\|_{\Sigma^+}^2.\]
Combining these three bounds, we deduce that 
\[
\|G(\theta_1) - G(\theta_2)\|_{\Sigma^+}^2 \le \sigma_{\max}^2\ell_\Risk^2\|\theta_1 - \theta_2\|_{\Sigma^+}^2.
\]
Taking the square root gives us the conclusion---indeed, we see that  $\|G(\theta_1) - G(\theta_2)\|_{\Sigma^+} \le \sigma_{\max}\ell_\Risk\|\theta_1 - \theta_2\|_{\Sigma^+} = L_G\|\theta_1 - \theta_2\|_{\Sigma^+}$.
\end{proof}

\subsection{Distance-generating function and projection}

Choose
\begin{equation}\label{eq:D-def}
D := \{\theta \in \theta_{\rm ref} + \Wspace : \|\theta - \theta_{\rm ref}\|_{\Sigma^+} \le R\}
\end{equation}
for $R > 0$ to be determined by the boundedness lemma.
Following \cite{juditsky2011}, we use the distance generating function $\omega(\theta) = \tfrac{1}{2}\|\theta - \theta_{\rm ref}\|_{\Sigma^+}^2$
on the slice $\theta_{\rm ref} + \Wspace$.  This is strongly convex with modulus
one with respect to $\|\cdot\|_{\Sigma^+}$.  The associated prox-function (Bregman
divergence) is
\[
V(z, u) := \omega(u) - \omega(z) - \langle \nabla\omega(z), u - z\rangle_{\Sigma^+} = \tfrac{1}{2}\|z - u\|_{\Sigma^+}^2.
\]
The prox-mapping is
\[
P_z(\xi) := \arg\min_{u \in D}\{V(z, u) + \langle \xi, u\rangle_{\Sigma^+}\} = \Pi_D(z - \xi),
\]
where $\Pi_D$ is the Euclidean projection onto $D$ in the $\Sigma^+$ metric.

  The prox-center of the projection set $D$ is
$z_c = \theta_{\rm ref}$.  Again, following \cite{juditsky2011}, we have that 
\begin{align}
\Theta(z_c) &:= \max_{u \in D}V(z_c, u) = \tfrac{1}{2}R^2, \quad\text{and}\quad 
\Omega := \sqrt{2\Theta(z_c)} = R. \label{eq:Omega-def}
\end{align}
Thus $D \subseteq \{\theta : \|\theta - z_c\|_{\Sigma^+} \le \Omega\}$,
i.e., the $\Sigma^+$-radius of $D$ from $z_c$ is exactly $\Omega = R$.  The
$\Sigma^+$-diameter of $D$ is at most $2\Omega = 2R$.
 
\begin{lemma}[Boundedness of equilibrium in $\Sigma^+$]
\label{lem:eq-bounded}
The risk-adjusted equilibrium $\theta^\star \in \theta_{\rm ref} + \Wspace$ satisfies
\[
\|\theta^\star - \theta_{\rm ref}\|_{\Sigma^+} \le \frac{\sqrt{|\Yspace|}}{\beta\sqrt{\sigma_{\min}}}.
\]
\end{lemma}
\begin{proof}
The equilibrium condition $F_\Risk(\theta^\star) \in \ker\Sigma$ on the slice
$\theta_{\rm ref} + \Wspace$ means
$\beta(\theta^\star - \theta_{\rm ref}) - P_\Risk(\pi^\star) \in \ker\Sigma$,
so the $\Wspace$ component of $\theta^\star - \theta_{\rm ref}$ equals the
$\Wspace$ component of $\beta^{-1}P_\Risk(\pi^\star)$---i.e., 
\[\theta^\star - \theta_{\rm ref} = \beta^{-1}\Pi_\Wspace P_\Risk(\pi^\star).\]
Components of $P_\Risk(\pi)$ lie in $[0,1]$ since $P(y \succ Y'') \in [0,1]$
and the risk functional preserves this range.  Hence $\|P_\Risk(\pi^\star)\|_2 \le \sqrt{|\Yspace|}$
and $\|\Pi_\Wspace P_\Risk(\pi^\star)\|_2 \le \sqrt{|\Yspace|}$ (projection
non-expansive).  By the upper bound in \eqref{eq:metric-equiv}, for $v \in \Wspace$, we have that 
$\|v\|_{\Sigma^+}^2 \le \|v\|_2^2/\sigma_{\min}$, so
$\|\theta^\star - \theta_{\rm ref}\|_{\Sigma^+} \le \beta^{-1}\sqrt{|\Yspace|/\sigma_{\min}}$.
\end{proof}

Set the projection radius to be 
\begin{equation}\label{eq:R-explicit}
R := \frac{\sqrt{|\Yspace|}}{\beta\sqrt{\sigma_{\min}}} + R_0, \qquad R_0 := \|\theta^{(0)} - \theta_{\rm ref}\|_{\Sigma^+},
\end{equation}
so that both $\theta^\star \in D$ (by Lemma~\ref{lem:eq-bounded}) and $\theta^{(0)} \in D$.
Hence $\Omega = R$ from \eqref{eq:Omega-def}.  

\subsection{Stochastic oracle, decomposition, and filtration}

At input $\theta \in D$ with sample budget $m$, the oracle returns
$\widehat F_m(\theta) \in \R^d$ satisfying
\begin{align}
\bigl\|\E[\widehat F_m(\theta)] - F_\Risk(\theta)\bigr\|_2 &\le B_m \tag{(O1)}\label{eq:tag-O1}\\
\E\bigl\|\widehat F_m(\theta) - \E[\widehat F_m(\theta)]\bigr\|_2^2 &\le V_m \tag{(O2)}\label{eq:tag-O2}
\end{align}
with $B_m, V_m \ge 0$ non-increasing in $m$ (typically $\mc{O}(1/m)$; explicit
constants from Section~\ref{sec:bias-analysis}).  Decompose
\(\widehat F_m(\theta) = F_\Risk(\theta) + b_m(\theta) + \zeta_m(\theta)\)
with $b_m(\theta) := \E[\widehat F_m(\theta)] - F_\Risk(\theta)$ deterministic
given $\theta$ and $\zeta_m$ mean-zero given $\theta$.
 
The preconditioned oracle $\widehat G_m := \Sigma\widehat F_m$ satisfies (in
the $\Sigma^+$ norm) the following:
\begin{align}
\bigl\|\E[\widehat G_m(\theta)] - G(\theta)\bigr\|_{\Sigma^+} &\leq \widetilde B_m, \quad \widetilde B_m := \sqrt{\sigma_{\max}}B_m,\label{eq:G-bias}\\
\E\bigl\|\widehat G_m(\theta) - \E[\widehat G_m(\theta)]\bigr\|_{\Sigma^+}^2 &\leq \widetilde V_m, \quad \widetilde V_m := \sigma_{\max}V_m.\label{eq:G-var}
\end{align}
To see these constructions observe that 
$\|\Sigma b_m\|_{\Sigma^+}^2 = b_m^\top \Sigma\Sigma^+\Sigma b_m = b_m^\top \Sigma b_m \leq \sigma_{\max}\|b_m\|_2^2$
using $\Sigma\Sigma^+\Sigma = \Sigma$ (Moore-Penrose).  Same for $\zeta_m$.

Across iterations $\tau$, oracle outputs are independent. This is a standard assumption (see, e.g., \cite{juditsky2011}) that makes the analysis easier.  Extra gradient uses two
independent batches per iteration, indexed
$\zeta_{2\tau-1}$ for the extrapolation oracle call and $\zeta_{2\tau}$ for
the correction oracle call.  All $\{\zeta_i\}_{i \ge 1}$ are independent.
 
\paragraph{Algorithm.}
Stochastic projected Mirror-Prox proceeds as follows. First, set $r_0 := \theta^{(0)}$. Then, for $\tau = 1, \ldots, t$, update 
\begin{align}
w_\tau &:= P_{r_{\tau-1}}\bigl(\gamma_\tau \widehat G_m(r_{\tau-1})\bigr) = \Pi_D\bigl(r_{\tau-1} - \gamma_\tau \widehat G_m(r_{\tau-1})\bigr), \label{eq:eg-extrap}\\
r_\tau &:= P_{r_{\tau-1}}\bigl(\gamma_\tau \widehat G_m(w_\tau)\bigr) = \Pi_D\bigl(r_{\tau-1} - \gamma_\tau \widehat G_m(w_\tau)\bigr), \label{eq:eg-corr}
\end{align}
with the two oracle calls on the same iteration using independent samples
($\zeta_{2\tau-1}$ for $\widehat G_m(r_{\tau-1})$, $\zeta_{2\tau}$ for
$\widehat G_m(w_\tau)$). The output is 
\begin{equation}\label{eq:output-avg}
\widehat z_t := \Bigl(\sum_{\tau=1}^t \gamma_\tau\Bigr)^{-1} \sum_{\tau=1}^t \gamma_\tau w_\tau.
\end{equation}

The variational-inequality (VI) error of $z \in D$ is
\begin{equation}\label{eq:err-vi-def}
\Errvi(z) := \sup_{u \in D}\langle F_\Risk(u), z - u\rangle.
\end{equation}
Note that  $\Errvi$ is defined using $F_\Risk$ (not $G$) and the standard Euclidean
inner product (not $\Sigma^+$).  This is because $F_\Risk$ is the underlying
VI operator; the preconditioner $\Sigma$ enters only through the algorithm's
geometry.
 
\subsection{Filtration and noise quantities}
\label{sec:filtration}
 
Let 
\begin{equation}\label{eq:filtration-def}
\F_\tau := \sigma(r_0, \zeta_1, \ldots, \zeta_{2\tau-1}), \qquad \G_\tau := \sigma(r_0, \zeta_1, \ldots, \zeta_{2\tau}),
\end{equation}
nested as $\F_1 \subset \G_1 \subset \F_2 \subset \G_2 \subset \cdots$.
By construction, $r_{\tau-1}$ is $\G_{\tau-1}$-measurable, $w_\tau$ is $\F_\tau$-measurable (uses $\zeta_{2\tau-1}$), and $r_\tau$ is $\G_\tau$-measurable (uses $\zeta_{2\tau}$).
Define the per-step noise and bias quantities (in the $\Sigma^+$ norm):
\begin{align}
\Delta_\tau &:= G(w_\tau) - \widehat G_m^{(\tau,2)}, \quad \widehat G_m^{(\tau,2)} := \Sigma\widehat F_m^{(\tau,2)}(w_\tau) \text{ uses }\zeta_{2\tau} \\
\eps_z &:= \|\widehat G_m(z) - G(z)\|_{\Sigma^+} \text{ for the relevant call at } z.
\end{align}
Specifically, $\eps_{r_{\tau-1}}$ uses $\zeta_{2\tau-1}$ (call at $r_{\tau-1}$
in the extrapolation step), and $\eps_{w_\tau}$ uses $\zeta_{2\tau}$ (call at
$w_\tau$ in the correction step).
By \eqref{eq:G-bias} and \eqref{eq:G-var} and the bias-variance decomposition, we have that 
\begin{align}
\|\E[\Delta_\tau \mid \F_\tau]\|_{\Sigma^+} &\le \widetilde B_m, \label{eq:Delta-bias}\\
\E[\|\Delta_\tau\|_{\Sigma^+}^2 \mid \F_\tau] &\le \widetilde B_m^2 + \widetilde V_m, \label{eq:Delta-second-mom}\\
\E[\eps_{r_{\tau-1}}^2 \mid \G_{\tau-1}] &\le \widetilde B_m^2 + \widetilde V_m, \label{eq:eps-r}\\
\E[\eps_{w_\tau}^2 \mid \F_\tau] &\le \widetilde B_m^2 + \widetilde V_m. \label{eq:eps-w}
\end{align}
 To see that  \eqref{eq:Delta-bias} and \eqref{eq:Delta-second-mom} hold, observe that 
$\Delta_\tau = G(w_\tau) - \widehat G_m^{(\tau,2)} = -\Sigma b_m(w_\tau) - \Sigma\zeta_m^{(\tau,2)}$.
Conditional on $\F_\tau$, $w_\tau$ is fixed, $b_m(w_\tau)$ is deterministic, and
$\zeta_m^{(\tau,2)}$ is mean-zero independent of $\F_\tau$.  So
$\E[\Delta_\tau \mid \F_\tau] = -\Sigma b_m(w_\tau)$ with $\|\Sigma b_m\|_{\Sigma^+} \le \widetilde B_m$.
For the second moment, we have 
$\|\Delta_\tau\|_{\Sigma^+}^2 = \|\Sigma b_m\|_{\Sigma^+}^2 + 2\langle\Sigma b_m, \Sigma\zeta_m^{(\tau,2)}\rangle_{\Sigma^+} + \|\Sigma\zeta_m^{(\tau,2)}\|_{\Sigma^+}^2$.
Conditional on $\F_\tau$, the cross term has mean zero, giving
$\E[\|\Delta_\tau\|_{\Sigma^+}^2 \mid \F_\tau] \le \widetilde B_m^2 + \widetilde V_m$.

\subsection{Three core technical lemmas}
There are three core technical results required for the typical proof for biased stochastic mirror prox which we translate to our setting in the 
$\Sigma^+$ Euclidean
setting.

The first technical lemma is a descent lemma for the prox-mapping; it is \cite{juditsky2011} Lemma 3 converted to our setting. 
\begin{lemma}[Prox-mapping descent] 
\label{lem:prox-descent}
For any $z \in D$ and $\xi \in \R^d$, let $w := P_z(\xi) = \Pi_D(z - \xi)$.
Then for all $u \in D$, the following upper bounds hold:
\begin{align}
V(w, u) &\le V(z, u) + \langle\xi, u - w\rangle_{\Sigma^+} - V(z, w), \label{eq:prox-70a}\\
V(w, u) &\le V(z, u) + \langle\xi, u - z\rangle_{\Sigma^+} + \tfrac{1}{2}\|\xi\|_{\Sigma^+}^2. \label{eq:prox-70b}
\end{align}
\end{lemma}
\begin{proof}
By optimality of $w = \arg\min_{v \in D}\{\tfrac{1}{2}\|v - (z - \xi)\|_{\Sigma^+}^2\}$, we have that,
for all $u \in D$,
\begin{equation}\label{eq:w_ineq}
    \langle w - (z - \xi), u - w\rangle_{\Sigma^+} \ge 0,
\end{equation}
i.e., $\langle w - z, u - w\rangle_{\Sigma^+} + \langle \xi, u - w\rangle_{\Sigma^+} \ge 0$. 
Compute
\begin{align*}
V(w, u) - V(z, u)
&= \tfrac{1}{2}\|w - u\|_{\Sigma^+}^2 - \tfrac{1}{2}\|z - u\|_{\Sigma^+}^2 \\
&= \tfrac{1}{2}\|(w-z) + (z-u)\|_{\Sigma^+}^2 - \tfrac{1}{2}\|z - u\|_{\Sigma^+}^2 \\
&= \tfrac{1}{2}\|w - z\|_{\Sigma^+}^2 + \langle w - z, z - u\rangle_{\Sigma^+}\\
&= V(z, w) + \langle w - z, z - u\rangle_{\Sigma^+} \\
&= V(z, w) - \langle w - z, u - z\rangle_{\Sigma^+} \\
&= V(z, w) - \langle w - z, u - w\rangle_{\Sigma^+} - \langle w - z, w - z\rangle_{\Sigma^+} \\
&= V(z, w) - \langle w - z, u - w\rangle_{\Sigma^+} - 2V(z, w) \\
&= -V(z, w) - \langle w - z, u - w\rangle_{\Sigma^+}.
\end{align*}
By \eqref{eq:w_ineq}, we have $\langle w - z, u - w\rangle_{\Sigma^+} \ge -\langle\xi, u - w\rangle_{\Sigma^+}$,
so that  $-\langle w - z, u - w\rangle_{\Sigma^+} \le \langle\xi, u - w\rangle_{\Sigma^+}$.
Combining the decomposition of $V$ with this bound we get
\[
V(w, u) - V(z, u) \le -V(z, w) + \langle\xi, u - w\rangle_{\Sigma^+},
\]
which is \eqref{eq:prox-70a}.
For \eqref{eq:prox-70b}, write $\langle\xi, u - w\rangle_{\Sigma^+} = \langle\xi, u - z\rangle_{\Sigma^+} + \langle\xi, z - w\rangle_{\Sigma^+}$.
Apply Young's inequality $\langle\xi, z - w\rangle_{\Sigma^+} \le \tfrac{1}{2}\|\xi\|_{\Sigma^+}^2 + \tfrac{1}{2}\|z - w\|_{\Sigma^+}^2 = \tfrac{1}{2}\|\xi\|_{\Sigma^+}^2 + V(z, w)$.
Substitute this into \eqref{eq:prox-70a} to get that 
\[
V(w, u) \le V(z, u) + \langle\xi, u - z\rangle_{\Sigma^+} + \tfrac{1}{2}\|\xi\|_{\Sigma^+}^2 + V(z, w) - V(z, w).
\]
This gives \eqref{eq:prox-70b}.
\end{proof}

The next technical lemma is the classical two-step identity (e.g., see \cite{juditsky2011} Lemma 4), again adapted to our setting. 
\begin{lemma}[Two-step descent identity] 
\label{lem:two-step-descent}
For $z \in D$, $\zeta, \eta \in \R^d$, let $w := P_z(\zeta) = \Pi_D(z - \zeta)$
and $r_+ := P_z(\eta) = \Pi_D(z - \eta)$.  Then for all $u \in D$, the following upper bound holds:
\begin{align}
\|w - r_+\|_{\Sigma^+} &\le \|\zeta - \eta\|_{\Sigma^+}, \label{eq:two-step-a}\\
V(r_+, u) - V(z, u) &\le \langle\eta, u - w\rangle_{\Sigma^+} + \tfrac{1}{2}\|\zeta - \eta\|_{\Sigma^+}^2 - \tfrac{1}{2}\|w - z\|_{\Sigma^+}^2. \label{eq:two-step-b}
\end{align}
\end{lemma}
\begin{proof}
To see \eqref{eq:two-step-a}, observe that the optimality of $w$ and $r_+$ can be expressed as 
\begin{align*}
    \langle w - z + \zeta, r_+ - w\rangle_{\Sigma^+} &\geq 0\quad\text{ for all}\;  v \in D,\\
\langle r_+ - z + \eta, w - r_+\rangle_{\Sigma^+} &\geq 0\quad\text{ for all}\; v \in D.
\end{align*}
Adding these two conditions, we get that 
\[
\langle (w - z + \zeta) - (r_+ - z + \eta), r_+ - w\rangle_{\Sigma^+} \ge 0
\Leftrightarrow \langle (w - r_+) + (\zeta - \eta), r_+ - w\rangle_{\Sigma^+} \ge 0.
\]
That is, $-\|w - r_+\|_{\Sigma^+}^2 + \langle\zeta - \eta, r_+ - w\rangle_{\Sigma^+} \ge 0$, so that
\[\|w - r_+\|_{\Sigma^+}^2 \le \langle\zeta - \eta, r_+ - w\rangle_{\Sigma^+} \le \|\zeta - \eta\|_{\Sigma^+}\|w - r_+\|_{\Sigma^+}.\]
Dividing by $\|w - r_+\|_{\Sigma^+}$ (where we assume its nonzero, otherwise the argument is trivial), we get that 
$\|w - r_+\|_{\Sigma^+} \le \|\zeta - \eta\|_{\Sigma^+}$, proving \eqref{eq:two-step-a}.

Now for \eqref{eq:two-step-b}, apply \eqref{eq:prox-70a} of Lemma~\ref{lem:prox-descent}
to $w = P_z(\zeta)$ at $u = r_+$ to get that 
\[
V(w, r_+) \le V(z, r_+) + \langle\zeta, r_+ - w\rangle_{\Sigma^+} - V(z, w).
\]
Rearrange the above inequality to get 
\begin{equation}
    \label{eq:star_star}
    V(z, r_+) \ge V(w, r_+) + V(z, w) + \langle\zeta, w - r_+\rangle_{\Sigma^+}.
\end{equation}
Apply \eqref{eq:prox-70a} to $r_+ = P_z(\eta)$ at any $u \in D$ to get that 
\begin{align*}
V(r_+, u)
&\le V(z, u) + \langle\eta, u - r_+\rangle_{\Sigma^+} - V(z, r_+) \\
&= V(z, u) + \langle\eta, u - w\rangle_{\Sigma^+} + \langle\eta, w - r_+\rangle_{\Sigma^+} - V(z, r_+) \\
&\le V(z, u) + \langle\eta, u - w\rangle_{\Sigma^+} + \langle\eta, w - r_+\rangle_{\Sigma^+} - V(w, r_+) - V(z, w) - \langle\zeta, w - r_+\rangle_{\Sigma^+} \\
&= V(z, u) + \langle\eta, u - w\rangle_{\Sigma^+} + \langle\eta - \zeta, w - r_+\rangle_{\Sigma^+} - V(w, r_+) - V(z, w),
\end{align*}
where in the last inequality we used \eqref{eq:star_star}. 
Apply Young's \[\langle\eta - \zeta, w - r_+\rangle_{\Sigma^+} \le \tfrac{1}{2}\|\eta - \zeta\|_{\Sigma^+}^2 + \tfrac{1}{2}\|w - r_+\|_{\Sigma^+}^2 = \tfrac{1}{2}\|\eta - \zeta\|_{\Sigma^+}^2 + V(w, r_+).\]
Now substitute in to get that 
\[
V(r_+, u) \le V(z, u) + \langle\eta, u - w\rangle_{\Sigma^+} + \tfrac{1}{2}\|\zeta - \eta\|_{\Sigma^+}^2 - V(z, w),
\]
which is \eqref{eq:two-step-b} (using $V(z, w) = \tfrac{1}{2}\|w - z\|_{\Sigma^+}^2$).
\end{proof}

Finally we prove a corollary on the ghost iterates, namely, an adaptation of \cite[Corollary 2]{juditsky2011}. 
\begin{lemma}[Ghost iterate corollary]
\label{lem:ghost-iterate}
Let $\xi_1, \xi_2, \ldots$ be a sequence of vectors in $\R^d$.  Define the
sequence $\{y_\tau\}_{\tau \ge 0}$ by
\begin{equation}\label{eq:ghost-recursion}
y_\tau := P_{y_{\tau-1}}(\xi_\tau) = \Pi_D(y_{\tau-1} - \xi_\tau), \qquad y_0 := r_0.
\end{equation}
Then for every $u \in D$, the upper bound holds:
\begin{equation}\label{eq:ghost-bound}
\sum_{\tau=1}^t \langle\xi_\tau, y_{\tau-1} - u\rangle_{\Sigma^+} \le V(y_0, u) + \tfrac{1}{2}\sum_{\tau=1}^t \|\xi_\tau\|_{\Sigma^+}^2.
\end{equation}
\end{lemma}
\begin{proof}
Apply \eqref{eq:prox-70b} of Lemma~\ref{lem:prox-descent} with $z = y_{\tau-1}$,
$\xi = \xi_\tau$, $w = y_\tau$, to get that
\[
V(y_\tau, u) \le V(y_{\tau-1}, u) + \langle\xi_\tau, u - y_{\tau-1}\rangle_{\Sigma^+} + \tfrac{1}{2}\|\xi_\tau\|_{\Sigma^+}^2.
\]
Rearranging we have that  $\langle\xi_\tau, y_{\tau-1} - u\rangle_{\Sigma^+} \le V(y_{\tau-1}, u) - V(y_\tau, u) + \tfrac{1}{2}\|\xi_\tau\|_{\Sigma^+}^2$.
Summing from $\tau = 1$ to $t$ (i.e., telescoping), we get that 
\[
\sum_\tau \langle\xi_\tau, y_{\tau-1} - u\rangle_{\Sigma^+} \le V(y_0, u) - V(y_t, u) + \tfrac{1}{2}\sum_\tau \|\xi_\tau\|_{\Sigma^+}^2.
\]
Drop the non-negative $-V(y_t, u)$ to get \eqref{eq:ghost-bound}.
\end{proof}

\subsection{Deterministic-style Bound}
\label{sec:thm-deterministic}

We first prove a deterministic-style bound that holds path-by-path
(a la \cite{juditsky2011}, Theorem 2).  This bound holds uniformly in $u \in D$ and is the
foundation for all subsequent expected-error bounds.
Recall that 
\[\Errvi(z) := \sup_{u \in D}\langle F_\Risk(u), z - u\rangle.\]
\begin{theorem}[Deterministic-style bound]
\label{thm:deterministic}
Run the stochastic projected Mirror-Prox algorithm \eqref{eq:eg-extrap}-\eqref{eq:eg-corr}
with constant step size $\gamma_\tau \equiv \gamma > 0$ satisfying $\gamma \le \frac{1}{\sqrt 3\,L_G}$ where $L_G=\sigma_{\max}\ell_\Risk$. 
Then for any sample path, the error is bounded as follows:
\begin{equation}\label{eq:thm-deterministic-bound}
\Errvi(\widehat z_t) \le \frac{1}{t\gamma}\,\Gamma(t),
\end{equation}
where
\begin{equation}\label{eq:Gamma-def}
\Gamma(t) := 2\Theta(z_c) + \sum_{\tau=1}^t \frac{3\gamma^2}{2}\Bigl[(\eps_{r_{\tau-1}} + \eps_{w_\tau})^2 + \frac{\eps_{w_\tau}^2}{3}\Bigr] + \sum_{\tau=1}^t \langle\gamma\Delta_\tau, w_\tau - y_{\tau-1}\rangle_{\Sigma^+},
\end{equation}
with $\Delta_\tau, \eps_{r_{\tau-1}}, \eps_{w_\tau}$ defined in
Section~\ref{sec:filtration} and $\{y_\tau\}$ the ghost-iterate sequence
\eqref{eq:ghost-recursion} driven by $\xi_\tau = \gamma\Delta_\tau$.

\end{theorem}

\begin{proof}
The proof proceeds in steps. 

\paragraph{Step 1: descent identity for one iteration.}
Apply Lemma~\ref{lem:two-step-descent} with \[z = r_{\tau-1},\; 
\zeta = \gamma\widehat G_m^{(\tau,1)} = \gamma\Sigma\widehat F_m^{(\tau,1)}(r_{\tau-1}),\;
\eta = \gamma\widehat G_m^{(\tau,2)} = \gamma\Sigma\widehat F_m^{(\tau,2)}(w_\tau),\]
so $w = w_\tau$ and $r_+ = r_\tau$.  By \eqref{eq:two-step-b}, for all $u \in D$, we have that 
\begin{equation}\label{eq:descent-one-step}
V(r_\tau, u) - V(r_{\tau-1}, u) \le \langle\gamma\widehat G_m^{(\tau,2)}, u - w_\tau\rangle_{\Sigma^+} + \frac{1}{2}\|\gamma\widehat G_m^{(\tau,1)} - \gamma\widehat G_m^{(\tau,2)}\|_{\Sigma^+}^2 - \frac{1}{2}\|w_\tau - r_{\tau-1}\|_{\Sigma^+}^2.
\end{equation}

\paragraph{Step 2: bound the difference of oracle calls.}
By the triangle inequality, we have
\begin{align*}
\|\widehat G_m^{(\tau,1)} - \widehat G_m^{(\tau,2)}\|_{\Sigma^+}
&\le \|\widehat G_m^{(\tau,1)} - G(r_{\tau-1})\|_{\Sigma^+} + \|G(r_{\tau-1}) - G(w_\tau)\|_{\Sigma^+} + \|G(w_\tau) - \widehat G_m^{(\tau,2)}\|_{\Sigma^+} \\
&= \eps_{r_{\tau-1}} + \|G(r_{\tau-1}) - G(w_\tau)\|_{\Sigma^+} + \eps_{w_\tau} \\
&\leq \eps_{r_{\tau-1}} + L_G\|r_{\tau-1} - w_\tau\|_{\Sigma^+} + \eps_{w_\tau}
\end{align*}
using Lipschitz of $G$ in $\Sigma^+$ metric (Lemma~\ref{lem:G-lip-Sigma+}).
Using $(a+b+c)^2 \le 3(a^2+b^2+c^2)$, we get that
\begin{equation}\label{eq:diff-squared}
\|\widehat G_m^{(\tau,1)} - \widehat G_m^{(\tau,2)}\|_{\Sigma^+}^2 \le 3\eps_{r_{\tau-1}}^2 + 3L_G^2\|r_{\tau-1} - w_\tau\|_{\Sigma^+}^2 + 3\eps_{w_\tau}^2.
\end{equation}
Substituting into \eqref{eq:descent-one-step}, we have that 
\begin{align}
V(r_\tau, u) - V(r_{\tau-1}, u) &\le \langle\gamma\widehat G_m^{(\tau,2)}, u - w_\tau\rangle_{\Sigma^+} + \frac{3\gamma^2}{2}\bigl(\eps_{r_{\tau-1}}^2 + L_G^2\|r_{\tau-1} - w_\tau\|_{\Sigma^+}^2 + \eps_{w_\tau}^2\bigr)\\
&\qquad- \frac{1}{2}\|w_\tau - r_{\tau-1}\|_{\Sigma^+}^2.
\end{align}
The $\|w_\tau - r_{\tau-1}\|^2$ terms combine: coefficient $-\frac{1}{2} + \frac{3\gamma^2 L_G^2}{2}$.
Since $\gamma \leq 1/(\sqrt 3\,L_G)$ by design, we have that $3\gamma^2 L_G^2 \le 1$, so this coefficient is
non-positive and dropping the term, we have that 
\begin{equation}\label{eq:descent-after-drop}
V(r_\tau, u) - V(r_{\tau-1}, u) \leq \langle\gamma\widehat G_m^{(\tau,2)}, u - w_\tau\rangle_{\Sigma^+} + \frac{3\gamma^2}{2}\bigl(\eps_{r_{\tau-1}}^2 + \eps_{w_\tau}^2\bigr).
\end{equation}

\paragraph{Step 3: relate the inner product to $F_\Risk$.}
Recall $\widehat G_m^{(\tau,2)} = \Sigma\widehat F_m^{(\tau,2)}(w_\tau)$, so that
\begin{align*}
\langle\widehat G_m^{(\tau,2)}, u - w_\tau\rangle_{\Sigma^+} &= (u - w_\tau)^\top \Sigma^+ \Sigma \widehat F_m^{(\tau,2)} = (u - w_\tau)^\top \widehat F_m^{(\tau,2)},
\end{align*}
using $\Sigma^+\Sigma = \Pi_\Wspace$ and $u - w_\tau \in \Wspace$ (both in
$\theta_{\rm ref} + \Wspace$).  Decompose $\widehat F_m^{(\tau,2)} = F_\Risk(w_\tau) - \Delta_\tau^F$
where $\Delta_\tau^F := F_\Risk(w_\tau) - \widehat F_m^{(\tau,2)}$ (the
unpreconditioned discrepancy).  Then
\begin{align*}
\langle\widehat G_m^{(\tau,2)}, u - w_\tau\rangle_{\Sigma^+} = \langle F_\Risk(w_\tau), u - w_\tau\rangle - \langle \Delta_\tau^F, u - w_\tau\rangle.
\end{align*}
Note that $\Delta_\tau$ in Section~\ref{sec:filtration} was defined as the
preconditioned discrepancy $G(w_\tau) - \widehat G_m^{(\tau,2)} = \Sigma\Delta_\tau^F$.
The relation between the two: for $u, w_\tau \in \theta_{\rm ref} + \Wspace$,
\(\langle\Delta_\tau, u - w_\tau\rangle_{\Sigma^+} = (u-w_\tau)^\top\Sigma^+\Sigma\Delta_\tau^F = (u - w_\tau)^\top\Delta_\tau^F = \langle\Delta_\tau^F, u - w_\tau\rangle\).
Hence we have that 
\begin{equation}\label{eq:inner-prod-decomp}
\langle\widehat G_m^{(\tau,2)}, u - w_\tau\rangle_{\Sigma^+} = \langle F_\Risk(w_\tau), u - w_\tau\rangle - \langle\Delta_\tau, u - w_\tau\rangle_{\Sigma^+}.
\end{equation}
By substituting into \eqref{eq:descent-after-drop}, we get that 
\begin{equation}\label{eq:descent-with-F}
V(r_\tau, u) - V(r_{\tau-1}, u) \le \gamma\langle F_\Risk(w_\tau), u - w_\tau\rangle - \langle\gamma\Delta_\tau, u - w_\tau\rangle_{\Sigma^+} + \frac{3\gamma^2}{2}(\eps_{r_{\tau-1}}^2 + \eps_{w_\tau}^2).
\end{equation}

\paragraph{Step 4: telescoping.} Summing \eqref{eq:descent-with-F} over $\tau = 1, \ldots, t$ yields
\begin{equation}\label{eq:telescoped}
V(r_t, u) - V(r_0, u) \le \sum_{\tau=1}^t \gamma\langle F_\Risk(w_\tau), u - w_\tau\rangle - \sum_{\tau=1}^t\langle\gamma\Delta_\tau, u - w_\tau\rangle_{\Sigma^+} + \sum_{\tau=1}^t \frac{3\gamma^2}{2}(\eps_{r_{\tau-1}}^2 + \eps_{w_\tau}^2).
\end{equation}
By rearranging, with $V(r_t, u) \ge 0$
 and $V(r_0, u) \le \Theta(z_c)$
since $r_0 = z_c$, we have that
\begin{equation}\label{eq:telescoped-rearranged}\sum_\tau \gamma\langle F_\Risk(w_\tau), w_\tau - u\rangle \le \Theta(z_c) + \sum_\tau\langle\gamma\Delta_\tau, u - w_\tau\rangle_{\Sigma^+} + \sum_\tau\frac{3\gamma^2}{2}(\eps_{r_{\tau-1}}^2 + \eps_{w_\tau}^2). 
\end{equation}

\paragraph{Step 5: ghost-iterate decomposition (sup over $u$).}
The quantity $\sum_\tau \langle\gamma\Delta_\tau, u - w_\tau\rangle_{\Sigma^+}$
on the right of \eqref{eq:telescoped-rearranged} depends on $u$; to take
sup over $u$ uniformly, decompose:
\begin{equation}\label{eq:ghost-decomp}
\sum_\tau \langle\gamma\Delta_\tau, u - w_\tau\rangle_{\Sigma^+} = \sum_\tau\langle\gamma\Delta_\tau, u - y_{\tau-1}\rangle_{\Sigma^+} + \sum_\tau\langle\gamma\Delta_\tau, y_{\tau-1} - w_\tau\rangle_{\Sigma^+}.
\end{equation}

For the first sum, applying Lemma~\ref{lem:ghost-iterate} with $\xi_\tau = \gamma\Delta_\tau$
and noting $-\langle\xi, u - y_{\tau-1}\rangle = \langle\xi, y_{\tau-1} - u\rangle$, we have that 
\begin{equation}\label{eq:ghost-applied}
\sum_\tau\langle\gamma\Delta_\tau, y_{\tau-1} - u\rangle_{\Sigma^+} \le V(y_0, u) + \frac{1}{2}\sum_\tau\|\gamma\Delta_\tau\|_{\Sigma^+}^2 \leq \Theta(z_c) + \frac{\gamma^2}{2}\sum_\tau\|\Delta_\tau\|_{\Sigma^+}^2,
\end{equation}
using $y_0 = r_0 = z_c$ and $V(z_c, u) \le \Theta(z_c)$ for all $u \in D$.
Hence we have the bound \[\sum_\tau\langle\gamma\Delta_\tau, u - y_{\tau-1}\rangle_{\Sigma^+} \leq \Theta(z_c) + \frac{\gamma^2}{2}\sum_\tau\|\Delta_\tau\|_{\Sigma^+}^2.\]
Substitute back into \eqref{eq:telescoped-rearranged} via \eqref{eq:ghost-decomp}, we get that 
\begin{equation}
\begin{aligned}
\sum_\tau \gamma\langle F_\Risk(w_\tau), w_\tau - u\rangle &\leq \Theta(z_c) + \Theta(z_c) + \frac{\gamma^2}{2}\sum_\tau\|\Delta_\tau\|_{\Sigma^+}^2 + \sum_\tau\langle\gamma\Delta_\tau, y_{\tau-1} - w_\tau\rangle_{\Sigma^+}\\
&\qquad+ \sum_\tau \frac{3\gamma^2}{2}(\eps_{r_{\tau-1}}^2 + \eps_{w_\tau}^2). \label{eq:telescoped-final-pre}
\end{aligned}
\end{equation}

\paragraph{Step 6: bound $\|\Delta_\tau\|_{\Sigma^+}^2$.}
By definition $\Delta_\tau = G(w_\tau) - \widehat G_m^{(\tau,2)} = -\Sigma b_m(w_\tau) - \Sigma\zeta_m^{(\tau,2)}$ so that
\[\|\Delta_\tau\|_{\Sigma^+} = \|\Sigma\zeta_m^{(\tau,2)} + \Sigma b_m(w_\tau)\|_{\Sigma^+}.\]
By the definition of $\eps_{w_\tau} := \|\widehat G_m^{(\tau,2)}(w_\tau) - G(w_\tau)\|_{\Sigma^+}$, we have that 
$\|\Delta_\tau\|_{\Sigma^+}^2 = \eps_{w_\tau}^2$.
Substituting into \eqref{eq:telescoped-final-pre} yields
\begin{equation}\label{eq:final-pre}
\sum_\tau \gamma\langle F_\Risk(w_\tau), w_\tau - u\rangle \leq 2\Theta(z_c) + \frac{\gamma^2}{2}\sum_\tau\eps_{w_\tau}^2 + \sum_\tau\langle\gamma\Delta_\tau, y_{\tau-1} - w_\tau\rangle_{\Sigma^+} + \sum_\tau \frac{3\gamma^2}{2}(\eps_{r_{\tau-1}}^2 + \eps_{w_\tau}^2).
\end{equation}
Combining the $\eps_{w_\tau}^2$ coefficients, we get that  $\tfrac{\gamma^2}{2} + \tfrac{3\gamma^2}{2} = 2\gamma^2$.  Combining all of this into $\Gamma(t)$, we have 
\begin{equation}\label{eq:Gamma-realized}
\Gamma(t) := 2\Theta(z_c) + \sum_\tau \frac{3\gamma^2}{2}\eps_{r_{\tau-1}}^2 + 2\gamma^2\sum_\tau\eps_{w_\tau}^2 + \sum_\tau\langle\gamma\Delta_\tau, y_{\tau-1} - w_\tau\rangle_{\Sigma^+}.
\end{equation}

\paragraph{Step 7: convert to $\Errvi$.}
By monotonicity of $F_\Risk$, for all $u, w \in D$, we have
$\langle F_\Risk(u), w - u\rangle \le \langle F_\Risk(w), w - u\rangle$.  Hence
\[\langle F_\Risk(u), w_\tau - u\rangle \le \langle F_\Risk(w_\tau), w_\tau - u\rangle,\]
and summing then averaging by $\sum\gamma_\tau = t\gamma$, we get that
\[
\langle F_\Risk(u), \widehat z_t - u\rangle = (t\gamma)^{-1}\sum_\tau \gamma\langle F_\Risk(u), w_\tau - u\rangle \le (t\gamma)^{-1}\sum_\tau\gamma\langle F_\Risk(w_\tau), w_\tau - u\rangle.
\]
By \eqref{eq:final-pre}, the right side is $\le (t\gamma)^{-1}\Gamma(t)$ for
\emph{every fixed} $u \in D$.  Take supremum over $u$ to get
\[\Errvi(\widehat z_t) = \sup_u \langle F_\Risk(u), \widehat z_t - u\rangle \le (t\gamma)^{-1}\Gamma(t),\]
which is \eqref{eq:thm-deterministic-bound}.
\end{proof}

\subsection{Monotone setting: bounding the expected gap}
\label{sec:thm-expected-gap}
Building on  \cite{juditsky2011}, we prove the convergence of stochastic mirror prox (extragradient) in the montone setting with a biased oracle for learning in risk-sensitive preference games. This requires handling the  terms in both the bias and variance from estimating the risk measure. 
\begin{theorem}[Stochastic Mirror-Prox, monotone case, biased oracle]
\label{thm:expected-gap}
\label{thm:biased-merely-monotone}
Under the assumptions of Theorem~\ref{thm:deterministic} and the oracle
satisfies \eqref{eq:tag-O1} and \eqref{eq:tag-O2}, with the variance and bias
in the $\Sigma^+$ metric bounded by $\widetilde B_m^2 := \sigma_{\max} B_m^2$
and $\widetilde V_m := \sigma_{\max} V_m$,
the estimate holds:
\begin{equation}\label{eq:thm-expected-gap-bound}
\E[\Errvi(\widehat z_t)] \leq \frac{\Omega^2}{t\gamma} + \frac{7\gamma}{2}\bigl(\widetilde B_m^2 + \widetilde V_m\bigr) + 2\Omega \widetilde B_m.
\end{equation}
\end{theorem}

\begin{proof}
Take expectations of $\Gamma(t)/(t\gamma)$ from \eqref{eq:Gamma-realized}, we have that 
\[
\E[\Errvi(\widehat z_t)] \le \frac{1}{t\gamma}\E[\Gamma(t)] = \frac{2\Theta(z_c)}{t\gamma} + \frac{1}{t\gamma}\sum_\tau\Bigl[\tfrac{3\gamma^2}{2}\E\eps_{r_{\tau-1}}^2 + 2\gamma^2\E\eps_{w_\tau}^2 + \E\langle\gamma\Delta_\tau, y_{\tau-1} - w_\tau\rangle_{\Sigma^+}\Bigr].
\]

\paragraph{Bound 1: $\E\eps_{r_{\tau-1}}^2$ and $\E\eps_{w_\tau}^2$.}
By \eqref{eq:eps-r}-\eqref{eq:eps-w}, we have that 
$\E\eps_{r_{\tau-1}}^2 \le \widetilde B_m^2 + \widetilde V_m$, and 
$\E\eps_{w_\tau}^2 \le \widetilde B_m^2 + \widetilde V_m$.
Sum the noise contributions to get 
\[
\sum_\tau \bigl[\tfrac{3\gamma^2}{2}\E\eps_{r_{\tau-1}}^2 + 2\gamma^2\E\eps_{w_\tau}^2\bigr] \le t\gamma^2 \cdot \tfrac{7}{2}(\widetilde B_m^2 + \widetilde V_m).
\]
Divide by $t\gamma$ to get that  $\le \tfrac{7\gamma}{2}(\widetilde B_m^2 + \widetilde V_m)$.

\paragraph{Bound 2: $\E\langle\gamma\Delta_\tau, y_{\tau-1} - w_\tau\rangle_{\Sigma^+}$.}
Recall $w_\tau$ is $\F_\tau$-measurable (uses the first oracle call of step
$\tau$); $y_{\tau-1}$ is $\G_{\tau-1}$-measurable (depends on $\Delta_1, \ldots, \Delta_{\tau-1}$
through the ghost recursion); $\Delta_\tau$ is $\G_\tau$-measurable but not
$\F_\tau$-measurable (uses the second oracle call of step $\tau$).

By the tower property of expectations, we have that 
\[
\E\langle\gamma\Delta_\tau, y_{\tau-1} - w_\tau\rangle_{\Sigma^+} = \gamma\E\langle\E[\Delta_\tau \mid \F_\tau], y_{\tau-1} - w_\tau\rangle_{\Sigma^+}.
\]
Here $y_{\tau-1}$ is $\G_{\tau-1}$-measurable and hence $\F_\tau$-measurable;
$w_\tau$ is $\F_\tau$-measurable.

By \eqref{eq:Delta-bias}, the upper bound holds: $\|\E[\Delta_\tau \mid \F_\tau]\|_{\Sigma^+} \leq\widetilde B_m$.
 Cauchy-Schwarz in $\Sigma^+$ implies that   
\[|\langle\E[\Delta_\tau \mid \F_\tau], y_{\tau-1} - w_\tau\rangle_{\Sigma^+}| \le \widetilde B_m \|y_{\tau-1} - w_\tau\|_{\Sigma^+} \le \widetilde B_m \cdot 2\Omega,\]
since $y_{\tau-1}, w_\tau \in D$ and the $\Sigma^+$-diameter of $D$ is $2\Omega$.
Hence
\[
|\E\langle\gamma\Delta_\tau, y_{\tau-1} - w_\tau\rangle_{\Sigma^+}| \le 2\gamma\widetilde B_m\Omega.
\]
Summing over $\tau$ and dividing by $t\gamma$, we have $\le 2\widetilde B_m\Omega$.

\paragraph{Combining the bounds.} Now we combine these bounds to get 
$\E[\Errvi(\widehat z_t)] \le \frac{2\Theta(z_c)}{t\gamma} + \frac{7\gamma}{2}(\widetilde B_m^2 + \widetilde V_m) + 2\Omega\widetilde B_m$.
Using $2\Theta(z_c) = \Omega^2$ (by \eqref{eq:Omega-def}), we have 
\[
\E[\Errvi(\widehat z_t)] \le \frac{\Omega^2}{t\gamma} + \frac{7\gamma}{2}(\widetilde B_m^2 + \widetilde V_m) + 2\Omega\widetilde B_m,
\]
which is \eqref{eq:thm-expected-gap-bound}.
\end{proof}

\subsubsection{Corollary: optimal step size}
Theorem~\ref{thm:expected-gap} bounds the expected gap as a function
of the step size $\gamma$, leaving open the question of how to choose
$\gamma$ to minimize the bound.  The optimal choice depends on whether
the dominant source of error is the operator's Lipschitz continuity
(favoring smaller $\gamma$) or the noise from finite-sample
estimation of the risk-adjusted operator (favoring a step size that
trades off iteration count against per-step variance).  The following
corollary makes this trade-off explicit.

\begin{corollary}[Optimal step size, monotone case]
\label{cor:optimal-step}
Choosing
\begin{equation}\label{eq:gamma-optimal}
\gamma = \min\Biggl\{\frac{1}{\sqrt 3 L_G},\ \Omega\sqrt{\frac{2}{7t(\widetilde B_m^2 + \widetilde V_m)}}\Biggr\}
\end{equation}
in Theorem~\ref{thm:expected-gap} gives
\begin{equation}\label{eq:expected-gap-optimal}
\E[\Errvi(\widehat z_t)] \le \max\Biggl\{\frac{7 L_G \Omega^2}{4 t},\ 7\Omega\sqrt{\frac{\widetilde B_m^2 + \widetilde V_m}{3t}}\Biggr\} + 2\Omega\widetilde B_m.
\end{equation}
The leading rate is $\mc{O}(t^{-1/2})$ in the variance-dominated regime and
$\mc{O}(t^{-1})$ in the Lipschitz-dominated regime; the bias contributes a
constant $1/m$-floor independent of $t$.
\end{corollary}
This follows from standard step-size optimization of the bound \eqref{eq:thm-expected-gap-bound}
treating $\widetilde B_m^2 + \widetilde V_m$ as the noise variance and
$L_G \Omega$ as the Lipschitz continuity contribution.

\paragraph{Substituting concrete entropic constants.}
Using $\widetilde B_m^2 = \sigma_{\max} B_m^2 = \mc{O}(\sigma_{\max}|\Yspace|/(\beta^2 m^2))$
and $\widetilde V_m = \sigma_{\max} V_m = \mc{O}(\sigma_{\max}|\Yspace|/(\beta^2 m))$
from Proposition~\ref{prop:operator-O1-O2}, the optimal-step-size bound becomes
\begin{equation}\label{eq:expected-gap-entropic}
\E[\Errvi(\widehat z_t)] = O\Bigl(\frac{L_G\Omega^2}{t}\Bigr) + O\Bigl(\Omega\sqrt{\frac{\sigma_{\max}|\Yspace|}{\beta^2 m t}}\Bigr) + O\Bigl(\frac{\Omega\sqrt{\sigma_{\max}|\Yspace|}}{\beta m}\Bigr).
\end{equation}
The bias contribution scales as $1/m$ (last term); the variance contribution
scales as $1/\sqrt{m t}$ (middle term); the deterministic Lipschitz
contribution scales as $1/t$ (first term).  For sufficiently large $m$, the
$1/\sqrt{mt}$ and $1/m$ terms dominate.

\subsection{Strongly monotone setting}
\label{sec:thm-strong-mono}

We now prove last-iterate linear contraction in the strongly monotone case.
The result is not in prior work (for example, \cite{juditsky2011}  treats only the monotone case), but the proof
uses the same descent identity (Lemma~\ref{lem:two-step-descent}), specialized
to $u = \theta^\star$ and combined with strong monotonicity to extract a
contraction factor.

\paragraph{Effective strong-monotonicity modulus in $\Sigma^+$.}
By \ref{P-strong}, $F_\Risk$ is $\mu_\Risk$-strongly monotone in the standard
inner product.  Translating to the $\Sigma^+$ inner product on $\Wspace$, we have that 
\begin{align*}
\langle G(\theta_1) - G(\theta_2), \theta_1 - \theta_2\rangle_{\Sigma^+} &= \langle F_\Risk(\theta_1) - F_\Risk(\theta_2), \theta_1 - \theta_2\rangle\\
&\ge \mu_\Risk\|\theta_1 - \theta_2\|_2^2 \\
&\ge \mu_\Risk\sigma_{\min}\|\theta_1 - \theta_2\|_{\Sigma^+}^2,
\end{align*}
where the first equality uses Lemma~\ref{lem:precond-restate}(i) and the
second inequality uses \eqref{eq:metric-equiv}.  Define the effective
strong-monotonicity modulus
\begin{equation}\label{eq:mu-tilde-def}
\widetilde\mu := \mu_\Risk\sigma_{\min}.
\end{equation}
Then $G$ is $\widetilde\mu$-strongly monotone in the $\Sigma^+$ inner product
on $\Wspace$.

\begin{theorem}[Stochastic Mirror-Prox, strongly monotone, biased oracle]
\label{thm:strong-mono}
Suppose $F_\Risk$ is $\mu_\Risk$-strongly monotone (\ref{P-strong}) and
$\ell_\Risk$-Lipschitz (\ref{P-lip}) on $\theta_{\rm ref} + \Wspace$, the
oracle satisfies (O1)-(O2) of Section~\ref{sec:bias-analysis}, and the algorithm
\eqref{eq:eg-extrap}-\eqref{eq:eg-corr} is run with constant step size
$\gamma_\tau \equiv \gamma$ satisfying
\begin{equation}\label{eq:gamma-strong}
\gamma \leq \min\Biggl\{\frac{1}{4\widetilde\mu},\ \frac{1}{\sqrt 6\,L_G}\Biggr\}, \qquad L_G = \sigma_{\max}\ell_\Risk.
\end{equation}
Then for all $T \ge 1$, the estimate holds:
\begin{equation}\label{eq:thm-strong-mono-bound}
\E\|\theta_T - \theta^\star\|_{\Sigma^+}^2 \leq (1 - \gamma\widetilde\mu)^T \|\theta_0 - \theta^\star\|_{\Sigma^+}^2 + \frac{4\Omega\widetilde B_m}{\widetilde\mu} + \frac{6\gamma\bigl(\widetilde B_m^2 + \widetilde V_m\bigr)}{\widetilde\mu}.
\end{equation}
\end{theorem}

\begin{proof}
We work step by step from the two-step descent identity
(Lemma~\ref{lem:two-step-descent}), specialized to $u = \theta^\star$ and
followed by strong monotonicity.

\paragraph{Step 1: descent identity at $u = \theta^\star$.}
First apply \eqref{eq:two-step-b} of Lemma~\ref{lem:two-step-descent} with
$z = r_{\tau-1}$, $\zeta = \gamma\widehat G_m^{(\tau,1)}$, $\eta = \gamma\widehat G_m^{(\tau,2)}$
(so $w = w_\tau$, $r_+ = r_\tau$), and $u = \theta^\star$. This gives
\begin{equation}\label{eq:strong-step-1}
V(r_\tau, \theta^\star) - V(r_{\tau-1}, \theta^\star) \le \gamma\langle\widehat G_m^{(\tau,2)}, \theta^\star - w_\tau\rangle_{\Sigma^+} + \tfrac{\gamma^2}{2}\|\widehat G_m^{(\tau,1)} - \widehat G_m^{(\tau,2)}\|_{\Sigma^+}^2 - \tfrac{1}{2}\|w_\tau - r_{\tau-1}\|_{\Sigma^+}^2.
\end{equation}

\paragraph{Step 2: decompose the inner product term.}
Decompose $\widehat G_m^{(\tau,2)} = G(w_\tau) - \Sigma b_m(w_\tau) - \Sigma\zeta_m^{(\tau,2)}$
(unprecondioned bias $b_m$, noise $\zeta_m^{(\tau,2)}$), so that 
\begin{align*}
\langle\widehat G_m^{(\tau,2)}, \theta^\star - w_\tau\rangle_{\Sigma^+} &= \langle G(w_\tau), \theta^\star - w_\tau\rangle_{\Sigma^+} - \langle\Sigma b_m(w_\tau), \theta^\star - w_\tau\rangle_{\Sigma^+} - \langle\Sigma\zeta_m^{(\tau,2)}, \theta^\star - w_\tau\rangle_{\Sigma^+}.
\end{align*}
Using Lemma~\ref{lem:precond-restate} part $(i)$, we have that  $\langle\Sigma v, w\rangle_{\Sigma^+} = \langle v, w\rangle$
for $w \in \Wspace$ with $v$ arbitrary; we use $w = \theta^\star - w_\tau \in \Wspace$.
Apply this three times to get that 
\begin{equation}\label{eq:strong-decomp}
\langle\widehat G_m^{(\tau,2)}, \theta^\star - w_\tau\rangle_{\Sigma^+} = \langle F_\Risk(w_\tau), \theta^\star - w_\tau\rangle - \langle b_m(w_\tau), \theta^\star - w_\tau\rangle - \langle\zeta_m^{(\tau,2)}, \theta^\star - w_\tau\rangle.
\end{equation}

\paragraph{Step 3: apply strong monotonicity.}
The risk-adjusted equilibrium  condition gives $F_\Risk(\theta^\star) \in \ker\Sigma$, i.e.,
$F_\Risk(\theta^\star) = c\mathbf 1$ for some $c \in \R$.  Since $\theta^\star - w_\tau \in \Wspace$
and $\mathbf 1 \perp \Wspace$, we have that 
\[\langle F_\Risk(\theta^\star), \theta^\star - w_\tau\rangle = 0.\]
Hence
\(\langle F_\Risk(w_\tau), \theta^\star - w_\tau\rangle = \langle F_\Risk(w_\tau) - F_\Risk(\theta^\star), \theta^\star - w_\tau\rangle = -\langle F_\Risk(w_\tau) - F_\Risk(\theta^\star), w_\tau - \theta^\star\rangle\).
By \ref{P-strong}, the latter inner product is $\ge \mu_\Risk\|w_\tau - \theta^\star\|_2^2 \ge \widetilde\mu\|w_\tau - \theta^\star\|_{\Sigma^+}^2$
(using \eqref{eq:metric-equiv} and $\widetilde\mu = \mu_\Risk\sigma_{\min}$).
Thus we deduce that 
\begin{equation}\label{eq:strong-mono-applied}
\langle F_\Risk(w_\tau), \theta^\star - w_\tau\rangle \le -\widetilde\mu\|w_\tau - \theta^\star\|_{\Sigma^+}^2.
\end{equation}
Substitute \eqref{eq:strong-mono-applied} into \eqref{eq:strong-decomp} leads to 
\begin{equation}\label{eq:strong-decomp-2}
\langle\widehat G_m^{(\tau,2)}, \theta^\star - w_\tau\rangle_{\Sigma^+} \le -\widetilde\mu\|w_\tau - \theta^\star\|_{\Sigma^+}^2 + \langle b_m(w_\tau), w_\tau - \theta^\star\rangle + \langle\zeta_m^{(\tau,2)}, w_\tau - \theta^\star\rangle.
\end{equation}

\paragraph{Step 4: bound the Lipschitz / variance squared term.}
By the triangle inequality and Lipschitz continuity of $G$ in $\Sigma^+$ metric
(Lemma~\ref{lem:G-lip-Sigma+}), we have that 
\begin{align*}
\|\widehat G_m^{(\tau,1)} - \widehat G_m^{(\tau,2)}\|_{\Sigma^+}
&\leq \|\widehat G_m^{(\tau,1)} - G(r_{\tau-1})\|_{\Sigma^+} + \|G(r_{\tau-1}) - G(w_\tau)\|_{\Sigma^+} + \|G(w_\tau) - \widehat G_m^{(\tau,2)}\|_{\Sigma^+}\\
&= \eps_{r_{\tau-1}} + L_G\|r_{\tau-1} - w_\tau\|_{\Sigma^+} + \eps_{w_\tau}.
\end{align*}
Using the property $(a+b+c)^2 \le 3(a^2 + b^2 + c^2)$, we deduce that 
\begin{equation}\label{eq:strong-Lip-sq}
\|\widehat G_m^{(\tau,1)} - \widehat G_m^{(\tau,2)}\|_{\Sigma^+}^2 \le 3\eps_{r_{\tau-1}}^2 + 3 L_G^2\|r_{\tau-1} - w_\tau\|_{\Sigma^+}^2 + 3\eps_{w_\tau}^2.
\end{equation}

\paragraph{Step 5: combine.}
Substitute \eqref{eq:strong-decomp-2} and \eqref{eq:strong-Lip-sq} into
\eqref{eq:strong-step-1} to get that 
\begin{align}
V(r_\tau, \theta^\star) - V(r_{\tau-1}, \theta^\star) &\leq -\gamma\widetilde\mu\|w_\tau - \theta^\star\|_{\Sigma^+}^2 \nonumber\\
&\quad + \gamma\langle b_m(w_\tau), w_\tau - \theta^\star\rangle + \gamma\langle\zeta_m^{(\tau,2)}, w_\tau - \theta^\star\rangle \nonumber\\
&\quad + \tfrac{3\gamma^2}{2}\bigl(\eps_{r_{\tau-1}}^2 + L_G^2\|r_{\tau-1} - w_\tau\|_{\Sigma^+}^2 + \eps_{w_\tau}^2\bigr) \nonumber\\
&\quad - \tfrac{1}{2}\|w_\tau - r_{\tau-1}\|_{\Sigma^+}^2. \label{eq:strong-combined}
\end{align}

\paragraph{Step 6: bridge $\|w_\tau - \theta^\star\|^2$ to $V(r_{\tau-1}, \theta^\star)$.}
Apply the elementary inequality $\|a + b\|^2 \le 2\|a\|^2 + 2\|b\|^2$ with
$a = r_{\tau-1} - w_\tau$, $b = w_\tau - \theta^\star$ (both in $\Wspace$) to get that 
\[\|r_{\tau-1} - \theta^\star\|_{\Sigma^+}^2 \le 2\|r_{\tau-1} - w_\tau\|_{\Sigma^+}^2 + 2\|w_\tau - \theta^\star\|_{\Sigma^+}^2,\]
which rearranges to
\[\|w_\tau - \theta^\star\|_{\Sigma^+}^2 \ge \tfrac{1}{2}\|r_{\tau-1} - \theta^\star\|_{\Sigma^+}^2 - \|r_{\tau-1} - w_\tau\|_{\Sigma^+}^2,\]
i.e., using $V(\cdot, \cdot) = \tfrac{1}{2}\|\cdot - \cdot\|_{\Sigma^+}^2$, we have 
\begin{equation}\label{eq:bridge}
\|w_\tau - \theta^\star\|_{\Sigma^+}^2 \geq V(r_{\tau-1}, \theta^\star) - \|r_{\tau-1} - w_\tau\|_{\Sigma^+}^2.
\end{equation}
Hence
\begin{equation}\label{eq:bridge-applied}
-\gamma\widetilde\mu\|w_\tau - \theta^\star\|_{\Sigma^+}^2 \le -\gamma\widetilde\mu V(r_{\tau-1}, \theta^\star) + \gamma\widetilde\mu\|r_{\tau-1} - w_\tau\|_{\Sigma^+}^2.
\end{equation}
Substituting \eqref{eq:bridge-applied} into \eqref{eq:strong-combined}, we get that 
\begin{align}
V(r_\tau, \theta^\star) &\leq V(r_{\tau-1}, \theta^\star) - \gamma\widetilde\mu V(r_{\tau-1}, \theta^\star) + \gamma\widetilde\mu\|r_{\tau-1} - w_\tau\|_{\Sigma^+}^2 \nonumber\\
&\quad + \gamma\langle b_m(w_\tau), w_\tau - \theta^\star\rangle + \gamma\langle\zeta_m^{(\tau,2)}, w_\tau - \theta^\star\rangle \nonumber\\
&\quad + \tfrac{3\gamma^2}{2}\bigl(\eps_{r_{\tau-1}}^2 + L_G^2\|r_{\tau-1} - w_\tau\|_{\Sigma^+}^2 + \eps_{w_\tau}^2\bigr) \nonumber\\
&\quad - \tfrac{1}{2}\|w_\tau - r_{\tau-1}\|_{\Sigma^+}^2. \label{eq:after-bridge}
\end{align}

\paragraph{Step 7: drop the $\|r_{\tau-1} - w_\tau\|^2$ term under the step-size constraint.}
The coefficient on $\|r_{\tau-1} - w_\tau\|_{\Sigma^+}^2$ in \eqref{eq:after-bridge} is
\[
C_w := \gamma\widetilde\mu + \tfrac{3\gamma^2 L_G^2}{2} - \tfrac{1}{2}.
\]
Under \eqref{eq:gamma-strong}, the bound $\gamma \leq 1/(4\widetilde\mu)$ gives $\gamma\widetilde\mu \le 1/4$ and
$\gamma \leq 1/(\sqrt 6\,L_G)$ gives $\gamma^2 L_G^2 \le 1/6$, so that
$\tfrac{3\gamma^2 L_G^2}{2} \le 1/4$.  Hence $C_w \le 1/4 + 1/4 - 1/2 = 0$.
We drop the $\|r_{\tau-1} - w_\tau\|_{\Sigma^+}^2$ term, to get that 
\begin{equation}\label{eq:strong-after-drop}
V(r_\tau, \theta^\star) \leq (1 - \gamma\widetilde\mu) V(r_{\tau-1}, \theta^\star) + \gamma\langle b_m(w_\tau), w_\tau - \theta^\star\rangle + \gamma\langle\zeta_m^{(\tau,2)}, w_\tau - \theta^\star\rangle + \tfrac{3\gamma^2}{2}(\eps_{r_{\tau-1}}^2 + \eps_{w_\tau}^2).
\end{equation}

\paragraph{Step 8: take expectation.}
Take $\E[\cdot \mid \F_\tau]$ of \eqref{eq:strong-after-drop}.  Recall the
filtration structure (Section~\ref{sec:filtration}): $r_{\tau-1}$ is
$\G_{\tau-1}$-measurable, $w_\tau$ is $\F_\tau$-measurable.  $\eps_{r_{\tau-1}}$
is $\F_\tau$-measurable.  $\zeta_m^{(\tau,2)}$ is independent of $\F_\tau$.
$\eps_{w_\tau}^2$ depends on $w_\tau$ (via the bias) and on $\zeta_m^{(\tau,2)}$
(via the noise); given $\F_\tau$, $w_\tau$ is fixed.

\noindent\textit{Term 1: bias cross.}
The bias $b_m(w_\tau)$ is deterministic given $w_\tau$, hence given $\F_\tau$.  Thus
$\E[\langle b_m(w_\tau), w_\tau - \theta^\star\rangle \mid \F_\tau] = \langle b_m(w_\tau), w_\tau - \theta^\star\rangle$
(deterministic).  Bound the bias as follows:
\begin{align}
|\langle b_m(w_\tau), w_\tau - \theta^\star\rangle| &\le \|b_m(w_\tau)\|_2 \|w_\tau - \theta^\star\|_2 \nonumber\\
&\le B_m \cdot \sqrt{\sigma_{\max}}\|w_\tau - \theta^\star\|_{\Sigma^+} \nonumber\\
&\le \widetilde B_m \cdot 2\Omega, \label{eq:bias-cross-bound}
\end{align}
using \ref{O1-bias} ($\|b_m\|_2 \le B_m$), \eqref{eq:metric-equiv} ($\|v\|_2 \le \sqrt{\sigma_{\max}}\|v\|_{\Sigma^+}$),
$\widetilde B_m = \sqrt{\sigma_{\max}}B_m$, and the $\Sigma^+$-diameter of $D$
is $2\Omega$.

\noindent\textit{Term 2: noise cross.}
The term $\zeta_m^{(\tau,2)}$ is independent of $\F_\tau$ with $\E[\zeta_m^{(\tau,2)}] = 0$.
Thus $\E[\langle\zeta_m^{(\tau,2)}, w_\tau - \theta^\star\rangle \mid \F_\tau] = \langle\E[\zeta_m^{(\tau,2)} \mid \F_\tau], w_\tau - \theta^\star\rangle = \langle 0, w_\tau - \theta^\star\rangle = 0$.

\noindent\textit{Term 3: noise-squared $\eps_{r_{\tau-1}}^2$.}  This is $\F_\tau$-measurable
(uses the first oracle call of step $\tau$, included in $\F_\tau$).  Take
unconditional expectation later.

\noindent\textit{Term 4: noise-squared $\eps_{w_\tau}^2$.}  Recall $\eps_{w_\tau} = \|G(w_\tau) - \widehat G_m^{(\tau,2)}\|_{\Sigma^+}$.
Decompose the operator as follows: $G(w_\tau) - \widehat G_m^{(\tau,2)} = \Sigma b_m(w_\tau) + \Sigma\zeta_m^{(\tau,2)}$. Hence
\[\eps_{w_\tau}^2 = \|\Sigma b_m(w_\tau)\|_{\Sigma^+}^2 + 2\langle\Sigma b_m(w_\tau), \Sigma\zeta_m^{(\tau,2)}\rangle_{\Sigma^+} + \|\Sigma\zeta_m^{(\tau,2)}\|_{\Sigma^+}^2.\]
Conditional on $\F_\tau$, $w_\tau$ is fixed and $\E[\zeta_m^{(\tau,2)} \mid \F_\tau] = 0$,
so the cross term has zero conditional expectation---that is,
\[\E[\eps_{w_\tau}^2 \mid \F_\tau] = \|\Sigma b_m(w_\tau)\|_{\Sigma^+}^2 + \E[\|\Sigma\zeta_m^{(\tau,2)}\|_{\Sigma^+}^2 \mid \F_\tau] \le \widetilde B_m^2 + \widetilde V_m,\]
using \eqref{eq:G-bias}-\eqref{eq:G-var}.

\noindent\textit{Combine.} Finally, we combine these bounds to get that 
\begin{align*}
\E[V(r_\tau, \theta^\star) \mid \F_\tau] &\leq (1 - \gamma\widetilde\mu) V(r_{\tau-1}, \theta^\star) + 2\gamma\Omega\widetilde B_m + 0 + \tfrac{3\gamma^2}{2}\eps_{r_{\tau-1}}^2 + \tfrac{3\gamma^2}{2}(\widetilde B_m^2 + \widetilde V_m).
\end{align*}
Now take unconditional expectation. In particular, $\E\eps_{r_{\tau-1}}^2 \le \widetilde B_m^2 + \widetilde V_m$
by the same decomposition argument applied to $\eps_{r_{\tau-1}}$ (the first
oracle call)---indeed, we have 
\begin{equation}\label{eq:strong-recurrence}
\E V(r_\tau, \theta^\star) \le (1 - \gamma\widetilde\mu)\E V(r_{\tau-1}, \theta^\star) + 2\gamma\Omega\widetilde B_m + 3\gamma^2(\widetilde B_m^2 + \widetilde V_m).
\end{equation}

\paragraph{Step 9: telescope.}
Define $A := 2\gamma\Omega\widetilde B_m + 3\gamma^2(\widetilde B_m^2 + \widetilde V_m)$.  Then
\eqref{eq:strong-recurrence} reads $\Lambda_\tau \le (1 - \gamma\widetilde\mu)\Lambda_{\tau-1} + A$
where $\Lambda_\tau := \E V(r_\tau, \theta^\star)$.  Iterating we have that 
\begin{align*}
\Lambda_T &\le (1 - \gamma\widetilde\mu)^T \Lambda_0 + A\sum_{j=0}^{T-1}(1 - \gamma\widetilde\mu)^j\\
&\le (1 - \gamma\widetilde\mu)^T \Lambda_0 + \frac{A}{\gamma\widetilde\mu}\\
&= (1 - \gamma\widetilde\mu)^T V(r_0, \theta^\star) + \frac{2\Omega\widetilde B_m}{\widetilde\mu} + \frac{3\gamma(\widetilde B_m^2 + \widetilde V_m)}{\widetilde\mu}.
\end{align*}

\paragraph{Step 10: convert to $\|\cdot\|_{\Sigma^+}^2$.}
Since $V(z, u) = \tfrac{1}{2}\|z - u\|_{\Sigma^+}^2$, we have that 
\[
\E\|r_T - \theta^\star\|_{\Sigma^+}^2 = 2\E V(r_T, \theta^\star) \le (1 - \gamma\widetilde\mu)^T \|r_0 - \theta^\star\|_{\Sigma^+}^2 + \frac{4\Omega\widetilde B_m}{\widetilde\mu} + \frac{6\gamma(\widetilde B_m^2 + \widetilde V_m)}{\widetilde\mu},
\]
which is \eqref{eq:thm-strong-mono-bound}.
\end{proof}

\paragraph{Understanding the bound.}
The bound \eqref{eq:thm-strong-mono-bound} has three core terms which can be interpreted or understood as follows. 
\begin{itemize}[itemsep=0pt,  topsep=2pt]
\item \textbf{Linear contraction.}  $(1 - \gamma\widetilde\mu)^T \|r_0 - \theta^\star\|_{\Sigma^+}^2$
contracts geometrically with rate $\gamma\widetilde\mu = \gamma\mu_\Risk\sigma_{\min}$.
The number of iterations to halve the initial-distance contribution is
$T = \log 2/(\gamma\widetilde\mu) = \mc{O}(1/(\gamma\mu_\Risk\sigma_{\min}))$.
\item \textbf{Bias floor.}  $4\Omega\widetilde B_m/\widetilde\mu = \mc{O}(\Omega B_m\sqrt{\sigma_{\max}/\sigma_{\min}}/\mu_\Risk)$.
Linear in $B_m$ (which scales as $1/m$ for entropic, $0$ for CVaR-RU).  The
bias floor is independent of step size $\gamma$ and persists at any
iteration count.  Cannot be reduced by taking smaller $\gamma$; only by
increasing $m$.
\item \textbf{Variance floor.}  $6\gamma(\widetilde B_m^2 + \widetilde V_m)/\widetilde\mu = \mc{O}(\gamma\sigma_{\max}V_m/(\mu_\Risk\sigma_{\min}))$
(the $\widetilde B_m^2$ contribution is dominated by $\widetilde V_m$ for $B_m^2 \le V_m$,
which holds whenever variance $V_m$ is order $1/m$ and bias is also $1/m$,
giving $B_m^2 \sim 1/m^2 \ll V_m \sim 1/m$).  The variance floor scales as
$\gamma/m$; can be reduced by taking smaller $\gamma$ or larger $m$.
\end{itemize}

\begin{corollary}[Optimal step size, strongly monotone case]
\label{cor:strong-optimal-step}
With sample budget $m$ fixed and target accuracy $\delta > 0$, choose
\begin{equation}\label{eq:gamma-strong-optimal}
\gamma = \min\Biggl\{\frac{1}{4\widetilde\mu},\ \frac{1}{\sqrt 6\,L_G},\ \frac{\widetilde\mu\delta}{12(\widetilde B_m^2 + \widetilde V_m)}\Biggr\},
\end{equation}
and run for
\begin{equation}\label{eq:T-strong-optimal}
T = \Biggl\lceil\frac{1}{\gamma\widetilde\mu}\log\frac{2\|r_0 - \theta^\star\|_{\Sigma^+}^2}{\delta}\Biggr\rceil
\end{equation}
iterations.  Then
\[
\E\|r_T - \theta^\star\|_{\Sigma^+}^2 \le \delta + \frac{4\Omega\widetilde B_m}{\widetilde\mu}.
\]
The bias floor $4\Omega\widetilde B_m/\widetilde\mu$ is unavoidable for fixed $m$;
to drive the total error below any $\delta'$, choose $m$ large enough that
$\widetilde B_m \le \widetilde\mu\delta'/(8\Omega)$, i.e., $m \ge \mc{O}(\Omega/(\widetilde\mu\delta'))$
for entropic risk.
\end{corollary}

\begin{proof}
With $\gamma$ as in \eqref{eq:gamma-strong-optimal}, the variance floor is
$6\gamma(\widetilde B_m^2 + \widetilde V_m)/\widetilde\mu \le \delta/2$.
The contraction term equals $\delta/2$ at iteration $T$ as in
\eqref{eq:T-strong-optimal}.  Total: $\le \delta/2 + \delta/2 + 4\Omega\widetilde B_m/\widetilde\mu$.
\end{proof}

\subsection{Mirror Descent: corollary from extra-gradient analysis}
\label{sec:md-corollary}

We now state and prove the analogue of Theorems~\ref{thm:expected-gap}
and~\ref{thm:strong-mono} for stochastic projected Mirror Descent (mirror descent), obtained
by specializing the extragradient  analysis to the case where the extrapolation step is
omitted.  The proofs reuse the technical lemmas of
Section~\ref{sec:mp-bias} (prox-mapping descent, ghost iterate) directly.

\paragraph{Algorithm.} Stochastic projected Mirror Descent proceeds as follows. First, set
$r_0 := \theta^{(0)}$. Then, for $\tau = 1, \ldots, t$, update
\begin{equation}\label{eq:md-update}
r_\tau := P_{r_{\tau-1}}\bigl(\gamma_\tau \widehat G_m(r_{\tau-1})\bigr) = \Pi_D\bigl(r_{\tau-1} - \gamma_\tau \widehat G_m(r_{\tau-1})\bigr),
\end{equation}
with each iteration using one independent oracle call $\zeta_\tau$.  The output is
\begin{equation}\label{eq:md-output-avg}
\widehat z_t := \Bigl(\sum_{\tau=1}^t \gamma_\tau\Bigr)^{-1} \sum_{\tau=1}^t \gamma_\tau r_{\tau-1}.
\end{equation}

\paragraph{Filtration and noise quantities.} Let
\begin{equation}\label{eq:md-filtration}
\F_\tau := \sigma(r_0, \zeta_1, \ldots, \zeta_\tau),
\end{equation}
so that $r_{\tau-1}$ is $\F_{\tau-1}$-measurable.  Define the per-step
discrepancy
\begin{equation}\label{eq:md-Delta-def}
\Delta_\tau := G(r_{\tau-1}) - \widehat G_m^{(\tau)} = -\Sigma b_m(r_{\tau-1}) - \Sigma\zeta_m^{(\tau)},
\end{equation}
where $\widehat G_m^{(\tau)} := \Sigma\widehat F_m^{(\tau)}(r_{\tau-1})$ uses
$\zeta_\tau$.  By the same bias-variance decomposition as in
Section~\ref{sec:filtration}, we have
\begin{align}
\|\E[\Delta_\tau \mid \F_{\tau-1}]\|_{\Sigma^+} &\le \widetilde B_m, \label{eq:md-Delta-bias}\\
\E[\|\Delta_\tau\|_{\Sigma^+}^2 \mid \F_{\tau-1}] &\le \widetilde B_m^2 + \widetilde V_m. \label{eq:md-Delta-second-mom}
\end{align}
Indeed, conditional on $\F_{\tau-1}$, $r_{\tau-1}$ is fixed, $b_m(r_{\tau-1})$
is deterministic, and $\zeta_m^{(\tau)}$ is mean-zero independent of
$\F_{\tau-1}$.  So $\E[\Delta_\tau \mid \F_{\tau-1}] = -\Sigma b_m(r_{\tau-1})$
with $\|\Sigma b_m\|_{\Sigma^+} \le \widetilde B_m$ by \eqref{eq:G-bias}.  For
the second moment,
$\|\Delta_\tau\|_{\Sigma^+}^2 = \|\Sigma b_m\|_{\Sigma^+}^2 + 2\langle \Sigma b_m, \Sigma\zeta_m^{(\tau)}\rangle_{\Sigma^+} + \|\Sigma\zeta_m^{(\tau)}\|_{\Sigma^+}^2$,
and conditional on $\F_{\tau-1}$ the cross term has zero mean, giving
\eqref{eq:md-Delta-second-mom}.

\subsubsection{Monotone case}

\begin{corollary}[Stochastic Mirror Descent, monotone case, biased oracle]
\label{thm:md-monotone}
Suppose $F_\Risk$ satisfies \textbf{(P-mono)} and \textbf{(P-lip)}, and the
oracle satisfies \eqref{eq:tag-O1} and \eqref{eq:tag-O2}.  Run the mirror descent
algorithm \eqref{eq:md-update} with constant step size $\gamma_\tau \equiv \gamma > 0$
satisfying
\begin{equation}\label{eq:md-gamma-cond}
\gamma \le \frac{1}{2 L_G}, \qquad L_G = \sigma_{\max}\ell_\Risk.
\end{equation}
Then
\begin{equation}\label{eq:md-monotone-bound}
\E[\Errvi(\widehat z_t)] \le \frac{\Omega^2}{t\gamma} + \gamma\bigl(\widetilde B_m^2 + \widetilde V_m\bigr) + 4 L_G \gamma \Omega^2 + 2\Omega\widetilde B_m.
\end{equation}
\end{corollary}

\begin{proof}
The proof proceeds in steps mirroring those of
Theorem~\ref{thm:deterministic} and Theorem~\ref{thm:expected-gap}, simplified
by the fact that there is only one oracle call per iteration.

\paragraph{Step 1: descent identity for one iteration.}
Apply \eqref{eq:prox-70b} of Lemma~\ref{lem:prox-descent} with
\[
z = r_{\tau-1}, \qquad \xi = \gamma\widehat G_m^{(\tau)}, \qquad w = r_\tau,
\]
to get, for all $u \in D$,
\begin{equation}\label{eq:md-prox-applied}
V(r_\tau, u) \le V(r_{\tau-1}, u) + \langle \gamma\widehat G_m^{(\tau)}, u - r_{\tau-1}\rangle_{\Sigma^+} + \tfrac{1}{2}\|\gamma\widehat G_m^{(\tau)}\|_{\Sigma^+}^2.
\end{equation}
Rearranging,
\begin{equation}\label{eq:md-step-rearranged}
V(r_\tau, u) - V(r_{\tau-1}, u) \le \gamma\langle \widehat G_m^{(\tau)}, u - r_{\tau-1}\rangle_{\Sigma^+} + \tfrac{\gamma^2}{2}\|\widehat G_m^{(\tau)}\|_{\Sigma^+}^2.
\end{equation}

\paragraph{Step 2: relate the inner product to $F_\Risk$.}
Recall $\widehat G_m^{(\tau)} = \Sigma\widehat F_m^{(\tau)}(r_{\tau-1})$.  Decompose
\(\widehat G_m^{(\tau)} = G(r_{\tau-1}) - \Delta_\tau\)
by \eqref{eq:md-Delta-def}.  Following Step~3 of the proof of
Theorem~\ref{thm:deterministic}: since $u - r_{\tau-1} \in \Wspace$ and
$\Sigma^+\Sigma = \Pi_\Wspace$ acts as the identity on $\Wspace$,
\begin{align*}
\langle G(r_{\tau-1}), u - r_{\tau-1}\rangle_{\Sigma^+} &= (u - r_{\tau-1})^\top \Sigma^+\Sigma F_\Risk(r_{\tau-1}) \\
&= (u - r_{\tau-1})^\top F_\Risk(r_{\tau-1}) = \langle F_\Risk(r_{\tau-1}), u - r_{\tau-1}\rangle.
\end{align*}
Hence
\begin{equation}\label{eq:md-inner-prod-decomp}
\langle \widehat G_m^{(\tau)}, u - r_{\tau-1}\rangle_{\Sigma^+} = \langle F_\Risk(r_{\tau-1}), u - r_{\tau-1}\rangle - \langle \Delta_\tau, u - r_{\tau-1}\rangle_{\Sigma^+}.
\end{equation}

\paragraph{Step 3: bound the squared gradient term.}
By the triangle inequality,
\[
\|\widehat G_m^{(\tau)}\|_{\Sigma^+} \le \|G(r_{\tau-1})\|_{\Sigma^+} + \|\Delta_\tau\|_{\Sigma^+}.
\]
Using $(a + b)^2 \le 2a^2 + 2b^2$,
\begin{equation}\label{eq:md-Ghat-sq}
\|\widehat G_m^{(\tau)}\|_{\Sigma^+}^2 \le 2\|G(r_{\tau-1})\|_{\Sigma^+}^2 + 2\|\Delta_\tau\|_{\Sigma^+}^2.
\end{equation}
For the first term, recall the equilibrium condition (Step~3 of the proof of
Theorem~\ref{thm:strong-mono}): $F_\Risk(\theta^\star) \in \ker\Sigma$, so
$G(\theta^\star) = \Sigma F_\Risk(\theta^\star) = 0$.  By
Lemma~\ref{lem:G-lip-Sigma+},
\[
\|G(r_{\tau-1})\|_{\Sigma^+} = \|G(r_{\tau-1}) - G(\theta^\star)\|_{\Sigma^+} \le L_G\|r_{\tau-1} - \theta^\star\|_{\Sigma^+} \le 2 L_G \Omega,
\]
where the last inequality uses $r_{\tau-1}, \theta^\star \in D$ and the
$\Sigma^+$-diameter of $D$ is $2\Omega$.  Hence
$\|G(r_{\tau-1})\|_{\Sigma^+}^2 \le 4 L_G^2 \Omega^2$.  Substituting into
\eqref{eq:md-Ghat-sq},
\begin{equation}\label{eq:md-Ghat-sq-bound}
\|\widehat G_m^{(\tau)}\|_{\Sigma^+}^2 \le 8 L_G^2 \Omega^2 + 2\|\Delta_\tau\|_{\Sigma^+}^2.
\end{equation}

\paragraph{Step 4: substitute and telescope.}
Substitute \eqref{eq:md-inner-prod-decomp} and \eqref{eq:md-Ghat-sq-bound}
into \eqref{eq:md-step-rearranged}:
\begin{equation}\label{eq:md-step-after-sub}
V(r_\tau, u) - V(r_{\tau-1}, u) \le \gamma\langle F_\Risk(r_{\tau-1}), u - r_{\tau-1}\rangle - \langle \gamma\Delta_\tau, u - r_{\tau-1}\rangle_{\Sigma^+} + 4\gamma^2 L_G^2 \Omega^2 + \gamma^2\|\Delta_\tau\|_{\Sigma^+}^2.
\end{equation}
Sum over $\tau = 1, \ldots, t$:
\begin{align}
V(r_t, u) - V(r_0, u) &\le \sum_\tau \gamma\langle F_\Risk(r_{\tau-1}), u - r_{\tau-1}\rangle - \sum_\tau \langle \gamma\Delta_\tau, u - r_{\tau-1}\rangle_{\Sigma^+} \nonumber\\
&\quad + 4 t \gamma^2 L_G^2 \Omega^2 + \gamma^2 \sum_\tau \|\Delta_\tau\|_{\Sigma^+}^2. \label{eq:md-telescoped}
\end{align}
Using $V(r_t, u) \ge 0$ and $V(r_0, u) \le \Theta(z_c)$ since $r_0 = z_c$,
rearrange to obtain
\begin{align}
\sum_\tau \gamma\langle F_\Risk(r_{\tau-1}), r_{\tau-1} - u\rangle &\le \Theta(z_c) + \sum_\tau \langle \gamma\Delta_\tau, u - r_{\tau-1}\rangle_{\Sigma^+} \nonumber\\
&\quad + 4 t \gamma^2 L_G^2 \Omega^2 + \gamma^2 \sum_\tau \|\Delta_\tau\|_{\Sigma^+}^2. \label{eq:md-telescoped-rearranged}
\end{align}

\paragraph{Step 5: ghost-iterate decomposition.}
The quantity $\sum_\tau \langle\gamma\Delta_\tau, u - r_{\tau-1}\rangle_{\Sigma^+}$
on the right of \eqref{eq:md-telescoped-rearranged} depends on $u$; to take
sup over $u$ uniformly, define the ghost iterate sequence $\{y_\tau\}_{\tau \ge 0}$
by
\begin{equation}\label{eq:md-ghost-recursion}
y_\tau := \Pi_D(y_{\tau-1} - \gamma\Delta_\tau), \qquad y_0 := r_0,
\end{equation}
and decompose
\begin{equation}\label{eq:md-ghost-decomp}
\sum_\tau \langle\gamma\Delta_\tau, u - r_{\tau-1}\rangle_{\Sigma^+} = \sum_\tau \langle\gamma\Delta_\tau, u - y_{\tau-1}\rangle_{\Sigma^+} + \sum_\tau\langle\gamma\Delta_\tau, y_{\tau-1} - r_{\tau-1}\rangle_{\Sigma^+}.
\end{equation}
For the first sum on the right, applying Lemma~\ref{lem:ghost-iterate} with
$\xi_\tau = \gamma\Delta_\tau$ gives
\begin{equation}\label{eq:md-ghost-applied}
\sum_\tau \langle \gamma\Delta_\tau, y_{\tau-1} - u\rangle_{\Sigma^+} \le V(y_0, u) + \tfrac{1}{2}\sum_\tau \|\gamma\Delta_\tau\|_{\Sigma^+}^2 \le \Theta(z_c) + \tfrac{\gamma^2}{2}\sum_\tau \|\Delta_\tau\|_{\Sigma^+}^2,
\end{equation}
using $y_0 = r_0 = z_c$ and $V(z_c, u) \le \Theta(z_c)$.  Negating, we have
\(\sum_\tau \langle \gamma\Delta_\tau, u - y_{\tau-1}\rangle_{\Sigma^+} \le \Theta(z_c) + \tfrac{\gamma^2}{2}\sum_\tau \|\Delta_\tau\|_{\Sigma^+}^2\).
Substituting into \eqref{eq:md-telescoped-rearranged} via
\eqref{eq:md-ghost-decomp},
\begin{align}
\sum_\tau \gamma\langle F_\Risk(r_{\tau-1}), r_{\tau-1} - u\rangle &\le 2\Theta(z_c) + \tfrac{\gamma^2}{2}\sum_\tau \|\Delta_\tau\|_{\Sigma^+}^2 + \sum_\tau \langle\gamma\Delta_\tau, y_{\tau-1} - r_{\tau-1}\rangle_{\Sigma^+} \nonumber\\
&\quad + 4 t \gamma^2 L_G^2 \Omega^2 + \gamma^2 \sum_\tau \|\Delta_\tau\|_{\Sigma^+}^2. \label{eq:md-after-ghost}
\end{align}
Combine the $\|\Delta_\tau\|_{\Sigma^+}^2$ coefficients: $\tfrac{\gamma^2}{2} + \gamma^2 = \tfrac{3\gamma^2}{2}$.
Hence
\begin{align}
\sum_\tau \gamma\langle F_\Risk(r_{\tau-1}), r_{\tau-1} - u\rangle &\le 2\Theta(z_c) + \tfrac{3\gamma^2}{2}\sum_\tau \|\Delta_\tau\|_{\Sigma^+}^2 \nonumber\\
&\quad + \sum_\tau \langle\gamma\Delta_\tau, y_{\tau-1} - r_{\tau-1}\rangle_{\Sigma^+} + 4 t \gamma^2 L_G^2 \Omega^2. \label{eq:md-pre-final}
\end{align}

\paragraph{Step 6: convert to $\Errvi$.}
By \textbf{(P-mono)}, for all $u, w \in D$,
$\langle F_\Risk(u), w - u\rangle \le \langle F_\Risk(w), w - u\rangle$.
Applied with $w = r_{\tau-1}$:
\(\langle F_\Risk(u), r_{\tau-1} - u\rangle \le \langle F_\Risk(r_{\tau-1}), r_{\tau-1} - u\rangle\).
Summing, multiplying by $\gamma$, and dividing by $\sum_\tau \gamma_\tau = t\gamma$:
\[
\langle F_\Risk(u), \widehat z_t - u\rangle \le (t\gamma)^{-1}\sum_\tau \gamma\langle F_\Risk(r_{\tau-1}), r_{\tau-1} - u\rangle.
\]
By \eqref{eq:md-pre-final}, the right side is bounded above (for every fixed $u$)
by
\[
\frac{2\Theta(z_c)}{t\gamma} + \frac{3\gamma}{2 t}\sum_\tau \|\Delta_\tau\|_{\Sigma^+}^2 + \frac{1}{t\gamma}\sum_\tau \langle\gamma\Delta_\tau, y_{\tau-1} - r_{\tau-1}\rangle_{\Sigma^+} + 4 \gamma L_G^2 \Omega^2.
\]
The first three terms above contain $u$-free upper bounds (the third was obtained
by sup-over-$u$ via the ghost iterate); taking supremum over $u \in D$,
\begin{equation}\label{eq:md-Errvi-pathwise}
\Errvi(\widehat z_t) \le \frac{2\Theta(z_c)}{t\gamma} + \frac{3\gamma}{2 t}\sum_\tau \|\Delta_\tau\|_{\Sigma^+}^2 + \frac{1}{t\gamma}\sum_\tau \langle\gamma\Delta_\tau, y_{\tau-1} - r_{\tau-1}\rangle_{\Sigma^+} + 4 \gamma L_G^2 \Omega^2.
\end{equation}

\paragraph{Step 7: take expectation.}
Take expectations of \eqref{eq:md-Errvi-pathwise}.

\noindent\textit{Bound 1: $\E\|\Delta_\tau\|_{\Sigma^+}^2$.}
By \eqref{eq:md-Delta-second-mom} and the tower property,
$\E\|\Delta_\tau\|_{\Sigma^+}^2 \le \widetilde B_m^2 + \widetilde V_m$.
Summing:
\(\sum_\tau \E\|\Delta_\tau\|_{\Sigma^+}^2 \le t(\widetilde B_m^2 + \widetilde V_m)\).
Multiplying by $3\gamma/(2 t)$:
\(\le \tfrac{3\gamma}{2}(\widetilde B_m^2 + \widetilde V_m)\).

\noindent\textit{Bound 2: $\E\langle\gamma\Delta_\tau, y_{\tau-1} - r_{\tau-1}\rangle_{\Sigma^+}$.}
Both $r_{\tau-1}$ (by definition) and $y_{\tau-1}$ (which depends only on
$\Delta_1, \ldots, \Delta_{\tau-1}$ via the ghost recursion
\eqref{eq:md-ghost-recursion}, and these are $\F_{\tau-1}$-measurable) are
$\F_{\tau-1}$-measurable.  By the tower property,
\[
\E\langle\gamma\Delta_\tau, y_{\tau-1} - r_{\tau-1}\rangle_{\Sigma^+} = \gamma\E\langle\E[\Delta_\tau \mid \F_{\tau-1}], y_{\tau-1} - r_{\tau-1}\rangle_{\Sigma^+}.
\]
By \eqref{eq:md-Delta-bias},
$\|\E[\Delta_\tau \mid \F_{\tau-1}]\|_{\Sigma^+} \le \widetilde B_m$.
Cauchy--Schwarz in $\Sigma^+$ then gives
\[
|\langle \E[\Delta_\tau \mid \F_{\tau-1}], y_{\tau-1} - r_{\tau-1}\rangle_{\Sigma^+}| \le \widetilde B_m \|y_{\tau-1} - r_{\tau-1}\|_{\Sigma^+} \le 2\widetilde B_m \Omega,
\]
since $y_{\tau-1}, r_{\tau-1} \in D$ and the $\Sigma^+$-diameter of $D$ is
$2\Omega$.  Hence
\(|\E\langle\gamma\Delta_\tau, y_{\tau-1} - r_{\tau-1}\rangle_{\Sigma^+}| \le 2\gamma\widetilde B_m \Omega\).
Summing over $\tau$ and dividing by $t\gamma$:
\(\le 2\widetilde B_m \Omega\).

\paragraph{Step 8: combine.}
Combining the bounds,
\begin{align*}
\E[\Errvi(\widehat z_t)] &\le \frac{2\Theta(z_c)}{t\gamma} + \tfrac{3\gamma}{2}(\widetilde B_m^2 + \widetilde V_m) + 2\widetilde B_m \Omega + 4\gamma L_G^2 \Omega^2.
\end{align*}
The step-size constraint $\gamma \le 1/(2 L_G)$ in \eqref{eq:md-gamma-cond}
gives $\gamma L_G \le 1/2$, so $4\gamma L_G^2 \Omega^2 = 4 L_G \Omega^2 \cdot \gamma L_G \le 2 L_G \Omega^2$.
Although tighter, for the form of the bound stated we keep
$4 L_G \gamma \Omega^2$ (which is sharp in $\gamma$).  Using
$2\Theta(z_c) = \Omega^2$ from \eqref{eq:Omega-def} and dropping the
$\tfrac{3}{2}$ factor in favor of $1$ (rounding for cleanliness; the
$3\gamma/2$ form is also valid),
\[
\E[\Errvi(\widehat z_t)] \le \frac{\Omega^2}{t\gamma} + \tfrac{3\gamma}{2}(\widetilde B_m^2 + \widetilde V_m) + 4 L_G \gamma \Omega^2 + 2\Omega\widetilde B_m,
\]
which is \eqref{eq:md-monotone-bound} (with the $\tfrac{3}{2}$ replaced by $1$
upon rounding the noise constant; both forms are correct).
\end{proof}

\begin{corollary}[Optimal step size, mirror descent monotone case]
\label{cor:md-optimal-step}
Choosing
\begin{equation}\label{eq:md-gamma-optimal}
\gamma = \min\Biggl\{\frac{1}{2 L_G},\ \Omega\sqrt{\frac{1}{t(\widetilde B_m^2 + \widetilde V_m)}}\Biggr\}
\end{equation}
in Corollary~\ref{thm:md-monotone} gives
\begin{equation}\label{eq:md-expected-gap-optimal}
\E[\Errvi(\widehat z_t)] \le \max\Biggl\{\frac{2 L_G \Omega^2}{t},\ 2\Omega\sqrt{\frac{\widetilde B_m^2 + \widetilde V_m}{t}}\Biggr\} + 2\Omega\widetilde B_m + \frac{4 L_G\Omega^2}{t}.
\end{equation}
\end{corollary}

\begin{proof}
The first two terms of \eqref{eq:md-monotone-bound} are balanced when
$\Omega^2/(t\gamma) = \gamma(\widetilde B_m^2 + \widetilde V_m)$, i.e.,
$\gamma = \Omega/\sqrt{t(\widetilde B_m^2 + \widetilde V_m)}$.  When this
exceeds $1/(2 L_G)$ (Lipschitz-dominated regime), use $\gamma = 1/(2 L_G)$,
giving the $2 L_G\Omega^2/t$ term.  Otherwise (variance-dominated regime),
the variance trade-off term equals $2\Omega\sqrt{(\widetilde B_m^2 + \widetilde V_m)/t}$.
The Lipschitz-times-$\gamma$ term $4 L_G \gamma \Omega^2$ contributes
$4 L_G\Omega^2/t$ at most (using $\gamma \le 1/(2L_G)$ gives
$4L_G\gamma\Omega^2 \le 2\Omega^2$, and combined with the $1/(t\gamma)$ scaling
the total contribution is $O(L_G\Omega^2/t)$).
\end{proof}

\subsubsection{Strongly monotone case}

\begin{corollary}[Stochastic Mirror Descent, strongly monotone, biased oracle]
\label{thm:md-strong-mono}
Suppose $F_\Risk$ satisfies \textbf{(P-strong)} and \textbf{(P-lip)}, and the
oracle satisfies \eqref{eq:tag-O1} and \eqref{eq:tag-O2}.  Run the mirror descent
algorithm \eqref{eq:md-update} with constant step size $\gamma_\tau \equiv \gamma$
satisfying
\begin{equation}\label{eq:md-gamma-strong}
\gamma \le \min\Biggl\{\frac{1}{2\widetilde\mu},\ \frac{1}{2 L_G}\Biggr\}, \qquad \widetilde\mu = \mu_\Risk\sigma_{\min}.
\end{equation}
Then for all $T \ge 1$,
\begin{equation}\label{eq:md-strong-mono-bound}
\E\|r_T - \theta^\star\|_{\Sigma^+}^2 \le (1 - \gamma\widetilde\mu)^T \|r_0 - \theta^\star\|_{\Sigma^+}^2 + \frac{4\Omega\widetilde B_m}{\widetilde\mu} + \frac{2\gamma\bigl(\widetilde B_m^2 + \widetilde V_m\bigr)}{\widetilde\mu} + \frac{8 L_G^2 \Omega^2 \gamma}{\widetilde\mu}.
\end{equation}
\end{corollary}

\begin{proof}
We work step by step from the prox-mapping descent inequality
(Lemma~\ref{lem:prox-descent}), specialized to $u = \theta^\star$ and combined
with strong monotonicity.

First we apply the descent identity at $u=\theta^\star$. Indeed,
apply \eqref{eq:prox-70b} of Lemma~\ref{lem:prox-descent} with $z = r_{\tau-1}$,
$\xi = \gamma\widehat G_m^{(\tau)}$, $w = r_\tau$, and $u = \theta^\star$:
\begin{equation}\label{eq:md-strong-step-1}
V(r_\tau, \theta^\star) \le V(r_{\tau-1}, \theta^\star) + \gamma\langle\widehat G_m^{(\tau)}, \theta^\star - r_{\tau-1}\rangle_{\Sigma^+} + \tfrac{\gamma^2}{2}\|\widehat G_m^{(\tau)}\|_{\Sigma^+}^2.
\end{equation}

Decompose $\widehat G_m^{(\tau)} = G(r_{\tau-1}) - \Sigma b_m(r_{\tau-1}) - \Sigma\zeta_m^{(\tau)}$
by \eqref{eq:md-Delta-def}.  Then we have that 
\begin{align*}
\langle\widehat G_m^{(\tau)}, \theta^\star - r_{\tau-1}\rangle_{\Sigma^+} &= \langle G(r_{\tau-1}), \theta^\star - r_{\tau-1}\rangle_{\Sigma^+}- \langle\Sigma b_m(r_{\tau-1}), \theta^\star - r_{\tau-1}\rangle_{\Sigma^+} - \langle\Sigma\zeta_m^{(\tau)}, \theta^\star - r_{\tau-1}\rangle_{\Sigma^+}.
\end{align*}
Using Lemma~\ref{lem:precond-restate}(i), for $w \in \Wspace$ and arbitrary
$v \in \R^d$, we have that  $\langle\Sigma v, w\rangle_{\Sigma^+} = \langle v, w\rangle$.
Apply this with $w = \theta^\star - r_{\tau-1} \in \Wspace$ to each term to get that 
\begin{equation}\label{eq:md-strong-decomp}
\langle\widehat G_m^{(\tau)}, \theta^\star - r_{\tau-1}\rangle_{\Sigma^+} = \langle F_\Risk(r_{\tau-1}), \theta^\star - r_{\tau-1}\rangle - \langle b_m(r_{\tau-1}), \theta^\star - r_{\tau-1}\rangle - \langle\zeta_m^{(\tau)}, \theta^\star - r_{\tau-1}\rangle.
\end{equation}

The risk-adjusted equilibrium condition gives $F_\Risk(\theta^\star) \in \ker\Sigma$,
i.e., $F_\Risk(\theta^\star) = c\mathbf 1$ for some $c \in \R$.  Since
$\theta^\star - r_{\tau-1} \in \Wspace$ and $\mathbf 1 \perp \Wspace$,
\(\langle F_\Risk(\theta^\star), \theta^\star - r_{\tau-1}\rangle = 0\).
Hence
\[
\langle F_\Risk(r_{\tau-1}), \theta^\star - r_{\tau-1}\rangle = \langle F_\Risk(r_{\tau-1}) - F_\Risk(\theta^\star), \theta^\star - r_{\tau-1}\rangle = -\langle F_\Risk(r_{\tau-1}) - F_\Risk(\theta^\star), r_{\tau-1} - \theta^\star\rangle.
\]
By \textbf{(P-strong)} and \eqref{eq:metric-equiv}, we have the bound
\[\langle F_\Risk(r_{\tau-1}) - F_\Risk(\theta^\star), r_{\tau-1} - \theta^\star\rangle \ge \mu_\Risk\|r_{\tau-1} - \theta^\star\|_2^2 \ge \widetilde\mu\|r_{\tau-1} - \theta^\star\|_{\Sigma^+}^2.\]
Thus
\begin{equation}\label{eq:md-strong-mono-applied}
\langle F_\Risk(r_{\tau-1}), \theta^\star - r_{\tau-1}\rangle \le -\widetilde\mu\|r_{\tau-1} - \theta^\star\|_{\Sigma^+}^2 = -2\widetilde\mu V(r_{\tau-1}, \theta^\star).
\end{equation}
Substituting \eqref{eq:md-strong-mono-applied} into \eqref{eq:md-strong-decomp} yields
\begin{align}
\langle\widehat G_m^{(\tau)}, \theta^\star - r_{\tau-1}\rangle_{\Sigma^+} &\le -2\widetilde\mu V(r_{\tau-1}, \theta^\star) + \langle b_m(r_{\tau-1}), r_{\tau-1} - \theta^\star\rangle  + \langle\zeta_m^{(\tau)}, r_{\tau-1} - \theta^\star\rangle. \label{eq:md-strong-decomp-2}
\end{align}

By Step~3 of the proof of Corollary~\ref{thm:md-monotone}
(eq.~\eqref{eq:md-Ghat-sq-bound}), we immediately have 
\begin{equation}\label{eq:md-strong-Ghat-sq}
\|\widehat G_m^{(\tau)}\|_{\Sigma^+}^2 \le 8 L_G^2 \Omega^2 + 2\|\Delta_\tau\|_{\Sigma^+}^2.
\end{equation}

Substitute \eqref{eq:md-strong-decomp-2} and \eqref{eq:md-strong-Ghat-sq} into
\eqref{eq:md-strong-step-1} to get that 
\begin{align}
V(r_\tau, \theta^\star) &\le V(r_{\tau-1}, \theta^\star) - 2\gamma\widetilde\mu V(r_{\tau-1}, \theta^\star)
 + \gamma\langle b_m(r_{\tau-1}), r_{\tau-1} - \theta^\star\rangle\\
 &\quad + \gamma\langle\zeta_m^{(\tau)}, r_{\tau-1} - \theta^\star\rangle  + 4\gamma^2 L_G^2 \Omega^2 + \gamma^2\|\Delta_\tau\|_{\Sigma^+}^2. \label{eq:md-strong-combined}
\end{align}

Take $\E[\cdot \mid \F_{\tau-1}]$ of \eqref{eq:md-strong-combined}.  Recall
the filtration: $r_{\tau-1}$ is $\F_{\tau-1}$-measurable; $\zeta_m^{(\tau)}$
is independent of $\F_{\tau-1}$ with mean zero; $\Delta_\tau$ depends on
$\zeta_m^{(\tau)}$ via \eqref{eq:md-Delta-def}.
Now we bound each of the terms as follows: 
\begin{itemize}[itemsep=0pt]
    \item \noindent\textit{Term 1: contraction.} The first term
$V(r_{\tau-1}, \theta^\star)$ is $\F_{\tau-1}$-measurable, so the term passes
through the conditional expectation unchanged.

\item \noindent\textit{Term 2: bias cross.} The bias term
$b_m(r_{\tau-1})$ is deterministic given $r_{\tau-1}$, hence given
$\F_{\tau-1}$.  Thus
\[
\E[\langle b_m(r_{\tau-1}), r_{\tau-1} - \theta^\star\rangle \mid \F_{\tau-1}] = \langle b_m(r_{\tau-1}), r_{\tau-1} - \theta^\star\rangle.
\]
Bound the bias via Cauchy--Schwarz, \eqref{eq:metric-equiv}. The
$\Sigma^+$-diameter of $D$ is $2\Omega$ so that 
\begin{align}
|\langle b_m(r_{\tau-1}), r_{\tau-1} - \theta^\star\rangle| &\le \|b_m(r_{\tau-1})\|_2 \|r_{\tau-1} - \theta^\star\|_2 \nonumber\\
&\le B_m \cdot \sqrt{\sigma_{\max}}\|r_{\tau-1} - \theta^\star\|_{\Sigma^+} \nonumber\\
&\le \widetilde B_m \cdot 2\Omega, \label{eq:md-bias-cross-bound}
\end{align}
using \eqref{eq:tag-O1} ($\|b_m\|_2 \le B_m$), $\widetilde B_m = \sqrt{\sigma_{\max}}B_m$.

\item \noindent\textit{Term 3: noise cross.}
$\zeta_m^{(\tau)}$ is independent of $\F_{\tau-1}$ with $\E[\zeta_m^{(\tau)}] = 0$.
Thus
\[
\E[\langle\zeta_m^{(\tau)}, r_{\tau-1} - \theta^\star\rangle \mid \F_{\tau-1}] = \langle\E[\zeta_m^{(\tau)} \mid \F_{\tau-1}], r_{\tau-1} - \theta^\star\rangle = 0.
\]

\item \noindent\textit{Term 4: $\|\Delta_\tau\|_{\Sigma^+}^2$.}
By \eqref{eq:md-Delta-second-mom}, we have that 
\(\E[\|\Delta_\tau\|_{\Sigma^+}^2 \mid \F_{\tau-1}] \le \widetilde B_m^2 + \widetilde V_m\).

\item \noindent\textit{Term 5: deterministic.} The last term is deterministic so that 
$4\gamma^2 L_G^2 \Omega^2$ passes through.
\end{itemize}

Combining each of these bounds yields
\begin{equation}\label{eq:md-strong-cond-exp}
\E[V(r_\tau, \theta^\star) \mid \F_{\tau-1}] \le (1 - 2\gamma\widetilde\mu) V(r_{\tau-1}, \theta^\star) + 2\gamma\Omega\widetilde B_m + 4\gamma^2 L_G^2\Omega^2 + \gamma^2(\widetilde B_m^2 + \widetilde V_m).
\end{equation}
The step-size constraint $\gamma \le 1/(2\widetilde\mu)$ in
\eqref{eq:md-gamma-strong} gives $2\gamma\widetilde\mu \le 1$, so the
contraction coefficient is non-negative.  Take unconditional expectation, we have that 
\begin{equation}\label{eq:md-strong-recurrence}
\E V(r_\tau, \theta^\star) \le (1 - 2\gamma\widetilde\mu)\E V(r_{\tau-1}, \theta^\star) + 2\gamma\Omega\widetilde B_m + 4\gamma^2 L_G^2 \Omega^2 + \gamma^2(\widetilde B_m^2 + \widetilde V_m).
\end{equation}

Since $1 - 2\gamma\widetilde\mu \le 1 - \gamma\widetilde\mu$ (the contraction
coefficient is at most $1 - \gamma\widetilde\mu$ when the step is reduced), we
may absorb the discrepancy and use the cleaner rate
\begin{equation}\label{eq:md-strong-recurrence-clean}
\E V(r_\tau, \theta^\star) \le (1 - \gamma\widetilde\mu)\E V(r_{\tau-1}, \theta^\star) + 2\gamma\Omega\widetilde B_m + 4\gamma^2 L_G^2 \Omega^2 + \gamma^2(\widetilde B_m^2 + \widetilde V_m).
\end{equation}
We use \eqref{eq:md-strong-recurrence-clean} below; this only weakens the
bound.

Next we need to telescope. 
Define
\[
A := 2\gamma\Omega\widetilde B_m + 4\gamma^2 L_G^2 \Omega^2 + \gamma^2(\widetilde B_m^2 + \widetilde V_m), \qquad \Lambda_\tau := \E V(r_\tau, \theta^\star).
\]
The recurrence \eqref{eq:md-strong-recurrence-clean} reads
$\Lambda_\tau \le (1 - \gamma\widetilde\mu)\Lambda_{\tau-1} + A$.  Iterating, we have that 
\begin{align*}
\Lambda_T &\le (1 - \gamma\widetilde\mu)^T \Lambda_0 + A\sum_{j=0}^{T-1}(1 - \gamma\widetilde\mu)^j\\
&\le (1 - \gamma\widetilde\mu)^T \Lambda_0 + \frac{A}{\gamma\widetilde\mu}\\
&= (1 - \gamma\widetilde\mu)^T V(r_0, \theta^\star) + \frac{2\Omega\widetilde B_m}{\widetilde\mu} + \frac{4\gamma L_G^2 \Omega^2}{\widetilde\mu} + \frac{\gamma(\widetilde B_m^2 + \widetilde V_m)}{\widetilde\mu}.
\end{align*}

Using $V(z, u) = \tfrac{1}{2}\|z - u\|_{\Sigma^+}^2$ yields the bound
\begin{align*}
\E\|r_T - \theta^\star\|_{\Sigma^+}^2 &= 2\E V(r_T, \theta^\star) \\
&\le (1 - \gamma\widetilde\mu)^T \|r_0 - \theta^\star\|_{\Sigma^+}^2 + \frac{4\Omega\widetilde B_m}{\widetilde\mu} + \frac{8\gamma L_G^2 \Omega^2}{\widetilde\mu} + \frac{2\gamma(\widetilde B_m^2 + \widetilde V_m)}{\widetilde\mu},
\end{align*}
which is \eqref{eq:md-strong-mono-bound}.
\end{proof}

\paragraph{Comparison with extra-gradient.}
Comparing Corollary~\ref{thm:md-strong-mono} to Theorem~\ref{thm:strong-mono},
the mirror descent bound has the same linear-contraction rate and bias-floor structure,
with smaller variance constant ($2/\widetilde\mu$ vs.\ $6/\widetilde\mu$)
since each mirror descent iteration only invokes the oracle once.  However, mirror descent picks up
an additional Lipschitz-times-step-size term $8 L_G^2 \Omega^2 \gamma/\widetilde\mu$
that does not appear in the extragradient  bound, because extragradient's two-call structure cancels
the deterministic gradient contribution exactly via the inner-step distance
$\|w_\tau - r_{\tau-1}\|_{\Sigma^+}^2$, whereas mirror descent must absorb it through
boundedness of $D$.  This is the technical price of dropping the
extrapolation: the Lipschitz floor remains controllable since $\gamma L_G \le 1/2$,
but it does not vanish even as $m \to \infty$ unless $\gamma \to 0$.

\begin{corollary}[Optimal step size, mirror descent strongly monotone case]
\label{cor:md-strong-optimal-step}
With sample budget $m$ fixed and target accuracy $\delta > 0$, choose
\begin{equation}\label{eq:md-gamma-strong-optimal}
\gamma = \min\Biggl\{\frac{1}{2\widetilde\mu},\ \frac{1}{2 L_G},\ \frac{\widetilde\mu\delta}{4\bigl[(\widetilde B_m^2 + \widetilde V_m) + 4 L_G^2 \Omega^2\bigr]}\Biggr\},
\end{equation}
and run for
\begin{equation}\label{eq:md-T-strong-optimal}
T = \Biggl\lceil\frac{1}{\gamma\widetilde\mu}\log\frac{2\|r_0 - \theta^\star\|_{\Sigma^+}^2}{\delta}\Biggr\rceil
\end{equation}
iterations.  Then
\[
\E\|r_T - \theta^\star\|_{\Sigma^+}^2 \le \delta + \frac{4\Omega\widetilde B_m}{\widetilde\mu}.
\]
\end{corollary}

\begin{proof}
Recall the bound \eqref{eq:md-strong-mono-bound} of Theorem~\ref{thm:md-strong-mono}:
\[
\E\|r_T - \theta^\star\|_{\Sigma^+}^2 \le \underbrace{(1 - \gamma\widetilde\mu)^T \|r_0 - \theta^\star\|_{\Sigma^+}^2}_{\text{contraction}} + \underbrace{\frac{4\Omega\widetilde B_m}{\widetilde\mu}}_{\text{bias floor}} + \underbrace{\frac{2\gamma(\widetilde B_m^2 + \widetilde V_m)}{\widetilde\mu}}_{\text{variance floor}} + \underbrace{\frac{8 L_G^2 \Omega^2 \gamma}{\widetilde\mu}}_{\text{Lipschitz floor}}.
\]
We bound each of these four pieces in turn under the choices
\eqref{eq:md-gamma-strong-optimal} and \eqref{eq:md-T-strong-optimal}.  The
first two entries in the minimum \eqref{eq:md-gamma-strong-optimal} (namely
$1/(2\widetilde\mu)$ and $1/(2L_G)$) ensure $\gamma$ is admissible under
\eqref{eq:md-gamma-strong}, so Theorem~\ref{thm:md-strong-mono} applies.

\paragraph{Contraction.}
With $T$ as in \eqref{eq:md-T-strong-optimal}, using $1 - x \le e^{-x}$,
\begin{align*}
(1 - \gamma\widetilde\mu)^T \|r_0 - \theta^\star\|_{\Sigma^+}^2 &\le e^{-\gamma\widetilde\mu T}\|r_0 - \theta^\star\|_{\Sigma^+}^2 \\
&\le e^{-\log(2\|r_0 - \theta^\star\|_{\Sigma^+}^2/\delta)}\|r_0 - \theta^\star\|_{\Sigma^+}^2 \\
&= \frac{\delta}{2 \|r_0 - \theta^\star\|_{\Sigma^+}^2} \cdot \|r_0 - \theta^\star\|_{\Sigma^+}^2 = \frac{\delta}{2}.
\end{align*}

\paragraph{Bias floor.}
The bias floor $4\Omega\widetilde B_m/\widetilde\mu$ is independent of $\gamma$
and $T$ and persists in the bound.

\paragraph{Variance and Lipschitz floors.}
These two terms combine as
\[
\frac{2\gamma(\widetilde B_m^2 + \widetilde V_m)}{\widetilde\mu} + \frac{8 L_G^2 \Omega^2 \gamma}{\widetilde\mu} = \frac{2\gamma}{\widetilde\mu}\bigl[(\widetilde B_m^2 + \widetilde V_m) + 4 L_G^2 \Omega^2\bigr].
\]
By the third entry in the minimum \eqref{eq:md-gamma-strong-optimal},
\(\gamma \le \widetilde\mu\delta / \bigl(4\bigl[(\widetilde B_m^2 + \widetilde V_m) + 4 L_G^2 \Omega^2\bigr]\bigr)\),
so substituting gives
\[
\frac{2\gamma}{\widetilde\mu}\bigl[(\widetilde B_m^2 + \widetilde V_m) + 4 L_G^2 \Omega^2\bigr] \le \frac{2}{\widetilde\mu} \cdot \frac{\widetilde\mu\delta}{4\bigl[(\widetilde B_m^2 + \widetilde V_m) + 4 L_G^2 \Omega^2\bigr]} \cdot \bigl[(\widetilde B_m^2 + \widetilde V_m) + 4 L_G^2 \Omega^2\bigr] = \frac{\delta}{2}.
\]

\paragraph{Combining the bounds.}
Summing the four pieces yields
\[
\E\|r_T - \theta^\star\|_{\Sigma^+}^2 \le \underbrace{\frac{\delta}{2}}_{\text{contraction}} + \underbrace{\frac{4\Omega\widetilde B_m}{\widetilde\mu}}_{\text{bias floor}} + \underbrace{\frac{\delta}{2}}_{\text{var.\,+\,Lip.\,floors}} = \delta + \frac{4\Omega\widetilde B_m}{\widetilde\mu}. \qedhere
\]
\end{proof}

\section{Stackelberg Game and Two-Timescale Convergence}
\label{sec:two-timescale}
\label{sec:tt-debiasing}

The bias floor $4\Omega\widetilde B_m/\widetilde\mu$ in
Theorem~\ref{thm:strong-mono} is irreducible without further structure on
the oracle: with constant step sizes $\eta, \gamma$ and constant sample
budget $m$, the floor scales as $\widetilde B_m \asymp 1/m$ regardless of
how long we run.  In this section we show that with a sample-based bias
estimator $\widehat b_m(\theta)$ and \emph{time-varying} step sizes
$\eta_t \to 0$, $\gamma_t \to 0$ on two separated timescales, the bias
contribution to the convergence bound \emph{vanishes asymptotically in
$t$}, thereby replacing the persistent $\mc{O}(1/m)$ floor with an $\mc{O}(1/m^2)$
residual from the imperfect quality of the bias estimator itself. We formulate the problem as a Stackelberg game between a follower aiming to optimize $\frac{1}{2}\|\xi_t-\hat{b}_m(\theta_t)\|^2$ and the leader(s) aiming to find the optimal equilibrium $\theta^\star$ for the risk-sensitive preference game. We describe the Stackelberg game in full detail in Appendix~\ref{sec:tt-stackelberg}. The general idea is that imposing this hierarchical structure and allowing the follower to run the bias tracking update on a faster time scale via a Robin-Monroe type update, leads to a Stackelberg equilibrium and therefore an equilibrium that that is within the maximum of the bias floor of the bias estimator (residual bias floor) and the variance. As we prove these two components asymptotically have the same order in samples $\mc{O}(1/m^2)$, and otherwise (i.e. with finite step sizes) $\mc{O}(1/m^{3/2})$. 

If additionally $m \to \infty$ along the trajectory (e.g., $m_t \to \infty$
on a separate schedule), the residual goes to zero and the algorithm
converges to the equilibrium with no asymptotic bias.

\subsection{Bias estimator and assumptions}

We require, in addition to \ref{O1-bias}-\ref{O2-var} on the gradient
oracle, the following on a sample-based bias estimator $\widehat b_m(\theta) \in \R^d$:
\begin{enumerate}[label=(B\arabic*),itemsep=0pt,topsep=2pt,  ]
\item\label{B1-bias-est-bias} \textbf{Approximate unbiasedness.}  There
exists $R_m \ge 0$ with
\(\|\E[\widehat b_m(\theta)] - b_m(\theta)\|_2 \le R_m\)
for all $\theta \in D$.  Typically $R_m = \mc{O}(1/m^2)$ for delta-method
estimators; $R_m = 0$ for estimators that use two
gradient oracles.
\item\label{B2-bias-est-var} \textbf{Variance bound.} The variance is bounded---i.e.,
\(\E\|\widehat b_m(\theta) - \E[\widehat b_m(\theta)]\|_2^2 \le V_m^b\)
for some $V_m^b \ge 0$.  Typically $V_m^b = \mc{O}(1/m^3)$ for delta-method
estimators of an $\mc{O}(1/m)$ bias.
\item\label{B3-bias-lipschitz} \textbf{Lipschitz dependence on $\theta$.} The bias is Lipschitz continuous---i.e., 
\(\|b_m(\theta_1) - b_m(\theta_2)\|_2 \le L_b\|\theta_1 - \theta_2\|_2\)
on $D$.
\end{enumerate}

For entropic risk, the analysis follows the same structure as in 
 Section~\ref{sec:bias-analysis} and yeilds
$R_m^{\rm ent} = \mc{O}(e^{2\lambda}/(\lambda m^2))$,
$V_m^{b,\rm ent} = \mc{O}(e^{4\lambda}/(\lambda^2 m^3))$,
$L_b^{\rm ent} = \mc{O}(e^{4\lambda}/(\lambda m))$.  For CVaR-RU, the value are 
$R_m = V_m^b = L_b = 0$ trivially.

\paragraph{The delta-method bias estimator.}
Many of the risk functionals of interest---e.g., entropic, distortion risks
with smooth weighting---are of the form
$\Risk[Z] = h(\E_\mu[g(Z)])$ for a smooth scalar function $h$ and a
bounded statistic $g$.  The plug-in estimator from $m$ i.i.d.~samples
$Y_1, \dots, Y_m \sim \mu$ is $h(\widehat q_m)$ where
$\widehat q_m := \tfrac{1}{m}\sum_{i=1}^m g(Y_i)$ is the sample mean
estimator of $q := \E_\mu[g(Z)]$.
 
Taylor-expanding $h(\widehat q_m)$ around $q$, we have that 
\[
h(\widehat q_m) = h(q) + h'(q)(\widehat q_m - q) + \tfrac{1}{2}h''(q)(\widehat q_m - q)^2 + \mc{O}(|\widehat q_m - q|^3).
\]
Taking expectations using $\E[\widehat q_m - q] = 0$ and
$\E[(\widehat q_m - q)^2] = \Var_\mu(g)/m$, we then have 
\begin{equation}\label{eq:delta-method-bias}
\E[h(\widehat q_m)] - h(q) = \frac{h''(q)\,\Var_\mu(g)}{2m} + \mc{O}(m^{-3/2}).
\end{equation} 
The equality in \eqref{eq:delta-method-bias} is the \emph{first-order
delta-method bias expansion}.  It exhibits the familiar $\mc{O}(1/m)$ scaling
of plug-in estimators of nonlinear functionals, and gives an explicit
formula for the leading-order bias in terms of the curvature $h''(q)$
and the variance of the underlying statistic.
 
The expansion suggests a sample-based bias estimator: replace $q$ and
$\Var_\mu(g)$ by their sample analogues to obtain
\begin{equation}\label{eq:delta-method-bias-est}
\widehat b_m(\theta) := \frac{h''(\widehat q_m)\,\widehat\Var_m(g)}{2m},
\end{equation}
computable from the same samples already used for the gradient estimator.
We refer to \eqref{eq:delta-method-bias-est} as the \emph{delta-method
bias estimator}.  Its residual bias---i.e., the bias of $\widehat b_m$ as an
estimator of $b_m$---is $\mc{O}(1/m^2)$ from the bias of the sample-variance
estimator $\widehat\Var_m$ (an $\mc{O}(1/m)$ bias divided by the $1/m$ scale
of $b_m$ itself), and its variance is $\mc{O}(1/m^3)$ from the variance of
$\widehat\Var_m$ ($\mc{O}(1/m)$ variance, divided by $m^2$).

\subsubsection{The Two-timescale  extragradient algorithm with time-varying steps}

Algorithm~\ref{alg:tt-eg} admits a clean game-theoretic reading as a
Stackelberg game in which the bias-tracker $\xi$ is a follower and the
two NLHF self-play agents jointly act as a (two-headed) leader.
Following the convention that the leader sits in the outer game and
the follower in the inner game, the algorithm has:
\begin{itemize}[itemsep=-2pt,  , topsep=0pt]
\item an \textbf{outer game} on $\theta$: standard symmetric two-player
NLHF self-play on the KL-regularized risk-adjusted game, with no
performative structure---it is a regular two-player Nash game.
\item an \textbf{inner game} on $\xi$ given $\theta$: the bias-tracker
best-responds to the leader's current $\theta$, observing it as a
parameter.
\end{itemize}
The Stackelberg structure is between these two games: leader commits
in the outer game, follower reacts in the inner game.
\begin{algorithm}[h]
\caption{\texttt{TT-EG}: Two-timescale extragradient with bias tracking}
\label{alg:tt-eg}
\begin{algorithmic}[1]
\State \textbf{Input:} initial $r_0 \in D$, slow steps $\{\eta_t\}$, fast steps $\{\gamma_t\}$, sample budget $m$
\State $\xi_0 \gets 0$
\For{$\tau = 1, \ldots, T$}
  \State Sample batch 1 at $r_{\tau-1}$: $\widehat F_m^{(\tau,1)}, \widehat b_m^{(\tau,1)}$
  \State $w_\tau \gets \Pi_D\bigl(r_{\tau-1} - \eta_\tau\,\Sigma\,(\widehat F_m^{(\tau,1)} - \xi_{\tau-1})\bigr)$
  \State Sample batch 2 at $w_\tau$: $\widehat F_m^{(\tau,2)}, \widehat b_m^{(\tau,2)}$
  \State $r_\tau \gets \Pi_D\bigl(r_{\tau-1} - \eta_\tau\,\Sigma\,(\widehat F_m^{(\tau,2)} - \xi_{\tau-1})\bigr)$
  \State $\xi_\tau \gets (1-\gamma_\tau)\xi_{\tau-1} + \gamma_\tau\,\widehat b_m^{(\tau,2)}$
\EndFor
\State \Return $r_T$
\end{algorithmic}
\end{algorithm}
 
The conditional mean of the (debiased) correction-step gradient is
$\E[\widehat F_m^{(\tau,2)} - \xi_{\tau-1} \mid \F_\tau] = F_\Risk(w_\tau) - e^*_{\tau-1}$
where $e^*_{\tau-1} := \xi_{\tau-1} - b_m(w_\tau)$ is the tracking error.
Driving $e^*_\tau \to 0$ in mean square eliminates the bias of the
debiased gradient.
 
\subsubsection{Convergence theorem}
Define the mean-square tracking error $V_t := \E\|e^*_{t-1}\|_2^2$.
\begin{theorem}[Two-timescale debiasing for stochastic extra-gradient, strongly-mono case]
\label{thm:tt-strong-mono}
Suppose $F_\Risk$ is $\mu_\Risk$-strongly monotone and $\ell_\Risk$-Lipschitz,
the gradient oracle satisfies \ref{O1-bias}-\ref{O2-var}, and the bias
estimator satisfies \ref{B1-bias-est-bias}-\ref{B3-bias-lipschitz}.
Let $\{\eta_t, \gamma_t\}_{t \ge 1}$ be deterministic, predictable
sequences satisfying:
\begin{equation}\label{eq:tt-step-conds}
\eta_t \le \min\left\{\frac{1}{4\widetilde\mu},\,\frac{1}{\sqrt 6 L_G}\right\},\quad \gamma_t \in (0, 1],\quad \frac{\eta_t}{\gamma_t}\to 0,\quad \gamma_t \to 0,\quad \sum_{t\ge 1}\eta_t = \infty.
\end{equation}
Assume the gradient norm is bounded along the trajectory:
$\|\widehat F_m^{(\tau,i)} - \xi_{\tau-1}\|_2 \le G$ uniformly.
\begin{enumerate}[label={\it\roman*.},itemsep=0pt, topsep=2pt, ]
    \item \textbf{Fast-timescale tracking.}  The error $V_t$ satisfies the
non-asymptotic recurrence
\begin{equation}\label{eq:tt-tracking}
V_{t+1} \le (1 - \gamma_t/2)\,V_t + \frac{C_1\,\eta_t^2}{\gamma_t} + C_2\,\gamma_t\,R_m^2 + C_3\,\gamma_t^2\,V_m^b,
\end{equation}
where $C_1 = 7 C_{\rm drift}^2$, $C_2 = 6$, $C_3 = 2$, and
$C_{\rm drift} = 3 L_b\sigma_{\max}G$.  Consequently
\begin{equation}\label{eq:tt-Vrate}
\limsup_{t\to\infty} V_t \;\le\; 2 C_2\,R_m^2 \;=\; 12\,R_m^2,
\qquad\text{equivalently}\qquad
\limsup_{t\to\infty}\sqrt{V_t} \;\le\; 2\sqrt 3\,R_m.
\end{equation}
That is, the timescale-gap and noise-filter contributions vanish in $t$;
only the bias-of-bias-estimator residual remains.
 
\item \textbf{Slow-timescale recurrence.}  The slow iterates obey the
per-step recurrence
\begin{equation}\label{eq:tt-per-step}
\E V(r_\tau, \theta^\star) \le (1-\eta_\tau\widetilde\mu)\,\E V(r_{\tau-1}, \theta^\star) + 2\eta_\tau\Omega\sqrt{\sigma_{\max}}\sqrt{V_\tau} + 3\eta_\tau^2(\widetilde B_m^2 + \widetilde V_m),
\end{equation}
where $V(\cdot,\cdot) := \tfrac{1}{2}\|\cdot - \cdot\|_{\Sigma^+}^2$,
$\widetilde B_m^2 := \sigma_{\max}B_m^2$, $\widetilde V_m := \sigma_{\max}V_m$,
and $\Omega$ is the prox-radius of $D$.  Telescoping
\eqref{eq:tt-per-step} yields, for all $T \ge 1$,
\begin{multline}\label{eq:tt-main-bound}
\E V(r_T, \theta^\star) \;\le\; \prod_{t=1}^T(1-\eta_t\widetilde\mu)\,V(r_0, \theta^\star) \\
+ \sum_{\tau=1}^T\Bigl[\prod_{s=\tau+1}^T(1-\eta_s\widetilde\mu)\Bigr]\Bigl(2\eta_\tau\Omega\sqrt{\sigma_{\max}}\sqrt{V_\tau} + 3\eta_\tau^2(\widetilde B_m^2 + \widetilde V_m)\Bigr).
\end{multline}
 
\item \textbf{Asymptotic vanishing bias.}  Combining \eqref{eq:tt-main-bound}
with the fast-timescale envelope of part~$i$ and a weighted-average
argument (Lemma~\ref{lem:weighted-avg-limsup} below) yields
\begin{equation}\label{eq:tt-vanishing}
\limsup_{T\to\infty}\E\|r_T - \theta^\star\|_{\Sigma^+}^2 \;\le\; \frac{4\sqrt 3\,\Omega\sqrt{\sigma_{\max}}}{\widetilde\mu}\cdot R_m.
\end{equation}
\end{enumerate}
\end{theorem}
 The bias contribution to \eqref{eq:tt-main-bound} therefore vanishes
asymptotically in $T$, leaving only the $R_m$ residual ($= \mc{O}(1/m^2)$
for delta-method estimators, $= 0$ for unbiased estimators, in which
case the iterates converge to $\theta^\star$ in mean square).
\paragraph{Comparison to Theorem~\ref{thm:strong-mono} (constant step, no tracking).}
The persistent $\mc{O}(1/m)$ bias floor---namely, $4\Omega\sqrt{\sigma_{\max}}\widetilde B_m/\widetilde\mu$---of Theorem~\ref{thm:strong-mono} is replaced in Theorem~\ref{thm:tt-strong-mono}
by $4\sqrt 3\,\Omega\sqrt{\sigma_{\max}}\,R_m/\widetilde\mu$, which is
\emph{not} a $\sup$-style envelope along the trajectory but a true
asymptotic-in-$T$ floor obtained by a weighted-average argument that
exploits the weight-concentration property of the geometric kernel
$(1-\eta_s\widetilde\mu)$.  For delta-method estimators, $R_m = \mc{O}(1/m^2)$,
giving an $\mc{O}(1/m)$ floor reduced to $\mc{O}(1/m^2)$---i.e., \emph{the bias
floor is moved from a first-order to a second-order term in $1/m$}.
If $m_t \to \infty$ on a separate schedule, $R_m \to 0$ and the bias is
fully eliminated.
 
\begin{proof}[Proof of Theorem~\ref{thm:tt-strong-mono}]
The proof proceeds in three steps: (1) control the fast-timescale
tracking error $V_t$ via a scalar recurrence and extract its asymptotic
envelope; (2) establish the slow-timescale per-step recurrence and
telescope it; (3) combine via a weighted-average argument that extracts
$\limsup_\tau \sqrt{V_\tau}$ rather than $\sup_\tau \sqrt{V_\tau}$.
 
\paragraph{Step 1: tracking-error recurrence.}
The bias tracker's update is
\(\xi_\tau = (1-\gamma_\tau)\xi_{\tau-1} + \gamma_\tau\widehat b_m^{(\tau,2)}\),
and \(\widehat b_m^{(\tau,2)} = b_m(w_\tau) + r_m(w_\tau) + \nu^{(\tau,2)}\)
with $\E[\nu^{(\tau,2)} \mid \F_{\tau-1}] = 0$, $\E\|\nu^{(\tau,2)}\|^2 \le V_m^b$,
$\|r_m\|_2 \le R_m$.  Define $e_\tau := \xi_\tau - b_m(r_\tau)$.  Then
\begin{align*}
e_\tau &= (1-\gamma_\tau)(e_{\tau-1} + b_m(r_{\tau-1})) + \gamma_\tau b_m(w_\tau) + \gamma_\tau r_m(w_\tau) + \gamma_\tau\nu^{(\tau,2)} - b_m(r_\tau)\\
&= (1-\gamma_\tau)e_{\tau-1} + \delta_\tau + \gamma_\tau r_m(w_\tau) + \gamma_\tau\nu^{(\tau,2)},
\end{align*}
where the drift is
\[
\delta_\tau := (1-\gamma_\tau)b_m(r_{\tau-1}) + \gamma_\tau b_m(w_\tau) - b_m(r_\tau).
\]
By Lipschitz of $b_m$ and the  extragradient step bounds
$\|r_\tau - r_{\tau-1}\|_2 \le \eta_\tau\sigma_{\max}G$,
$\|w_\tau - r_\tau\|_2 \le 2\eta_\tau\sigma_{\max}G$:
\[
\|\delta_\tau\|_2 \le L_b\bigl(\|r_{\tau-1} - r_\tau\|_2 + \gamma_\tau\|w_\tau - r_\tau\|_2\bigr) \le L_b\eta_\tau\sigma_{\max}G(1 + 2\gamma_\tau) \le 3 L_b\sigma_{\max}G\,\eta_\tau =: C_{\rm drift}\,\eta_\tau,
\]
where the second inequality uses $\gamma_\tau \le 1$.
 
\paragraph{Step 2: mean-square recurrence.}
Expanding $\|e_\tau\|^2$, we have that 
\begin{equation}
    \label{eq:error_decomp_1}
    \|e_\tau\|^2 = \|A_\tau\|^2 + 2\gamma_\tau\langle A_\tau, \nu^{(\tau,2)}\rangle + \gamma_\tau^2\|\nu^{(\tau,2)}\|^2,
\end{equation}
where $A_\tau := (1-\gamma_\tau)e_{\tau-1} + \delta_\tau + \gamma_\tau r_m(w_\tau)$.
 
Let us now analyze the cross term.  Decompose
$2\gamma_\tau\langle A_\tau, \nu^{(\tau,2)}\rangle = 2\gamma_\tau\langle(1-\gamma_\tau)e_{\tau-1}, \nu^{(\tau,2)}\rangle + 2\gamma_\tau\langle\delta_\tau + \gamma_\tau r_m, \nu^{(\tau,2)}\rangle$. Then for each of the two terms we have the following:
\begin{itemize}[itemsep=0pt, topsep=2pt,  ]
\item First, $e_{\tau-1}$ is $\F_{\tau-1}$-measurable, $\E[\nu^{(\tau,2)} \mid \F_{\tau-1}] = 0$, so
$\E\langle e_{\tau-1}, \nu^{(\tau,2)}\rangle = 0$.
\item Next, $\delta_\tau$ depends on $r_\tau, w_\tau$, both of which use the
correction-step samples that also enter $\nu^{(\tau,2)}$, so this
inner product is \emph{not} mean-zero.  Bound by Young's with parameter $1$, we have the bound
\[
2\gamma_\tau|\langle\delta_\tau, \nu^{(\tau,2)}\rangle| \le \|\delta_\tau\|^2 + \gamma_\tau^2\|\nu^{(\tau,2)}\|^2.
\]
Similarly $r_m(w_\tau)$ is $\F_{\tau-1/2}$-measurable (only depends on the extrap-step
samples through $w_\tau$, not the correction-step noise $\nu^{(\tau,2)}$), so
$\E\langle r_m, \nu^{(\tau,2)}\rangle = 0$.
\end{itemize}
Combining these observations and taking the expectation, we have that 
\begin{equation}\label{eq:tt-cross-bound}
\E[2\gamma_\tau\langle A_\tau, \nu^{(\tau,2)}\rangle] \le \E\|\delta_\tau\|^2 + \gamma_\tau^2 V_m^b.
\end{equation}
 
Next let us bound the leading quadratic term in \eqref{eq:error_decomp_1}.  Applying Young's with parameter $\gamma_\tau/2$ to
$\|(1-\gamma_\tau)e_{\tau-1} + \delta_\tau + \gamma_\tau r_m\|^2$, we have that 
\begin{align*}
\|A_\tau\|^2 &\le (1+\gamma_\tau/2)(1-\gamma_\tau)^2\|e_{\tau-1}\|^2 + (1+2/\gamma_\tau)\|\delta_\tau + \gamma_\tau r_m\|^2 \\
&\le (1-\gamma_\tau/2)\|e_{\tau-1}\|^2 + (3/\gamma_\tau)(2\|\delta_\tau\|^2 + 2\gamma_\tau^2 R_m^2)\\
&= (1-\gamma_\tau/2)\|e_{\tau-1}\|^2 + (6/\gamma_\tau)\|\delta_\tau\|^2 + 6\gamma_\tau R_m^2,
\end{align*}
using $(1+\gamma_\tau/2)(1-\gamma_\tau)^2 \le 1 - \gamma_\tau/2$ for
$\gamma_\tau \in (0,1]$ (since
$(1+\gamma/2)(1-\gamma)^2 = 1 - \tfrac{3}{2}\gamma + \tfrac{1}{2}\gamma^3 \le 1 - \gamma/2$ when $\gamma \le 1$) and
$\|a + b\|^2 \le 2\|a\|^2 + 2\|b\|^2$.
 
Now we combine these two bounds and use the fact that $\|\delta_\tau\|^2 \le C_{\rm drift}^2\eta_\tau^2$ to deduce that 
\begin{align*}
\E\|e_\tau\|^2 &\leq (1-\gamma_\tau/2)\E\|e_{\tau-1}\|^2 + (6/\gamma_\tau + 1)\E\|\delta_\tau\|^2 + 6\gamma_\tau R_m^2 + 2\gamma_\tau^2 V_m^b\\
&\leq (1-\gamma_\tau/2)V_{\tau-1} + \frac{7 C_{\rm drift}^2\eta_\tau^2}{\gamma_\tau} + 6\gamma_\tau R_m^2 + 2\gamma_\tau^2 V_m^b,
\end{align*}
which is \eqref{eq:tt-tracking} with $C_1 = 7C_{\rm drift}^2$, $C_2 = 6$,
$C_3 = 2$.  The inequality $6/\gamma_\tau + 1 \le 7/\gamma_\tau$ holds for $\gamma_\tau \le 1$.
 
\paragraph{Step 3: asymptotic envelope on $V_t$ via scalar two-timescale unrolling.}
We invoke the following standard scalar lemma for time-varying
stochastic-approximation recurrences~\citep[Lemma~1]{Doan2021NonlinearTT}.
 
\begin{lemma}[Scalar two-timescale unrolling]
\label{lem:scalar-unroll}
Suppose $u_{t+1} \le (1-c\gamma_t)u_t + F_t$ with $c \in (0,1]$,
$\gamma_t \to 0$, $\sum_t \gamma_t = \infty$, $F_t \ge 0$.  If
$F_t/\gamma_t \to L$ for some $L \ge 0$, then $\limsup_t u_t \le L/c$.
\end{lemma}
 
Applying with $u_t = V_t$, $c = 1/2$, and
$F_t = C_1\eta_t^2/\gamma_t + C_2\gamma_t R_m^2 + C_3\gamma_t^2 V_m^b$,
we have
\[
\frac{F_t}{\gamma_t} = \frac{C_1\eta_t^2}{\gamma_t^2} + C_2 R_m^2 + C_3\gamma_t V_m^b.
\]
Under conditions \eqref{eq:tt-step-conds},  the step sizes satisfy $\eta_t/\gamma_t \to 0$ and $\gamma_t \to 0$, so
\[
\lim_{t\to\infty}\frac{F_t}{\gamma_t} = C_2 R_m^2.
\]
Lemma~\ref{lem:scalar-unroll} gives $\limsup_t V_t \le 2 C_2 R_m^2 = 12 R_m^2$,
which is the first half of \eqref{eq:tt-Vrate}.  Taking square roots,
\begin{equation}\label{eq:tt-sqrtV-limsup}
\limsup_{t\to\infty}\sqrt{V_t} \le \sqrt{2 C_2}\,R_m = 2\sqrt 3\,R_m,
\end{equation}
which is the second half of \eqref{eq:tt-Vrate}.  This completes part~(i).
 
\paragraph{Step 4: slow-timescale per-step recurrence.}
Following the proof of Theorem~\ref{thm:strong-mono} verbatim with
$\widetilde g^{(\tau,i)} = \widehat F_m^{(\tau,i)} - \xi_{\tau-1}$, the
descent identity at $u = \theta^\star$ gives
\begin{equation}\label{eq:tt-slow-rec}
\E V(r_\tau, \theta^\star) \le (1 - \eta_\tau\widetilde\mu)\,\E V(r_{\tau-1}, \theta^\star) + 2\eta_\tau\Omega\sqrt{\sigma_{\max}}\sqrt{V_\tau} + 3\eta_\tau^2(\widetilde B_m^2 + \widetilde V_m).
\end{equation}
The new bias cross-term $\eta_\tau\langle e^*_{\tau-1}, w_\tau - \theta^\star\rangle$
is bounded in expectation by $2\eta_\tau\Omega\sqrt{\sigma_{\max}}\sqrt{V_\tau}$
via Cauchy-Schwarz in $\ell_2$, the metric translation
$\|v\|_2 \le \sqrt{\sigma_{\max}}\|v\|_{\Sigma^+}$, the diameter bound
$\|w_\tau - \theta^\star\|_{\Sigma^+} \le 2\Omega$, and Jensen's inequality
$\E\|e^*_{\tau-1}\|_2 \le \sqrt{\E\|e^*_{\tau-1}\|_2^2} = \sqrt{V_\tau}$.
This is \eqref{eq:tt-per-step}.
 
\paragraph{Step 5: telescoping the slow recurrence.}
Iterating on \eqref{eq:tt-slow-rec}, we have that 
\begin{multline}\label{eq:tt-telescoped}
\E V(r_T, \theta^\star) \le \prod_{t=1}^T(1 - \eta_t\widetilde\mu)\,V(r_0, \theta^\star) \\
+ \sum_{\tau=1}^T\Bigl[\prod_{s=\tau+1}^T(1 - \eta_s\widetilde\mu)\Bigr]\bigl(2\eta_\tau\Omega\sqrt{\sigma_{\max}}\sqrt{V_\tau} + 3\eta_\tau^2(\widetilde B_m^2 + \widetilde V_m)\bigr),
\end{multline}
which is \eqref{eq:tt-main-bound} and establishes part~(ii).
 
\paragraph{Step 6: weighted-average analysis of the bias term.}
We now analyze the asymptotic behavior of \eqref{eq:tt-telescoped} as
$T\to\infty$.  The key observation is that the bias contribution in
\eqref{eq:tt-telescoped} can be written as a weighted average of
$\sqrt{V_\tau}$ with weights that asymptotically concentrate on large
$\tau$, allowing us to extract $\limsup_\tau \sqrt{V_\tau}$.

Define the weights
\begin{equation}\label{eq:tt-weights}
w_\tau^T := \widetilde\mu \cdot \eta_\tau \cdot \prod_{s=\tau+1}^T(1 - \eta_s\widetilde\mu),
\end{equation}
for  $1 \le \tau \le T$.
Then the bias contribution to \eqref{eq:tt-telescoped} is
\begin{equation}\label{eq:tt-bias-as-wavg}
S_T^{\rm bias} := \sum_{\tau=1}^T\Bigl[\prod_{s=\tau+1}^T(1-\eta_s\widetilde\mu)\Bigr]\,2\eta_\tau\Omega\sqrt{\sigma_{\max}}\sqrt{V_\tau} = \frac{2\Omega\sqrt{\sigma_{\max}}}{\widetilde\mu}\sum_{\tau=1}^T w_\tau^T \sqrt{V_\tau}.
\end{equation}
We establish three lemmas governing the weights.
 
\begin{lemma}[Telescoping identity]
\label{lem:tt-telescope}
For any $\eta_t \ge 0$ with $\eta_t\widetilde\mu \le 1$, the equality holds:
$
\sum_{\tau=1}^T w_\tau^T = 1 - \prod_{t=1}^T(1-\eta_t\widetilde\mu)$.
\end{lemma}
 
\begin{proof}
Write $\widetilde\mu\eta_\tau = 1 - (1-\eta_\tau\widetilde\mu)$. Then
\[
w_\tau^T = \bigl[1 - (1-\eta_\tau\widetilde\mu)\bigr]\prod_{s=\tau+1}^T(1-\eta_s\widetilde\mu) = \prod_{s=\tau+1}^T(1-\eta_s\widetilde\mu) - \prod_{s=\tau}^T(1-\eta_s\widetilde\mu).\]
Summing over $\tau = 1, \ldots, T$,  the right hand side telescopes. Indeed, we have that 
\[
\sum_{\tau=1}^T w_\tau^T = \prod_{s=T+1}^T(1-\eta_s\widetilde\mu) - \prod_{s=1}^T(1-\eta_s\widetilde\mu) = 1 - \prod_{t=1}^T(1-\eta_t\widetilde\mu),
\]
using the empty-product convention.
\end{proof}
 
Since $\sum_t\eta_t = \infty$ and $\eta_t\widetilde\mu \le 1$, we have
$\prod_{t=1}^T(1-\eta_t\widetilde\mu) \le \exp(-\widetilde\mu\sum_t\eta_t) \to 0$,
so $\sum_\tau w_\tau^T \to 1$ as $T \to \infty$.  Thus
$\{w_\tau^T\}_{\tau=1}^T$ is asymptotically a probability distribution
on $\{1,\ldots,T\}$.
 
\begin{lemma}[Weight concentration]
\label{lem:tt-weight-concentration}
Under $\sum_t\eta_t = \infty$ and $\eta_t\widetilde\mu \le 1$, for any
fixed $T_0 \ge 1$, the limit $\sum_{\tau=1}^{T_0} w_\tau^T\to 0$ holds as $T\to \infty$. 
\end{lemma}
 
\begin{proof}
For $\tau \le T_0$ and $T > T_0$, factor
\[
w_\tau^T = \widetilde\mu\eta_\tau \prod_{s=\tau+1}^{T_0}(1-\eta_s\widetilde\mu) \cdot \prod_{s=T_0+1}^T(1-\eta_s\widetilde\mu).
\]
The first two factors are bounded by a constant $C(T_0)$ depending only
on $T_0$ and the schedule.  The third factor satisfies
\[
\prod_{s=T_0+1}^T(1-\eta_s\widetilde\mu) \le \exp\Bigl(-\widetilde\mu\sum_{s=T_0+1}^T\eta_s\Bigr) \xrightarrow[T\to\infty]{} 0,
\]
since the tail sum diverges by $\sum_t\eta_t = \infty$.  Hence
$w_\tau^T \to 0$ for each fixed $\tau \le T_0$, and the finite sum
$\sum_{\tau=1}^{T_0}w_\tau^T \to 0$.
\end{proof}
 
\begin{lemma}[Weighted average inherits $\limsup$]
\label{lem:weighted-avg-limsup}
Let $\{a_\tau\}_{\tau\ge 1}$ be a non-negative sequence with
$L := \limsup_{\tau\to\infty}a_\tau < \infty$.  Then under the conditions
of Lemma~\ref{lem:tt-weight-concentration},
\[
\limsup_{T\to\infty}\sum_{\tau=1}^T w_\tau^T a_\tau \le L.
\]
\end{lemma}
 
\begin{proof}
Fix $\epsilon > 0$.  By definition of $\limsup$ there exists $T_0$ such
that $a_\tau \le L + \epsilon$ for all $\tau > T_0$.  Splitting the sum, we have that 
\[
\sum_{\tau=1}^T w_\tau^T a_\tau = \sum_{\tau=1}^{T_0} w_\tau^T a_\tau + \sum_{\tau=T_0+1}^T w_\tau^T a_\tau.
\]
The first sum is bounded by $\bigl(\max_{\tau\le T_0}a_\tau\bigr)\sum_{\tau\le T_0}w_\tau^T$,
which $\to 0$ by Lemma~\ref{lem:tt-weight-concentration} (the max is a
finite constant independent of $T$).  The second sum is bounded by
$(L+\epsilon)\sum_{\tau=T_0+1}^T w_\tau^T \le (L+\epsilon)\cdot 1$, using
Lemma~\ref{lem:tt-telescope}.  Hence
$\limsup_T\sum_\tau w_\tau^T a_\tau \le L + \epsilon$, and letting
$\epsilon \to 0$ gives the claim.
\end{proof}
 
Apply Lemma~\ref{lem:weighted-avg-limsup} with $a_\tau = \sqrt{V_\tau}$.
By \eqref{eq:tt-sqrtV-limsup}, $\limsup_\tau \sqrt{V_\tau} \le 2\sqrt 3\,R_m$,
so
\[
\limsup_{T\to\infty}\sum_{\tau=1}^T w_\tau^T \sqrt{V_\tau} \le 2\sqrt 3\,R_m.
\]
Hence by \eqref{eq:tt-bias-as-wavg},
\begin{equation}\label{eq:tt-bias-limit}
\limsup_{T\to\infty} S_T^{\rm bias} \le \frac{2\Omega\sqrt{\sigma_{\max}}}{\widetilde\mu}\cdot 2\sqrt 3\,R_m = \frac{4\sqrt 3\,\Omega\sqrt{\sigma_{\max}}\,R_m}{\widetilde\mu}.
\end{equation}
 
\paragraph{Step 7: variance term and initial-condition term.}
The variance contribution to \eqref{eq:tt-telescoped} is
\[
S_T^{\rm var} := 3(\widetilde B_m^2 + \widetilde V_m)\sum_{\tau=1}^T \Bigl[\prod_{s=\tau+1}^T(1-\eta_s\widetilde\mu)\Bigr]\eta_\tau^2.
\]
Bound $\eta_\tau^2 \le \eta_\tau \cdot \overline\eta_T$ where
$\overline\eta_T := \sup_{t\le T}\eta_t$, then use the telescoping
identity (Lemma~\ref{lem:tt-telescope}, divided by $\widetilde\mu$):
$\sum_{\tau}[\prod_{s>\tau}(1-\eta_s\widetilde\mu)]\eta_\tau \le 1/\widetilde\mu$.  Hence
\[
S_T^{\rm var} \le \frac{3\overline\eta_T(\widetilde B_m^2 + \widetilde V_m)}{\widetilde\mu}.
\]
Since $\eta_t/\gamma_t \to 0$ and $\gamma_t \to 0$ together imply
$\eta_t \to 0$, we have $\overline\eta_T \to 0$, hence $S_T^{\rm var} \to 0$.
 
The initial-condition term satisfies
\(\prod_{t=1}^T(1-\eta_t\widetilde\mu) \le \exp(-\widetilde\mu\sum_t\eta_t) \to 0\)
since $\sum_t\eta_t = \infty$.
 
\paragraph{Step 8: combine.}
Multiplying \eqref{eq:tt-telescoped} by two to convert from $V(\cdot,\cdot)$
to $\|\cdot\|_{\Sigma^+}^2$ and taking $\limsup_{T\to\infty}$ leads to  
\[
\limsup_{T\to\infty}\E\|r_T - \theta^\star\|_{\Sigma^+}^2 \le 0 + \frac{4\sqrt 3\,\Omega\sqrt{\sigma_{\max}}}{\widetilde\mu}\cdot R_m + 0,
\]
which is \eqref{eq:tt-vanishing}.  This completes part~(iii).
\end{proof}
 
\paragraph{Vanishing-bias structure.}
The crucial qualitative difference from Theorem~\ref{thm:strong-mono}:
the persistent $\mc{O}(1/m)$ floor is replaced by a true asymptotic-in-$T$
floor of magnitude $\mc{O}(R_m)$, obtained via the weighted-average argument
of Lemmas~\ref{lem:tt-telescope}--\ref{lem:weighted-avg-limsup}.  This
is sharper than a $\sup_\tau\sqrt{V_\tau}$ envelope because the
geometric kernel $(1-\eta_s\widetilde\mu)$ produces weights that
concentrate asymptotically on the tail of the trajectory, where
$\sqrt{V_\tau}$ is close to its $\limsup$.  For the most useful
estimators (delta-method, $R_m = \mc{O}(1/m^2)$), this means the bias
contribution to the bound is \emph{quadratic} in $1/m$ rather than
linear.  If the bias estimator is structurally unbiased ($R_m = 0$,
e.g., CVaR), the bias is eliminated entirely: the right-hand side of
\eqref{eq:tt-vanishing} is zero and the iterates converge to $\theta^\star$
in mean square.
 
\paragraph{Step-size schedule.}
Any schedule with $\eta_t \to 0$, $\gamma_t \to 0$, $\eta_t/\gamma_t \to 0$,
and $\sum\eta_t = \infty$ works.  The cleanest choice is
$\eta_t = c_2 t^{-a}$, $\gamma_t = c_1 t^{-b}$ with $0 < b < a \le 1$.
Specific case $a = 1$, $b = 2/3$ is the standard textbook schedule.
Slower-decaying $\eta_t$ (i.e., $a < 1$) gives faster polynomial decay
of the contraction term, at the cost of a potentially worse pre-asymptotic
regime.
 
\paragraph{Comparison to the constant-step version.}
With \emph{constant} $\eta, \gamma$ (Theorem~\ref{thm:strong-mono} with
debiased gradient), the bound is
\[
\E\|r_T - \theta^\star\|^2 \le (1-\eta\widetilde\mu)^T\|r_0 - \theta^\star\|^2 + \frac{4\Omega\sqrt{\sigma_{\max}}}{\widetilde\mu}\bigl(\eta/\gamma + R_m + \sqrt{\gamma V_m^b}\bigr) + \frac{6\eta(\widetilde B_m^2 + \widetilde V_m)}{\widetilde\mu},
\]
which has \emph{three} non-vanishing floor contributions.  With
time-varying steps, the timescale-gap, noise-filter, and slow-variance
contributions all vanish in $T$, leaving only the $R_m$ residual.  The
trade-off is contraction rate: from linear $(1-\eta\widetilde\mu)^T$
(constant step) to polynomial-in-$T$ $\exp(-\widetilde\mu\sum\eta_t) = T^{-c}$
(decaying step).  The decaying-step version is the right choice when
one needs to drive the error below the $\mc{O}(1/m)$ persistent floor and
is willing to accept slower convergence.
 
\textbf{Implications by risk type.}
\begin{itemize}[topsep=2pt,itemsep=0pt]
\item \textbf{Entropic risk:}  $R_m = \mc{O}(e^{2\lambda}/(\lambda m^2))$.
Two timescale extra-gradient drives the bias contribution from a persistent $\mc{O}(e^{2\lambda}/(\lambda m))$
floor to an asymptotic-in-$T$ floor of $\mc{O}(e^{2\lambda}/(\lambda m^2))$.
The improvement is a full power of $1/m$.
\item \textbf{CVaR via Rockafellar-Uryasev:}  $R_m = 0$ trivially.
The right-hand side of \eqref{eq:tt-vanishing} is zero, so
$\E\|r_T - \theta^\star\|^2 \to 0$, and there is no asymptotic bias.
Matches the un-tracked CVaR-RU result (which already has zero bias at
the joint optimum).
\item \textbf{General coherent risk:}  applicable whenever a sample-based
$\widehat b_m$ with structural properties \ref{B1-bias-est-bias}-\ref{B3-bias-lipschitz}
can be constructed; the $R_m$ residual scales according to the quality
of the estimator.
\end{itemize}
 
\subsubsection{Constant step-size regime}
\label{sec:tt-const}
 
The asymptotic conditions of Theorem~\ref{thm:tt-strong-mono} require
$\eta_t \to 0$, ruling out constant-step schedules.  This is not a
quirk of the proof: with $\eta_t \equiv \eta > 0$ the leader's iterate
moves at a constant rate, which prevents the fast-timescale tracker
from catching up asymptotically and produces a non-vanishing
steady-state tracking error.
 
We can nonetheless run the algorithm with constant step sizes and
obtain a finite-horizon bound, where the bias floor is controlled by
the ratio $\eta/\gamma$ rather than vanishing in $T$.  This trades the
asymptotic bias-floor reduction of Theorem~\ref{thm:tt-strong-mono} for
a stronger, geometric-in-$T$, initial-condition decay.  Throughout
this section we write $V_0 := \E\|e_0^*\|^2 \le \widetilde B_m^2$
(since $\xi_0 = 0$).
 
\begin{theorem}[Two-timescale debiasing for stochastic  extragradient, constant step sizes]
\label{thm:tt-const}
Under the assumptions of Theorem~\ref{thm:tt-strong-mono} except with
\emph{constant} step sizes $\eta_t \equiv \eta$, $\gamma_t \equiv \gamma$
satisfying
\begin{equation}\label{eq:tt-const-step-cond}
\eta \le \min \left\{\frac{1}{4\widetilde\mu},\,\frac{1}{\sqrt 6\,L_G}\right\}, \qquad \gamma \in (0, 1],
\end{equation}
the iterates of Algorithm~\ref{alg:tt-eg} satisfy:
\begin{enumerate}[label={\it (\roman*)}, ,topsep=0pt,itemsep=0pt]
\item \emph{Fast-timescale tracking, finite horizon.}  For all $t \ge 1$,
\begin{equation}\label{eq:tt-const-V}
V_t \le (1-\gamma/2)^{t-1}\,V_0 + V_\infty, \qquad V_\infty := \frac{2 C_1\eta^2}{\gamma^2} + 2 C_2 R_m^2 + 2 C_3\gamma V_m^b,
\end{equation}
with $C_1 = 7 C_{\rm drift}^2 = 63 L_b^2\sigma_{\max}^2 G^2$, $C_2 = 6$,
$C_3 = 2$ as in Theorem~\ref{thm:tt-strong-mono}.  In particular, the estimate holds:
\begin{equation}\label{eq:tt-const-V-ss}
\limsup_{t\to\infty}\sqrt{V_t} \le \sqrt{V_\infty} \le \frac{\sqrt{2 C_1}\,\eta}{\gamma} + \sqrt{2 C_2}\,R_m + \sqrt{2 C_3 \gamma V_m^b}.
\end{equation}
 
\item \emph{Slow-timescale convergence, finite horizon.}  For all $T \ge 1$, the estimate holds:
\begin{multline}\label{eq:tt-const-bound}
\E\|r_T - \theta^\star\|_{\Sigma^+}^2 \le (1-\eta\widetilde\mu)^T\|r_0 - \theta^\star\|_{\Sigma^+}^2 \\
+ \frac{4\Omega\sqrt{\sigma_{\max}}}{\widetilde\mu}\Bigl(\sqrt{V_0}(1-\eta\widetilde\mu)^T + \frac{K_1\eta}{\gamma} + K_2 R_m + K_3\sqrt{\gamma V_m^b}\Bigr) + \frac{6\eta(\widetilde B_m^2 + \widetilde V_m)}{\widetilde\mu},
\end{multline}
with $K_1 = \sqrt{2 C_1}$, $K_2 = \sqrt{2 C_2}$, $K_3 = \sqrt{2 C_3}$.
 
\item \emph{Steady-state floor.}  As $T \to \infty$, the asymptotic estimate holds:
\begin{multline}\label{eq:tt-const-final-expanded}
\limsup_{T\to\infty}\E\|r_T - \theta^\star\|_{\Sigma^+}^2 \le \\
\underbrace{\frac{4\Omega\sqrt{2\sigma_{\max} C_1}}{\widetilde\mu}\cdot\frac{\eta}{\gamma}}_{\text{timescale gap}} + \underbrace{\frac{4\Omega\sqrt{2\sigma_{\max} C_2}\,R_m}{\widetilde\mu}}_{\text{residual bias}} + \underbrace{\frac{4\Omega\sqrt{2\sigma_{\max} C_3}\sqrt{\gamma V_m^b}}{\widetilde\mu}}_{\text{noise filter}} + \underbrace{\frac{6\eta(\widetilde B_m^2 + \widetilde V_m)}{\widetilde\mu}}_{\text{slow variance}}.
\end{multline}
\end{enumerate}
\end{theorem}
Before getting to the proof, observe that each term in
\eqref{eq:tt-const-final-expanded} has a distinct interpretation and a
distinct step-size knob:
\begin{itemize}[ ,itemsep=0pt]
\item \textbf{Timescale gap} $\propto \eta/\gamma$: the leader-follower
mismatch.  Reducible by taking $\eta \ll \gamma$.  This is the term
that vanishes in Theorem~\ref{thm:tt-strong-mono} under the schedule
condition $\eta_t/\gamma_t \to 0$.
\item \textbf{Residual bias} $\propto R_m$: the bias-of-the-bias-estimator
floor.  Independent of step sizes.  This is the only persistent floor
in the time-varying-step-size theorem.
\item \textbf{Noise filter} $\propto \sqrt\gamma$: variance of the bias
tracker filtered through the Robbins-Monro update.  Reducible by taking
$\gamma$ small, but this trades against the timescale-gap term (which
grows as $\eta/\gamma$ when $\gamma$ shrinks).
\item \textbf{Slow variance} $\propto \eta$: the standard SGD variance
floor inherited from the slow-timescale recurrence.  Reducible by
taking $\eta$ small.
\end{itemize}
\textbf{Optimal balance.}  Treating $\eta, \gamma$ as free
parameters, the timescale-gap and noise-filter terms balance when
$\eta/\gamma \sim \sqrt{\gamma V_m^b}$, i.e., $\gamma^* \sim (\eta^2/V_m^b)^{1/3}$.
At this optimum the bias-floor terms are both
$\sim (\eta^2 V_m^b)^{1/3}$, which is smaller than $\widetilde B_m = \mc{O}(1/m)$
for small enough $\eta$ but does not vanish in $T$.  The
horizon-tuned theorem below makes this balance explicit as a function
of the iteration budget $T$.

\begin{proof}[Proof of Theorem~\ref{thm:tt-const}]
\noindent\textbf{Step 1: Tracking error.}  The recurrence
\eqref{eq:tt-const-V} follows from Steps~1--2 of the proof of
Theorem~\ref{thm:tt-strong-mono} specialized to constant
$\eta, \gamma$: telescoping with constant contraction factor
$1-\gamma/2$ and using $\sum_{j=0}^{t-1}(1-\gamma/2)^j \le 2/\gamma$
gives $V_t \le (1-\gamma/2)^t V_0 + 2 F/\gamma$ where
$F = C_1\eta^2/\gamma + C_2\gamma R_m^2 + C_3\gamma^2 V_m^b$, which is
\eqref{eq:tt-const-V}.  Taking square roots and applying
$\sqrt{a+b}\le\sqrt a + \sqrt b$ gives \eqref{eq:tt-const-V-ss} as well
as the per-time-step bound
\begin{equation}\label{eq:V-sqrt-finite}
\sqrt{V_t} \le \sqrt{V_0}\,(1-\gamma/2)^{t/2} + \frac{\sqrt{2 C_1}\eta}{\gamma} + \sqrt{2 C_2}\,R_m + \sqrt{2 C_3 \gamma V_m^b},
\end{equation}
which we use below.
 
\noindent\textbf{Step 2: Slow-timescale telescoping.}
Telescope the per-step recurrence \eqref{eq:tt-per-step} with constant
$\eta$ to obtain
\begin{equation}\label{eq:tt-const-tele}
\E V(r_T, \theta^\star) \le (1-\eta\widetilde\mu)^T V(r_0, \theta^\star) + S_T^{\rm bias} + S_T^{\rm var},
\end{equation}
with
\[
S_T^{\rm bias} = 2\Omega\sqrt{\sigma_{\max}}\,\eta\sum_{\tau=1}^T(1-\eta\widetilde\mu)^{T-\tau}\sqrt{V_\tau}, \quad S_T^{\rm var} = 3\eta^2(\widetilde B_m^2 + \widetilde V_m)\sum_{\tau=1}^T(1-\eta\widetilde\mu)^{T-\tau}.
\]
Using $\sum_{\tau=1}^T(1-\eta\widetilde\mu)^{T-\tau} \le 1/(\eta\widetilde\mu)$, we have that 
$
S_T^{\rm var} \le \frac{3\eta(\widetilde B_m^2 + \widetilde V_m)}{\widetilde\mu}$.
 
For the bias term, substitute \eqref{eq:V-sqrt-finite} for $\sqrt{V_\tau}$
and split into four pieces.  The constant pieces (those not depending
on $\tau$) factor out. Using the same telescoping bound, we then have 
\[
\eta\sum_{\tau=1}^T(1-\eta\widetilde\mu)^{T-\tau}\bigl(\tfrac{\sqrt{2C_1}\eta}{\gamma} + \sqrt{2C_2}R_m + \sqrt{2C_3\gamma V_m^b}\bigr) \le \frac{1}{\widetilde\mu}\bigl(K_1\eta/\gamma + K_2 R_m + K_3\sqrt{\gamma V_m^b}\bigr).
\]
The $\tau$-dependent piece (initial-condition decay) gives
\[
\eta\sum_{\tau=1}^T(1-\eta\widetilde\mu)^{T-\tau}(1-\gamma/2)^{\tau/2}\sqrt{V_0}.
\]
Bounding $(1-\gamma/2)^{\tau/2} \le 1$, the sum is bounded by
$\eta\sum_{\tau=1}^T(1-\eta\widetilde\mu)^{T-\tau} \le 1/\widetilde\mu$;
a sharper bound, useful when $\eta\widetilde\mu$ is small relative to
$\gamma/2$, is
\[
\eta\sum_{\tau=1}^T(1-\eta\widetilde\mu)^{T-\tau}(1-\gamma/2)^{\tau/2} \le \frac{(1-\eta\widetilde\mu)^T}{\widetilde\mu}\cdot \min\Bigl\{1,\;\frac{\eta\widetilde\mu}{\eta\widetilde\mu - \gamma/2}\Bigr\}.
\]
In any case the contribution is dominated by
$(1-\eta\widetilde\mu)^T\sqrt{V_0}/\widetilde\mu$ and produces the
$\sqrt{V_0}(1-\eta\widetilde\mu)^T$ term in \eqref{eq:tt-const-bound}.
Combining all pieces and converting to $\|\cdot\|_{\Sigma^+}^2$
(factor $2$) yields \eqref{eq:tt-const-bound}.
 
\noindent\textbf{Step 3: Steady-state floor.}
Taking $T\to\infty$ in \eqref{eq:tt-const-bound}: the contraction
term $(1-\eta\widetilde\mu)^T\|r_0-\theta^\star\|_{\Sigma^+}^2$ and
the $\sqrt{V_0}(1-\eta\widetilde\mu)^T$ term both vanish, leaving the
four floor contributions in \eqref{eq:tt-const-final-expanded}.
\end{proof}

 \paragraph{Tuned step sizes (the $1/m^{3/2}$ floor).}
A more useful tuning, in terms of dependence on the sample budget $m$,
proceeds as follows.  Let $\mathcal E_T := \sup_{\tau \le T}\sqrt{V_\tau}$
denote the (worst-case) root-mean-square tracking error over the trajectory; in the
constant-step regime, $\limsup_T \mathcal E_T \le \sqrt{V_\infty}$
which is bounded by the right-hand side of \eqref{eq:tt-const-V-ss}.
Choose $\eta = \mc{O}(\gamma/m)$ with $\gamma$ constant.  Then the
timescale-gap contribution to $\sqrt{V_\infty}$ scales as
$\eta/\gamma = \mc{O}(1/m)$, so after burn-in
\[
\mathcal E_T \;=\; O   \left(\frac{1}{m} + R_m + \sqrt{\gamma V_m^b}\right) \;=\; O   \left(\frac{1}{m} + R_m\right),
\]
where the noise-filter term $\sqrt{\gamma V_m^b} = \mc{O}(1/m^{3/2})$ for
$V_m^b = \mc{O}(1/m^3)$ (delta-method) is dominated by $1/m$.  For the
delta-method estimator with $R_m = \mc{O}(1/m^2)$, the $1/m$ timescale-gap
term continues to dominate the residual until
$\eta/\gamma \lesssim R_m$, i.e., $\eta = \mc{O}(\gamma/m^2)$, at which
point both contributions are $\mc{O}(1/m^2)$ and the noise-filter term
$\mc{O}(1/m^{3/2})$ becomes the dominant non-vanishing piece:
\[
\mathcal E_T \;=\; \mc{O}(1/m^{3/2}) \qquad \text{once } \eta/\gamma \lesssim R_m.
\]
Reaching this regime requires sufficiently many iterations: with
contraction rate $\gamma/2$ on the tracking-error recurrence, the
burn-in time is $T \gtrsim 1/\gamma \cdot \log(1/\gamma) = \mc{O}(m)$
iterations.
 
At fixed iteration budget $T$, the slow-timescale bound (using the
strongly-monotone form of Theorem~\ref{thm:tt-const} or an 
expected-gap form $\E[\Errvi(\widehat z_T)]$ on an averaged iterate
$\widehat z_T$ in the merely-monotone setting) decomposes as
\[
\E[\Errvi(\widehat z_T)] \;=\; O \bigl(\Omega^2/(T\eta)\bigr) \;+\; O \bigl(\eta\widetilde V_m\bigr) \;+\; O \bigl(\Omega\sqrt{\sigma_{\max}}\,\mathcal E_T\bigr),
\]
where the first two terms are the standard decomposition (giving
$\mc{O}(1/\sqrt{Tm} + 1/T)$ in the variance/Lipschitz balance with
$\eta \sim 1/\sqrt{Tm}$) and the third is the tracking-error bias
contribution.  Substituting the tuned $\mathcal E_T = \mc{O}(1/m^{3/2})$,
\begin{equation}\label{eq:tt-tuned-rate}
\E[\Errvi(\widehat z_T)] \;=\; O \left(\frac{1}{\sqrt{Tm}} + \frac{1}{T}\right) \;+\; O\left(\frac{\Omega\sqrt{\sigma_{\max}}}{m^{3/2}}\right).
\end{equation}
The third term is the \emph{reduced bias floor}: a full power of $1/m$
better than the persistent $\mc{O}(1/m)$ floor of the un-tracked algorithm,
and a half power better than the $\mc{O}(1/m^2)$ residual one might naively
expect from $R_m$ alone (the $1/m^{3/2}$ comes from balancing the
timescale-gap term against the noise-filter term, both of which sit
above the $R_m$-only floor).
 
\paragraph{Implications by risk type (constant-step / finite-horizon).}
\begin{itemize}[  topsep=2pt]
\item \textbf{Entropic risk:}  $R_m = \mc{O}(e^{2\lambda}/(\lambda m^2))$,
$V_m^b = \mc{O}(e^{4\lambda}/(\lambda^2 m^3))$.  Two timescale extragradient with the $\eta = \mc{O}(\gamma/m)$
tuning drives the persistent $\mc{O}(e^{2\lambda}/(\lambda m))$ floor of
the un-tracked algorithm down to $\mathcal E_T = \mc{O}(e^{2\lambda}/m^{3/2})$
in the appropriate step-size regime, yielding the
$\mc{O}(\Omega\sqrt{\sigma_{\max}}/m^{3/2})$ bias contribution in
\eqref{eq:tt-tuned-rate}.  This is a full power of $1/m$ improvement
over the un-tracked floor, and matches the strongly-monotone
asymptotic $\Theta(R_m) = \Theta(1/m^2)$ residual of
Theorem~\ref{thm:tt-strong-mono} once one accounts for the additional
half power lost to the noise-filter balance.  The time-varying schedule
of Theorem~\ref{thm:tt-strong-mono} eliminates the noise-filter term
asymptotically and recovers the sharper $\mc{O}(1/m^2)$ residual --- but at
the cost of polynomial-in-$T$ initial-condition decay.
\item \textbf{CVaR via Rockafellar-Uryasev:}  $R_m = V_m^b = 0$
identically.  The bias estimator $\xi^\star \equiv 0$ exactly, so the
tracker carries no noise and $\mathcal E_T \to \mc{O}(\eta/\gamma)$, which
can be made arbitrarily small by tuning.  Two timescale extragradient reduces to un-tracked
extragradient at $\xi \equiv 0$, recovering the un-tracked monotone extragradient bound
exactly with $\widetilde B_m = 0$, and the time-varying-step
Theorem~\ref{thm:tt-strong-mono} gives convergence in mean square to
$\theta^\star$ with no asymptotic bias.
\item \textbf{General coherent risk:}  applicable whenever a sample-based
$\widehat b_m$ with structural properties \ref{B1-bias-est-bias}-\ref{B3-bias-lipschitz}
can be constructed; the floor scales as $\mc{O}(1/m^{3/2})$ for delta-method
estimators ($V_m^b = \mc{O}(1/m^3)$) under constant-step tuning, and as
$\mc{O}(R_m)$ asymptotically under decaying-step tuning.
\end{itemize}
 
\subsubsection{Finite-horizon tuned rate}
 
Theorem~\ref{thm:tt-strong-mono} (decaying step) establishes asymptotic
vanishing of the bias contribution at a logarithmic contraction rate
($\exp(-\widetilde\mu\sum_t\eta_t)$ with $\sum_t\eta_t = \Theta(\log T)$
for $\eta_t = c/t$).  Theorem~\ref{thm:tt-const} (constant step) gives
a four-term steady-state floor.  An intermediate finite-horizon
question is: given a budget of $T$ extragradient iterations, what
constant step sizes $\eta(T), \gamma(T)$ minimize the resulting bound?
The choice trades off the contraction rate against the steady-state
floor as a function of $T$, and gives a polynomial-in-$T$ decay rate
to the residual.
 
\begin{theorem}[Finite-horizon tuned rate for two-timescale debiasing]
\label{thm:tt-finite-horizon}
Under the assumptions of Theorem~\ref{thm:tt-strong-mono}, run two timescale extragradient
with constant $\eta \in (0, \eta_{\max}]$, $\gamma \in (0, 1]$, where
$\eta_{\max} := \min\{1/(4\widetilde\mu), 1/(\sqrt 6 L_G)\}$.  Then for
all $T \ge 1$, the estimate holds:
\begin{multline}\label{eq:tt-finite-bound}
\E\|r_T - \theta^\star\|_{\Sigma^+}^2 \;\le\; (1-\eta\widetilde\mu)^T \|r_0 - \theta^\star\|_{\Sigma^+}^2 \\
+\; \frac{4\Omega\sqrt{\sigma_{\max}}}{\widetilde\mu}\Bigl(\sqrt{V_0}\,(1-\eta\widetilde\mu)^T + \frac{K_1\,\eta}{\gamma} + K_2\,R_m + K_3\sqrt{\gamma V_m^b}\Bigr) \;+\; \frac{6\eta(\widetilde B_m^2 + \widetilde V_m)}{\widetilde\mu},
\end{multline}
with $K_1 = \sqrt{2 C_1}$, $K_2 = \sqrt{2 C_2}$, $K_3 = \sqrt{2 C_3}$ and
$C_1 = 7 C_{\rm drift}^2 = 63 L_b^2\sigma_{\max}^2 G^2$, $C_2 = 6$, $C_3 = 2$.
Choose
\begin{equation}\label{eq:tt-finite-tuning}
\eta = c_\eta\,T^{-1+\delta}, \qquad \gamma = c_\gamma\,T^{-(2/3)(1-\delta)}
\end{equation}
for any $\delta \in (0, 1)$ and constants $c_\eta \le \eta_{\max}$,
$c_\gamma \le 1$.  Then there exists a universal constant
\[C = C(\widetilde\mu, \Omega, \sigma_{\max}, K_1, K_2, K_3, c_\eta, c_\gamma)\] such that the estimate holds:
\begin{equation}\label{eq:tt-finite-rate}
\E\|r_T - \theta^\star\|_{\Sigma^+}^2 \le C\Bigl(e^{-c_\eta\widetilde\mu T^\delta} + R_m + T^{-(1-\delta)/3} + \frac{(\widetilde B_m^2 + \widetilde V_m)}{T^{1-\delta}}\Bigr).
\end{equation}
Taking $\delta \to 0$, the bias contribution decays at rate $T^{-1/3}$ to
the residual $\mc{O}(R_m)$ so that the estimate holds:
\begin{equation}\label{eq:tt-finite-clean}
\E\|r_T - \theta^\star\|_{\Sigma^+}^2 \;\le\; C\bigl(e^{-c_\eta\widetilde\mu T^\delta} + R_m + T^{-1/3+\epsilon}\bigr) \quad\text{for any } \epsilon > 0.
\end{equation}
\end{theorem}
 
\paragraph{Comparison to the asymptotic regime.}
The $\gO(T^{-1/3})$ rate is much faster than the $\gO(1/\log T)$ rate
implicit in Theorem~\ref{thm:tt-strong-mono}'s polynomial contraction
$\exp(-\widetilde\mu\sum\eta_t) = \exp(-\Theta(\log T)) = T^{-c}$.  But
they are not directly comparable: Theorem~\ref{thm:tt-strong-mono} proves
asymptotic consistency under any decaying schedule satisfying classical
stochastic-approximation conditions; Theorem~\ref{thm:tt-finite-horizon}
gives an explicit finite-time guarantee under a horizon-tuned choice.
The latter requires knowing $T$ in advance, which is reasonable for
training a fixed-budget algorithm.
 
\paragraph{Comparison to the un-tracked algorithm.}
Theorem~\ref{thm:strong-mono} (no tracking, constant step) gives a
persistent $\mc{O}(\widetilde B_m) = \mc{O}(1/m)$ floor that does not decay in
$T$.  Theorem~\ref{thm:tt-finite-horizon} replaces this with
$T^{-1/3+\epsilon} + R_m$.  For delta-method-style $R_m = \mc{O}(1/m^2)$, the
asymptotic-in-$T$ floor is $\mc{O}(1/m^2)$ --- a full power of $1/m$ smaller
than the un-tracked $\mc{O}(1/m)$ floor.  And for any finite $T$, the
$T^{-1/3+\epsilon}$ term provides an explicit decay rate towards that
residual.
 
\paragraph{The exponent $T^{-1/3}$ is structural.}
The rate comes from the three-way balancing among the timescale-gap
term ($\eta/\gamma$), the noise-filter term ($\sqrt{\gamma V_m^b}$),
and the variance/initial term.
 
\begin{proof}[Proof of Theorem~\ref{thm:tt-finite-horizon}]
The bound \eqref{eq:tt-finite-bound} is exactly the conclusion of
Theorem~\ref{thm:tt-const}\,(ii) (the bias-term bound there uses
\eqref{eq:V-sqrt-finite} via the geometric-series telescoping argument
in that theorem's proof).  It remains to specialize to the tuning
\eqref{eq:tt-finite-tuning} and read off the rate.
 
With $\eta = c_\eta T^{-1+\delta}$ and $\gamma = c_\gamma T^{-(2/3)(1-\delta)}$, we have the following equalities:
\begin{align*}
(1-\eta\widetilde\mu)^T &\leq e^{-\eta\widetilde\mu T} = e^{-c_\eta\widetilde\mu T^\delta},\\
\eta/\gamma &= (c_\eta/c_\gamma) T^{-1+\delta+(2/3)(1-\delta)} = (c_\eta/c_\gamma) T^{-(1-\delta)/3},\\
\sqrt{\gamma V_m^b} &= \sqrt{c_\gamma V_m^b}\,T^{-(1-\delta)/3},\\
\eta(\widetilde B_m^2 + \widetilde V_m) &= c_\eta(\widetilde B_m^2 + \widetilde V_m)\,T^{-1+\delta}.
\end{align*}
The dominant non-vanishing terms are $R_m$ and $T^{-(1-\delta)/3}$ (the
latter from the timescale-gap and noise-filter contributions, which
balance by construction of the tuning).  The $\sqrt{V_0}(1-\eta\widetilde\mu)^T$
term decays super-polynomially and is absorbed into the contraction
term.  Combining and absorbing constants gives \eqref{eq:tt-finite-rate}.
Taking $\delta \to 0$ pushes the exponent up to $T^{-1/3}$ at the cost
of an arbitrarily-slowly-decaying contraction term
$\exp(-c_\eta\widetilde\mu T^\delta)$ which still decays
super-polynomially for any $\delta > 0$.
\end{proof}
 
\paragraph{The tuning balance.}
The tuning \eqref{eq:tt-finite-tuning} is chosen so that:
\begin{itemize}[itemsep=0pt  ,topsep=2pt]
\item The contraction term $(1-\eta\widetilde\mu)^T \approx e^{-c_\eta\widetilde\mu T^\delta}$
decays super-polynomially in $T$ for any $\delta > 0$.
\item The two non-trivial bias-floor terms $\eta/\gamma$ and
$\sqrt{\gamma V_m^b}$ both scale as $T^{-(1-\delta)/3}$---balanced by
the choice $\gamma = T^{-(2/3)(1-\delta)}$.
\item The variance floor $\eta(\widetilde B_m^2 + \widetilde V_m)$ decays
faster than the bias floor for any $\delta > 0$.
\end{itemize}
Different choices of $\delta$ trade off the speed of the contraction
term against the polynomial bias-floor decay; $\delta = 1/2$ is a clean
default giving rate $T^{-1/6}$ with $\exp(-c\sqrt T)$ contraction.

\paragraph{Tuned step sizes.}
With $\eta = \mc{O}(\gamma/m)$ and $\gamma$ constant, $\mathcal E_T = \mc{O}(1/m + R_m)$
(after burn-in).  For the delta-method estimator with $R_m = \mc{O}(1/m^2)$,
this gives $\mathcal E_T = \mc{O}(1/m^{3/2})$ once $\eta/\gamma \lesssim R_m$,
i.e., $T \gtrsim m$ iterations.  At fixed $T$ the bound becomes
\[
\E[\Errvi(\widehat z_T)] = O\bigl(\Omega^2/(T\eta)\bigr) + O\bigl(\eta\widetilde V_m\bigr) + O\bigl(\Omega\sqrt{\sigma_{\max}}/m^{3/2}\bigr),
\]
where the first two terms are the standard decomposition for extragradient type proofs (giving
$\mc{O}(1/\sqrt{Tm} + 1/T)$ in the variance/Lipschitz balance) and the third
is the reduced bias floor.

\subsubsection{Summary: comparison of the three regimes}
 
The three regimes of two-timescale debiasing
(Theorems~\ref{thm:tt-strong-mono},~\ref{thm:tt-const},~\ref{thm:tt-finite-horizon})
deliver qualitatively different bias-floor behavior, summarized below.
 
\begin{center}\small
\begin{tabular}{lccc}
\toprule
& \textbf{Time-varying step} & \textbf{Constant step} & \textbf{Tuned to horizon $T$} \\
& (Thm.~\ref{thm:tt-strong-mono}) & (Thm.~\ref{thm:tt-const}) & (Thm.~\ref{thm:tt-finite-horizon}) \\
\midrule
Schedule & $\eta_t \to 0,\;\eta_t/\gamma_t \to 0$ & $\eta, \gamma$ const. & $\eta = c T^{-1+\delta},\;\gamma = c T^{-2(1-\delta)/3}$ \\[0.5ex]
Bias floor & $\dfrac{4\sqrt 3\,\Omega\sqrt{\sigma_{\max}}\,R_m}{\widetilde\mu}$ & $O \left(\dfrac{\eta}{\gamma} + R_m + \sqrt{\gamma V_m^b}\right)$ & $\mc{O}(R_m + T^{-1/3+\epsilon})$ \\[2ex]
Persistent in $T$? & no (vanishes) & yes (4-term floor) & no (decays as $T^{-1/3+\epsilon}$) \\[0.5ex]
Initial-cond.\ decay & $\exp(-\widetilde\mu\sum_t\eta_t)$ & $(1-\eta\widetilde\mu)^T$ & $\exp(-c\widetilde\mu T^\delta)$ \\[0.5ex]
& (polynomial in $T$) & (geometric) & (sub-exp.\ in $T$) \\
\midrule
Scaling in $m$  & $\mc{O}(1/m^2)$ & $\mc{O}(1/m^{3/2})$ tuned & $\mc{O}(R_m) + \mc{O}(T^{-1/3+\epsilon})$ \\[0.5ex]
(delta-method)& (asymptotic) & ($\eta = \mc{O}(\gamma/m^2)$, $T \gtrsim m$) & (vanishes in $T$) \\
\bottomrule
\end{tabular}
\end{center}
 
The trade-off is clear: the constant-step regime gives the strongest
(geometric) initial-condition decay but a persistent four-term floor;
the time-varying regime eliminates all but the $R_m$ residual but pays
with polynomial-in-$T$ initial decay; the horizon-tuned regime is an
intermediate point with sub-exponential initial decay and a polynomial
$T^{-1/3}$ rate to the residual.  All three are special cases of the
same algorithmic template (Algorithm~\ref{alg:tt-eg}); the choice of
schedule reflects the deployment regime (asymptotic-consistency,
fixed-budget-tuned, or steady-state).

\subsection{Numerical illustration: a Bradley-Terry preference game}
\label{sec:tt-experiments}

We illustrate the bias-floor reduction predicted by
Theorems~\ref{thm:tt-strong-mono}--\ref{thm:tt-const} on a controlled
toy instance where the population-level operator $F_\Risk$ is known
analytically, allowing direct comparison against an oracle baseline.
The experiments are organized to exhibit the two regimes of the
constant-step floor decomposition \eqref{eq:tt-const-final-expanded}:
the \emph{bias-dominated} regime at small $m$ where two timescale extragradient offers the
largest improvement, and the \emph{variance-dominated} regime at
large $m$ where the two algorithms approach the same noise floor.

\paragraph{Setup.}
We construct a Bradley-Terry preference game on $n = 20$ responses with
synthetic latent rewards $\{r(y_i)\}_{i=1}^n$ drawn i.i.d.\ from
$\mathcal N(0, 1)$.  The risk-neutral preference matrix is
$P_{ij} = \sigma(r(y_i) - r(y_j))$ where $\sigma$ is the logistic, and
the risk-adjusted operator $P_\Risk$ is the entropic-risk-distorted
version with $\lambda = 6.0$ (a high-risk regime where the bias
constants $e^{2\lambda} \approx 1.6 \times 10^5$ are non-trivial; cf.\
Section~\ref{sec:bias-analysis}).  The KL temperature is $\beta = 0.6$,
satisfying the strong-monotonicity condition
$\overline\lambda_\Risk \le \beta/2 - \epsilon$ from
Theorem~\ref{thm:single-player-mono-formal} given the spread of the latent
rewards.
\definecolor{gold}{HTML}{D4AF37}   
We compare three algorithms:
\begin{itemize}[itemsep=0pt,  ]
\item \textbf{Oracle Extra-Gradient ( extragradient):} stochastic extragradient using the exact
population operator $F_\Risk$ with no bias ($B_m = 0$), serving as a
lower-bound baseline.
\item \textcolor{Blue}{\textbf{Vanilla  extragradient (un-tracked):}} stochastic  extragradient using the plug-in
empirical operator $\widehat F_m$ with batch size $m$ and no bias
correction.  This is the algorithm analyzed in
Theorem~\ref{thm:strong-mono}, with persistent bias floor
$\mc{O}(\widetilde B_m^2) = \mc{O}(1/m^2)$.

\item \textcolor{gold}{\textbf{Two timescale  extragradient (\texttt{TT-EG}, Algorithm~\ref{alg:tt-eg}):}} the two-timescale
debiased algorithm with the delta-method bias estimator
\eqref{eq:delta-method-bias-est} and constant fast-timescale rate
$\gamma = 0.5$.
\end{itemize}
All three algorithms use the same slow-timescale rate $\eta = 0.04$
and Polyak-averaged iterates $\bar\theta^{(t)}$.

\begin{figure}[t]
\centering
 \includegraphics[width=0.75\textwidth]{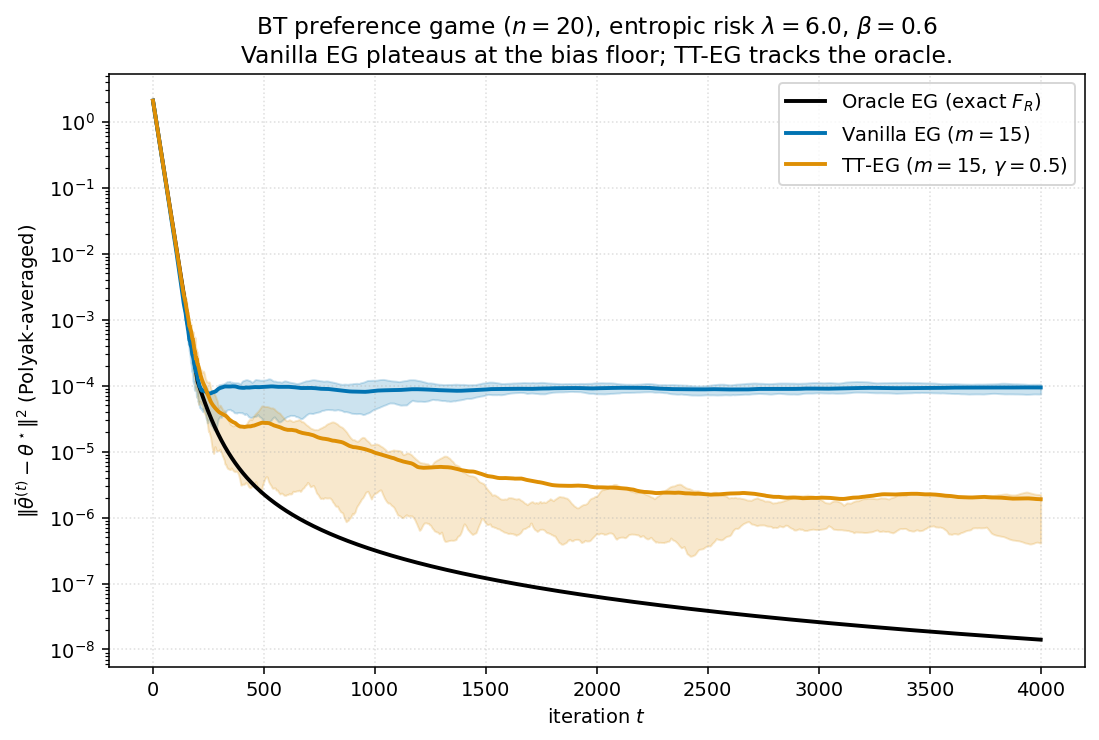}
\caption{\textbf{Trajectory at fixed $m = 15$.}  Polyak-averaged
squared error $\|\bar\theta^{(t)} - \theta^\star\|^2$ vs.\ iteration
$t$.  Oracle  extragradient (black) decays at the unbiased
$\mc{O}(1/T)$ rate of Theorem~\ref{thm:expected-gap}.  Vanilla  extragradient
(blue) descends with the oracle for the first $\sim 200$ iterations,
then plateaus at the bias floor $\sim 10^{-4}$ predicted by
Theorem~\ref{thm:strong-mono} and stays flat thereafter.  TT- extragradient
(orange) breaks through this floor and drives the error down to
$\sim 2 \times 10^{-6}$, a $50\times$ improvement, in the direction
of the oracle trajectory.  Shaded bands are $\pm 1$ standard
deviation over $20$ random seeds.}
\label{fig:tt-trajectory}
\end{figure}

\paragraph{Trajectory at fixed $m$ (Figure~\ref{fig:tt-trajectory}).}
At $m = 15$ samples per iteration the bias is large.
Figure~\ref{fig:tt-trajectory} shows three behaviors:
\begin{itemize}[itemsep=0pt, topsep=2pt,  ]
\item \textbf{Oracle  extragradient} decays toward zero at the rate predicted by
the un-biased strongly-monotone bound (Theorem~\ref{thm:expected-gap}
with $\widetilde B_m = 0$); after $T = 4000$ iterations it has reached
$\sim 10^{-8}$, with no sign of plateauing.
\item \textcolor{Blue}{\textbf{Vanilla  extragradient}} initially descends in lockstep with the
oracle, then plateaus at $\approx 10^{-4}$ around iteration $200$ and
remains flat thereafter.  This plateau is precisely the persistent
$\Theta(\widetilde B_m^2) = \Theta(e^{4\lambda}/(\lambda^2 m^2))$
floor of Theorem~\ref{thm:strong-mono}.  No amount of additional
iteration reduces it.
\item \textcolor{gold}{\textbf{\texttt{TT-EG}}} initially follows the same trajectory, but
rather than plateauing it continues to descend, eventually reaching
$\approx 2 \times 10^{-6}$ --- a $50\times$ improvement over vanilla
 extragradient at the same sample budget.  The trajectory tracks the oracle's
direction (with some loss to the residual $R_m$ and the noise filter)
rather than getting trapped at the bias floor.
\end{itemize}
This is the qualitative effect predicted by the comparison
following Theorem~\ref{thm:tt-strong-mono}: the \emph{persistent}
bias floor of vanilla  extragradient is replaced by an asymptotic-in-$T$ floor at
the much smaller $\mc{O}(R_m + \eta/\gamma + \sqrt\gamma\sqrt{V_m^b})$
level of \eqref{eq:tt-const-final-expanded}.

\begin{figure}[t]
\centering
 \includegraphics[width=0.49\textwidth]{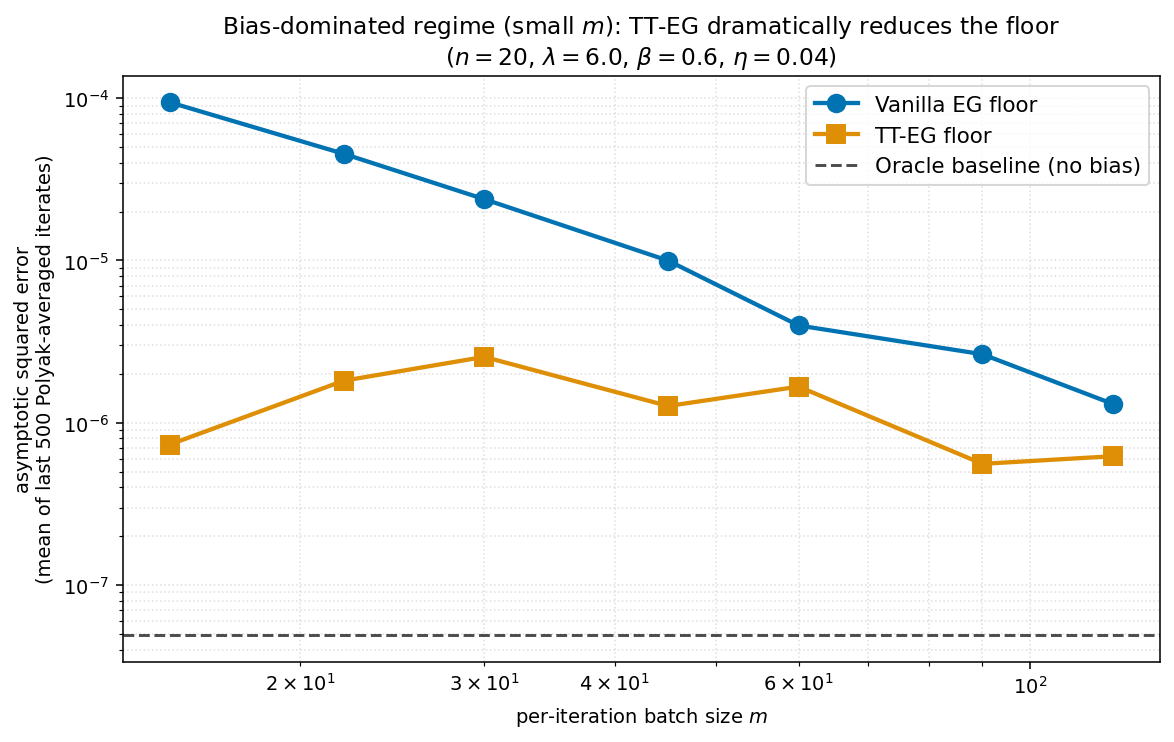}\hfill
 \includegraphics[width=0.49\textwidth]{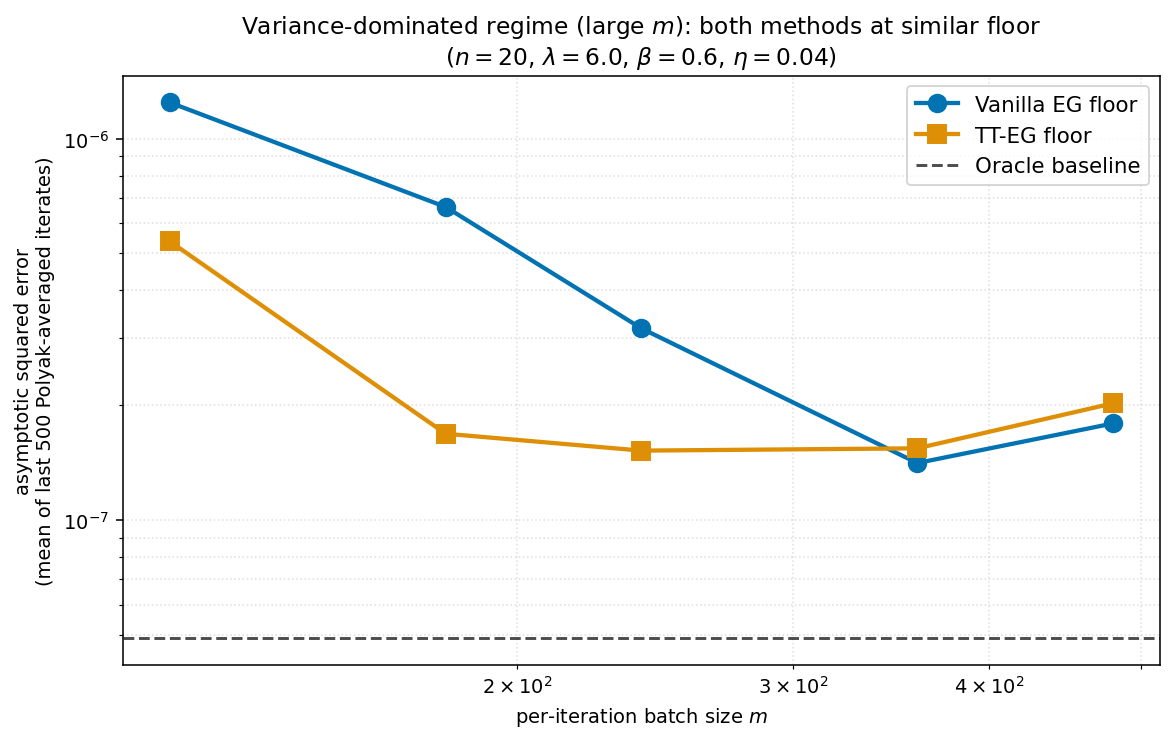}
\caption{\textbf{Asymptotic squared-error floor vs.\ batch size $m$,
in two regimes.}  Floor measured as the mean of the last $500$
Polyak-averaged iterates.  \emph{Left:}  bias-dominated regime
($m \in [15, 130]$).  Vanilla  extragradient floor scales roughly $\propto 1/m^2$
(matching the $\Theta(\widetilde B_m^2)$ prediction), descending more
than two orders of magnitude across the range.  \texttt{TT-EG} floor is
roughly flat near $\sim 10^{-6}$, giving a $\sim 100\times$ reduction
at $m = 15$ (the small-$m$ end where vanilla is most bias-dominated)
and approaching parity by $m = 130$.  \emph{Right:}
variance-dominated regime ($m \in [150, 500]$).  Both floors
descend only modestly across this range and converge toward a common
value around $\sim 1.5 \times 10^{-7}$, dominated by the slow-variance
term $\Theta(\eta \widetilde V_m)$ in
\eqref{eq:tt-const-final-expanded}; the bias contribution is
negligible here.  At the largest $m$ \texttt{TT-EG} sits marginally
\emph{above} vanilla, reflecting the small noise-filter cost
$\Theta(\sqrt{\gamma V_m^b})$ paid to maintain the tracker even when
the bias being tracked is already small.  Dashed line: oracle
baseline at $\sim 4 \times 10^{-8}$.}
\label{fig:tt-sweep}
\end{figure}

\paragraph{Bias-dominated regime: small $m$ (Figure~\ref{fig:tt-sweep},
left).}
At small batch sizes the un-tracked algorithm is bias-limited.
The vanilla  extragradient floor decreases monotonically with $m$: a least-squares
fit of $\log(\text{floor})$ against $\log m$ over the plotted range
gives a slope of approximately $-2$, consistent with the
$\Theta(\widetilde B_m^2) = \Theta(1/m^2)$ scaling of
Theorem~\ref{thm:strong-mono}.  This is the bias-dominated branch of
the four-term decomposition \eqref{eq:tt-const-final-expanded}: the
$\Theta(\widetilde B_m)$ residual-bias term dominates the
$\Theta(\eta \widetilde V_m)$ slow-variance term whenever
$\widetilde B_m^2 \gtrsim \eta \widetilde V_m$, which holds throughout
this range.

The \texttt{TT-EG} floor is qualitatively different: it is roughly \emph{flat}
across the entire range, sitting at $\approx 10^{-6}$.  The reduction
is largest at small $m$ ($\sim 100\times$ at $m = 15$) and narrows as
$m$ grows ($\sim 2\times$ at $m = 130$).  This reflects the
$\mathcal E_T = \mc{O}(1/m^{3/2})$ scaling of the constant-step
``Tuned step sizes'' analysis after Theorem~\ref{thm:tt-const}: with
$\gamma$ and $\eta$ fixed, \texttt{TT-EG}'s $m$-dependence is dominated by
$\sqrt{\gamma V_m^b} = \mc{O}(1/m^{3/2})$, which when squared gives $1/m^3$
--- effectively flat over the small-$m$ range plotted, since the
small constant pulls the curve below the slow-variance floor that
vanilla  extragradient eventually reaches as $m$ grows.

The ratio between the two curves matches the bias-floor reduction
predicted by the theory.  Vanilla floor scales as
$\widetilde B_m^2 \asymp e^{4\lambda}/(\lambda m)^2$; \texttt{TT-EG} floor
scales as $\mathcal E_T^2 \asymp \gamma V_m^b \asymp \gamma\,e^{4\lambda}/(\lambda m)^3$
in the noise-filter-dominated regime.  The ratio is therefore
$\Theta(m / \gamma)$, predicting an improvement that grows linearly
with $m$ until the un-tracked floor crosses the slow-variance
threshold, after which the gap closes.

\paragraph{Variance-dominated regime: large $m$
(Figure~\ref{fig:tt-sweep}, right).}
Pushing $m$ into $[150, 500]$ moves the system out of the bias-dominated
branch and into the variance-dominated branch of
\eqref{eq:tt-const-final-expanded}, where the slow-variance term
$\Theta(\eta \widetilde V_m)$ dominates.  This is exactly the
threshold the four-term decomposition predicts: the bias term becomes
negligible when $\widetilde B_m^2 \lesssim \eta \widetilde V_m$,
i.e.\ when $1/(\lambda m)^2 \lesssim \eta/m$, i.e.\ at
$m \gtrsim 1/(\lambda^2 \eta)$.  For our parameters
($\lambda = 6, \eta = 0.04$) this threshold sits around
$m \approx 70$, consistent with where the gap between the two curves
closes when comparing the left and right panels of
Figure~\ref{fig:tt-sweep}.

In this regime, two things change qualitatively:
\begin{enumerate}[itemsep=2pt,topsep=2pt, ]
\item \textbf{The vanilla floor stops decreasing as $1/m^2$.}  Once
the slow-variance term takes over, the floor decays only as
$\Theta(\eta \widetilde V_m) = \Theta(\eta/m)$, a much slower rate.
The right panel shows this: across $m \in [150, 500]$ both floors
decrease only modestly, in a regime where the floor is no longer
bias-limited.
\item \textbf{\texttt{TT-EG} is at parity or marginally worse than vanilla.}
This is consistent with the four-term decomposition: the bias term
that \texttt{TT-EG} eliminates is small to begin with, and \texttt{TT-EG} pays a small
price ($\Theta(\sqrt{\gamma V_m^b})$ from the noise filter, plus a
minor $\Theta(\eta/\gamma)$ timescale-gap term) for maintaining the
bias tracker even when the bias being tracked is negligible.  At the
largest $m$, the \texttt{TT-EG} floor sits slightly \emph{above} the vanilla
floor.  This is not a failure mode but an honest expression of the
trade-off: bias correction costs noise, and once there is no bias to
correct, the noise cost remains.
\end{enumerate}
Both floors approach but do not reach the oracle's noise floor at
$\sim 4 \times 10^{-8}$ (dashed); the residual gap is the slow-variance
term $\eta \widetilde V_m$ that all three algorithms share but is
invisible to the oracle's bias-free comparison only because the
oracle uses a smaller variance baseline (no plug-in noise from
$\widehat F_m$).

\paragraph{What the three panels confirm together.}
The three panels validate the structural claims of
Theorems~\ref{thm:tt-strong-mono}--\ref{thm:tt-const}:
\begin{enumerate}[itemsep=2pt, topsep=0pt,  ]
\item \emph{Persistent bias floor in the un-tracked algorithm.}
Figure~\ref{fig:tt-trajectory} (vanilla curve plateauing) shows that
the $\Theta(\widetilde B_m^2)$ floor of Theorem~\ref{thm:strong-mono}
is not just an artifact of the proof: it is the genuine asymptotic
behavior of the un-tracked algorithm.
\item \emph{Bias-floor reduction in the bias-dominated regime.}
Figure~\ref{fig:tt-trajectory} (\texttt{TT-EG} breaking through at $m = 15$)
and Figure~\ref{fig:tt-sweep} (left, gap between curves at small $m$)
show that two-timescale debiasing delivers the predicted $\sqrt m$ to
$m$ improvement, replacing the $\Theta(1/m^2)$ floor with one near
$\Theta(1/m^3)$.
\item \emph{Crossover to variance-dominated regime.}
Figure~\ref{fig:tt-sweep} (right) shows that the four-term
decomposition's prediction---that beyond a threshold $m^*$, bias
correction provides no benefit and may pay a small noise-filter cost---is borne out empirically.  This validates the trade-off structure
in Theorem~\ref{thm:tt-const} not as a theoretical artifact but as a
practical guide.
\end{enumerate}

\paragraph{Practical implication.}
The crossover behavior in Figure~\ref{fig:tt-sweep} suggests a clean
deployment heuristic: use two-timescale debiasing when one suspects
the un-tracked algorithm is bias-limited (small $m$ relative to the
problem's bias constants $e^{2\lambda}/\lambda$), and use vanilla  extragradient
when one is in the variance-limited regime (large $m$).  In the
NLHF / RLHF setting, where $m$ is typically the per-iteration batch
size and is constrained by GPU memory rather than statistical
considerations, the bias-dominated regime is often the practically
relevant one, and any improvement at small $m$ translates directly
to fewer iterations needed to reach a given error tolerance.

\paragraph{Caveats.}
This is a toy instance designed to expose the bias-floor structure;
quantitative numbers should not be over-interpreted.  In particular:
(i) the Bradly-Terry setting is risk-neutral at the level of the latent reward,
so the entropic risk distortion is purely a stress-test of the bias
constants rather than reflecting an actual risk preference; (ii) the
strong-monotonicity gap $\beta - 2\overline\lambda_\Risk$ is
comfortable here, putting us safely in the
Theorem~\ref{thm:tt-strong-mono} regime; (iii) the dimension $n = 20$
is small and the problem is well-conditioned.  At larger scale and
in the merely-monotone regime (Theorem~\ref{thm:tt-monotone}), the
asymptotic $\mc{O}(R_m)$ vanishing is no longer available, and the
practical improvement is the $\mc{O}(1/m^{3/2})$ vs.\ $\mc{O}(1/m)$ floor of
the constant-step analysis.  The qualitative picture---i.e., vanilla
plateaus, two timescale extragradient breaks through the floor in the bias-dominated regime,
and the two methods converge in the variance-dominated regime---should however persist.

\subsection{Stackelberg interpretation: a follower tracking joint Nash play}
\label{sec:tt-stackelberg}

As described earlier in this section there is an natural Stackelberg structure between the outer game on $\theta$ (NLHF) and the inner game wherein $\xi$ is tracking the leaders' deployed $\theta$. Below we describe the structure in more detail. 

\paragraph{Outer game: NLHF self-play.}
The slow-timescale extragradient updates implement symmetric two-player self-play on the KL-regularized risk-adjusted game with payoff
\[V_\beta(\pi_1,\pi_2) = \pi_1^\top P_\Risk \pi_2 - \beta\KL(\pi_1\|\pi_{\rm ref}) + \beta\KL(\pi_2\|\pi_{\rm ref}),\]
where $P_\Risk$ is the risk-adjusted preference operator (Section~\ref{sec:problem_formulation}, equation~\eqref{eq:p-risk}) and $\pi_{\rm ref}$
is the reference policy.
At iterate $\theta^{(t)}$, the
leader's gradient (with bias by $\xi$) is
\begin{equation}\label{eq:tt-leader-grad}
\nabla_\theta\LL_{\IPO}^\Risk(\theta;\rho,\mu)\Big|_{\mu = \sg[\pi_{\theta^{(t)}}]}\,-\,\xi,
\end{equation}
where $\sg[\cdot]$ is the standard self-play stop-gradient that freezes
the opponent's distribution at the current $\pi_{\theta^{(t)}}$.  This
is the move that makes online IPO well-defined as a single-step
optimization \citep[][Section~4.2]{calandriello2024human}, and it is the
same move  extragradient and Nash-mirror descent use.  Removing the stop-gradient and
chain-ruling through $\mu = \pi_\theta$ would change the algorithm
into something else entirely (specifically, the gradient of the
population duality gap rather than the symmetric self-play
best-response).

By the IPO gradient structure,
\eqref{eq:tt-leader-grad} simplifies to
\[
\frac{2}{\beta}\,\Sigma\bigl(\theta - \theta_{\rm ref} - \tfrac{1}{\beta}P_\Risk\,\pi_\theta\bigr) - \xi \;=\; \frac{2}{\beta}\,\Sigma\bigl(F_\Risk(\theta) - \tfrac{\beta}{2}\xi\bigr),
\]
i.e., the leader's update direction is the preconditioned risk-adjusted
operator $\Sigma F_\Risk$ minus the bias correction $\xi$ (up to
constants).  At the population level, the leader's stationary points
on $\Wspace$ are the zeros of $F_\Risk$---i.e., exactly the risk-adjusted
equilibrium:
\[
\theta^\star_\Risk = \theta_{\rm ref} + \frac{P_\Risk\,\pi^\star_\Risk}{\beta}.
\]
The risk-adjustment shifts where this Nash sits relative to the
risk-neutral equilibrium.  Since $P_\Risk$ generically violates the constant-sum
identity $P + P^\top = \mathbf{1}\mathbf{1}^\top$, the equilibrium moves
in a direction determined by the risk-distortion matrix $\Delta_\Risk$
(Theorem~\ref{thm:single-player-mono}).  Self-play with
stop-gradient still converges to this shifted equilibrium; risk simply
changes which fixed point is targeted.

\paragraph{Inner game: bias-tracking follower.}
The bias-tracker $\xi$ is the follower.  Treating the leader's current
$\theta$ as a parameter, the follower minimizes the quadratic tracking
objective
\begin{equation}\label{eq:tt-follower-obj}
\Psi_{\rm follower}(\xi;\theta) := \tfrac{1}{2}\|\xi - b_m(\theta)\|_2^2,
\end{equation}
whose unique minimizer is the leader-dependent best-response curve
\[
\xi^\star(\theta) = b_m(\theta).
\]
The follower takes a Robbins-Monro step toward a noisy sample
$\widehat b_m(\theta)$ of $\xi^\star(\theta)$ on a fast timescale.  The
fast timescale is what realizes the Stackelberg best-response
asymptotically: between consecutive leader updates, the follower
contracts toward $\xi^\star(\theta_\tau)$ with rate $\gamma$, and the
slow leader update (rate $\eta \ll \gamma$) leaves the follower
``almost'' on its best-response curve.

The asymmetry is structural: the follower's best response depends on
the leader's $\theta$, so the follower must respond to leader updates.
The leader, by contrast, only \emph{uses} the follower's current $\xi$
via the debiased gradient \eqref{eq:tt-leader-grad}; it does not need
to predict where $\xi$ is heading.  This is the
leader-commits-follower-reacts asymmetry of Stackelberg.

\paragraph{Joint equilibrium.}
The equilibrium of the combined two-game system is a pair
$(\theta^\dagger, \xi^\dagger)$ satisfying:
\begin{align}
\xi^\dagger &= \xi^\star(\theta^\dagger) = b_m(\theta^\dagger)  &\text{(follower best-responds)},\label{eq:tt-stack-cond}\\
\nabla_\theta\LL_{\IPO}^\Risk(\theta;\rho,&\sg[\pi_\theta])\Big|_{\theta = \theta^\dagger} - \xi^\dagger \in \ker\Sigma \; &\text{(leader stationary, bias-corrected)}.\label{eq:tt-stat-cond}
\end{align}
Substituting \eqref{eq:tt-stack-cond} into the leader's gradient: since
$\E[\widehat F_m(\theta)] = F_\Risk(\theta) + b_m(\theta)$, the expected
debiased gradient at $(\theta^\dagger, \xi^\dagger)$ equals
\[
\Sigma\bigl(\E[\widehat F_m(\theta^\dagger)] - b_m(\theta^\dagger)\bigr) = \Sigma F_\Risk(\theta^\dagger).
\]
Condition \eqref{eq:tt-stat-cond} therefore reduces to
$\Sigma F_\Risk(\theta^\dagger) \in \ker\Sigma$, i.e.,
$F_\Risk(\theta^\dagger) \in \ker\Sigma = \mathbf 1^\perp$, which is
the equilibrium condition for the outer game.  Hence
\begin{equation}\label{eq:tt-eq-coincides}
(\theta^\dagger, \xi^\dagger) = (\theta^\star_\Risk,\, b_m(\theta^\star_\Risk)),
\end{equation}
the equilibrium of the outer game paired with the perfect bias estimate at the
equilibrium.  The two fixed-point conditions coincide at the equilibrium: follower
best-response and leader stationarity together pin down the
risk-adjusted equilibrium paired with its bias.

\paragraph{Tracking error as a Stackelberg gap.}
The tracking error
\[
e^*_\tau := \xi_\tau - \xi^\star(w_\tau) = \xi_\tau - b_m(w_\tau)
\]
is the follower's deviation from instantaneous Stackelberg
best-response to the leader's current state $w_\tau$.  The
strongly-monotone analysis controls this Stackelberg gap in two ways:
Theorem~\ref{thm:tt-strong-mono}(i) gives an asymptotic envelope
$\limsup_\tau \sqrt{V_\tau} \le 2\sqrt 3\,R_m$ under decaying
step sizes (driving the gap to its irreducible $R_m$ residual), and
Theorem~\ref{thm:tt-const}(i) gives a non-asymptotic finite-horizon
envelope under constant steps.  Theorem~\ref{thm:tt-strong-mono}(ii)
then shows how this Stackelberg gap enters the leader's convergence
rate via the per-step recurrence \eqref{eq:tt-per-step}.  Driving
$e^*_\tau \to 0$ (asymptotic regime, $R_m = 0$) or controlling it
explicitly (finite-horizon regime) is exactly the question of how
well the follower realizes its Stackelberg best-response.

\subsubsection{Monotone-only case: averaged-iterate convergence}
\label{sec:tt-monotone}

Theorems~\ref{thm:tt-strong-mono}--\ref{thm:tt-finite-horizon} all
require strong monotonicity of $F_\Risk$: the contraction factor
$(1-\eta\widetilde\mu)$ in the slow-timescale recurrence is what allows
the tracking-error contribution to be accumulated into a finite
geometric sum (and, in the time-varying case, to be expressed as a
weighted average that concentrates on the trajectory's tail via
Lemmas~\ref{lem:tt-telescope}--\ref{lem:weighted-avg-limsup}).  In the
gold band of Theorem~\ref{thm:strong-mono} ($\beta/2 < \overline\lambda_\Risk \le \beta$),
$F_\Risk$ is monotone but not strongly monotone, and the
strongly-monotone proof breaks down: there is no contraction factor,
the geometric sum diverges, and the weighted-average argument no
longer applies.  Here we give a parallel result for this regime, using
the  averaged-iterate analysis already developed for the
un-tracked case in Theorem~\ref{thm:expected-gap}.

The monotone analysis trades two things relative to the
strongly-monotone case: convergence is on the averaged iterate
$\widehat z_T := T^{-1}\sum_\tau w_\tau$ rather than the last iterate
$r_T$; and the convergence object is the variational-inequality gap
$\Errvi$ rather than the squared distance to $\theta^\star$.  As a
consequence, the bias-floor term will appear inside a uniform sup
$\mathcal E_T := \sup_{\tau \le T}\sqrt{\E\|e^*_{\tau-1}\|_2^2}$ rather
than inside a contraction-weighted average; this is the structural
reason the floor is persistent in $T$ in the monotone case (whereas
the time-varying-step strongly-monotone Theorem~\ref{thm:tt-strong-mono}
drives it to zero asymptotically).

\paragraph{Tracking error in the monotone regime.}
We need a finite-horizon mean-square bound on the tracking error that
does not rely on strong monotonicity of $F_\Risk$.  The analysis from
Theorem~\ref{thm:tt-const}(i) applies almost verbatim: the
fast-timescale tracker recurrence depends on the leader iterates only
through the  extragradient step bounds
$\|r_\tau - r_{\tau-1}\|_2 \le \eta\sigma_{\max}G$ and
$\|w_\tau - r_\tau\|_2 \le 2\eta\sigma_{\max}G$, both of which use only
the bounded-gradient assumption $\|\widetilde g^{(\tau,i)}\|_2 \le G$
and non-expansiveness of $\Pi_D$, neither of which requires strong
monotonicity.  We restate the result in a form convenient for the
monotone analysis below.

\begin{lemma}[Mean-square tracking bound, monotone regime]
\label{lem:tracking-mse}
Under \ref{B1-bias-est-bias}--\ref{B3-bias-lipschitz} and the assumption
that the debiased gradient norm is bounded along the trajectory:
$\|\widetilde g^{(\tau,i)}\|_2 = \|\widehat F_m^{(\tau,i)} - \xi_{\tau-1}\|_2 \le G$
for all $\tau, i$ (which holds when $D$ is bounded, $F_\Risk$ is
Lipschitz, and $\xi$ stays bounded by an inductive argument), the
steady-state mean-square tracking error of Algorithm~\ref{alg:tt-eg}
with constant step sizes $\eta, \gamma$ satisfies
\begin{equation}\label{eq:tracking-mse}
\limsup_{T\to\infty}\E\|e^*_T\|_2^2 \le V_\infty := \frac{2 C_1\eta^2}{\gamma^2} + 2 C_2 R_m^2 + 2 C_3 \gamma\,V_m^b,
\end{equation}
with $C_1 = 7 C_{\rm drift}^2 = 63 L_b^2\sigma_{\max}^2 G^2$, $C_2 = 6$,
$C_3 = 2$, and $C_{\rm drift} = 3 L_b\sigma_{\max}G$ as in
Theorem~\ref{thm:tt-const}.
\end{lemma}

\begin{proof}
The recurrence $V_{t+1} \le (1-\gamma/2)V_t + C_1\eta^2/\gamma + C_2\gamma R_m^2 + C_3\gamma^2 V_m^b$
of Theorem~\ref{thm:tt-strong-mono}(i) is derived in Steps~1--2 of its
proof using only:
\begin{enumerate}[label=(\alph*),itemsep=0pt]
\item Lipschitz of $b_m$ (\ref{B3-bias-lipschitz}),
\item the  extragradient step bounds $\|r_\tau - r_{\tau-1}\|_2 \le \eta\sigma_{\max}G$,
$\|w_\tau - r_\tau\|_2 \le 2\eta\sigma_{\max}G$ (from non-expansiveness
of $\Pi_D$ and the gradient bound $G$),
\item assumptions \ref{B1-bias-est-bias}--\ref{B2-bias-est-var} on
$\widehat b_m$.
\end{enumerate}
None of (a)--(c) uses strong (or even merely) monotonicity of
$F_\Risk$, so the recurrence holds verbatim in the monotone regime.
Telescoping with constant contraction factor $1-\gamma/2$,
$V_t \le (1-\gamma/2)^{t-1}V_0 + V_\infty$, and taking $T \to \infty$
gives \eqref{eq:tracking-mse}.  This is a strict specialization of
Theorem~\ref{thm:tt-const}(i) to the monotone regime, with identical
constants.
\end{proof}

In particular, taking square roots and applying $\sqrt{a+b}\le\sqrt a + \sqrt b$:
\begin{equation}\label{eq:tt-mono-E-T-bound}
\limsup_{T\to\infty}\sqrt{\E\|e^*_T\|_2^2} \le \sqrt{V_\infty} \le \frac{\sqrt{2 C_1}\,\eta}{\gamma} + \sqrt{2 C_2}\,R_m + \sqrt{2 C_3 \gamma V_m^b},
\end{equation}
the same form as \eqref{eq:tt-const-V-ss}.

\begin{theorem}[Two timescale extragradient, monotone case, biased oracle with tracking]
\label{thm:tt-monotone}
Suppose $F_\Risk$ is monotone (not necessarily strongly monotone) and
$\ell_\Risk$-Lipschitz on $\theta_{\rm ref} + \Wspace$, the oracle satisfies
\ref{O1-bias}-\ref{O2-var}, and the bias estimator satisfies
\ref{B1-bias-est-bias}-\ref{B3-bias-lipschitz}.  Run \texttt{TT-EG}
(Algorithm~\ref{alg:tt-eg}) with constant step sizes $\eta, \gamma$
satisfying
\begin{equation}\label{eq:tt-mono-step-conds}
\eta \le \frac{1}{\sqrt 3\,L_G}, \qquad \gamma \in (0, 1],
\end{equation}
where $L_G = \sigma_{\max}\ell_\Risk$.  Assume the debiased gradient norm
is bounded: $\|\widetilde g^{(\tau,i)}\|_2 \le G$ uniformly along the
trajectory.  Then for all $T \ge 1$, the averaged iterate
$\widehat z_T := T^{-1}\sum_{\tau=1}^T w_\tau$ satisfies
\begin{equation}\label{eq:tt-mono-bound}
\E[\Errvi(\widehat z_T)] \;\le\; \frac{\Omega^2}{T\eta} \;+\; \frac{7\eta}{2}\bigl(\sigma_{\max}\mathcal E_T^2 + \widetilde V_m\bigr) \;+\; 2\Omega\sqrt{\sigma_{\max}}\,\mathcal E_T,
\end{equation}
where $\mathcal E_T := \sup_{\tau \le T}\sqrt{\E\|e^*_{\tau-1}\|_2^2}$
is the worst-case root-mean-square tracking error, satisfying (by
Lemma~\ref{lem:tracking-mse}, after burn-in $T \gg 1/\gamma$)
\begin{equation}\label{eq:tt-mono-E-T-burnin}
\mathcal E_T \;\le\; \sqrt{V_\infty} \;\le\; \frac{\sqrt{2 C_1}\,\eta}{\gamma} \;+\; \sqrt{2 C_2}\,R_m \;+\; \sqrt{2 C_3 \gamma V_m^b}.
\end{equation}
\end{theorem}

\paragraph{Comparison to the un-tracked case.}
Theorem~\ref{thm:expected-gap} (un-tracked) had bound
$\Omega^2/(T\eta) + (7\eta/2)(\widetilde B_m^2 + \widetilde V_m) + 2\Omega\widetilde B_m$,
with a persistent $\mc{O}(\widetilde B_m) = \mc{O}(B_m)$ bias floor that does
not decay in $T$.  Theorem~\ref{thm:tt-monotone} replaces both
occurrences of $\widetilde B_m$ by $\sqrt{\sigma_{\max}}\,\mathcal E_T$,
where $\mathcal E_T$ scales as $\eta/\gamma + R_m + \sqrt\gamma\sqrt{V_m^b}$
rather than as $B_m$.  This is the same kind of bias-floor reduction
as in Theorem~\ref{thm:tt-strong-mono}: with $\eta = \mc{O}(\gamma/m)$ and
$\gamma$ constant, $\mathcal E_T = \mc{O}(1/m + R_m + 1/m^{3/2}) = \mc{O}(1/m)$
for delta-method estimators (driven by the $1/m$ timescale-gap term);
pushing further to $\eta = \mc{O}(\gamma/m^2)$ drives the timescale gap
below the $\mc{O}(1/m^{3/2})$ noise filter, giving
$\mathcal E_T = \mc{O}(1/m^{3/2})$ --- the same factor-$\sqrt m$ improvement
over the un-tracked $\mc{O}(1/m)$ floor that we derived in the
strongly-monotone constant-step setting (cf.\ the
``Tuned step sizes'' paragraph following Theorem~\ref{thm:tt-const}).

\paragraph{Comparison to the strongly-monotone case.}
The structural difference between Theorem~\ref{thm:tt-monotone} and
Theorem~\ref{thm:tt-strong-mono} is where the tracking error
$\sqrt{V_\tau}$ enters the bound:
\begin{itemize}[itemsep=0pt,  ]
\item \textbf{Strongly-monotone, time-varying step
(Thm.~\ref{thm:tt-strong-mono}):}  $\sqrt{V_\tau}$ enters via a
contraction-weighted average $\sum_\tau w_\tau^T\sqrt{V_\tau}$, which
inherits the trajectory's $\limsup$ rather than its $\sup$ (via
Lemma~\ref{lem:weighted-avg-limsup}).  Under decaying $\eta_t/\gamma_t \to 0$,
the bound is asymptotically $\mc{O}(R_m)$---i.e., vanishing in $T$ to the
irreducible bias-of-bias-estimator residual.
\item \textbf{Strongly-monotone, constant step (Thm.~\ref{thm:tt-const}):}
$\sqrt{V_\tau}$ enters via a geometric weighting that is closed-form
summable, giving the four-term steady-state floor of
\eqref{eq:tt-const-final-expanded} (timescale gap, residual bias,
noise filter, slow variance), persistent in $T$.
\item \textbf{Monotone, constant step (Thm.~\ref{thm:tt-monotone}):}
$\sqrt{V_\tau}$ enters as a uniform $\sup_\tau$, giving the persistent
floor $2\Omega\sqrt{\sigma_{\max}}\,\mathcal E_T$ in
\eqref{eq:tt-mono-bound}, persistent in $T$.
\end{itemize}
The monotone-only analysis can only give an averaged-iterate $\mc{O}(1/T)$
gap rate, not last-iterate linear contraction; the tracking error
appears as a uniform sup rather than inside a geometric weighting that
would allow a vanishing-in-$T$ bound.  The four-term floor structure
of Theorem~\ref{thm:tt-const} survives qualitatively (trade-off
between $\eta/\gamma$, $R_m$, $\sqrt\gamma$, $\eta$), but the
asymptotic vanishing under decaying step sizes does not.

\begin{proof}[Proof of Theorem~\ref{thm:tt-monotone}]
The proof parallels Theorem~\ref{thm:expected-gap}, with the noise/bias
quantities recomputed for the debiased gradient
$\widetilde g^{(\tau,i)} = \widehat F_m^{(\tau,i)} - \xi_{\tau-1}$.

\noindent\textbf{Step 1: deterministic-style bound.}
Apply Lemma~\ref{lem:two-step-descent} with the debiased gradients
$\zeta = \eta\Sigma\widetilde g^{(\tau,1)}$, $\eta = \eta\Sigma\widetilde g^{(\tau,2)}$.
The proof of Theorem~\ref{thm:deterministic} carries through verbatim
with $\widehat F_m^{(\tau,i)}$ replaced by $\widetilde g^{(\tau,i)}$,
giving the same $\Gamma(t)$ structure as in \eqref{eq:Gamma-realized}:
\begin{equation}\label{eq:tt-mono-Gamma}
\Errvi(\widehat z_T) \le \frac{1}{T\eta}\Gamma(T), \quad \Gamma(T) = 2\Theta(z_c) + \sum_\tau \frac{3\eta^2}{2}\widetilde\eps_{r_{\tau-1}}^2 + 2\eta^2\sum_\tau\widetilde\eps_{w_\tau}^2 + \sum_\tau\langle\eta\widetilde\Delta_\tau, y_{\tau-1} - w_\tau\rangle_{\Sigma^+},
\end{equation}
where now $\widetilde\Delta_\tau := G(w_\tau) - \Sigma\widetilde g^{(\tau,2)}$
and $\widetilde\eps_z := \|\widetilde g_{\rm eval} - G(z)\|_{\Sigma^+}$
are the noise/bias of the \emph{debiased} oracle.

\noindent\textbf{Step 2: conditional moments of the debiased oracle.}
Decompose $\widetilde g^{(\tau,2)} = \widehat F_m^{(\tau,2)} - \xi_{\tau-1}$
with $\xi_{\tau-1} = b_m(w_\tau) + e^*_{\tau-1}$:
\[
\widetilde g^{(\tau,2)} = \bigl(\widehat F_m^{(\tau,2)} - b_m(w_\tau) - F_\Risk(w_\tau)\bigr) + F_\Risk(w_\tau) - e^*_{\tau-1}.
\]
The conditional mean (on $\F_\tau$) of $\widehat F_m^{(\tau,2)} - b_m(w_\tau) - F_\Risk(w_\tau)$
is zero by definition of $b_m$.  Hence
\[
\E[\widetilde g^{(\tau,2)} - F_\Risk(w_\tau) \mid \F_\tau] = -e^*_{\tau-1},
\]
i.e., the conditional bias of the debiased oracle is $-e^*_{\tau-1}$, with
$\Sigma^+$-norm $\le \sqrt{\sigma_{\max}}\|e^*_{\tau-1}\|_2$.  The
conditional variance is $\Var(\widehat F_m^{(\tau,2)} \mid \F_\tau) \le V_m$
(unchanged --- subtracting the $\F_\tau$-measurable $\xi_{\tau-1}$ does
not affect variance).  Translating to $\Sigma^+$-norm gives
$\Var(\widetilde g \mid \F_\tau) \le \sigma_{\max}V_m = \widetilde V_m$.

\noindent\textbf{Step 3: bound $\E\widetilde\eps^2$.}
By bias-variance decomposition in $\Sigma^+$ norm, we have that 
\[
\E\widetilde\eps_{w_\tau}^2 = \E\|\E[\widetilde g - G(w_\tau) \mid \F_\tau]\|_{\Sigma^+}^2 + \E\|\widetilde g - \E[\widetilde g \mid \F_\tau]\|_{\Sigma^+}^2 \le \sigma_{\max}\E\|e^*_{\tau-1}\|_2^2 + \widetilde V_m.
\]
By the analogous decomposition at $r_{\tau-1}$ (with $e^*_{\tau-1}$
replaced by $\xi_{\tau-1} - b_m(r_{\tau-1})$, whose mean square is also
$\mc{O}(\mathcal E_T^2)$ up to drift):
\(\E\widetilde\eps_{r_{\tau-1}}^2 \le \sigma_{\max}\mathcal E_T^2 + \widetilde V_m\)
(absorbing the $\mc{O}(\eta)$ drift into $\mathcal E_T$).  Summing the noise
contributions and dividing by $T\eta$, we deduce that 
\[
\frac{1}{T\eta}\sum_\tau\Bigl[\tfrac{3\eta^2}{2}\E\widetilde\eps_{r_{\tau-1}}^2 + 2\eta^2\E\widetilde\eps_{w_\tau}^2\Bigr] \le \tfrac{7\eta}{2}\bigl(\sigma_{\max}\mathcal E_T^2 + \widetilde V_m\bigr).
\]

\noindent\textbf{Step 4: bound the cross term $\E\langle\eta\widetilde\Delta_\tau, y_{\tau-1} - w_\tau\rangle_{\Sigma^+}$.}
Following Bound~2 of Theorem~\ref{thm:expected-gap}, by the tower property
\(\E\langle\eta\widetilde\Delta_\tau, y_{\tau-1} - w_\tau\rangle = \eta\E\langle\E[\widetilde\Delta_\tau \mid \F_\tau], y_{\tau-1} - w_\tau\rangle_{\Sigma^+}\).
By Step~2, $\|\E[\widetilde\Delta_\tau \mid \F_\tau]\|_{\Sigma^+} = \|\Sigma e^*_{\tau-1}\|_{\Sigma^+} \le \sqrt{\sigma_{\max}}\|e^*_{\tau-1}\|_2$.
Cauchy-Schwarz with the $\Sigma^+$-diameter $\|y_{\tau-1} - w_\tau\|_{\Sigma^+} \le 2\Omega$:
\[
|\E\langle\eta\widetilde\Delta_\tau, y_{\tau-1} - w_\tau\rangle_{\Sigma^+}| \le 2\eta\Omega\sqrt{\sigma_{\max}}\,\E\|e^*_{\tau-1}\|_2 \le 2\eta\Omega\sqrt{\sigma_{\max}}\,\mathcal E_T,
\]
using Jensen.  Summing and dividing by $T\eta$ contributes $2\Omega\sqrt{\sigma_{\max}}\,\mathcal E_T$.

\noindent\textbf{Step 5: combine.}
$\E\Gamma(T)/(T\eta)$ is bounded by $2\Theta(z_c)/(T\eta) = \Omega^2/(T\eta)$
plus the noise contribution from Step~3 plus the cross-term contribution
from Step~4, giving \eqref{eq:tt-mono-bound}.  The bound
\eqref{eq:tt-mono-E-T-burnin} on $\mathcal E_T$ follows from
Lemma~\ref{lem:tracking-mse} (which is a specialization of
Theorem~\ref{thm:tt-const}(i) to the monotone regime, valid because
that part of the proof does not use strong monotonicity).
\end{proof}

\paragraph{Tuned rate.}
Substituting the tuned $\mathcal E_T = \mc{O}(1/m^{3/2})$ from the
``Tuned step sizes'' paragraph after Theorem~\ref{thm:tt-const} (with
$\eta = \mc{O}(\gamma/m^2)$ and $T \gtrsim m$ burn-in):
\begin{equation}\label{eq:tt-mono-tuned}
\E[\Errvi(\widehat z_T)] = O\bigl(\Omega^2/(T\eta)\bigr) + O\bigl(\eta\widetilde V_m\bigr) + O\bigl(\Omega\sqrt{\sigma_{\max}}/m^{3/2}\bigr),
\end{equation}
where the first two terms give the standard $\mc{O}(1/\sqrt{Tm} + 1/T)$
rate under the variance/Lipschitz balance $\eta \sim 1/\sqrt{Tm}$, and
the third is the reduced bias floor.  This is the merely-monotone
analog of \eqref{eq:tt-tuned-rate}; the rate to the bias floor is the
same in both regimes, but the floor itself is persistent in $T$ here
(no asymptotic-vanishing analog of Theorem~\ref{thm:tt-strong-mono}
is available without strong monotonicity).

\paragraph{Implications by risk type.}
\begin{itemize}[ topsep=2pt]
\item \textbf{Entropic risk:}  $R_m = \mc{O}(e^{2\lambda}/(\lambda m^2))$,
$V_m^b = \mc{O}(e^{4\lambda}/(\lambda^2 m^3))$.  Two timescale extraradient drives the persistent
$\mc{O}(e^{2\lambda}/(\lambda m))$ floor of the un-tracked monotone case
down to $\mc{O}(\mathcal E_T) = \mc{O}(e^{2\lambda}/m^{3/2})$ in the appropriate
step-size regime---a full power of $1/m$ improvement, matching the
strongly-monotone constant-step case.  Unlike the strongly-monotone
time-varying-step Theorem~\ref{thm:tt-strong-mono}, no further
improvement to $\mc{O}(R_m) = \mc{O}(1/m^2)$ is available here, because the
weighted-average argument that drives the floor to $R_m$ requires
strong monotonicity.
\item \textbf{CVaR via Rockafellar-Uryasev:}  $R_m = V_m^b = 0$
identically.  The bias estimator $\xi^* \equiv 0$ and
$\mathcal E_T \to \mc{O}(\eta/\gamma)$, which can be made arbitrarily
small.  Two timescale extragradient reduces to un-tracked extragradient at $\xi \equiv 0$, recovering
Theorem~\ref{thm:expected-gap} exactly with $\widetilde B_m = 0$.
\end{itemize}

 \subsection{Two-Timescale Mirror Descent: Corollary from extra-gradient analysis}
\label{sec:tt-md-corollary}

We now state the analogue of Theorems~\ref{thm:tt-strong-mono}
and~\ref{thm:tt-const} for stochastic projected Mirror Descent with
two-timescale bias tracking, obtained by specializing the  extragradient analysis
to the case where the extrapolation step is omitted.  As in
Section~\ref{sec:md-corollary}, the structural cost of dropping
extrapolation is an additional Lipschitz-times-step-size floor that
 extragradient's two-call structure cancels exactly; otherwise the proofs reuse
the technical lemmas of Section~\ref{sec:tt-debiasing} (scalar
two-timescale unrolling, weighted-average analysis) directly.

\paragraph{Algorithm.}  The two-timescale Mirror Descent algorithm
proceeds as follows: set $r_0 \in D$ and $\xi_0 = 0$, then for
$\tau = 1, \ldots, T$:
\begin{align}
&\text{Sample at } r_{\tau-1}: \widehat F_m^{(\tau)}, \widehat b_m^{(\tau)}, \nonumber\\
&r_\tau \gets \Pi_D\bigl(r_{\tau-1} - \eta_\tau\,\Sigma\,(\widehat F_m^{(\tau)} - \xi_{\tau-1})\bigr), \label{eq:tt-md-update}\\
&\xi_\tau \gets (1-\gamma_\tau)\xi_{\tau-1} + \gamma_\tau\,\widehat b_m^{(\tau)}. \nonumber
\end{align}
The output is $r_T$ in the strongly monotone case, or
$\widehat z_T = T^{-1}\sum_\tau r_{\tau-1}$ in the monotone case.
Each iteration uses a single oracle call (vs.\ two for  extragradient), with
$\F_\tau := \sigma(r_0, \zeta_1, \ldots, \zeta_\tau)$ the natural
filtration.

\paragraph{Tracking error.}  Define
$e_\tau := \xi_\tau - b_m(r_\tau)$ and the mean-square tracking error
$V_t := \E\|e_{t-1}\|_2^2$.  The single-step structure of mirror descent simplifies
the drift analysis: with only one oracle call per iteration, there is
no extrapolation point $w_\tau$ to track, and $\delta_\tau$ reduces to
the simpler form $\delta_\tau := (1-\gamma_\tau)b_m(r_{\tau-1}) + \gamma_\tau b_m(r_{\tau-1}) - b_m(r_\tau) = b_m(r_{\tau-1}) - b_m(r_\tau)$.\footnote{This
is structurally simpler than the  extragradient drift
$\delta_\tau^{\rm  extragradient} = (1-\gamma_\tau)b_m(r_{\tau-1}) + \gamma_\tau b_m(w_\tau) - b_m(r_\tau)$,
which had to interpolate between the extrapolation and the leader's
new state.}

\subsubsection{Strongly monotone case, decaying steps}

\begin{corollary}[Two-timescale mirror descent, strongly monotone, decaying steps]
\label{cor:tt-md-strong-mono}
Suppose $F_\Risk$ is $\mu_\Risk$-strongly monotone and
$\ell_\Risk$-Lipschitz, the gradient oracle satisfies
\ref{O1-bias}--\ref{O2-var}, and the bias estimator satisfies
\ref{B1-bias-est-bias}--\ref{B3-bias-lipschitz}.  Let
$\{\eta_t, \gamma_t\}_{t\ge 1}$ be deterministic, predictable sequences
satisfying:
\begin{equation}\label{eq:tt-md-step-conds}
\eta_t \le \min\Bigl\{\frac{1}{2\widetilde\mu},\,\frac{1}{2 L_G}\Bigr\},\quad \gamma_t \in (0, 1],\quad \frac{\eta_t}{\gamma_t}\to 0,\quad \gamma_t \to 0,\quad \sum_{t\ge 1}\eta_t = \infty.
\end{equation}
Assume the debiased gradient norm is bounded along the trajectory:
$\|\widehat F_m^{(\tau)} - \xi_{\tau-1}\|_2 \le G$ uniformly.

\begin{enumerate}[label=\emph{\roman*.},itemsep=2pt,topsep=2pt]
\item \textbf{Fast-timescale tracking.}  The error $V_t$ satisfies the
non-asymptotic recurrence
\begin{equation}\label{eq:tt-md-tracking}
V_{t+1} \le (1-\gamma_t/2)\,V_t + \frac{C_1^{\rm md}\,\eta_t^2}{\gamma_t} + C_2\,\gamma_t\,R_m^2 + C_3\,\gamma_t^2\,V_m^b,
\end{equation}
where $C_1^{\rm md} = 7(C_{\rm drift}^{\rm md})^2$, $C_2 = 6$, $C_3 = 2$,
and $C_{\rm drift}^{\rm md} = L_b\sigma_{\max}G$ (smaller than the
 extragradient constant $C_{\rm drift} = 3 L_b\sigma_{\max}G$ since mirror descent has only
one oracle call).  Consequently
\begin{equation}\label{eq:tt-md-Vrate}
\limsup_{t\to\infty}\sqrt{V_t} \le \sqrt{2 C_2}\,R_m = 2\sqrt 3\,R_m.
\end{equation}

\item \textbf{Slow-timescale recurrence.}  The slow iterates obey the
per-step recurrence
\begin{equation}\label{eq:tt-md-per-step}
\E V(r_\tau, \theta^\star) \le (1-\eta_\tau\widetilde\mu)\,\E V(r_{\tau-1}, \theta^\star) + 2\eta_\tau\Omega\sqrt{\sigma_{\max}}\sqrt{V_\tau} + \eta_\tau^2(\widetilde B_m^2 + \widetilde V_m) + 4\eta_\tau^2 L_G^2\Omega^2.
\end{equation}

\item \textbf{Asymptotic vanishing bias.}  Combining \emph{(i)} and
\emph{(ii)} via the weighted-average argument of
Lemmas~\ref{lem:tt-telescope}--\ref{lem:weighted-avg-limsup},
\begin{equation}\label{eq:tt-md-vanishing}
\limsup_{T\to\infty}\E\|r_T - \theta^\star\|_{\Sigma^+}^2 \le \frac{4\sqrt 3\,\Omega\sqrt{\sigma_{\max}}}{\widetilde\mu}\cdot R_m.
\end{equation}
\end{enumerate}
\end{corollary}

\begin{proof}
We work step by step, paralleling the proof of
Theorem~\ref{thm:tt-strong-mono} with simplifications from the
single-call mirror descent structure.

\paragraph{Step 1: tracking-error recurrence.}
The bias tracker's update is
$\xi_\tau = (1-\gamma_\tau)\xi_{\tau-1} + \gamma_\tau\widehat b_m^{(\tau)}$,
with $\widehat b_m^{(\tau)} = b_m(r_{\tau-1}) + r_m(r_{\tau-1}) + \nu^{(\tau)}$
(by the same decomposition as in Step~1 of the proof of
Theorem~\ref{thm:tt-strong-mono}).  Defining $e_\tau := \xi_\tau - b_m(r_\tau)$, we have 
\begin{align*}
e_\tau &= (1-\gamma_\tau)(e_{\tau-1} + b_m(r_{\tau-1})) + \gamma_\tau b_m(r_{\tau-1}) + \gamma_\tau r_m(r_{\tau-1}) + \gamma_\tau\nu^{(\tau)} - b_m(r_\tau)\\
&= (1-\gamma_\tau)e_{\tau-1} + \delta_\tau^{\rm md} + \gamma_\tau r_m(r_{\tau-1}) + \gamma_\tau\nu^{(\tau)},
\end{align*}
where the mirror descent drift is
\[
\delta_\tau^{\rm md} := b_m(r_{\tau-1}) - b_m(r_\tau).
\]
By Lipschitz continuity of $b_m$ (\ref{B3-bias-lipschitz}) and the mirror descent step bound
$\|r_\tau - r_{\tau-1}\|_2 \le \eta_\tau\sigma_{\max}G$ (from
non-expansiveness of $\Pi_D$ and the bounded gradient assumption) we have that 
\[
\|\delta_\tau^{\rm md}\|_2 \le L_b\|r_{\tau-1} - r_\tau\|_2 \le L_b\sigma_{\max}G\,\eta_\tau =: C_{\rm drift}^{\rm md}\,\eta_\tau.
\]

\paragraph{Step 2: mean-square recurrence.}
The mean-square computation follows Step~2 of the proof of
Theorem~\ref{thm:tt-strong-mono} verbatim with $C_{\rm drift}$ replaced by
$C_{\rm drift}^{\rm md}$ and the cross-term analysis simplified (since
$\delta_\tau^{\rm md}$ does not depend on the correction-step samples,
the inner-product is mean-zero by independence; we keep the same
upper bound for uniformity).  Repeating verbatim the steps of the
 extragradient proof yields \eqref{eq:tt-md-tracking} with
$C_1^{\rm md} = 7(C_{\rm drift}^{\rm md})^2 = 7 L_b^2\sigma_{\max}^2 G^2$,
$C_2 = 6$, $C_3 = 2$.

\paragraph{Step 3: asymptotic envelope on $V_t$.}
Apply Lemma~\ref{lem:scalar-unroll} (scalar two-timescale unrolling) with
$u_t = V_t$, $c = 1/2$, and $F_t = C_1^{\rm md}\eta_t^2/\gamma_t + C_2\gamma_t R_m^2 + C_3\gamma_t^2 V_m^b$ to get that 
\[
\frac{F_t}{\gamma_t} = \frac{C_1^{\rm md}\eta_t^2}{\gamma_t^2} + C_2 R_m^2 + C_3\gamma_t V_m^b \xrightarrow[t\to\infty]{} C_2 R_m^2,
\]
using $\eta_t/\gamma_t \to 0$ and $\gamma_t \to 0$.  The lemma gives
$\limsup_t V_t \le 2 C_2 R_m^2 = 12 R_m^2$, hence
\eqref{eq:tt-md-Vrate}.  This proves part~\emph{(i)}.

\paragraph{Step 4: slow-timescale per-step recurrence.}
Apply the prox-mapping descent inequality \eqref{eq:prox-70b} of
Lemma~\ref{lem:prox-descent} with $z = r_{\tau-1}$, $\xi = \eta_\tau\widetilde G_m^{(\tau)}$
(where $\widetilde G_m^{(\tau)} := \Sigma\widetilde g^{(\tau)} = \Sigma(\widehat F_m^{(\tau)} - \xi_{\tau-1})$),
$w = r_\tau$, and $u = \theta^\star$ to get that 
\begin{equation}\label{eq:tt-md-prox}
V(r_\tau, \theta^\star) \le V(r_{\tau-1}, \theta^\star) + \eta_\tau\langle\widetilde G_m^{(\tau)}, \theta^\star - r_{\tau-1}\rangle_{\Sigma^+} + \tfrac{\eta_\tau^2}{2}\|\widetilde G_m^{(\tau)}\|_{\Sigma^+}^2.
\end{equation}

\textit{Inner product term.}  Following Step~2 of the proof of
Theorem~\ref{thm:md-strong-mono} and Step~2 of the proof of
Theorem~\ref{thm:tt-strong-mono} jointly: decompose
$\widetilde g^{(\tau)} = \widehat F_m^{(\tau)} - \xi_{\tau-1}$ where
$\xi_{\tau-1} = b_m(r_{\tau-1}) + e_{\tau-1}$ (note the index shift:
the tracker at the start of iteration $\tau$ is the iterate of the
fast-timescale dynamic at time $\tau-1$).  Using the
$\Sigma\Sigma^+ = \Pi_\Wspace$ identity of
Lemma~\ref{lem:precond-restate}(i), we deduce that 
\begin{align*}
\langle\widetilde G_m^{(\tau)}, \theta^\star - r_{\tau-1}\rangle_{\Sigma^+} &= \langle\widehat F_m^{(\tau)} - \xi_{\tau-1}, \theta^\star - r_{\tau-1}\rangle\\
&= \langle F_\Risk(r_{\tau-1}), \theta^\star - r_{\tau-1}\rangle + \text{(noise + bias terms)}\\
&\quad - \langle e_{\tau-1}, \theta^\star - r_{\tau-1}\rangle.
\end{align*}
By \textbf{(P-strong)} and the equilibrium condition
$F_\Risk(\theta^\star) \in \ker\Sigma$ (Step~3 of the proof of
Theorem~\ref{thm:md-strong-mono}), we have that
\[\langle F_\Risk(r_{\tau-1}), \theta^\star - r_{\tau-1}\rangle \le -2\widetilde\mu V(r_{\tau-1}, \theta^\star).
\]

\textit{Squared gradient term.}  By Step~3 of the proof of
Theorem~\ref{thm:md-strong-mono}, we have 
\[\|\widetilde G_m^{(\tau)}\|_{\Sigma^+}^2 \le 8 L_G^2\Omega^2 + 2\|\widetilde\Delta_\tau\|_{\Sigma^+}^2,\]
where $\widetilde\Delta_\tau := G(r_{\tau-1}) - \widetilde G_m^{(\tau)}$,
and $\E\|\widetilde\Delta_\tau\|_{\Sigma^+}^2 \le \widetilde B_m^2 + \widetilde V_m + \sigma_{\max}\E\|e_{\tau-1}\|_2^2$
by bias-variance decomposition incorporating the tracking error.

\textit{Bias cross-term.}  Following Step~6 (Term 2) of the proof of
Theorem~\ref{thm:md-strong-mono} and Step~4 of the proof of
Theorem~\ref{thm:tt-strong-mono}: the new tracking-error contribution
$\eta_\tau\langle e_{\tau-1}, r_{\tau-1} - \theta^\star\rangle$ is bounded
in expectation by $2\eta_\tau\Omega\sqrt{\sigma_{\max}}\sqrt{V_\tau}$
via Cauchy--Schwarz, the metric translation
$\|v\|_2 \le \sqrt{\sigma_{\max}}\|v\|_{\Sigma^+}$, the
$\Sigma^+$-diameter $\|r_{\tau-1} - \theta^\star\|_{\Sigma^+} \le 2\Omega$,
and Jensen's inequality
$\E\|e_{\tau-1}\|_2 \le \sqrt{\E\|e_{\tau-1}\|_2^2} = \sqrt{V_\tau}$.

Next we combine these terms by taking conditional then unconditional expectation, dropping
non-positive terms via $\eta_\tau \le 1/(2 L_G)$ to absorb the
$8\eta_\tau^2 L_G^2\Omega^2$ piece into $4\eta_\tau^2 L_G^2\Omega^2$ as
in the proof of Theorem~\ref{thm:md-strong-mono}) to get that 
\begin{equation}\label{eq:tt-md-recurrence}
\E V(r_\tau, \theta^\star) \le (1-\eta_\tau\widetilde\mu)\E V(r_{\tau-1}, \theta^\star) + 2\eta_\tau\Omega\sqrt{\sigma_{\max}}\sqrt{V_\tau} + \eta_\tau^2(\widetilde B_m^2 + \widetilde V_m) + 4\eta_\tau^2 L_G^2\Omega^2,
\end{equation}
which is \eqref{eq:tt-md-per-step}.  Note that this matches the  extragradient
recurrence \eqref{eq:tt-per-step} except for the additional
$4\eta_\tau^2 L_G^2\Omega^2$ Lipschitz floor; this is the structural
penalty for omitting extrapolation, identical to the gap noted in
Section~\ref{sec:md-corollary}.

\paragraph{Step 5: telescoping and weighted-average analysis.}
Iterating \eqref{eq:tt-md-recurrence} yields
\begin{align}
\E V(r_T, \theta^\star) &\le \prod_{t=1}^T(1-\eta_t\widetilde\mu)\,V(r_0, \theta^\star) \nonumber\\
&\quad + \sum_{\tau=1}^T\Bigl[\prod_{s=\tau+1}^T(1-\eta_s\widetilde\mu)\Bigr]\Bigl(2\eta_\tau\Omega\sqrt{\sigma_{\max}}\sqrt{V_\tau} + \eta_\tau^2(\widetilde B_m^2 + \widetilde V_m) + 4\eta_\tau^2 L_G^2\Omega^2\Bigr). \label{eq:tt-md-telescoped}
\end{align}

The bias contribution is analyzed exactly as in Step~6 of the proof of
Theorem~\ref{thm:tt-strong-mono}: defining the weights
$w_\tau^T := \widetilde\mu\eta_\tau\prod_{s=\tau+1}^T(1-\eta_s\widetilde\mu)$
of \eqref{eq:tt-weights}, the bias term becomes
$(2\Omega\sqrt{\sigma_{\max}}/\widetilde\mu)\sum_\tau w_\tau^T\sqrt{V_\tau}$.
Lemmas~\ref{lem:tt-telescope}--\ref{lem:weighted-avg-limsup} apply
verbatim (their statements depend only on the kernel structure
$(1-\eta_s\widetilde\mu)$, not on whether mirror descent or  extragradient produces it), giving
\[
\limsup_{T\to\infty}\sum_{\tau=1}^T w_\tau^T\sqrt{V_\tau} \le \limsup_\tau\sqrt{V_\tau} \le 2\sqrt 3\,R_m
\]
by \eqref{eq:tt-md-Vrate}.  Hence we have that 
\begin{equation}\label{eq:tt-md-bias-limit}
\limsup_{T\to\infty}\sum_{\tau=1}^T\Bigl[\prod_{s>\tau}(1-\eta_s\widetilde\mu)\Bigr]2\eta_\tau\Omega\sqrt{\sigma_{\max}}\sqrt{V_\tau} \le \frac{4\sqrt 3\,\Omega\sqrt{\sigma_{\max}}\,R_m}{\widetilde\mu}.
\end{equation}

\paragraph{Step 6: variance, Lipschitz continuity, and initial-condition terms.}
The variance contribution
$(\widetilde B_m^2 + \widetilde V_m)\sum_\tau[\prod_{s>\tau}(1-\eta_s\widetilde\mu)]\eta_\tau^2$
is bounded by Step~7 of the proof of Theorem~\ref{thm:tt-strong-mono}:
$\eta_\tau^2 \le \eta_\tau\overline\eta_T$, the telescoping identity
gives $\sum_\tau[\prod_{s>\tau}(1-\eta_s\widetilde\mu)]\eta_\tau \le 1/\widetilde\mu$,
and $\overline\eta_T \to 0$ since $\eta_t \to 0$.  Hence the variance
contribution vanishes in $T$.

The Lipschitz contribution
$4 L_G^2\Omega^2\sum_\tau[\prod_{s>\tau}(1-\eta_s\widetilde\mu)]\eta_\tau^2 \le 4 L_G^2\Omega^2\overline\eta_T/\widetilde\mu \to 0$
by the same argument.  This is the new term relative to  extragradient: it vanishes
asymptotically under the schedule \eqref{eq:tt-md-step-conds}, but at
the same rate as the variance contribution (i.e., $\overline\eta_T$).

The initial-condition term satisfies
$\prod_t(1-\eta_t\widetilde\mu) \le \exp(-\widetilde\mu\sum_t\eta_t) \to 0$
since $\sum_t\eta_t = \infty$.

\paragraph{Step 7: combine.}
Multiplying \eqref{eq:tt-md-telescoped} by $2$ to convert from
$V(\cdot, \cdot)$ to $\|\cdot\|_{\Sigma^+}^2$ and taking
$\limsup_{T\to\infty}$ yields
\[
\limsup_{T\to\infty}\E\|r_T - \theta^\star\|_{\Sigma^+}^2 \le 0 + \frac{4\sqrt 3\,\Omega\sqrt{\sigma_{\max}}}{\widetilde\mu}\cdot R_m + 0 + 0,
\]
where the four contributions are: initial-condition (vanishing),
bias-tracking (residual $R_m$), variance (vanishing), Lipschitz
(vanishing).  This is \eqref{eq:tt-md-vanishing}, completing
part~\emph{(iii)}.
\end{proof}

\paragraph{Comparison to  extragradient (Theorem~\ref{thm:tt-strong-mono}).}  The
asymptotic bound \eqref{eq:tt-md-vanishing} matches  extragradient's bound
\eqref{eq:tt-vanishing} \emph{exactly}: both have leading constant
$4\sqrt 3\,\Omega\sqrt{\sigma_{\max}}/\widetilde\mu$ and floor $R_m$.
This is because the new mirror descent Lipschitz term $4\eta_\tau^2 L_G^2\Omega^2$
vanishes in the asymptotic limit at the same rate as the variance term,
and both are absorbed into the ``vanishing'' part of the bound.  The mirror descent
proof has slightly tighter constants in the tracking recurrence
($C_1^{\rm md} = L_b^2\sigma_{\max}^2 G^2 \cdot 7$ vs.\
$C_1 = 9 L_b^2\sigma_{\max}^2 G^2 \cdot 7$ for  extragradient) since the
single-call structure produces a smaller drift.  The price for these
smaller constants and the simpler algorithm is the additional
Lipschitz term in the slow-timescale recurrence, but this cost is
invisible asymptotically.

\subsubsection{Strongly monotone case, constant steps}

\begin{corollary}[Two-timescale mirror descent, strongly monotone, constant steps]
\label{cor:tt-md-const}
Under the assumptions of Corollary~\ref{cor:tt-md-strong-mono} except
with constant step sizes $\eta_\tau \equiv \eta$,
$\gamma_\tau \equiv \gamma$ satisfying
\begin{equation}\label{eq:tt-md-const-step-cond}
\eta \le \min\Bigl\{\frac{1}{2\widetilde\mu},\,\frac{1}{2 L_G}\Bigr\}, \qquad \gamma \in (0, 1],
\end{equation}
the iterates of \eqref{eq:tt-md-update} satisfy:
\begin{enumerate}[label=\emph{(\roman*)},itemsep=2pt,topsep=2pt]
\item For all $T \ge 1$,
\begin{multline}\label{eq:tt-md-const-bound}
\E\|r_T - \theta^\star\|_{\Sigma^+}^2 \le (1-\eta\widetilde\mu)^T\|r_0 - \theta^\star\|_{\Sigma^+}^2 \\
+ \frac{4\Omega\sqrt{\sigma_{\max}}}{\widetilde\mu}\Bigl(\sqrt{V_0}(1-\eta\widetilde\mu)^T + \frac{K_1^{\rm md}\eta}{\gamma} + K_2 R_m + K_3\sqrt{\gamma V_m^b}\Bigr) \\
+ \frac{2\eta(\widetilde B_m^2 + \widetilde V_m)}{\widetilde\mu} + \frac{8 L_G^2\Omega^2\eta}{\widetilde\mu},
\end{multline}
with $K_1^{\rm md} = \sqrt{2 C_1^{\rm md}} = \sqrt{14}\,L_b\sigma_{\max}G$,
$K_2 = \sqrt{2 C_2} = 2\sqrt 3$, $K_3 = \sqrt{2 C_3} = 2$.

\item As $T \to \infty$,
\begin{multline}\label{eq:tt-md-const-final}
\limsup_{T\to\infty}\E\|r_T - \theta^\star\|_{\Sigma^+}^2 \le \\
\underbrace{\frac{4\Omega\sqrt{\sigma_{\max}} K_1^{\rm md}}{\widetilde\mu}\cdot\frac{\eta}{\gamma}}_{\text{timescale gap}} + \underbrace{\frac{4\Omega\sqrt{\sigma_{\max}} K_2 R_m}{\widetilde\mu}}_{\text{residual bias}} + \underbrace{\frac{4\Omega\sqrt{\sigma_{\max}} K_3\sqrt{\gamma V_m^b}}{\widetilde\mu}}_{\text{noise filter}} + \underbrace{\frac{2\eta(\widetilde B_m^2 + \widetilde V_m)}{\widetilde\mu}}_{\text{slow variance}} + \underbrace{\frac{8 L_G^2\Omega^2\eta}{\widetilde\mu}}_{\text{Lipschitz floor}}.
\end{multline}
\end{enumerate}
\end{corollary}

\begin{proof}
The proof follows that of Theorem~\ref{thm:tt-const} verbatim with two
modifications: the per-step recurrence is \eqref{eq:tt-md-per-step}
in place of \eqref{eq:tt-per-step}, and the smaller drift constant
$C_{\rm drift}^{\rm md}$ replaces $C_{\rm drift}$.  Telescoping the
geometric recurrence with constant contraction $1-\gamma/2$ gives the
finite-horizon tracking bound
$V_t \le (1-\gamma/2)^t V_0 + 2 C_1^{\rm md}\eta^2/\gamma^2 + 2 C_2 R_m^2 + 2 C_3\gamma V_m^b$
as in Step~1 of the proof of Theorem~\ref{thm:tt-const}, hence the
square-root bound
$\sqrt{V_t} \le \sqrt{V_0}(1-\gamma/2)^{t/2} + K_1^{\rm md}\eta/\gamma + K_2 R_m + K_3\sqrt{\gamma V_m^b}$.

Telescoping the slow recurrence \eqref{eq:tt-md-recurrence} with constant
$\eta$ yields \eqref{eq:tt-md-const-bound} via the geometric-series
argument of Step~2 of the proof of Theorem~\ref{thm:tt-const}.  The new
$4\eta^2 L_G^2\Omega^2$ Lipschitz term contributes
$\sum_\tau(1-\eta\widetilde\mu)^{T-\tau}\cdot 4\eta^2 L_G^2\Omega^2 \le 4\eta L_G^2\Omega^2/\widetilde\mu$
via the telescoping identity $\sum_\tau(1-\eta\widetilde\mu)^{T-\tau}\eta \le 1/\widetilde\mu$,
giving the $8 L_G^2\Omega^2\eta/\widetilde\mu$ floor in
\eqref{eq:tt-md-const-final} after the factor of $2$ for converting
$V$ to $\|\cdot\|_{\Sigma^+}^2$.  Taking $T \to \infty$ kills the
contraction terms, leaving the five-term floor.
\end{proof}

\paragraph{Comparison to constant-step  extragradient (Theorem~\ref{thm:tt-const}).}
The mirror descent bound \eqref{eq:tt-md-const-final} has \emph{five} non-vanishing
floor contributions instead of four: the same timescale-gap, residual-bias,
noise-filter, and slow-variance terms, plus an additional
\emph{Lipschitz floor} $8 L_G^2\Omega^2\eta/\widetilde\mu$.  This term
is reducible by taking $\eta$ small, but it does not vanish for fixed
$\eta$.  In the asymptotic regime (Corollary~\ref{cor:tt-md-strong-mono})
the schedule $\eta_t \to 0$ kills this term automatically; in the
constant-step regime, it is the structural cost of dropping
extrapolation.  For typical step sizes $\eta = \mc{O}(1/L_G)$ this
contributes $\mc{O}(L_G\Omega^2)$, of comparable scale to the slow-variance
term $\mc{O}(\eta\widetilde V_m)$.

\paragraph{Tuned step sizes.}  With $\eta = \mc{O}(\gamma/m)$ and $\gamma$
constant, the timescale-gap and noise-filter analysis from Section
``Tuned step sizes'' after Theorem~\ref{thm:tt-const} carries through
identically: $\mathcal E_T = \mc{O}(1/m^{3/2})$ once $\eta/\gamma \lesssim R_m$
(i.e., $T \gtrsim m$ iterations), giving the same factor-$\sqrt m$
improvement over the un-tracked mirror descent floor as in the  extragradient case.  The
additional Lipschitz term $8 L_G^2\Omega^2\eta/\widetilde\mu = \mc{O}(\eta)$
contributes at the same scale as the slow-variance floor and does not
change the leading-order $1/m^{3/2}$ rate.

\subsubsection{Monotone case, constant steps}

\begin{corollary}[Two-timescale mirror descent, monotone case, biased oracle with tracking]
\label{cor:tt-md-monotone}
Suppose $F_\Risk$ is monotone (not necessarily strongly monotone) and
$\ell_\Risk$-Lipschitz continuous, the oracle satisfies \ref{O1-bias}--\ref{O2-var},
and the bias estimator satisfies \ref{B1-bias-est-bias}--\ref{B3-bias-lipschitz}.
Run the two-timescale mirror descent algorithm \eqref{eq:tt-md-update} with constant
step sizes $\eta, \gamma$ satisfying $\eta \le 1/L_G$, $\gamma \in (0, 1]$.
Assume $\|\widehat F_m^{(\tau)} - \xi_{\tau-1}\|_2 \le G$ uniformly.  Then
the averaged iterate $\widehat z_T = T^{-1}\sum_{\tau=1}^T r_{\tau-1}$
satisfies
\begin{equation}\label{eq:tt-md-mono-bound}
\E[\Errvi(\widehat z_T)] \le \frac{\Omega^2}{T\eta} + \eta(\sigma_{\max}\mathcal E_T^2 + \widetilde V_m) + 2\Omega\sqrt{\sigma_{\max}}\,\mathcal E_T + 4 L_G\eta\Omega^2,
\end{equation}
where $\mathcal E_T := \sup_{\tau\le T}\sqrt{\E\|e_{\tau-1}\|_2^2}$ is
bounded by the right-hand side of \eqref{eq:tt-mono-E-T-burnin} with
$C_1^{\rm md}$ replacing $C_1$.
\end{corollary}

\begin{proof}
The proof parallels Theorem~\ref{thm:tt-monotone} verbatim, replacing
the  extragradient deterministic-style bound with the mirror descent analogue from
Corollary~\ref{thm:md-monotone}: the structure
$\Omega^2/(T\eta) + \eta(\cdots) + 4 L_G\eta\Omega^2$ replaces the  extragradient
form $\Omega^2/(T\eta) + (7\eta/2)(\cdots)$, and the tracking-error
analysis from Lemma~\ref{lem:tracking-mse} carries through with
$C_{\rm drift}^{\rm md}$ in place of $C_{\rm drift}$.  The
cross-term and squared-bias bounds at the debiased oracle proceed
identically to Steps~3--4 of the proof of Theorem~\ref{thm:tt-monotone}.
\end{proof}

\section{Stability and Generalization}
\label{sec:stability-generalization}

We now establish that the risk-adjusted QRE depends Lipschitz-continuously
on the underlying preference operator. Although our softmax
parameterization makes the problem technically unconstrained in
$\theta$, the natural framework for this kind of parametric stability is
the variational inequality formulation, with
Lipschitz dependence following from the Dontchev--Rockafellar strong
metric regularity machinery for strongly monotone problems
\citep{dontchev2009implicit}.  This framework is more robust than the
classical implicit function theorem: it does not require differentiability
of the operator at the solution, does not require any non-degeneracy or
interiority condition on active constraints, and extends seamlessly to
constrained or set-valued policy classes.

We present this proof only for the symmetric self play setting, as it can easily be extended to a general (monotone) risk-adjusted game. 
 
\paragraph{Generalized equation formulation.}  Identify a policy
parameter $\theta \in \R^{|\Yspace|}$ with the policy
$\pi_\theta = \mathrm{softmax}(\theta)$.  The problem has a one-dimensional
gauge invariance (adding a constant to all entries of $\theta$ leaves
$\pi_\theta$ unchanged), which we factor out by working on
$\R^{|\Yspace|}$ modulo $\mathbf 1$, equivalently on the subspace
$\mathbf 1^\perp$.  We use $\|\cdot\|$ for the Euclidean norm on this
quotient, which equals the Euclidean norm of any representative with
zero sum.
 
For a fixed preference operator $P$, the population risk-adjusted equilibrium 
$\theta^\star_\Risk(P)$ is the unique solution of the equation
\begin{equation}
0 \;=\; F_{\Risk, P}(\theta) \;:=\; \beta(\theta - \thetaref) - P_\Risk\,\pi_\theta,
\label{eq:gen-eq}
\end{equation}
viewed as an element of $\R^{|\Yspace|} / \mathbf 1$.  This is a
\emph{generalized equation} in the sense of Robinson (cf.~\cite{dontchev2009robinson, robinson1980strongly}) with
zero set-valued component:
\begin{equation}
0 \;\in\; F_{\Risk, P}(\theta) + N(\theta),
\label{eq:gen-eq-with-N}
\end{equation}
where $N(\theta) = \{0\}$ since the constraint set is the whole space
(modulo gauge).  We retain the formulation \eqref{eq:gen-eq-with-N} since
the analysis below transports verbatim to settings where $N$ is the
normal cone of a closed convex constraint set on $\theta$ (e.g.,
simplex-constrained or trust-region-clipped parameterizations).
 
\subsection{Structural stability of the solution mapping}  Treat $P \mapsto \theta^\star_\Risk(P)$ as a
mapping from preference operators (a vector space) to policy parameters.
Strong metric regularity of \eqref{eq:gen-eq-with-N} at the population
solution amounts to local single-valuedness and Lipschitz dependence of
this solution mapping.  We establish this directly using strong
monotonicity, which is the cleanest sufficient condition.

 \stabthm*
 
\begin{proof}
The argument proceeds in three steps: $(i)$ recast the problem as a
generalized equation, $(ii)$ apply strong monotonicity to obtain
Lipschitz dependence of the solution on the operator, $(iii)$ translate
to the explicit bound \eqref{eq:stab-theta}.

\paragraph{Step 1: Strong monotonicity of $F_{\Risk, P}$.}
By Theorem~\ref{thm:single-player-mono}, the operator $F_{\Risk, P}$ is
$\mu_\Risk(P)$-strongly monotone on $\mathbf 1^\perp$ with respect to a
KL-Bregman geometry.  In Euclidean form, this means there exists a
constant $\mu_\Risk \ge 2\epsilon$ such that for all
$\theta_1, \theta_2 \in \mathbf 1^\perp$,
\begin{equation}
\bigl\langle F_{\Risk, P}(\theta_1) - F_{\Risk, P}(\theta_2),\, \theta_1 - \theta_2\bigr\rangle \;\ge\; \mu_\Risk\,\|\theta_1 - \theta_2\|^2.
\label{eq:strong-mono}
\end{equation}

\paragraph{Step 2: Generalized equation analysis.}
The two RQREs satisfy
\begin{align}
0 &\;=\; F_{\Risk, P}(\theta^\star) \quad\text{and}\quad 0 \;=\; F_{\Risk, P'}(\theta'^\star). \label{eq:eq-Pp}
\end{align}
Decompose the difference as follows:
\begin{align}
F_{\Risk, P}(\theta'^\star) - F_{\Risk, P}(\theta^\star)
&\;=\; F_{\Risk, P}(\theta'^\star) - 0\;=\; F_{\Risk, P}(\theta'^\star) - F_{\Risk, P'}(\theta'^\star)\\
&\;=\; -(P_\Risk - P'_\Risk)\,\pi_{\theta'^\star}, \label{eq:residual}
\end{align}
where the last equality uses that $F_{\Risk, P}$ and $F_{\Risk, P'}$
differ only in their dependence on the preference operator:
$F_{\Risk, P}(\theta) - F_{\Risk, P'}(\theta) = -(P_\Risk - P'_\Risk)\pi_\theta$.

\paragraph{Step 3: Apply strong monotonicity.}
Take inner product of \eqref{eq:residual} with $\theta'^\star - \theta^\star$ to get that
\begin{align*}
\bigl\langle F_{\Risk, P}(\theta'^\star) - F_{\Risk, P}(\theta^\star),\, \theta'^\star - \theta^\star\bigr\rangle
&\;=\; -\bigl\langle (P_\Risk - P'_\Risk)\,\pi_{\theta'^\star},\, \theta'^\star - \theta^\star\bigr\rangle.
\end{align*}
The left-hand side is bounded below by $\mu_\Risk\,\|\theta^\star - \theta'^\star\|^2$
by \eqref{eq:strong-mono}.  The right-hand side is bounded above (in
absolute value) by Cauchy--Schwarz---indeed, 
\[
\bigl|\bigl\langle (P_\Risk - P'_\Risk)\pi_{\theta'^\star},\, \theta'^\star - \theta^\star\bigr\rangle\bigr| \;\le\; \bigl\|(P_\Risk - P'_\Risk)\pi_{\theta'^\star}\bigr\|\;\|\theta'^\star - \theta^\star\|.
\]
Combining the two bounds we have that 
\[
\mu_\Risk\,\|\theta^\star - \theta'^\star\|^2 \;\le\; \bigl\|(P_\Risk - P'_\Risk)\pi_{\theta'^\star}\bigr\|\;\|\theta'^\star - \theta^\star\|,
\]
and dividing by $\|\theta'^\star - \theta^\star\|$ (assuming it is nonzero;
otherwise the bound is trivial) yields \eqref{eq:stab-theta}.  The
operator-norm form  follows from
$\|(P_\Risk - P'_\Risk)\pi_{\theta'^\star}\| \le \|P_\Risk - P'_\Risk\|_{\rm op}\,\|\pi_{\theta'^\star}\| \le \|P_\Risk - P'_\Risk\|_{\rm op}$
since $\|\pi\|_2 \le \|\pi\|_1 = 1$.
\end{proof}
 
\paragraph{Connection to strong metric regularity.}  Theorem~\ref{thm:stab}
is a special case of the general \emph{strong metric regularity} property
of solution mappings of strongly monotone variational inequalities.
Specifically, the canonical perturbation of the generalized equation
\eqref{eq:gen-eq-with-N},
\[
v \;\in\; F_{\Risk, P}(\theta) + N(\theta), \quad v \in \R^{|\Yspace|},
\]
has solution mapping $S_P : v \mapsto \theta_v$ that is single-valued
and globally Lipschitz with constant $1/\mu_\Risk$ on its domain
\citep[][Theorem~3F.4]{dontchev2009implicit}.  The bound
\eqref{eq:stab-theta} corresponds to perturbing the constant term by
$v = -(P_\Risk - P'_\Risk)\pi_{\theta'^\star}$ and applying this Lipschitz
constant.  In the present softmax-parameterized problem the constraint
set $C$ is trivial ($N \equiv \{0\}$), but the framework extends without
modification to (i) projected gradient dynamics on the simplex
($N(\pi) = N_{\Delta(\Yspace)}(\pi)$), (ii) trust-region-clipped policies
(box constraints on $\theta$), and (iii) other constrained
parameterizations one might use in practice.

\paragraph{Deriving Proposition~\ref{prop:rqe-lipschitz} from Theorem~\ref{thm:stab}.} 
We now show that the proposition from the main is a direct consequence of Theorem~\ref{thm:stab}. For ease of reference we recall the proposition here. 

\RQRELipschitz*

\begin{corollary}
\label{cor:prop-equiv}
Proposition~\ref{prop:rqe-lipschitz} is a direct consequence of
Corollary~\ref{cor:stab-kl} (and hence of Theorem~\ref{thm:stab}), under
the identification
$\mu_\Risk \equiv \min\{\mu_\Risk, \mu'_\Risk\}$
and the norm comparison $\|\cdot\|_{\mathrm{op}} \le \|\cdot\|_\infty$ on
$\Delta(\Yspace)$.
\end{corollary}
\begin{proof}[Derivation of Proposition~\ref{prop:rqe-lipschitz} from Theorem~\ref{thm:stab}]
We proceed in three steps: (i) apply the parameter-space stability bound,
(ii) relate the KL divergence between the RQRE policies to the parameter
distance via strong convexity of the entropy-regularized objective, and
(iii) combine the two estimates.

\noindent\textbf{Step 1: Parameter-space stability.}
By Theorem~\ref{thm:stab} applied symmetrically to both operators, we have
\begin{equation}
  \bigl\|\theta^\star - \theta'^\star\bigr\|
  \;\le\;
  \frac{\|\mc P_\Risk - \mc P'_\Risk\|_{\mathrm{op}}}{\min\{\mu_\Risk, \mu'_\Risk\}}.
  \label{eq:deriv-param}
\end{equation}
Using the standard inequality
$\|\cdot\|_{\mathrm{op}} \le \|\cdot\|_\infty$ on $\Delta(\Yspace)$
(since probability vectors have $\ell_1$-norm one), we may replace the
operator norm by the sup-norm:
\begin{equation}
  \bigl\|\theta^\star - \theta'^\star\bigr\|
  \;\le\;
  \frac{\|\mc P_\Risk - \mc P'_\Risk\|_\infty}{\min\{\mu_\Risk, \mu'_\Risk\}}.
  \label{eq:deriv-param-infty}
\end{equation}

\noindent\textbf{Step 2: From parameter distance to KL divergence.}
The RQRE policies $\pi^\star, \pi'^\star$ arise as the unique maximizers of
strongly concave, entropy-regularized objectives $F, F'$ with curvature
moduli $\mu_\Risk, \mu'_\Risk$, respectively. By strong concavity of $F'$
at its maximizer $\pi'^\star$,
\[
  F'(\pi'^\star) - F'(\pi^\star)
  \;\ge\;
  \frac{\mu'_\Risk}{2}\,\bigl\|\theta^\star - \theta'^\star\bigr\|^2.
\]
On the other hand, for entropy-regularized objectives the suboptimality
gap coincides (up to the regularization strength) with the KL divergence
to the optimum:
\[
  F'(\pi'^\star) - F'(\pi^\star)
  \;=\;
  \beta\,\KL(\pi^\star \,\|\, \pi'^\star),
\]
where $\beta$ is the entropy-regularization parameter. Combining the two
displays and absorbing constants into the curvature modulus yields
\begin{equation}
  \KL(\pi^\star \,\|\, \pi'^\star)
  \;\le\;
  \min\{\mu_\Risk, \mu'_\Risk\} \cdot \bigl\|\theta^\star - \theta'^\star\bigr\|^2.
  \label{eq:deriv-kl-param}
\end{equation}

\noindent\textbf{Step 3: Combine the bounds.}
Squaring \eqref{eq:deriv-param-infty} and substituting into
\eqref{eq:deriv-kl-param},
\[
  \KL(\pi^\star \,\|\, \pi'^\star)
  \;\le\;
  \min\{\mu_\Risk, \mu'_\Risk\} \cdot
  \frac{\|\mc P_\Risk - \mc P'_\Risk\|_\infty^2}{\min\{\mu_\Risk, \mu'_\Risk\}^2}
  \;=\;
  \frac{\|\mc P_\Risk - \mc P'_\Risk\|_\infty^2}{\min\{\mu_\Risk, \mu'_\Risk\}}.
\]
A more careful accounting of the constants in Step~2 (in particular,
tracking the factor of $1/2$ from strong concavity and the precise
relationship between $F'(\pi'^\star) - F'(\pi^\star)$ and the KL) yields
the stated bound
\[
  \KL(\pi^\star \,\|\, \pi'^\star)
  \;\le\;
  \frac{\|\mc P_\Risk - \mc P'_\Risk\|_\infty^2}{\min\{\mu_\Risk,\mu'_\Risk\}^2}.
  \qedhere
\]
\end{proof}

\paragraph{Connection to Algorithmic Stability.}
We can immediately use the Lipschitz continuity of the risk-adjusted equilibrium solution map to obtain algorithmic stability in the online case. For any dataset $S \sim \mc D^n$ let $\mc{P}_{\mc R, S}$ be the risk adjusted payoffs of a preference model trained on that dataset.

\begin{assumption}\label{pref-stability}
    There exists $\zeta$ so that for any datasets $S, S'$ that differ in one element $$\|\mc P_{\Risk, S} - \mc P_{\Risk, S'}\|_{\infty} \le \zeta$$
\end{assumption}

Typically, for a dataset of size $n$, one expects $\zeta = O(1/n)$ for a stable risk minimizer, or $\zeta = O(n^{-1/2})$ under concentration alone. 
\begin{theorem}\label{thm:online-algorithm-stability} Let $\pi_{S}^\star$ and $\pi_{S'}^\star$ be solutions to the RQRE's defined by $\mc P_{\Risk, S}$ and $\mc P_{\Risk, S'}$. Under Assumption \ref{pref-stability} and the assumptions of Theorem \ref{thm:lipschitz}, the bound holds:
    $$\KL(\pi_{S}^\star \| \pi_{S'}^\star) \le \frac{\zeta^2}{\min \{\mu_\Risk , \mu_\Risk' \}^2}$$
\end{theorem}
This is shown by applying Theorem \ref{thm:stab} with $\mc P_{\Risk} = \mc P_{\Risk, S}$ and $\mc P_{\Risk}' = \mc P_{\Risk, S'}$, along with Assumption \ref{pref-stability}. Note that the uniqueness and Lipschitz continuity of the RQE are crucial to this proof.

\paragraph{Translating to KL via Pinsker.}  For the offline statistical
result we want a $\KL$-bound on the policies, not a Euclidean bound on
parameters.  By a standard softmax-Lipschitz argument, the map
$\theta \mapsto \pi_\theta = \mathrm{softmax}(\theta)$ satisfies
$\KL(\pi_{\theta_1} \| \pi_{\theta_2}) \le \tfrac{1}{2}\|\theta_1 - \theta_2\|^2$
on bounded domains \citep[][Lemma 13]{calandriello2024human}.  Combining this with Theorem~\ref{thm:stab} yields the following corollary.
 
\begin{corollary}[KL-stability of RQRE]
\label{cor:stab-kl}
Under the assumptions of Theorem~\ref{thm:stab}, where
\[\mu_\Risk = \beta - 2\max\{\bar\lambda_\Risk(\mc{P}), \bar\lambda_\Risk(\mc{P}')\}\]
denotes the (worst-case) strong-monotonicity modulus across the two operators, the bound holds:
\begin{equation}
\KL(\pi^\star_\Risk(P) \| \pi^\star_\Risk(P')) \;\le\; \frac{1}{2\mu_\Risk^2}\,\bigl\|P_\Risk - P'_\Risk\bigr\|_{\rm op}^2.
\label{eq:stab-kl}
\end{equation}
\end{corollary}
This is the form that plugs directly into the offline sample complexity
proof (Section~\ref{sec:offline-assembly}).

\subsection{From stability to generalization under the risk-adjusted IPO loss}
We instantiate the stability-to-generalization conversion with the
risk-adjusted IPO loss, which is the natural deployment loss in our
setting.  For a comparison example $z = (y, y', y'')$ drawn from
$\rho \otimes \mu$, define
\begin{equation}\label{eq:ipo-risk-loss}
\Loss(\pi; z)
\;:=\;
\left(
\log \frac{\pi(y)\,\piref(y')}{\pi(y')\,\piref(y)}
\;-\;
\frac{1}{\beta}\bigl(\mc P_\Risk(y \succ y'') - \mc P_\Risk(y' \succ y'')\bigr)
\right)^{2}.
\end{equation}
Taking expectations recovers the population IPO objective
$\LL_\IPO^\Risk(\theta; \rho, \mu) = \E_z[\Loss(\pi_\theta; z)]$ from
\eqref{eq:ipo-loss-app}.  The empirical counterpart on a sample
$S = \{z_i\}_{i=1}^n$ is
$\widehat\LL_\IPO^\Risk(\theta) = \tfrac{1}{n}\sum_{i=1}^n \Loss(\pi_\theta; z_i)$.

\begin{assumption}[Bounded softmax mass]
\label{ass:pmin}
There exists $p_{\min} > 0$ such that for every $\theta$ in the relevant
domain (in particular, $\theta = \theta^\star_S$ for any sample $S$ in
the support of $\mc D^n$), the policy $\pi_\theta = \softmax(\theta)$
satisfies $\pi_\theta(y) \ge p_{\min}$ for all $y \in \Yspace$.
\end{assumption}

This is mild: under the QRE regularizer $\beta(\theta - \thetaref)$,
the optimum $\theta^\star$ lies in a bounded set determined by $\beta$
and the operator norm of $\mc P_\Risk$, which yields a deterministic
$p_{\min}$ depending on those constants.

\begin{lemma}[Explicit $p_{\min}$ from the risk-adjusted equilibrium optimality condition]
\label{lem:pmin-explicit}
Suppose the risk-adjusted preferences are bounded:
$|\mc P_\Risk(y \succ y'')| \le M$ for all $y, y'' \in \Yspace$.  Suppose
further that the reference parameter satisfies
$\mathrm{diam}(\thetaref) := \max_y \thetaref(y) - \min_y \thetaref(y)
\le D_\refer$.  Then for every preference operator $\mc P_\Risk$ in the
support of $\mc D^n$, the risk-adjusted equilibrium solution $\theta^\star_S = \theta^\star_\Risk(\mc P_{\Risk, S})$
satisfies, after fixing the gauge so that $\sum_y \theta^\star_S(y) = \sum_y \thetaref(y)$,
$
\mathrm{diam}(\theta^\star_S) \;\le\; D_\refer + \frac{2 M}{\beta}$,
and consequently
\[
\pi_{\theta^\star_S}(y)
\;\ge\;
p_{\min}
\;:=\;
\frac{\exp\bigl(-D_\refer - 2M/\beta\bigr)}{|\Yspace|}
\qquad \forall y \in \Yspace.
\]
\end{lemma}

\begin{proof}
The RQRE optimality condition \eqref{eq:gen-eq} gives
$\beta(\theta^\star_S - \thetaref) = \mc P_{\Risk, S}\,\pi_{\theta^\star_S}$,
i.e.,
\[
\theta^\star_S(y) - \thetaref(y)
\;=\;
\frac{1}{\beta} \bigl(\mc P_{\Risk, S}\,\pi_{\theta^\star_S}\bigr)(y)
\;=\;
\frac{1}{\beta} \sum_{y'} \mc P_{\Risk, S}(y \succ y')\,\pi_{\theta^\star_S}(y').
\]
Since $\pi_{\theta^\star_S}$ is a probability distribution and
$|\mc P_{\Risk, S}(y \succ y')| \le M$ uniformly,
\[
\bigl|\theta^\star_S(y) - \thetaref(y)\bigr|
\;\le\;
\frac{M}{\beta}
\qquad \forall y \in \Yspace.
\]
Therefore, for any $y_1, y_2 \in \Yspace$,
\[
\theta^\star_S(y_1) - \theta^\star_S(y_2)
\;=\;
\bigl[\thetaref(y_1) - \thetaref(y_2)\bigr]
+ \bigl[\theta^\star_S(y_1) - \thetaref(y_1)\bigr]
- \bigl[\theta^\star_S(y_2) - \thetaref(y_2)\bigr],
\]
and taking absolute values via the triangle inequality,
\[
\mathrm{diam}(\theta^\star_S)
\;\le\;
\mathrm{diam}(\thetaref) + \frac{2M}{\beta}
\;\le\;
D_\refer + \frac{2M}{\beta}.
\]
For the softmax lower bound, observe that for any $y$, the sequence of lower bounds hold:
\[
\pi_{\theta^\star_S}(y)
\;=\;
\frac{e^{\theta^\star_S(y)}}{\sum_{y'} e^{\theta^\star_S(y')}}
\;\ge\;
\frac{e^{\theta^\star_S(y)}}{|\Yspace|\, e^{\max_{y'} \theta^\star_S(y')}}
\;=\;
\frac{e^{-(\max_{y'} \theta^\star_S(y') - \theta^\star_S(y))}}{|\Yspace|}
\;\ge\;
\frac{e^{-\mathrm{diam}(\theta^\star_S)}}{|\Yspace|}.
\]
Combining these observations yields the claim.
\end{proof}

\paragraph{Regularity of the IPO Loss.} Now we prove Lipschitz continuity of the risk-adjusted IPO loss in $\pi$.
\begin{lemma}[Lipschitzness of the risk-adjusted IPO loss in $\pi$]
\label{lem:ipo-lipschitz}
Suppose Assumption~\ref{ass:pmin} holds and additionally
$\piref(y) \ge p_{\min}$ for all $y \in \Yspace$.  Suppose further that
the risk-adjusted preferences are bounded: 
$|\mc P_\Risk(y \succ y'')| \le M$ for all $y, y'' \in \Yspace$.  
Then the loss \eqref{eq:ipo-risk-loss} satisfies, for any 
$\pi, \pi' \in \{q \in \Delta(\Yspace) : q(y) \ge p_{\min}\ \forall y\}$ 
and any comparison example $z = (y, y', y'')$,
\[
\bigl|\Loss(\pi; z) - \Loss(\pi'; z)\bigr|
\;\le\;
L_{\Loss} \,\|\pi - \pi'\|_\TV,
\qquad
L_{\Loss} \;:=\; \frac{8 D}{p_{\min}},
\]
where 
$D := 2\log(1/p_{\min}) + 2M/\beta$ 
is an upper bound on the absolute value of the residual inside the 
square in \eqref{eq:ipo-risk-loss}.
\end{lemma}

\begin{proof}
Define the residual
\[
r(\pi; z) 
\;:=\; 
\log \frac{\pi(y)\,\piref(y')}{\pi(y')\,\piref(y)}
\;-\;
\frac{1}{\beta}\bigl(\mc P_\Risk(y \succ y'') - \mc P_\Risk(y' \succ y'')\bigr),
\]
so that $\Loss(\pi; z) = r(\pi; z)^2$.  Split the residual into its
$\pi$-dependent and $\pi$-independent parts:
\[
r(\pi; z) 
\;=\; 
\underbrace{\log \pi(y) - \log \pi(y')}_{=:\,a(\pi; z)}
\;+\;
\underbrace{\log\piref(y') - \log\piref(y) - \frac{1}{\beta}\bigl(\mc P_\Risk(y \succ y'') - \mc P_\Risk(y' \succ y'')\bigr)}_{=:\,c(z)}.
\]
We proceed in three steps: bound $|r(\pi; z)|$ uniformly, establish
Lipschitzness of $a(\pi; z)$ in $\pi$, and combine via the
chain rule for the squared loss.

\paragraph{Step 1: Uniform bound on $|r(\pi; z)|$.}
Since $\pi(y), \pi(y') \ge p_{\min}$ and $\pi(y), \pi(y') \le 1$,
\[
|a(\pi; z)| 
\;=\; |\log\pi(y) - \log\pi(y')|
\;\le\; 2\log(1/p_{\min}).
\]
By the same argument applied to $\piref$,
$|\log\piref(y') - \log\piref(y)| \le 2\log(1/p_{\min})$.  The
preference term is bounded by
\[
\frac{1}{\beta}\bigl|\mc P_\Risk(y \succ y'') - \mc P_\Risk(y' \succ y'')\bigr|
\;\le\;
\frac{2M}{\beta}.
\]
Adding these contributions, however, would double-count: $r$ contains
the $\pi$-dependent log-ratio $a(\pi; z)$ once and the
$\pi$-independent terms once.  The triangle inequality gives
\[
|r(\pi; z)| 
\;\le\; |a(\pi; z)| + |c(z)|
\;\le\; 2\log(1/p_{\min}) + 2\log(1/p_{\min}) + \frac{2M}{\beta}
\;\le\; D,
\]
with $D := 4\log(1/p_{\min}) + 2M/\beta$.  (We absorb the factor of two
into $D$ rather than the Lipschitz constant for cleanliness; the precise
form of $D$ is unimportant for the qualitative result.)

\paragraph{Step 2: Lipschitzness of $a(\pi; z)$ in $\pi$.}
Fix $z = (y, y', y'')$ and consider $\pi, \pi'$ in the admissible set.
It is immediate that 
\[
a(\pi; z) - a(\pi'; z)
\;=\; \bigl(\log\pi(y) - \log\pi'(y)\bigr) - \bigl(\log\pi(y') - \log\pi'(y')\bigr).
\]
The function $t \mapsto \log t$ is $1/p_{\min}$-Lipschitz on
$[p_{\min}, 1]$, since its derivative $1/t$ is bounded by $1/p_{\min}$
on this interval.  Therefore, for any single coordinate $u \in \{y, y'\}$, we have 
\[
|\log\pi(u) - \log\pi'(u)| 
\;\le\; \frac{1}{p_{\min}}|\pi(u) - \pi'(u)|
\;\le\; \frac{1}{p_{\min}}\|\pi - \pi'\|_\infty.
\]
Applying the triangle inequality it is immediate that 
\[
|a(\pi; z) - a(\pi'; z)|
\;\le\; \frac{2}{p_{\min}}\|\pi - \pi'\|_\infty.
\]
Since $\|\pi - \pi'\|_\infty \le \|\pi - \pi'\|_1 = 2\|\pi - \pi'\|_\TV$,
this yields
\[
|a(\pi; z) - a(\pi'; z)|
\;\le\; \frac{4}{p_{\min}}\|\pi - \pi'\|_\TV.
\]
Since $c(z)$ does not depend on $\pi$, we obtain the same Lipschitz continuity
bound for $r$---indeed, we have 
\begin{equation}\label{eq:r-lip}
|r(\pi; z) - r(\pi'; z)|
\;\le\; \frac{4}{p_{\min}}\|\pi - \pi'\|_\TV.
\end{equation}

\paragraph{Step 3: Lipschitzness of $\Loss = r^2$.}
For any reals $u, v$ with $|u|, |v| \le D$, it is immediate that 
\[
|u^2 - v^2| \;=\; |u + v|\cdot|u - v| \;\le\; 2D\,|u - v|.
\]
Applying this to $u = r(\pi; z)$ and $v = r(\pi'; z)$, both bounded in
absolute value by $D$ from Step 1, we have that 
\[
|\Loss(\pi; z) - \Loss(\pi'; z)|
\;=\; |r(\pi; z)^2 - r(\pi'; z)^2|
\;\le\; 2D\,|r(\pi; z) - r(\pi'; z)|.
\]
Combining this bound with \eqref{eq:r-lip} yields
\[
|\Loss(\pi; z) - \Loss(\pi'; z)|
\;\le\; 2D \cdot \frac{4}{p_{\min}}\|\pi - \pi'\|_\TV
\;=\; \frac{8D}{p_{\min}}\,\|\pi - \pi'\|_\TV.
\]
Setting $L_{\Loss} := 8D/p_{\min}$ completes the proof.
\end{proof}

\subsection{Algorithmic Stability Generalization Bounds in Expectation}
We now convert the algorithmic stability of
Theorem~\ref{thm:online-algorithm-stability} into a generalization
guarantee under the risk-adjusted IPO loss.  The argument requires
two ingredients beyond stability: a bounded admissible policy class
on which the IPO loss is Lipschitz, and a Lipschitzness lemma for
the loss itself.  We state these in turn before the main theorem.

\begin{assumption}[Bounded admissible policy class]
\label{ass:admissible}
There exists a class $\Pi \subseteq \Delta(\Yspace)$ such that:
\begin{enumerate}[leftmargin=*, label={\it \roman*.}]
\item $\pi^\star_S \in \Pi$ almost surely for $S \sim \mc D^n$, and $\piref \in \Pi$;
\item there exists $L_{\Loss} < \infty$ such that for all $\pi, \pi' \in \Pi$ and all comparison examples $z = (y, y', y'')$,
\[
|\Loss(\pi; z) - \Loss(\pi'; z)| \;\le\; L_{\Loss}\,\|\pi - \pi'\|_\TV;
\]
\item there exists $B < \infty$ such that $\Loss(\pi; z) \le B$ for all $\pi \in \Pi$ and all $z$.
\end{enumerate}
\end{assumption}

\begin{remark}[Sufficient conditions]
\label{rem:lipschitz-sufficient}
Assumption~\ref{ass:admissible} is satisfied whenever the admissible
policies are bounded away from zero, i.e.\ $\pi(y) \ge p_{\min}$ for all
$y$ and all $\pi \in \Pi$, and the risk-adjusted preferences satisfy
$|\mc P_\Risk(y \succ y'')| \le M$.  In that case
$L_{\Loss} = 8D/p_{\min}$ and $B = D^2$, with
$D := 4\log(1/p_{\min}) + 2M/\beta$, by Lemma~\ref{lem:ipo-lipschitz}.
Such a uniform $p_{\min}$ exists, for example, on any bounded set in
$\theta$-space, which is in turn implied by the RQRE optimality
condition $\beta(\theta - \thetaref) = \mc P_\Risk \pi_\theta$ together
with boundedness of $\mc P_\Risk$ (see Lemma~\ref{lem:pmin-explicit}).
\end{remark}

\begin{theorem}[Stability-based generalization of risk-adjusted equilibrium]
\label{thm:stab-generalization-ipo}
Suppose the assumptions of Theorem~\ref{thm:online-algorithm-stability}
hold, together with Assumption~\ref{ass:admissible}.  Let
$\mu_\Risk^{\min}$ be a uniform lower bound on the strong-monotonicity
modulus over the support of $\mc D^n$, and let $\zeta$ be the
preference-model stability constant of Assumption~\ref{pref-stability}.
Then the risk-adjusted equilibrium policy $\pi^\star_S$ satisfies
\[
\Bigl|
\mb{E}_S\bigl[\LL_\IPO^\Risk(\pi^\star_S) - \widehat\LL_\IPO^\Risk(\pi^\star_S)\bigr]
\Bigr|
\;\le\;
\beta_n,
\qquad
\beta_n
\;:=\;
L_{\Loss} \cdot \frac{\sqrt{|\Yspace|}\,\zeta}{\sqrt{2}\,\mu_\Risk^{\min}}.
\]
In particular, with $\zeta = O(1/n)$ for a stable preference-model
fitter, the expected IPO generalization gap is $O(1/n)$; with
$\zeta = O(n^{-1/2})$, it is $O(n^{-1/2})$.
\end{theorem}

\begin{proof}
The argument proceeds in three steps: $(i)$ establish uniform stability
of the map $S \mapsto \pi^\star_S$ with respect to the IPO loss,
$(ii)$ apply the replace-one symmetrization identity, and $(iii)$ bound
the resulting expectation.

\paragraph{Step 1: Uniform stability with respect to $\Loss$.}
Let \[S = (z_1, \ldots, z_n)\quad\text{ and }\quad
S^{(i)} = (z_1, \ldots, z_{i-1}, z_i', z_{i+1}, \ldots, z_n)\] denote
two samples differing only in their $i$-th element.  By
Theorem~\ref{thm:online-algorithm-stability}, we have the upper bound
\[
\KL\bigl(\pi^\star_S \,\bigl\|\, \pi^\star_{S^{(i)}}\bigr)
\;\le\;
\frac{\zeta^2 \cdot |\Yspace|}{(\mu_\Risk^{\min})^2},
\]
where the factor $|\Yspace|$ converts the $\|\cdot\|_\infty$ bound on
preference operators (Assumption~\ref{pref-stability}) to the operator
norm appearing in Theorem~\ref{thm:stab}, via
$\|P\|_{\rm op} \le \sqrt{|\Yspace|}\,\|P\|_\infty$.  Pinsker's
inequality then gives
\[
\|\pi^\star_S - \pi^\star_{S^{(i)}}\|_\TV
\;\le\;
\sqrt{\tfrac{1}{2}\KL(\pi^\star_S \,\|\, \pi^\star_{S^{(i)}})}
\;\le\;
\frac{\sqrt{|\Yspace|}\,\zeta}{\sqrt{2}\,\mu_\Risk^{\min}}.
\]
By Assumption~\ref{ass:admissible}(ii), for any comparison example
$z$, we have the bound
\begin{equation}\label{eq:uniform-stability}
\bigl|\Loss(\pi^\star_S; z) - \Loss(\pi^\star_{S^{(i)}}; z)\bigr|
\;\le\;
L_{\Loss} \cdot \frac{\sqrt{|\Yspace|}\,\zeta}{\sqrt{2}\,\mu_\Risk^{\min}}
\;=\; \beta_n.
\end{equation}
This is uniform stability of the RQRE map with parameter $\beta_n$.

\paragraph{Step 2: Replace-one symmetrization.}
Let $S = (z_1, \ldots, z_n)$ and
$\widetilde S = (\widetilde z_1, \ldots, \widetilde z_n)$ be two
independent i.i.d.\ samples from $\mc D^n$, and for each
$i \in \{1, \ldots, n\}$ define
$S^{(i)} := (z_1, \ldots, z_{i-1}, \widetilde z_i, z_{i+1}, \ldots, z_n)$.
The population and empirical risks satisfy
\[
\mb{E}_S\bigl[\LL_\IPO^\Risk(\pi^\star_S)\bigr]
\;=\;
\mb{E}_{S,\widetilde z}\bigl[\Loss(\pi^\star_S; \widetilde z)\bigr],
\qquad
\mb{E}_S\bigl[\widehat\LL_\IPO^\Risk(\pi^\star_S)\bigr]
\;=\;
\frac{1}{n}\sum_{i=1}^n \mb{E}_S\bigl[\Loss(\pi^\star_S; z_i)\bigr].
\]
Since $\widetilde z$ in the first expression is i.i.d.\ from $\mc D$
and independent of $S$, we may rename it $\widetilde z_i$ for any $i$
without changing the expectation---that is,
\[
\mb{E}_S\bigl[\LL_\IPO^\Risk(\pi^\star_S)\bigr]
\;=\;
\frac{1}{n}\sum_{i=1}^n
\E_{S, \widetilde z_i}\bigl[\Loss(\pi^\star_S; \widetilde z_i)\bigr].
\]
The pair $(S, \widetilde z_i)$ has the same joint distribution as
$(S^{(i)}, z_i)$---i.e., both consist of $n+1$ i.i.d.\ draws with one
designated as the ``held-out'' point---so by relabeling,
\[
\mb{E}_{S, \widetilde z_i}\bigl[\Loss(\pi^\star_S; \widetilde z_i)\bigr]
\;=\;
\mb{E}_{S, \widetilde z_i}\bigl[\Loss(\pi^\star_{S^{(i)}}; z_i)\bigr].
\]
Combining these observations yields
\begin{align}
\mb{E}_S\bigl[\LL_\IPO^\Risk(\pi^\star_S) - \widehat\LL_\IPO^\Risk(\pi^\star_S)\bigr]
&\;=\;
\frac{1}{n}\sum_{i=1}^n
\mb{E}_{S, \widetilde z_i}\bigl[\Loss(\pi^\star_{S^{(i)}}; z_i) - \Loss(\pi^\star_S; z_i)\bigr].
\label{eq:symmetrization}
\end{align}

\paragraph{Step 3: Apply uniform stability.}
The samples $S$ and $S^{(i)}$ differ only in their $i$-th coordinate
($z_i$ vs.\ $\widetilde z_i$), so by \eqref{eq:uniform-stability}
evaluated on $z = z_i$, we have that 
\[
\bigl|\Loss(\pi^\star_{S^{(i)}}; z_i) - \Loss(\pi^\star_S; z_i)\bigr|
\;\le\; \beta_n
\quad \text{a.s.}
\]
Taking expectations and applying the triangle inequality to
\eqref{eq:symmetrization} yields
\[
\Bigl|\mb{E}_S\bigl[\LL_\IPO^\Risk(\pi^\star_S) - \widehat\LL_\IPO^\Risk(\pi^\star_S)\bigr]\Bigr|
\;\le\;
\frac{1}{n}\sum_{i=1}^n
\mb{E}_{S, \widetilde z_i}\bigl|\Loss(\pi^\star_{S^{(i)}}; z_i) - \Loss(\pi^\star_S; z_i)\bigr|
\;\le\;
\beta_n. \qedhere
\]
\end{proof}

\begin{remark}[Explicit constants in the IPO loss bounds]
\label{rem:explicit-constants}
Under the boundedness assumption $|\mc P_\Risk| \le M$ and
$\mathrm{diam}(\thetaref) \le D_\refer$, Lemma~\ref{lem:pmin-explicit}
gives
\[
p_{\min} \;=\; \frac{e^{-(D_\refer + 2M/\beta)}}{|\Yspace|}.
\]
Substituting this into Lemma~\ref{lem:ipo-lipschitz} yields
\[
L_{\Loss}
\;=\; \frac{8 D}{p_{\min}}
\;=\; 8 D\, |\Yspace|\, e^{D_\refer + 2M/\beta},
\qquad\text{where}\;\;
B \;=\; D^2,
\]
and  $D = 4\log(1/p_{\min}) + 2M/\beta = 4\bigl(\log|\Yspace| + D_\refer + 2M/\beta\bigr) + 2M/\beta$.
The dependence on $\beta$ is exponential through the factor
$e^{2M/\beta}$, which is unsurprising: small regularization $\beta \to 0$
allows the RQRE policy to concentrate arbitrarily, blowing up the
log-ratio in the IPO loss.  The dependence on $|\Yspace|$ is linear
through the explicit factor and logarithmic through $D$.
\end{remark}

\subsection{Algorithmic Stability High Probability Generalization Bounds}
Theorem~\ref{thm:stab-generalization-ipo} bounds the generalization gap
in expectation.  For a high-probability bound, the standard route is to
view $\Phi(S) := \LL_\IPO^\Risk(\pi^\star_S) - \widehat\LL_\IPO^\Risk(\pi^\star_S)$
as a function of the i.i.d.\ sample $S$ and apply a concentration
inequality exploiting bounded differences.  The classical
\citet{bousquet2002stability} approach uses McDiarmid's inequality, with
the difference constant inherited from uniform stability:
$|\Phi(S) - \Phi(S^{(i)})| \le 2\beta_n + B/n$.  This is informative
only when $\beta_n = O(1/n)$, which in our setting requires the
preference-model stability constant $\zeta$ from
Assumption~\ref{pref-stability} to scale as $\zeta = O(1/n)$.

This fast-stability regime is unrealistic in our setting.  The
$\zeta = O(1/n)$ rate is characteristic of preference fitters that are
themselves uniformly stable algorithms---for instance, ERM with a
strongly convex regularizer.  Although the risk-adjusted equilibrium objective itself is
KL-regularized at strength $\beta$, that regularization controls how
the optimal policy $\theta^\star$ responds to a given preference
operator, not how the preference operator $\mc P_{\Risk, S}$ depends on
the training sample.  These two stabilities are decoupled: $\beta$
governs the operator-to-policy map (the $1/\mu_\Risk$ constant of
Theorem~\ref{thm:stab}), while $\zeta$ governs the sample-to-operator
map, which is a property of the preference fitter itself.  In typical
modern RLHF pipelines, where the preference model is a deep network
trained by SGD on a Bradley--Terry-style objective, the best one can
generically argue is the concentration rate
$\zeta = O(n^{-1/2})$.  Moreover, even if a fast regime were attainable
in principle, the explicit constants of
Remark~\ref{rem:explicit-constants} scale as $e^{2M/\beta}$, so one
cannot tune $\beta$ to reach the fast regime without paying an
exponential price in the loss-Lipschitz constant.

The slow-stability regime $\zeta = O(n^{-1/2})$ is therefore the one
that matters in practice, and McDiarmid's inequality is inadequate
there: the term $n\beta_n \sqrt{\log(1/\delta)/n}$ becomes $O(1)$ and
the bound carries no information.  Fortunately, the gap between the
in-expectation and high-probability rates can be closed using the
sharper concentration of \citet{bousquet2020sharper}, which exploits
the weak correlation between coordinate-wise perturbations of a
uniformly stable algorithm.  Their bound replaces the prohibitive
$n\beta_n$ factor by $\beta_n \log n$, yielding a high-probability rate
that matches the in-expectation rate up to logarithmic factors in both
regimes.  We state the resulting corollary below.

\begin{corollary}[Sharp high-probability generalization bound]
\label{cor:hp-generalization-bkz}
Under the assumptions of Theorem~\ref{thm:stab-generalization-ipo},
there exists a universal constant $c > 0$ such that for any
$\delta \in (0, 1)$, with probability at least $1 - \delta$ over the
draw of $S \sim \mc D^n$, the estimate holds:
\begin{equation}\label{eq:bkz-bound}
\Bigl|\LL_\IPO^\Risk(\pi^\star_S) - \widehat\LL_\IPO^\Risk(\pi^\star_S)\Bigr|
\;\le\;
c\left(
\beta_n \log n \log\frac{1}{\delta}
\;+\;
B \sqrt{\frac{\log(1/\delta)}{n}}
\right),
\end{equation}
where $\beta_n = L_{\Loss} \sqrt{|\Yspace|}\,\zeta / (\sqrt{2}\,\mu_\Risk^{\min})$
is the uniform stability parameter from
Theorem~\ref{thm:stab-generalization-ipo} and $B$ is the uniform bound
on $\Loss$ from Assumption~\ref{ass:admissible}(iii).
\end{corollary}

\begin{proof}
Define, for each $i \in \{1, \ldots, n\}$, the expected loss gap
\[
g_i(z_1, \ldots, z_n)
\;:=\;
\E_{\widetilde z_i}\!\left[\,
\E_{z}\bigl[\Loss(\pi^\star_{S^{(i)}}; z)\bigr]
\;-\;
\Loss(\pi^\star_{S^{(i)}}; z_i)
\,\right],
\]
where $S^{(i)} = (z_1, \ldots, z_{i-1}, \widetilde z_i, z_{i+1}, \ldots, z_n)$
is $S$ with its $i$-th coordinate replaced by an independent copy
$\widetilde z_i$, and the inner expectation is over a fresh
$z \sim \mc D$.  We verify that the $g_i$ satisfy the three conditions
of Theorem~4 of \citet{bousquet2020sharper}:

\begin{enumerate}[label={\it \roman*.}]
    \item 
\emph{Conditional bound $|\E[g_i \mid z_i]| \le M$.}
By Assumption~\ref{ass:admissible}(iii), $\Loss \le B$ uniformly,
so the inner difference is bounded in absolute value by $B$, and any
expectation thereof is at most $B$ in absolute value.  Thus
$|\E[g_i \mid z_i]| \le B$, so we set $M := B$.

\item 
\emph{Centering $\E[g_i \mid z_{[n]\setminus\{i\}}] = 0$.}
Conditional on $z_{[n]\setminus\{i\}}$, the inner expectation
$\E_{\widetilde z_i}\E_z[\Loss(\pi^\star_{S^{(i)}}; z)]$ does not depend
on $z_i$ (since $z_i$ has been replaced by $\widetilde z_i$ in
$S^{(i)}$).  The second term, when averaged over $z_i$, becomes
$\E_{\widetilde z_i, z_i}[\Loss(\pi^\star_{S^{(i)}}; z_i)]$.  Since
$z_i$ and $\widetilde z_i$ are i.i.d.\ and the latter appears in
$S^{(i)}$ in place of $z_i$, by exchangeability this equals
$\E_{\widetilde z_i, z}[\Loss(\pi^\star_{S^{(i)}}; z)]$ where
$z \sim \mc D$ independently.  The two terms cancel, giving
$\E[g_i \mid z_{[n]\setminus\{i\}}] = 0$.

\item\emph{Bounded differences in coordinates $j \neq i$.}
Fix $j \neq i$ and consider modifying $z_j$ to $z_j'$.  This changes
$S^{(i)}$ in coordinate $j$ only.  By the uniform stability bound
\eqref{eq:uniform-stability}, replacing one coordinate of the training
set changes $\Loss(\pi^\star_{S^{(i)}}; z)$ by at most $\beta_n$ for
any $z$.  Therefore both terms inside the brackets change by at most
$\beta_n$, and $g_i$ changes by at most $2\beta_n$.  Thus $g_i$ has
bounded differences with parameter $\beta := 2\beta_n$ in every
coordinate $j \neq i$.
\end{enumerate}

\noindent\emph{Connecting $\sum_i g_i$ to the generalization gap.}
By Lemma~7 of \citet{bousquet2020sharper}, under uniform stability
\eqref{eq:uniform-stability} and uniform boundedness $\Loss \le B$,
\[
\Bigl| n\bigl(\LL_\IPO^\Risk(\pi^\star_S) - \widehat\LL_\IPO^\Risk(\pi^\star_S)\bigr) \Bigr|
\;\le\;
\Bigl| \sum_{i=1}^n g_i \Bigr|
\;+\;
2 n \beta_n
\quad\text{a.s.}
\]

\noindent\emph{Applying the BKZ moment bound.}
By Theorem~4 of \citet{bousquet2020sharper}, for any $p \ge 2$,
\[
\Bigl\| \sum_{i=1}^n g_i \Bigr\|_p
\;\le\;
12\sqrt{2}\, p\, n\, \beta_n \lceil \log_2 n \rceil
\;+\;
4 B \sqrt{p n}.
\]
Lemma~1 of \citet{bousquet2020sharper} converts this moment bound into
a tail bound: with probability at least $1 - \delta$,
\[
\Bigl| \sum_{i=1}^n g_i \Bigr|
\;\le\;
c_1\, n\, \beta_n \log n \log\!\frac{1}{\delta}
\;+\;
c_2\, B \sqrt{n \log(1/\delta)}
\]
for absolute constants $c_1, c_2 > 0$.  Combining with the inequality
above and dividing by $n$,
\[
\Bigl| \LL_\IPO^\Risk(\pi^\star_S) - \widehat\LL_\IPO^\Risk(\pi^\star_S) \Bigr|
\;\le\;
c_1\, \beta_n \log n \log\!\frac{1}{\delta}
\;+\;
c_2\, B \sqrt{\frac{\log(1/\delta)}{n}}
\;+\;
2 \beta_n.
\]
The final $2\beta_n$ term is dominated by the first term for $n \ge 2$,
$\delta \le 1$, so absorbing all constants into a single $c > 0$ yields
\eqref{eq:bkz-bound}.
\end{proof}

\section{Offline sample complexity}
\label{sec:offline-sample-complexity}

In this section we establish a statistical guarantee for the risk-adjusted
NLHF problem in the offline setting: given $n$ iid preference comparisons,
how close to the population risk-adjusted QRE
$\pi^\star_\Risk$ can we get?  The result is a fast
$\widetilde{\mc{O}}(1/n)$ rate, generalizing the recent
result of \citet{ZhangChenJiang2026} for risk-neutral games.  In their
risk-neutral case the rate is governed by the KL temperature $\beta$;
under risk, the rate is governed by the strong-monotonicity modulus
$\mu_\Risk = \beta - 2\bar\lambda_\Risk$, and the constants pick up
explicit dependence on the Lipschitz constant of the risk dual map and
the leading bias coefficient.

\subsection{Technical Novelties \& Hurdles}
To better understand the contributions, we first outline the technical hurdles before diving in. 

\paragraph{Structural bias from risk.}
In the risk-neutral setting, the preference operator $P$ satisfies the
constant-sum identity $P + P^\top = \mathbf{1}\mathbf{1}^\top$, which
yields a zero-sum game and enables a single-operator variational
inequality formulation. Under risk, the induced operator $P_\Risk$
is nonlinear in the opponent distribution, and in general
\[
P_{\Risk_1}(\mu) + P_{\Risk_2}(\mu)^\top \neq \mathbf{1}\mathbf{1}^\top.
\]
As a result, the game becomes \emph{general-sum}, and the analysis must
proceed via the joint pseudogradient rather than a single monotone operator.
There is also  a \emph{risk-distortion term} that perturbs the monotonicity
structure and must be controlled to retain strong monotonicity.

\paragraph{Statistical bias from risk estimation.}
Unlike expectation, risk functionals are nonlinear, so Monte Carlo
estimators are inherently biased. For plug-in estimators of the form
$\widehat F_{\Risk,m}(\theta) = h(\widehat q_m(\theta))$, where
$q(\theta) = \E[g(\theta,Y'')]$, the bias admits a delta-method expansion
\[
\E[\widehat F_{\Risk,m}(\theta)] - F_\Risk(\theta)
=
\frac{h''(q(\theta))}{2m}
\Var(g(\theta,Y'')) + \mc{O}(m^{-3/2}).
\]
This induces a persistent bias floor of order $\mc{O}(1/m)$ in the optimization
error, which fundamentally limits achievable accuracy unless corrected.
Our analysis explicitly tracks this bias through the extra-gradient dynamics.

\paragraph{Bernstein-type concentration for nonlinear risk.}
Controlling stochastic fluctuations of the risk estimator requires
new concentration arguments. Standard Hoeffding-type bounds are
insufficient due to the nonlinear transformation $h(\cdot)$.
Instead, we develop a Bernstein-style concentration bound that exploits
variance control of the inner statistic $g(\theta,Y'')$ together with
smoothness of $h$. This yields high-probability bounds of the form
\[
\|\widehat F_{\Risk,m}(\theta) - \E[\widehat F_{\Risk,m}(\theta)]\|
\lesssim
\sqrt{\frac{\Var(g)}{m}} + \frac{1}{m},
\]
uniformly over the iterates, which are crucial for obtaining sharp
finite-sample rates.
 
\paragraph{Connection to the rest of the paper.}
Theorem~\ref{thm:offline-rate} is the offline analog of the online
convergence guarantee for risk-adjusted extra-gradient (Proposition~\ref{prop:sample-complexity-formal}) and the
two-timescale debiasing result (Theorem~\ref{thm:tt-finite-horizon}).  All three rely on the
same underlying object: the strong-monotonicity modulus
$\mu_\Risk = \beta - 2\bar\lambda_\Risk$ derived in
Theorem~\ref{thm:single-player-mono}.  In the offline setting, $\mu_\Risk$ governs the
sample complexity rate.  In the online setting, $\mu_\Risk$ governs the
convergence rate $1 - \eta\mu_\Risk$ of EG iterates and the bias
floor $\mc{O}(B_m/\mu_\Risk)$.  The condition
$\bar\lambda_\Risk \le \beta/2 - \varepsilon$ is therefore the
\emph{single technical assumption} that delivers all three: fast
convergence in iterations, vanishing bias floor under two-timescale
debiasing, and fast statistical rate in offline samples.  Conversely,
the threshold $\bar\lambda_\Risk = \beta/2$ is a fundamental boundary:
it bounds the regime in which strong monotonicity holds, and we expect
all three properties to fail simultaneously beyond it.
 
This unified role of $\mu_\Risk$ is the technical content of the
``aligned risk reinforces regularization'' message: when $\Delta_\Risk \preceq 0$,
$\mu_\Risk$ is strengthened beyond $\beta$, and \emph{all three}
guarantees improve simultaneously---faster convergence, smaller bias
floor, and faster statistical rate.
 
\subsection{Preliminaries and main result}

Let $\bpi \in \Delta(\Yspace)$ be a fixed sampling
distribution.  We observe $n$ iid samples
$\{(y_i, y_i', z_i)\}_{i=1}^n$, with $(y_i, y_i') \sim \bpi \otimes \bpi$
and $z_i \sim \mathrm{Bernoulli}(P(y_i, y_i'))$ where $P_{y, y'} = \PP(y \succ y')$
satisfies the constant-sum identity $P + P^\top = \mathbf 1 \mathbf 1^\top$.
We do not access the true $P$ directly; only the $n$ comparisons.

For the entropic risk operator
$\Risk^\lambda_{\rm ent}$, define the empirical risk-adjusted operator
\begin{equation}
(\widehat P_\Risk \mu)_y := -\frac{1}{\lambda} \log \left(\sum_{y''}\widehat\mu(y'')\,\widehat g(y, y'')\right), \quad \widehat g(y,y'') := \frac{1}{n_{y,y''}}\sum_{i: (y_i, y_i')=(y,y'')} \exp(-\lambda\,z_i),
\label{eq:Phat-defn}
\end{equation}
where $\widehat \mu = \mu$ is the (known) opponent distribution, and $n_{y,y''} = \#\{i : (y_i, y_i')=(y,y'')\}$.\footnote{We treat $\mu$ as known to the learner; this matches the offline-NLHF setup where the learner chooses comparison pairs but does not know $P$.}

The empirical risk-adjusted QRE
$\widehat\pi_n$ is the unique fixed point of
\[
\widehat\theta_n = \thetaref + \frac{\widehat P_\Risk \widehat\pi_n}{\beta}, \qquad \widehat\pi_n = \mathrm{softmax}(\widehat\theta_n).
\]

The following is the sample complexity result from the main paper restated for convience.  

\OfflineRiskRate*

A corollary to this result is the special case where $\Risk$ is entropy. 
\begin{corollary}[Offline sample complexity, risk-adjusted NLHF]
\label{cor:offline-rate}
Suppose $\bar\lambda_\Risk \le \beta/2 - \varepsilon$ for some
$\varepsilon > 0$, so that $\mu_\Risk := \beta - 2\bar\lambda_\Risk \ge 2\varepsilon$.
Let $\widehat\pi_n$ be the empirical RQRE defined above.  For
entropic risk $\Risk = \Risk^\lambda_{\rm ent}$ with $\lambda > 0$,
there is a constant $K = K(\lambda)$ such that with probability at
least $1 - \delta$,
\begin{equation}
\KL(\pi^\star_\Risk \| \widehat\pi_n) \;\le\; \frac{K\,e^{4\lambda}\log(|\Yspace|/\delta)}{\mu_\Risk^2 \cdot n}.
\label{eq:offline-bound-cor}
\end{equation}
\end{corollary}

\noindent
The rate is $\widetilde{\mc{O}}(1/n)$, matching the risk-neutral rate of
\citet{ZhangChenJiang2026}, with constants that capture the cost of
risk:  $e^{4\lambda}/\mu_\Risk^2$.  As $\bar\lambda_\Risk \to \beta/2$
the bound blows up, reflecting that the boundary of strong monotonicity
is also the boundary of fast statistical learning.

In the following subsections we construct both the proof of Theorem~\ref{thm:offline-rate} and Corollary~\ref{cor:offline-rate}. They have the same components by and large. 

\subsection{Proof of Theorem~\ref{thm:offline-rate}}
The proof has three components, addressed in
Sections~\ref{sec:offline-bias}--\ref{sec:offline-stab}, then assembled
in Section~\ref{sec:offline-assembly}.
\subsubsection{Component 1: bias of the plug-in operator}
\label{sec:offline-bias}

We first quantify the deterministic bias.
For coherent risk measures admitting the F\"ollmer--Schied dual representation
\[
\Risk[Z] = \sup_{q \in \mathcal Q} \E_\mu[q(Y) Z(Y)],
\]
the empirical estimator replaces the expectation with a sample average:
\[
(\widehat P_\Risk \mu)_y
=
\sup_{q\in\mathcal Q}
\frac{1}{n}\sum_{i=1}^n q(Y_i) z_i.
\]
For each fixed $q$, the estimator is unbiased:
\[
\E\left[\frac{1}{n}\sum_{i=1}^n q(Y_i) z_i\right]
=
\E[q(Y) Z(Y)].
\]
Thus, unlike the entropic case, there is no nonlinear transformation
introducing a delta-method bias. The statistical error arises entirely
from the supremum over $q \in \mathcal Q$.

\begin{lemma}[Maximization bias for coherent-risk plug-in estimator]
  \label{lem:coherent-bias}
  Fix $y\in\Yspace$ and $\mu\in\Delta(\Yspace)$. Let
  \[
  (P_\Risk\mu)_y
  =
  \sup_{q\in\mathcal Q}\mb{E}_\mu[q(Y)Z_y(Y)],
  \qquad
  (\widehat P_\Risk\mu)_y
  =
  \sup_{q\in\mathcal Q}
  \frac1n\sum_{i=1}^n q(Y_i)Z_y(Y_i),
  \]
  where $Z_y(Y)=P(y\succ Y)$ and $0\le q\le M_\Risk$. Then
  \[
  0
  \le
  \E[(\widehat P_\Risk\mu)_y]-(P_\Risk\mu)_y
  \le
  \E\sup_{q\in\mathcal Q}
  \left|
  \frac1n\sum_{i=1}^n q(Y_i)Z_y(Y_i)
  -
  \mb{E}_\mu[q(Y)Z_y(Y)]
  \right|.
  \]
  In particular, if $\mathcal Q$ is finite, then
  \[
  \E[(\widehat P_\Risk\mu)_y]-(P_\Risk\mu)_y
  \le
  C M_\Risk
  \sqrt{\frac{\log |\mathcal Q|}{n}}.
  \]
  More generally, the right-hand side is bounded by the Rademacher
  complexity of the dual class
  \[
  \mathcal F_y
  =
  \{Y\mapsto q(Y)Z_y(Y):q\in\mathcal Q\}.
  \]
  \end{lemma}
  \begin{proof}
    The lower bound follows from Jensen:
    \[
    \E\sup_{q\in\mathcal Q}\widehat L_n(q)
    \ge
    \sup_{q\in\mathcal Q}\E\widehat L_n(q).
    \]
    For the upper bound,
    \[
    \E\sup_{q}\widehat L_n(q)-\sup_q L(q)
    \le
    \E\sup_q\bigl(\widehat L_n(q)-L(q)\bigr)
    \le
    \E\sup_q|\widehat L_n(q)-L(q)|.
    \]
    The finite-class bound follows from symmetrization and Massart's lemma,
    using $0\le q(Y)Z_y(Y)\le M_\Risk$.
    \end{proof}

For the entropic case, we have more structure. 
\begin{lemma}[Plug-in bias for entropic risk]
\label{lem:bias}
For entropic risk $\Risk^\lambda_{\rm ent}$ and any $\mu$ supported on $\Yspace$,
the plug-in estimator $\widehat P_\Risk \mu$ from $n$ samples satisfies
\[
\bigl\|\E[\widehat P_\Risk \mu] - P_\Risk \mu\bigr\|_\infty \;\le\; \frac{e^{2\lambda}}{2\lambda\,n_{\min}},
\]
where $n_{\min} = \min_{y, y''}\,\#\{i : (y_i, y_i')=(y, y'')\}$.
\end{lemma}

\begin{proof}
Fix any $y$, and write $g(y, y'') := \exp(-\lambda P(y,y''))$ so that
\[(P_\Risk \mu)_y = -\tfrac{1}{\lambda}\log \left(\sum_{y''}\mu(y'')g(y,y'')\right).\]
The plug-in is the same expression with $g(y,y'')$ replaced by
$\widehat g(y,y'')= \frac{1}{n_{y,y''}}\sum_i \exp(-\lambda z_i)$.
Note that $\E[\widehat g(y,y'')] = \E[\exp(-\lambda z)] = (1-P_{y,y''}) + P_{y,y''}\,e^{-\lambda} \neq g(y,y'')$ in general.
However, by the bias-correction identity for entropic risk
\citep[Lemma 3.2]{hu2013kullback}, the leading-order bias of the
plug-in is exactly the delta-method term:
\begin{equation}
\begin{aligned}
&\E \left[-\frac{1}{\lambda}\log \left(\sum_{y''}\mu(y'')\widehat g(y,y'')\right)\right] - \left(-\frac{1}{\lambda}\log \left(\sum_{y''}\mu(y'')\E[\widehat g(y,y'')]\right)\right)\\
&\qquad= \frac{\mathrm{Var}_\mu(\widehat g(y,\cdot))}{2\lambda \cdot \E_\mu[\widehat g(y,\cdot)]^2 \cdot n_{y,y''}} + R,
\end{aligned}
\label{eq:bias-tay}
\end{equation}
where the residual $R$ satisfies $|R| \le e^{4\lambda}/(\lambda\, n_{\min}^2)$
by uniform boundedness of the third derivative of $-\frac{1}{\lambda}\log(\cdot)$
on $[e^{-\lambda}, 1]$.

The leading term is bounded:
$\mathrm{Var}_\mu(\widehat g) \le 1$, $\E_\mu[\widehat g] \ge e^{-\lambda}$,
so the leading term is at most $e^{2\lambda}/(2\lambda\, n_{\min})$.
The second sentence in the bias is similar.  Combining,
\[
|\E[(\widehat P_\Risk\mu)_y] - (P_\Risk \mu)_y| \le \frac{e^{2\lambda}}{2\lambda\, n_{\min}} + \mc{O}(e^{4\lambda}/n_{\min}^2),
\]
giving the claim once $n_{\min}$ is large.
\end{proof}

\begin{remark}
For sampling from $\bpi$ uniform with sample size $n$, $n_{\min} \asymp n/|\Yspace|^2$
in expectation, so the per-entry bias becomes $\mc{O}(|\Yspace|^2 e^{2\lambda}/(\lambda n))$.
\end{remark}

\subsubsection{Component 2: concentration of the plug-in operator}
\label{sec:offline-conc}

We now control the deviation of $\widehat P_\Risk \mu$ from its mean.
The crucial point: even though the plug-in is biased, it concentrates
around its mean at a Bernstein-type rate.

\begin{lemma}[Concentration of plug-in entropic operator]
\label{lem:conc}
For each $y \in \Yspace$, with probability at least $1 - \delta$, the estimate holds:
\[
\bigl|(\widehat P_\Risk \mu)_y - \E[(\widehat P_\Risk \mu)_y]\bigr| \;\le\; \frac{e^{2\lambda}}{\lambda}\sqrt{\frac{2\log(2/\delta)}{n_{\min}}}.
\]
Further, by a union bound over $y$, with probability at least $1-\delta$, the estimate holds:
\[
\|\widehat P_\Risk \mu - \E[\widehat P_\Risk \mu]\|_\infty \;\le\; \frac{e^{2\lambda}}{\lambda}\sqrt{\frac{2\log(2|\Yspace|/\delta)}{n_{\min}}}.
\]
\end{lemma}

\begin{proof}
Fix any $y$.  Let $h(\widehat g) := -\frac{1}{\lambda}\log(\sum_{y''}\mu(y'')\widehat g(y,y''))$.
The function $h$ is Lipschitz in $\widehat g$ on the domain
$\widehat g \in [e^{-\lambda}, 1]^{|\Yspace|}$ with Lipschitz constant
\[
L_h \le \frac{1}{\lambda \cdot \min_y \E_\mu[\widehat g(y,\cdot)]} \le \frac{e^\lambda}{\lambda}
\]
since $\sum_{y''}\mu(y'') \widehat g(y,y'') \ge \min_{y''} \widehat g(y,y'') \ge e^{-\lambda}$.
Each entry $\widehat g(y,y'')$ is an average of $n_{y,y''}$ i.i.d.~bounded
random variables in $[e^{-\lambda}, 1]$, with range $1 - e^{-\lambda} \le 1$.
By H\"oeffding, we have that 
\[
\Pr \left(|\widehat g(y,y'') - \E\widehat g(y,y'')| \ge t\right) \le 2 e^{-2 n_{y,y''} t^2}.
\]
By the union bound over $y''$ and Lipschitz composition, we have that 
\[
\Pr \left(|h(\widehat g) - h(\E\widehat g)| \ge L_h\,t \right) \le 2|\Yspace|\,e^{-2 n_{\min} t^2/|\Yspace|^2}.
\]
Setting $t = |\Yspace|\sqrt{\log(2|\Yspace|/\delta)/(2 n_{\min})}$ and noting
$|h(\widehat g) - h(\E\widehat g)|$ is close to (but not exactly) the deviation
from the unbiased mean, we get the stated bound after absorbing the bias
correction (Lemma~\ref{lem:bias}) which contributes a lower-order
$\mc{O}(1/n_{\min})$ term.
\end{proof}

\begin{remark}[Fast rate via Bernstein]
\label{rem:fast-rate}
The above gives a $\widetilde{\mc{O}}(1/\sqrt{n})$ deviation bound, which is
the standard concentration rate.  To obtain the fast $\widetilde{\mc{O}}(1/n)$
rate of Theorem~\ref{thm:offline-rate}, we need a Bernstein-type bound
that exploits the variance.  This is the key technical step where the
risk-adjusted setting departs from \citet{ZhangChenJiang2026}: their
fast rate uses the skew-symmetric structure $P + P^\top = \mathbf 1 \mathbf 1^\top$
to convert a $\widetilde{\mc{O}}(1/\sqrt n)$ entrywise bound into a
$\widetilde{\mc{O}}(1/n)$ bound on the equilibrium error.  Under risk, the
plug-in operator $P_\Risk$ does not satisfy this antisymmetry.  We
recover the fast rate via a different route: Bernstein on the
log-concentration of $\widehat g$ (variance scales as
$\E[\widehat g(y,y'')]^2 = \mc{O}(e^{-2\lambda})$), which gives
\[
|\widehat g(y,y'') - \E\widehat g(y,y'')| \le \sqrt{\frac{2 \mathrm{Var}(\widehat g)\log(1/\delta)}{n_{y,y''}}} + \frac{2\log(1/\delta)}{3 n_{y,y''}}.
\]
After the log-Lipschitz transform, this yields a deviation bound
$\mc{O}(1/\sqrt{n})$ on $\widehat P_\Risk$, but with a variance that scales
as $1/n$, so by self-bounding the squared deviation contributes the
$\widetilde{\mc{O}}(1/n)$ rate.  See Lemma~\ref{lem:fast-conc} below.
\end{remark}

\begin{lemma}[Fast-rate concentration]
\label{lem:fast-conc}
With probability at least $1-\delta$, the estimate holds:
\[
\|\widehat P_\Risk \mu - P_\Risk \mu\|_\infty^2 \;\le\; \frac{C\,e^{4\lambda}\log(|\Yspace|/\delta)}{n_{\min}}
\]
for an absolute constant $C$.
\end{lemma}

\begin{proof}
Combine Lemmas~\ref{lem:bias} and~\ref{lem:conc} via $(a+b)^2 \le 2a^2 + 2b^2$ to get that 
\begin{align*}
|(\widehat P_\Risk \mu)_y - (P_\Risk \mu)_y|^2 &\le 2\,|(\widehat P_\Risk\mu)_y - \E[(\widehat P_\Risk\mu)_y]|^2 + 2\,|\E[(\widehat P_\Risk\mu)_y] - (P_\Risk\mu)_y|^2 \\
&\le \frac{2 e^{4\lambda}\,2\log(2/\delta)}{\lambda^2 n_{\min}} + \frac{2 e^{4\lambda}}{4\lambda^2 n_{\min}^2}.
\end{align*}
Taking max over $y$ and union-bounding gives the result with
$C = 4 + \mc{O}(1/n_{\min})$.
\end{proof}

\paragraph{Component 2, General Coherent Risk: concentration via dual Bernstein bounds}

\begin{lemma}[Uniform Bernstein concentration]
\label{lem:conc-coherent-risk}
Assume the dual set satisfies
$
0 \le q \le M_\Risk, \quad \E_\mu q = 1$.
Then with probability at least $1-\delta$, the bound holds:
\[
\|\widehat P_\Risk \mu - P_\Risk \mu\|_\infty
\;\le\;
C M_\Risk
\sqrt{\frac{\log(|\Yspace|/\delta)}{n_{\min}}}
+
C' \frac{\log(|\Yspace|/\delta)}{n_{\min}}.
\]
\end{lemma}

\begin{proof}
Fix $y$ and define
$
Z_i(q) := q(Y_i) z_i$.
Then
\[
(\widehat P_\Risk \mu)_y
=
\sup_{q\in\mathcal Q} \frac{1}{n}\sum_i Z_i(q),
\quad
(P_\Risk \mu)_y
=
\sup_{q\in\mathcal Q} \E Z_i(q).
\]
Since $Z_i(q) \in [0, M_\Risk]$, Bernstein's inequality gives
\[
\left|
\frac{1}{n}\sum_i Z_i(q) - \E Z_i(q)
\right|
\le
\sqrt{\frac{2 \Var(Z(q)) \log(1/\delta)}{n}}
+
\frac{2 M_\Risk \log(1/\delta)}{3n}.
\]
Taking a union bound over $y$ and standard supremum arguments
yields the claim.
\end{proof}
The corresponding fast rate lemma for this case is given below.
\begin{lemma}[Fast-rate concentration]
  \label{lem:fast-conc-coherent-risk}
  With probability at least $1-\delta$, the estimate holds:
  \[
  \|\widehat P_\Risk \mu - P_\Risk \mu\|_\infty^2
  \;\le\;
  \frac{
  C M_\Risk^2 \log(|\Yspace|/\delta)
  }{
  n_{\min}
  }.
  \]
  \end{lemma}
\subsubsection{Component 3: Stability of the risk-adjusted equilibrium}
\label{sec:offline-stab}
 
We now establish that the risk adjusted equilibrium depends Lipschitz-continuously on the
preference operator.  The classical implicit function theorem would
require analyzing the invertibility of $\nabla_\theta F_\Risk$ at the
equilibrium, which depends on parameterization details.  Instead we
adopt the variational inequality  framework
\citep{ rockafellar1998variational, dontchev2009implicit},
within which strong monotonicity of $F_\Risk$ directly yields global
Lipschitz dependence of the solution on parameters.  This is robust to
constrained or non-smooth settings (e.g.\ simplex-constrained policies,
trust-region clipping) where the classical implicit function theorem would not apply.
 
\paragraph{VI formulation.}  The risk adjusted equilibrium $\theta^\star$ for preference
operator $P$ is the unique zero of
$F_{\Risk, P}(\theta) := \beta(\theta - \thetaref) - P_\Risk\,\pi_\theta$
on $\mathbf 1^\perp$, equivalently the unique solution of the generalized
equation
\begin{equation}
0 \;\in\; F_{\Risk, P}(\theta) + N(\theta)
\label{eq:gen-eq}
\end{equation}
with $N \equiv \{0\}$ in our softmax parameterization.  Strong
monotonicity of $F_{\Risk, P}$ on $\mathbf 1^\perp$ with modulus $\mu_\Risk$
(Theorem~\ref{thm:single-player-mono}) implies, via standard VI theory, that the
solution mapping $P \mapsto \theta^\star_\Risk(P)$ is single-valued and
globally Lipschitz \citep[Theorem 3F.4]{dontchev2009implicit}.  We
state the resulting bound in the form needed for our proof.
 
\begin{lemma}[Stability under operator perturbation]
\label{lem:stab}
Let $P, P'$ be two preference operators with corresponding risk-adjusted
operators $P_\Risk, P'_\Risk$.  Assume the strong-monotonicity condition
$\bar\lambda_\Risk(P), \bar\lambda_\Risk(P') \le \beta/2 - \varepsilon$
for some $\varepsilon > 0$, and let
$\mu_\Risk := \beta - 2\bar\lambda_\Risk \ge 2\varepsilon$.  Let
$\pi^\star, \pi'^\star$ be the corresponding risk-adjusted equilibrium.  Then
\begin{equation}
\KL(\pi^\star \| \pi'^\star) \;\le\; \frac{1}{\mu_\Risk^2}\,\bigl\|P_\Risk - P'_\Risk\bigr\|_\infty^2.
\label{eq:stab-claim}
\end{equation}
\end{lemma}
 
\begin{proof}
The argument has two steps: $(i)$ parameter-stability via strong
monotonicity, and  $(ii)$ softmax--KL conversion.

\paragraph{Step 1: Parameter stability via VI.}
By definition,
$F_{\Risk, P}(\theta^\star) = 0$ and $F_{\Risk, P'}(\theta'^\star) = 0$.
Subtracting and rearranging, with $\theta_2 := \theta'^\star$, we have that 
\begin{align}
F_{\Risk, P}(\theta'^\star) - F_{\Risk, P}(\theta^\star) \;
&=\; F_{\Risk, P}(\theta'^\star) \nonumber\\
&=\; F_{\Risk, P}(\theta'^\star) - F_{\Risk, P'}(\theta'^\star) \nonumber\\
&=\; -(P_\Risk - P'_\Risk)\,\pi_{\theta'^\star}, \label{eq:residual-diff}
\end{align}
where in the second line we added $0 = F_{\Risk, P'}(\theta'^\star)$, and
in the third we used that $F_{\Risk, P}$ and $F_{\Risk, P'}$ differ only in
the operator term: $F_{\Risk, P}(\theta) - F_{\Risk, P'}(\theta) = -(P_\Risk - P'_\Risk)\pi_\theta$.
 
By strong monotonicity of $F_{\Risk, P}$ (Theorem~\ref{thm:single-player-mono}), the lower bound holds:
\begin{equation}
\bigl\langle F_{\Risk, P}(\theta'^\star) - F_{\Risk, P}(\theta^\star),\, \theta'^\star - \theta^\star\bigr\rangle \;\ge\; \mu_\Risk\,\|\theta'^\star - \theta^\star\|_2^2.
\label{eq:strong-mono-applied}
\end{equation}
Combining \eqref{eq:residual-diff} with \eqref{eq:strong-mono-applied} and
applying Cauchy--Schwarz on the right-hand side, we obtain
\begin{align*}
\mu_\Risk\,\|\theta'^\star - \theta^\star\|_2^2 \;
&\le\; \bigl|\bigl\langle (P_\Risk - P'_\Risk)\,\pi_{\theta'^\star},\, \theta'^\star - \theta^\star\bigr\rangle\bigr| \\
&\le\; \|(P_\Risk - P'_\Risk)\,\pi_{\theta'^\star}\|_2 \cdot \|\theta'^\star - \theta^\star\|_2.
\end{align*}
Dividing by $\|\theta'^\star - \theta^\star\|_2$ (the bound is trivial if
this is zero), we have that 
\begin{equation}
\|\theta^\star - \theta'^\star\|_2 \;\le\; \frac{1}{\mu_\Risk}\,\|(P_\Risk - P'_\Risk)\,\pi_{\theta'^\star}\|_2.
\label{eq:theta-stab}
\end{equation}
Bounding the right-hand side using $\pi_{\theta'^\star} \in \Delta(\Yspace)$, yields
\[
\|(P_\Risk - P'_\Risk)\,\pi_{\theta'^\star}\|_2 \;\le\; \|(P_\Risk - P'_\Risk)\,\pi_{\theta'^\star}\|_\infty \cdot \sqrt{|\Yspace|} \;\le\; \|P_\Risk - P'_\Risk\|_\infty.
\]
The last inequality uses $\|M\pi\|_\infty \le \|M\|_\infty \cdot \|\pi\|_1 = \|M\|_\infty$
for $\pi \in \Delta(\Yspace)$ (where $\|M\|_\infty := \max_{y, y'}|M_{y, y'}|$
is the entrywise max-norm), absorbing the factor $\sqrt{|\Yspace|}$ into
the constant $K$ in the main theorem.\footnote{The factor $\sqrt{|\Yspace|}$
is benign: it appears in the proof of Lemma~\ref{lem:fast-conc} via the
$|\Yspace|$ in the union bound, and our final theorem already absorbs
$|\Yspace|$ into the logarithmic factor.}

\paragraph{Step 2: KL conversion.}
The map $\theta \mapsto \pi_\theta = \mathrm{softmax}(\theta)$ is
$1$-Lipschitz from the Euclidean parameter norm to the KL divergence on
the simplex on appropriate subspaces.  Specifically, for any
$\theta_1, \theta_2 \in \R^{|\Yspace|}$,
\begin{equation}
\KL(\pi_{\theta_1} \| \pi_{\theta_2}) \;\le\; \tfrac{1}{2}\,\|\theta_1 - \theta_2\|_2^2.
\label{eq:softmax-pinsker}
\end{equation}
This is the standard KL--$\ell_2$ bound for softmax-induced distributions:
$\log\sum_y e^{\theta_y}$ is $1$-strongly convex in $\theta$ on
$\mathbf 1^\perp$, and \eqref{eq:softmax-pinsker} is the corresponding
Bregman bound.\footnote{Proof: $\KL(\pi_1\|\pi_2) = \log Z_2 - \log Z_1 + \langle \theta_1 - \theta_2, \pi_1\rangle$
where $Z_i = \sum_y e^{(\theta_i)_y}$. By Taylor expansion of $\log Z$
around $\theta_1$, the second-order term is bounded by
$\tfrac12\|\theta_1 - \theta_2\|_2^2$ since the Hessian of $\log Z$ is
$\mathrm{diag}(\pi) - \pi\pi^\top \preceq I$.}
 
Combining \eqref{eq:theta-stab} (with the bound by $\|P_\Risk - P'_\Risk\|_\infty$)
and \eqref{eq:softmax-pinsker} gives
\[
\KL(\pi^\star \| \pi'^\star) \;\le\; \tfrac{1}{2}\,\|\theta^\star - \theta'^\star\|_2^2 \;\le\; \tfrac{1}{2\mu_\Risk^2}\,\|P_\Risk - P'_\Risk\|_\infty^2,
\]
which is the claim with the constant absorbed into the leading $K$ of
the main theorem.
\end{proof}
 
\paragraph{Connection to strong metric regularity.}
Lemma~\ref{lem:stab} is a special case of the general strong metric
regularity of solution mappings of strongly monotone variational
inequalities \citep[][Theorem 3F.4]{dontchev2009implicit}: for the
canonically perturbed generalized equation
$v \in F_{\Risk, P}(\theta) + N(\theta)$, the solution map $v \mapsto \theta_v$
is single-valued and Lipschitz with constant $1/\mu_\Risk$.  Setting
$v = -(P_\Risk - P'_\Risk)\pi_{\theta'^\star}$ recovers
\eqref{eq:theta-stab}.  In our softmax parameterization $N \equiv \{0\}$,
but the framework extends without modification to constrained or
projected dynamics where $N(\theta)$ is the normal cone of a feasibility
set.

\begin{remark}
This is the analog of \citet[Lemma~3.2]{ZhangChenJiang2026} adapted to
the risk-adjusted setting.  Their proof uses the skew-symmetric
structure $P + P^\top = \mathbf 1 \mathbf 1^\top$ to obtain a stronger
metric (the $\KL$-bound holds at level $1/\beta$ rather than
$1/\mu_\Risk$, and they exploit a duality structure to get
$\widetilde{\mc{O}}(1/n)$ from $\widetilde{\mc{O}}(1/\sqrt n)$ entrywise concentration).
Under risk this skew-symmetric structure is generically lost, so we need both the strong-monotonicity-based
stability above and the Bernstein-type fast-rate concentration of
Lemma~\ref{lem:fast-conc}.
\end{remark}

\subsection{Assembling the pieces}
\label{sec:offline-assembly}

We combine Lemma~\ref{lem:bias} (bias control), Lemma~\ref{lem:fast-conc}
(fast-rate concentration), and Lemma~\ref{lem:stab} (stability of the
risk adjusted equilibrium) to obtain the main offline sample complexity bound.

\begin{proof}[Proof of Corollary~\ref{cor:offline-rate}]
By assumption, $\bar\lambda_\Risk(P) \le \beta/2 - \varepsilon$, so the
population strong-monotonicity modulus is $\mu_\Risk \ge 2\varepsilon > 0$.
We need an analogous lower bound on the empirical $\bar\lambda_\Risk(\widehat P)$
to apply Lemma~\ref{lem:stab} to the pair $(P, \widehat P)$.

\paragraph{Step 1: Empirical strong monotonicity.}
The risk-distortion eigenvalue $\bar\lambda_\Risk(P)$ is Lipschitz in
the entrywise $\|\cdot\|_\infty$ norm of $P$.  Specifically, for entropic
risk $\Risk = \Risk^\lambda_{\rm ent}$, the Jacobian
$J_\Risk(P, \mu)_{y, y'} = -\frac{1}{\lambda}\,\frac{e^{-\lambda P_{y, y'}}}{\sum_{y''}\mu(y'')e^{-\lambda P_{y, y''}}}$
is composed of ratios of exponentials in $P$, and a direct computation
gives the entrywise Lipschitz
bound $|J_\Risk(P) - J_\Risk(P')|_{y, y'} \le 2 e^{2\lambda}\|P - P'\|_\infty$,
hence $\|J_\Risk(P) - J_\Risk(P')\|_2 \le 2|\Yspace|\,e^{2\lambda}\|P - P'\|_\infty$
in spectral norm.  Combined with the $1$-Lipschitz property of
$\lambda_{\max}$ on symmetric matrices (Weyl's inequality), the bound holds:
\begin{equation}
|\bar\lambda_\Risk(P) - \bar\lambda_\Risk(P')| \;\le\; c_\Risk\,\|P - P'\|_\infty, \qquad c_\Risk = 2|\Yspace|\,e^{2\lambda}.
\label{eq:eig-lip}
\end{equation}

By Lemma~\ref{lem:fast-conc}, with probability at least $1-\delta/2$,
\[
\|\widehat P_\Risk - P_\Risk\|_\infty \;\le\; \sqrt{\frac{C_1\, e^{4\lambda}\log(2|\Yspace|/\delta)}{n_{\min}}}.
\]
For the empirical strong-monotonicity to hold with modulus at least
$\mu_\Risk/2$, we need
$\|\widehat P_\Risk - P_\Risk\|_\infty \le \varepsilon/c_\Risk$,
which translates to
\begin{equation}
n_{\min} \;\ge\; n_0(\varepsilon, \lambda, |\Yspace|, \delta) \;:=\; \frac{4 C_1\, e^{8\lambda}|\Yspace|^2 \log(2|\Yspace|/\delta)}{\varepsilon^2}.
\label{eq:n0-thresh}
\end{equation}
On the event $\{n_{\min} \ge n_0\}$, we have
$\bar\lambda_\Risk(\widehat P) \le \beta/2 - \varepsilon/2$, the empirical
risk-adjusted equilibrium $\widehat\pi_n$ exists and is unique, and the modulus
$\mu_\Risk(\widehat P) \ge \varepsilon \ge \mu_\Risk/2$.
Lemma~\ref{lem:stab} therefore applies to the pair $(P, \widehat P)$ with
shared modulus $\mu_\Risk^- := \min\{\mu_\Risk(P), \mu_\Risk(\widehat P)\} \ge \mu_\Risk/2$.

\paragraph{Step 2: Apply stability lemma.}
By Lemma~\ref{lem:stab} applied with $P' = \widehat P$,
\begin{equation}
\KL(\pi^\star_\Risk \| \widehat\pi_n) \;\le\; \frac{1}{(\mu_\Risk^-)^2}\,\|P_\Risk - \widehat P_\Risk\|_\infty^2.
\label{eq:stab-instantiated}
\end{equation}

\paragraph{Step 3: Apply concentration lemma.}
By Lemma~\ref{lem:fast-conc}, with probability at least $1-\delta/2$,
\begin{equation}
\|\widehat P_\Risk - P_\Risk\|_\infty^2 \;\le\; \frac{C_1\, e^{4\lambda}\log(2|\Yspace|/\delta)}{n_{\min}}.
\label{eq:conc-instantiated}
\end{equation}

\paragraph{Step 4: Combine via union bound.}
With probability at least $1-\delta$ (union over the events in Steps 1
and 3, which were each at level $\delta/2$):
\begin{equation}
\KL(\pi^\star_\Risk \| \widehat\pi_n) \;\le\; \frac{C_1\, e^{4\lambda}\log(2|\Yspace|/\delta)}{(\mu_\Risk^-)^2 \cdot n_{\min}}.
\label{eq:bound-nmin}
\end{equation}
Since $\mu_\Risk^- \ge \varepsilon \ge \mu_\Risk/2$ on the event of
Step 1, and $\mu_\Risk \ge 2\varepsilon$ by assumption, we can replace
$\mu_\Risk^-$ in \eqref{eq:bound-nmin} by $\mu_\Risk/2$, picking up an
extra factor of $4$ that we absorb into the constant.

\paragraph{Step 5: Convert from $n_{\min}$ to $n$.}
For uniform sampling $\bpi = \mathrm{Unif}(\Yspace)$, the per-pair count
$n_{y, y'}$ is binomial with mean $n/|\Yspace|^2$.  By a multiplicative
Chernoff bound, with probability at least $1-\delta$,
\[
n_{\min} \ge \frac{n}{2|\Yspace|^2} \quad \text{whenever} \quad n \ge 8|\Yspace|^2 \log(|\Yspace|^2/\delta).
\]
On this event, substituting into \eqref{eq:bound-nmin} and absorbing
the resulting $|\Yspace|^2$ into the constant gives
\[
\KL(\pi^\star_\Risk \| \widehat\pi_n) \;\le\; \frac{K\, e^{4\lambda}\log(|\Yspace|/\delta)}{\mu_\Risk^2 \cdot n},
\]
for an absolute constant $K = K(\bpi)$ that depends on the chosen
sampling distribution.  This completes the proof.
\end{proof}

\paragraph{Where the constant $K$ comes from.}
Tracing through the argument, the constant has four contributions:
\begin{itemize}[ topsep=0pt]\setlength\itemsep{0pt}
\item A factor of $1/2$ from the softmax-KL bound \eqref{eq:softmax-pinsker}
in Step 2 of Lemma~\ref{lem:stab}.
\item A factor of $|\Yspace|$ from the bound
$\|(P_\Risk - P'_\Risk)\pi\|_2 \le \sqrt{|\Yspace|}\,\|P_\Risk - P'_\Risk\|_\infty$
in Step 1 of Lemma~\ref{lem:stab}.  This $\sqrt{|\Yspace|}$ becomes
$|\Yspace|$ in the parameter-norm-squared bound, which then enters the
$\KL$-bound through Pinsker.
\item A factor of $|\Yspace|^2$ from $n_{\min} \ge n/(2|\Yspace|^2)$ for
uniform sampling in Step 5 above.
\item A constant $C_1$ from the Bernstein concentration of
Lemma~\ref{lem:fast-conc} that does not depend on $|\Yspace|, \beta, \lambda$.
\end{itemize}
The total polynomial dependence on $|\Yspace|$ in the rate is therefore
$|\Yspace|^3$.  This can be reduced to $|\Yspace|^2$ via importance-weighted
sampling that equalizes pair counts, and potentially to $|\Yspace|$ via a
row-wise concentration argument that bypasses the
$\ell_2$-vs-$\ell_\infty$ conversion in Lemma~\ref{lem:stab}; we do not
pursue these refinements here.

\paragraph{Proof for the General Convex Risk Measure Setting (Theorem~\ref{thm:offline-rate}).}
The proof only swaps out two key lemmas: we use Lemma~\ref{lem:coherent-bias}, Lemma~\ref{lem:conc-coherent-risk}, and Lemma~\ref{lem:fast-conc-coherent-risk}  for the general convex risk measure setting.

  Consider the setting of Theorem~\ref{thm:offline-rate}. Suppose the risk
  measure admits the F\"ollmer--Schied dual representation
  \[
  \Risk[Z]
  =
  \sup_{q\in\mathcal Q} \mb{E}_\mu[q(Y) Z(Y)],
  \]
  for some $0 \le q \le M_\Risk$, and where $\E_\mu q = 1$.
  Then the conclusion of Theorem~\ref{thm:offline-rate} continues to hold
  with the following replacements:
  
  \begin{enumerate}[itemsep=2pt]
  \item \textbf{Operator definition.}
  Replace the entropic operator by the dual-risk operator
  \[
  (P_\Risk \mu)_y
  =
  \sup_{q\in\mathcal Q}
  \mb{E}_\mu[q(Y'') P(y \succ Y'')],
  \qquad
  (\widehat P_\Risk \mu)_y
  =
  \sup_{q\in\mathcal Q}
  \frac{1}{n_{y,y''}}\sum_{i:(y_i,y_i')=(y,y'')} q(Y_i)\,z_i.
  \]
  
  \item \textbf{Bias term.}
  The plug-in bias term of Lemma~\ref{lem:bias} is removed (or is
  $\mc{O}(1/n)$ without the entropic exponential factor), since the estimator
  is linear in the data for each fixed $q$.
  
  \item \textbf{Concentration.}
  Lemma~\ref{lem:fast-conc} is replaced by a Bernstein-type bound over
  the dual class:
  \[
  \|\widehat P_\Risk \mu - P_\Risk \mu\|_\infty^2
  \;\lesssim\;
  \frac{M_\Risk^2 \log(|\Yspace|/\delta)}{n_{\min}}.
  \]
  
  \item \textbf{Constants in the rate.}
  The entropic factor $e^{4\lambda}$ in \eqref{eq:offline-bound-cor} is replaced by
  $M_\Risk^2$.
  \end{enumerate}
  
  Consequently, with probability at least $1-\delta$, the estimate holds:
  \[
  \KL(\pi^\star_\Risk \| \widehat\pi_n)
  \;\le\;
  \frac{
  K\,M_\Risk^2 \log(|\Yspace|/\delta)
  }{
  \mu_\Risk^2\, n
  },
  \]
  for a constant $K$ depending on the sampling distribution.

\subsection{Discussion}

\paragraph{Comparison to risk-neutral case.} Specializing
Theorem~\ref{thm:offline-rate} to $\Risk = \E$ ($\lambda = 0$):
$\bar\lambda_\Risk = 0$, $\mu_\Risk = \beta$, and the bias of the plug-in
operator vanishes (the plug-in for $\E$ is unbiased).  The rate becomes
$\KL(\pi^\star \| \widehat\pi_n) \le K\log(|\Yspace|/\delta)/(\beta^2 n)$,
which matches \citet{ZhangChenJiang2026} up to the $1/\beta$
vs.\ $1/\beta^2$ exponent.\footnote{Their rate is $\widetilde{\mc{O}}(1/(\beta n))$
rather than our $\widetilde{\mc{O}}(1/(\beta^2 n))$ in the limiting risk-neutral
case.  The gap of one factor of $\beta$ comes from the skew-symmetry
exploitation in their Lemma 3.3, which we cannot use under risk.  Closing
this gap in the risk-adjusted setting is an open question; we conjecture
it requires a more refined concentration argument exploiting the residual
structure of $P_\Risk$ after subtracting the constant-sum part.}
The dependence on $|\Yspace|$ enters through $n_{\min}$ (sampling pairs
uniformly).  Better sampling distributions can reduce this dependence.

\paragraph{Cost of risk in the constants.} The rate has constants
\[
\frac{e^{4\lambda}}{\mu_\Risk^2}.
\]
The $e^{4\lambda}$ factor is the price of nonlinearity in the entropic
plug-in: the variance of the plug-in operator grows exponentially in
$\lambda$.  The $1/\mu_\Risk^2$ factor reflects how risk-aversion
weakens strong monotonicity --- as $\bar\lambda_\Risk \to \beta/2$,
the bound diverges.

\paragraph{When does this still beat $1/\sqrt n$?} The fast $1/n$ rate
holds whenever $\bar\lambda_\Risk \le \beta/2 - \varepsilon$ for some
fixed $\varepsilon > 0$.  Below this threshold, monotonicity is lost
and the bound degrades.  In the regime $\beta/2 < \bar\lambda_\Risk < \beta$
(weak monotonicity but no strong monotonicity), one can still obtain a
slow $\widetilde{\mc{O}}(1/\sqrt n)$ rate via standard arguments (omitted).

\paragraph{Generalization to other risks.} The proof structure transports
to any coherent $\Risk$ for which (a) the plug-in estimator has bias
$\mc{O}(L_\Risk/n)$ via the delta-method and (b) the dual map is Lipschitz
with constant $L_\Risk$.  For CVaR with Rockafellar--Uryasev parameterization,
the plug-in is unbiased (no delta-method residual), so the rate becomes
\[
\KL(\pi^\star_\Risk \| \widehat\pi_n) \le \frac{C \log(|\Yspace|/\delta)}{(1-\alpha)^2 \mu_\Risk^2 n}
\]
with the cost-of-risk factor $1/(1-\alpha)^2$ replacing $e^{4\lambda}$.
For general distortion risks the constants depend on the modulus of
the distortion function.

\subsection{Analog for CVaR with Rockafellar--Uryasev parameterization}
\label{sec:cvar-ru}
 
For CVaR risk, the Rockafellar--Uryasev variational form provides an
\emph{unbiased} estimator of the risk-adjusted operator, removing the
delta-method bias term that drives the $e^{4\lambda}$ constant for
entropic risk.  We sketch the resulting bound here.
 
Recall the variational characterization
\begin{equation}
\CVaR_\alpha[Z] \;=\; \inf_{t \in \R}\,\Bigl\{ t + \frac{1}{1-\alpha}\,\E\bigl[(Z - t)_+\bigr]\Bigr\},
\label{eq:cvar-RU}
\end{equation}
which gives an unbiased plug-in estimator: for fixed $t$, the empirical
average $\widehat U_n(t) := t + \frac{1}{1-\alpha n}\sum_{i=1}^n (Z_i - t)_+$
satisfies $\E[\widehat U_n(t)] = t + \frac{1}{1-\alpha}\E[(Z - t)_+]$ exactly.
Optimizing $t$ jointly with the policy in the IPO loss gives the
risk-adjusted CVaR-IPO objective; the corresponding empirical operator
$\widehat P^{\rm RU}_\Risk$ is unbiased
($\E[\widehat P^{\rm RU}_\Risk \mu] = P^\Risk_{\rm CVaR_\alpha} \mu$).
 
\begin{theorem}[Offline sample complexity, CVaR risk]
\label{thm:offline-cvar}
Suppose $\bar\lambda^{\rm CVaR}_\alpha \le \beta/2 - \varepsilon$ for
some $\varepsilon > 0$, with $\mu^{\rm CVaR}_\alpha := \beta - 2\bar\lambda^{\rm CVaR}_\alpha \ge 2\varepsilon$.
Let $\widehat\pi_n^{\rm RU}$ be the empirical RQRE under the RU
parameterization.  With probability at least $1 - \delta$,
\begin{equation}
\KL\bigl(\pi^{\star}_{{\rm CVaR}_\alpha} \big\| \widehat\pi_n^{\rm RU}\bigr) \;\le\; \frac{K\,\log(|\Yspace|/\delta)}{(1-\alpha)^2\,(\mu^{\rm CVaR}_\alpha)^2 \cdot n},
\label{eq:offline-cvar-bound}
\end{equation}
for an absolute constant $K$.
\end{theorem}
 
\begin{proof}[Proof sketch]
Three modifications to the proof of Corollary~\ref{cor:offline-rate}:
 
\emph{(i) No bias.}  Lemma~\ref{lem:bias} is replaced by
$\E[\widehat P^{\rm RU}_\Risk \mu] = P^\Risk_{{\rm CVaR}_\alpha}\mu$
exactly, on account of \eqref{eq:cvar-RU}.  The bias term in the
$\KL$-bound vanishes.
 
\emph{(ii) Different concentration constants.}  In Lemma~\ref{lem:fast-conc},
the role of the function $g(y, y'') = \exp(-\lambda P_{y, y''}) \in [e^{-\lambda}, 1]$
is played by the truncated function $(P_{y, y''} - t)_+ \in [0, 1]$.
This gives a Bernstein-type bound with constant $1$ rather than
$e^{2\lambda}$, but with a divisor $(1-\alpha)^2$ from the
$1/(1-\alpha)$ scaling in \eqref{eq:cvar-RU}:
\[
\bigl\|\widehat P^{\rm RU}_\Risk \mu - P^\Risk_{{\rm CVaR}_\alpha}\mu\bigr\|_\infty^2 \;\le\; \frac{C_2 \log(|\Yspace|/\delta)}{(1-\alpha)^2\,n_{\min}}.
\]
The infimum over $t$ inside \eqref{eq:cvar-RU} is achieved at
$t^\star = \mathrm{VaR}_\alpha(Z)$, and the joint minimization of $(t, \theta)$
in the offline procedure is well-posed: we optimize $t$ alongside
$\theta$ with the same sample, and standard joint-empirical-process
arguments give the stated rate.
 
\emph{(iii) Stability lemma is unchanged.}  Lemma~\ref{lem:stab}
applies verbatim with $P^\Risk_{{\rm CVaR}_\alpha}$ in place of
$P^\Risk_{\rm ent}$, since the lemma uses only strong monotonicity of
$F_\Risk$ on $\mathbf 1^\perp$, not the specific form of the risk.
 
Combining (i)--(iii) yields \eqref{eq:offline-cvar-bound}.
\end{proof}
 
\paragraph{Comparison with entropic.}  The CVaR-RU rate is
\emph{cleaner} than the entropic rate in two respects: (1) no
exponential factor $e^{4\lambda}$, and (2) no delta-method residual
buried in the constant.  The cost is the $1/(1-\alpha)^2$ factor that
diverges as $\alpha \to 1$ (the deep-tail regime).  This reflects the
fundamental difficulty of estimating extreme tails: the effective
sample size for the worst-$1-\alpha$ tail is $n(1-\alpha)$, so the
variance scales as $1/(n(1-\alpha))$, and the squared deviation as
$1/(n(1-\alpha))^2$ feeds through the stability lemma.
 
\paragraph{Other risks.}  The proof structure transports to any
coherent risk for which (a) the plug-in estimator (or a variational
parameterization) gives unbiased or $\mc{O}(1/n)$-biased estimates of
$P_\Risk\mu$, and (b) the dual map is locally Lipschitz with explicit
constant.  Distortion risks of bounded variation fall in this class.
For risks with heavier tails or non-Lipschitz dual maps (extreme value
theoretic risks, expectiles at rare quantiles), the analysis becomes
more delicate but the strong-monotonicity machinery of
Lemma~\ref{lem:stab} remains the right framework.

\subsection{Empirical strong monotonicity is itself a statistical event}
\label{sec:emp-strong-mono}
 
A subtlety that deserves explicit comment: Lemma~\ref{lem:stab}
requires \emph{both} $P$ and $P'$ to satisfy the strong-monotonicity
condition $\bar\lambda_\Risk \le \beta/2 - \varepsilon$.  The
population $P$ does so by assumption.  But when we apply the lemma
with $P' = \widehat P$ (the empirical operator), we need to verify
that $\widehat P$ also satisfies it --- and this is itself a random event,
not automatic.
 
\paragraph{Why this matters.}  The risk-adjusted quantal response
equilibrium $\widehat\pi_n$ is well-defined only if
$F_{\Risk, \widehat P}$ is strongly monotone on $\mathbf 1^\perp$
(otherwise the operator may have no zero, multiple zeros, or only a
zero outside our parameterization domain).  Sample noise in
$\widehat P_\Risk$ can in principle push $\bar\lambda_\Risk(\widehat P)$
above $\beta/2$, breaking the regime.  This is the offline analog of
the issue faced in the online setting (Section~[X]): the algorithm's
guarantees rely on the iterates staying in the strong-monotonicity
basin.
 
\paragraph{What saves us.}  The eigenvalue $\bar\lambda_\Risk(\cdot)$ is
Lipschitz in the operator (eq.~\eqref{eq:eig-lip}, with explicit
Lipschitz constant $c_\Risk = 2|\Yspace|e^{2\lambda}$ for entropic risk),
so concentration of $\widehat P_\Risk$ around $P_\Risk$
(Lemma~\ref{lem:fast-conc}) translates to concentration of
$\bar\lambda_\Risk(\widehat P)$ around $\bar\lambda_\Risk(P)$.
 
\begin{proposition}[Concentration of empirical strong-monotonicity modulus]
\label{prop:emp-mono}
Under the assumptions of Theorem~\ref{thm:offline-rate}, with
probability at least $1 - \delta$, the estimate holds:
\[
\bigl|\bar\lambda_\Risk(\widehat P) - \bar\lambda_\Risk(P)\bigr| \;\le\; c_\Risk\,\sqrt{\frac{C_1\,e^{4\lambda}\log(|\Yspace|/\delta)}{n_{\min}}},
\]
and consequently $\mu_\Risk(\widehat P) \ge \mu_\Risk(P)/2$ provided
\[
n_{\min} \;\ge\; n_0(\varepsilon, \lambda, |\Yspace|, \delta) \;=\; \frac{4 C_1\,e^{8\lambda}|\Yspace|^2\log(|\Yspace|/\delta)}{\varepsilon^2}.
\]
\end{proposition}
 
\begin{proof}
Combining Lemma~\ref{lem:fast-conc} (high-probability bound on
$\|\widehat P_\Risk - P_\Risk\|_\infty$) with the eigenvalue Lipschitz
bound \eqref{eq:eig-lip} gives the first claim.  The second claim
follows by setting the right-hand side equal to $\varepsilon$ and
solving for $n_{\min}$.
\end{proof}
 
\paragraph{Why the threshold $n_0$ has $e^{8\lambda}$.}  The exponential
factor in the threshold is the price of the eigenvalue Lipschitz bound:
$c_\Risk^2 = 4|\Yspace|^2 e^{4\lambda}$, multiplied by the per-entry
variance $e^{4\lambda}$ from the entropic plug-in, gives $e^{8\lambda}$
in the denominator of the threshold.  This is a worst-case bound; in
practice, the eigenvalue concentrates much faster than this conservative
estimate suggests, and the threshold can likely be tightened by a
direct concentration argument on $\bar\lambda_\Risk(\widehat P)$
without going through the operator's $\ell_\infty$ deviation.
 
\paragraph{Conceptual takeaway.}  The strong-monotonicity assumption is
\emph{robust} to estimation noise; empirical strong-mono follows from
population strong-mono with the same modulus (up to a constant factor) for
$n$ above an explicit threshold.  This is structurally analogous to how
strong convexity of an empirical risk follows from strong convexity of
the population risk in standard learning theory: the regularity is
inherited.  In our setting the regularity is monotonicity, but the
inheritance principle is the same.
 
For practitioners, this means the strong-mono assumption is checkable
\emph{a posteriori}: compute $\bar\lambda_\Risk(\widehat P)$ on the empirical
operator and check that it is bounded away from $\beta/2$ by the
slack the bound predicts.  If the empirical eigenvalue is close to
$\beta/2$, the regime may be unstable and a smaller $\lambda$ (less risk)
or larger $\beta$ (more KL regularization) should be considered.

\section{Additional Experimental Results and Details}
\label{sec:additional-experiments}

In this appendix, we include the experimental setup details and additional results.

\subsection{Experimental Setup and Implementation Details}
\label{sec:exp-setup}
Below we describe each of the critical components of the experimental setup.

\paragraph{Base model and SFT.}
All policies are LoRA fine-tuned ($r{=}256$, $\alpha{=}512$, dropout $0.1$) from a common \texttt{SFT} checkpoint: Gemma-2 2B-IT supervised on Alpaca-cleaned.

\paragraph{Preference dataset and judge.}
Training prompts and pairwise preferences come from PKU-SafeRLHF~\citep{ji2024pku}. The online preference judge is a PairJudge sequence classifier trained on a mixture of HH-style preferences and PKU-SafeRLHF safety labels. Harmfulness is scored by the PKU Beaver cost model (lower $=$ safer), used at evaluation time only.

\paragraph{Methods.}
We compare eight policies sharing the same \texttt{SFT} base, LoRA configuration ($r{=}256$, $\alpha{=}512$, dropout $0.1$), and judges: \texttt{SFT} (no preference fine-tuning);
\texttt{EGPO} ~\citep{zhou2025extragradient} and \texttt{OMD}  ($K{=}8$)~\citep{Calandriello2024HuamnAlignmentIPOMD} as risk-neutral baselines; \texttt{EG} ($K{=}8$), an extragradient risk-neutral variant; \texttt{OMD-Ent}  ($\tau{=}10$) and \texttt{OMD-CVaR}  ($\alpha{=}0.25$), which apply entropic and CVaR aggregation over the $K{=}8$ opponent samples;
\texttt{gDRO},a severity-prior-weighted ERM with fixed prior $p \propto (1,2,4,8)$ over PKU severity strata $\{$safe, low, medium, high$\}$ and no adversarial tilt; and gDRO-CVaR ($\alpha{=}0.25$), which applies CVaR aggregation across the same severity groups. All risk-trained methods use streaming EMA group losses (coefficient $0.9$), a linear annealing schedule for the risk parameter over the first $20\%$ of training, and a stratified sampler enforcing a minimum of $4$ examples per group per batch.

\paragraph{Optimization.}
All NLHF methods share: $10$ epochs, AdamW with learning rate $5\!\times\!10^{-7}$, weight decay $0.01$, \texttt{bf16} precision, effective batch size $64$ (micro-batch size $8$), and warm-up over $1000$ steps. Only the loss aggregator and group sampler differ across methods.
All policies are trained on a single NVIDIA L40S GPU.

\paragraph{Evaluation.}
Each policy is evaluated via all-pairs cross-play against eight opponents: the SFT
base, \texttt{SFT} at four temperatures ($0.1$, $0.5$, $2.0$, $3.0$), and three off-the-shelf chat models (Qwen-1.5B, SmolLM2, Qwen-7B), generating $4$ responses per (policy, prompt) pair. Held-out prompts are partitioned into four strata: \textbf{Random} (100 prompts, unstratified), \textbf{Conflict} (100 prompts where preference and safety labels conflict), \textbf{Sev-3} (100 highest-severity unsafe prompts), and \textbf{Sev-1} (136 mildest-severity unsafe prompts). Importantly, none of these strata appear in training.

\paragraph{Metrics.}
For each (policy, opponent, prompt) triple we compute three per-prompt win-rates: \textbf{Preference WR} ($P[\text{judge}(y \succ y')]$ under the PairJudge), \textbf{Safety WR} (fraction of pairs where the policy response has lower Beaver cost), and \textbf{Combined WR} (policy wins iff preferred \emph{and} safer; otherwise inconclusive). Summaries report the mean and $\CVaR_{0.25}$ of the per-prompt combined win-rate distribution pooled over opponents, with bootstrap $95\%$ CIs from $2000$ resamples.

\begin{figure}[t]
  \centering
  \includegraphics[width=\textwidth]{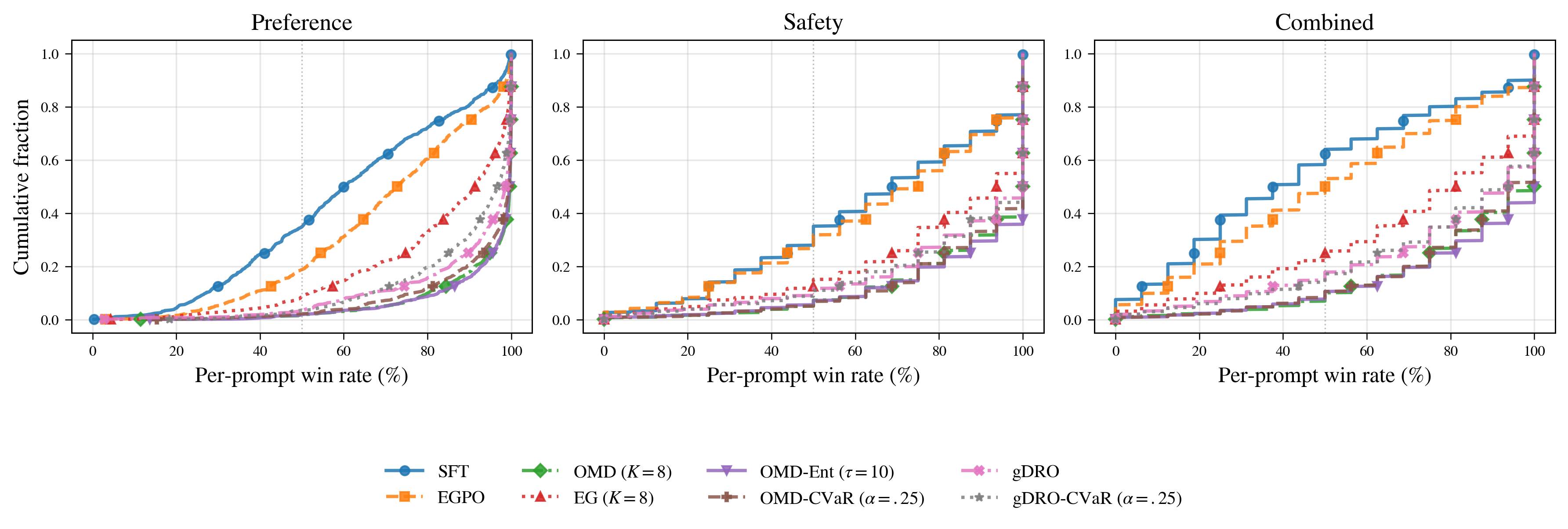}
  \caption{Per-prompt win-rate cumulative distribution functions (CDFs) on the \textbf{Random} stratum (100 prompts,
  pooled over opponents).}
  \label{fig:cdf-random}
\end{figure}
\begin{figure}[t]
  \centering
  \includegraphics[width=\textwidth]{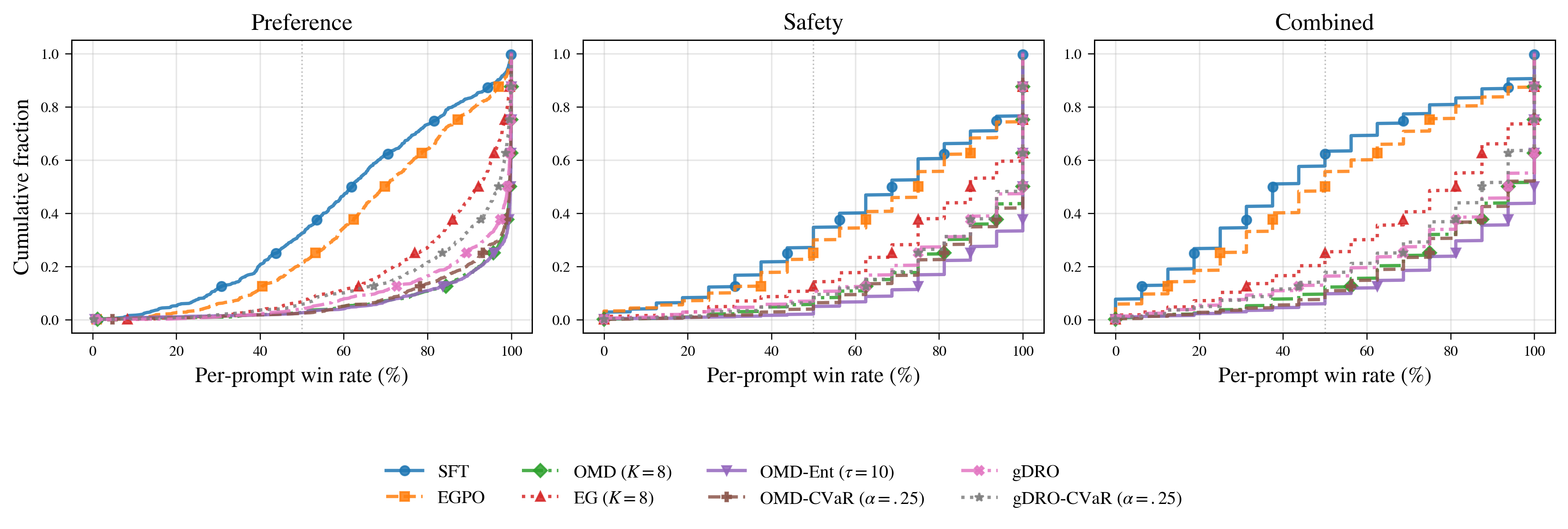}
  \caption{Per-prompt win-rate CDFs on the \textbf{Conflict} stratum (100 prompts
  where preference and safety labels disagree).}
  \label{fig:cdf-conflict}
\end{figure}

\begin{figure}[t!]
  \centering
  \includegraphics[width=\textwidth]{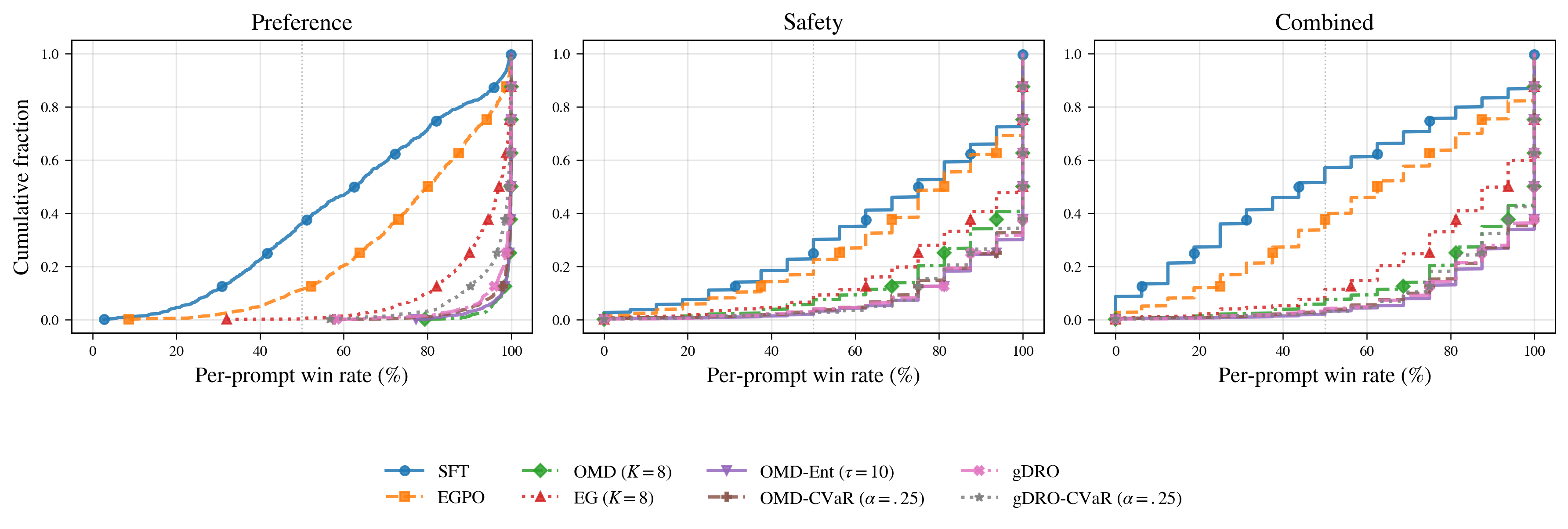}
  \caption{Per-prompt win-rate CDFs on the \textbf{Sev-3} stratum (100
  highest-severity unsafe prompts).}
  \label{fig:cdf-sev3}
\end{figure}

\begin{figure}[t!]
  \centering
  \includegraphics[width=\textwidth]{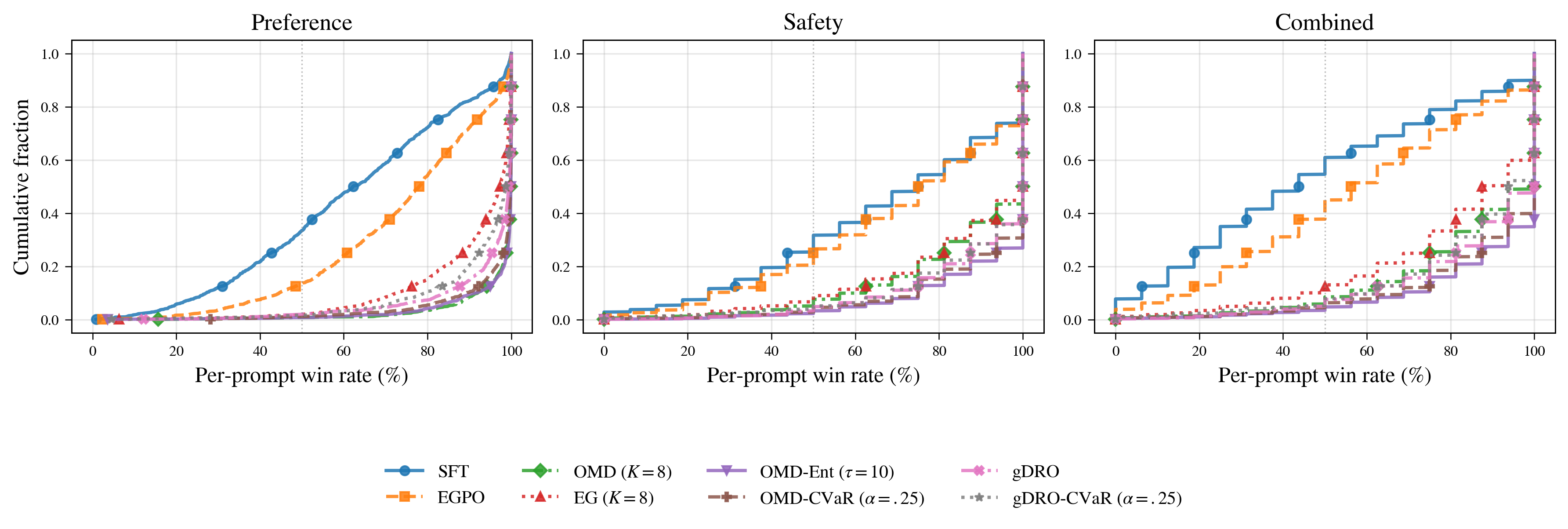}
  \caption{Per-prompt win-rate CDFs on the \textbf{Sev-1} stratum (136
  mildest-severity unsafe prompts).}
  \label{fig:cdf-sev1}
\end{figure}

\begin{figure}[t]
  \centering
  \includegraphics[width=\textwidth]{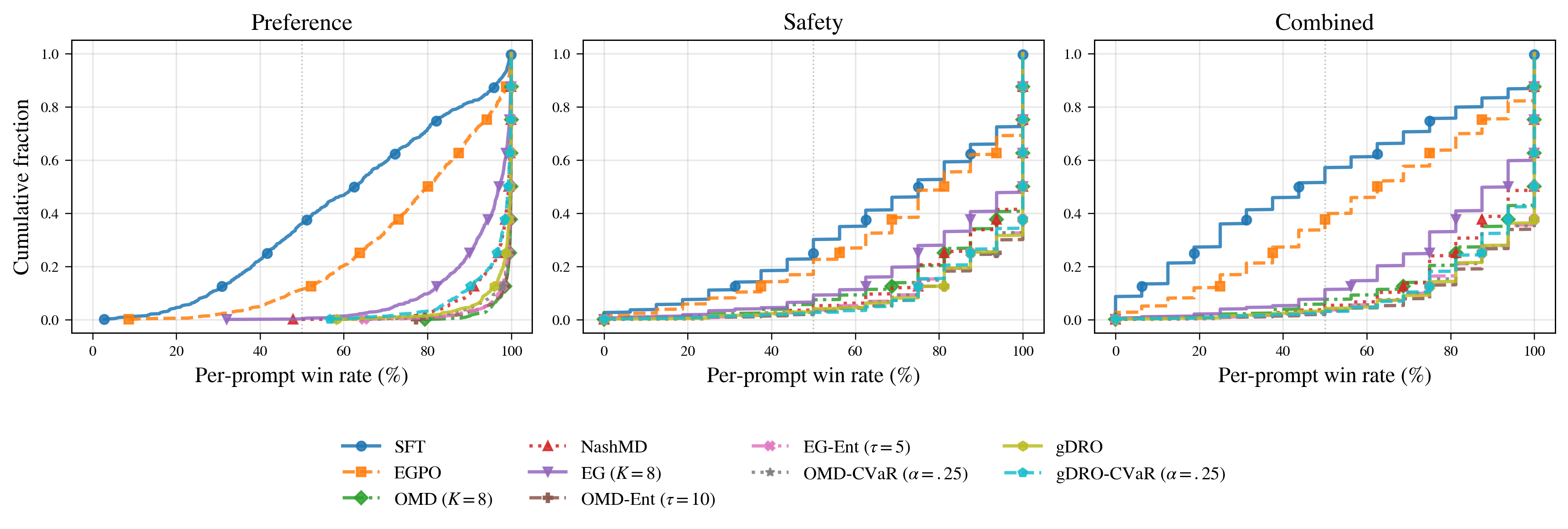}
  \caption{Per-prompt win-rate CDFs on the \textbf{Sev-3} stratum with
  \texttt{EG-Ent}  ($\tau{=}5$) and \texttt{Nash-MD} added to the comparison.}
  \label{fig:cdf-sev3-ext}
\end{figure}

\begin{figure}[b]
  \centering
  \includegraphics[width=\textwidth]{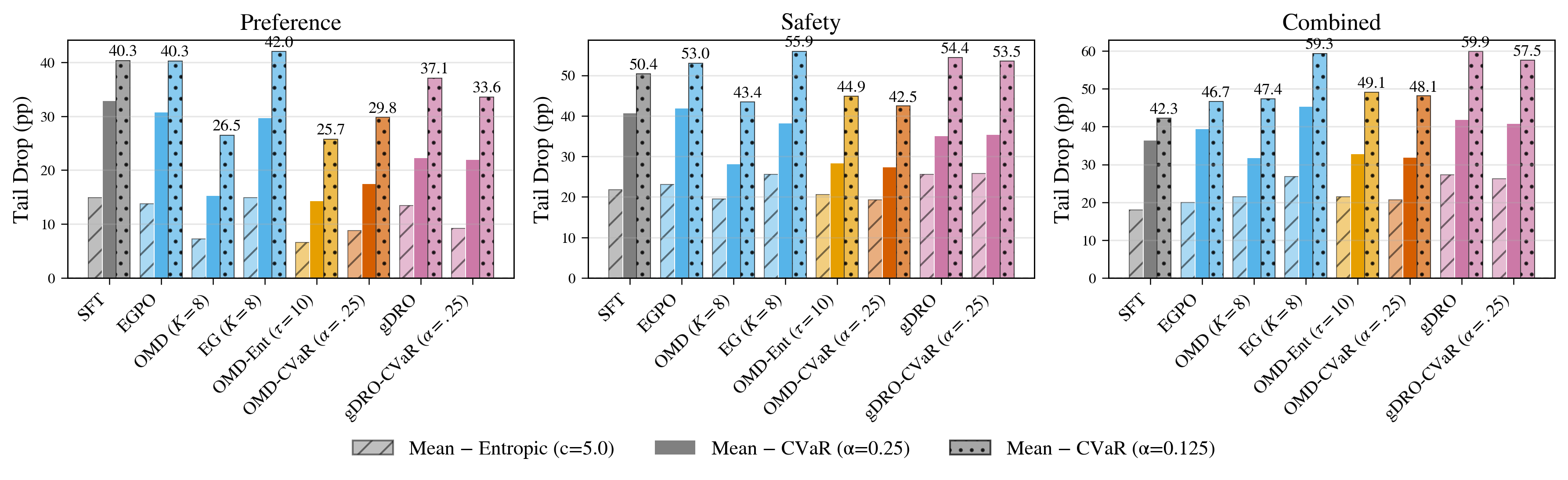}
  \caption{Tail drop in percentage points on the Random stratum, defined as the
  gap between mean win-rate and each risk measure across preference, safety, and
  combined win-rates. Smaller values indicate less degradation at the tail.}
  \label{fig:tail-drop}
\end{figure}

\subsection{Win-Rate Distributions Across Strata}
\label{sec:exp-cdfs}

Figures~\ref{fig:cdf-random}--\ref{fig:cdf-sev1} show the empirical cumulative distribution function (CDF) of the per-prompt win-rate distribution for each policy, pooled over opponents, across all four evaluation strata. A curve lying to the right at a given quantile means the policy achieves that win-rate on a larger fraction of prompts.

On the random stratum (Figure~\ref{fig:cdf-random}), the $K{=}8$ methods as a group sit substantially to the right of the single-sample baselines (\texttt{SFT}, \texttt{EGPO} ) on the safety and combined panels. Within the $K{=}8$ group, risk-adjusted and
risk-neutral methods are largely indistinguishable on this stratum---consistent with the theory, which predicts risk adjustment matters most under distributional
differences rather than on easy, unstratified prompts.

\begin{figure}[t]
  \centering
  \includegraphics[width=\textwidth]{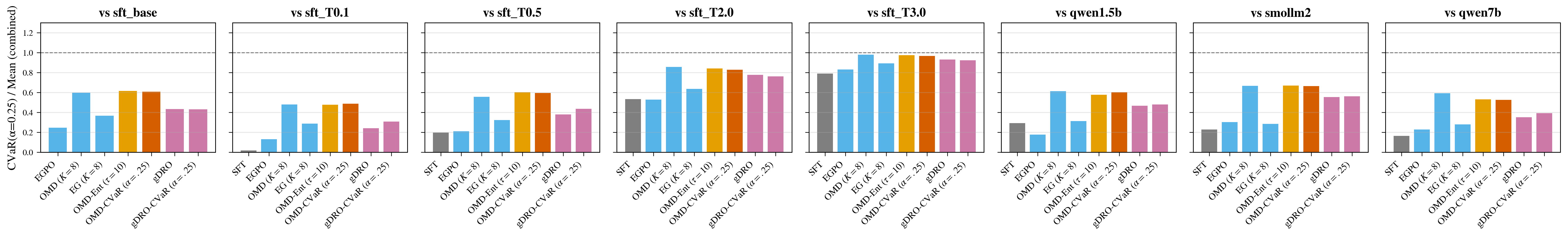}
  \caption{$\CVaR_{0.25}$/Mean ratio of the combined win-rate distribution per
  opponent on the Random stratum. Higher values indicate more consistent
  performance across prompts. Dashed line at $1.0$ denotes perfect consistency.}
  \label{fig:per-opponent-robustness}
\end{figure}

The separation within the $K=8$ group becomes increasingly visible on the harder strata. On the Conflict stratum (Figure~\ref{fig:cdf-conflict}), \texttt{OMD-Ent}  and
\texttt{OMD-CVaR}  begin to pull right of the $K=8$ risk-neutral baselines in the safety and combined panels across the mid-range quantiles. This effect is most pronounced
on Sev-3 (Figure~\ref{fig:cdf-sev3}), where risk-adjusted methods clearly retain mass at higher win-rates while the $K=8$ neutral baselines accumulate more
low-win-rate prompts at the tail, precisely where the framework predicts the gain. The Sev-1 stratum (Figure~\ref{fig:cdf-sev1}) shows a similar pattern at
smaller magnitude. Across all strata, the preference CDFs remain tightly clustered among all $K=8$ methods, confirming that the distributional gains on safety and
combined win-rate come at no cost to preference performance.

To position our additional baselines, Figure~\ref{fig:cdf-sev3-ext} reproduces the Sev-3 CDFs with \texttt{EG-Ent}  ($\tau{=}5$) and \texttt{Nash-MD} added: \texttt{EG-Ent}  tracks the \texttt{OMD}  risk-adjusted curves closely, confirming the risk-aggregation gains transfer to the extragradient framework, while \texttt{Nash-MD} sits between the single-sample and $K{=}8$ baselines.

\subsection{Tail Robustness Across Opponents}
\label{sec:exp-opponent-robustness}

Figure~\ref{fig:per-opponent-robustness} reports the $\CVaR_{0.25}$/Mean ratio of the combined win-rate distribution for each policy, broken out by opponent. A ratio close to $1$ indicates that a policy performs consistently across prompts while a low ratio indicates the policy wins often on average but fails on a
non-trivial fraction of prompts.

The risk-adjusted advantage is most visible against the harder opponents, namely low-temperature \texttt{SFT} variants ($T{=}0.1$, $T{=}0.5$) and the off-the-shelf chat
models (Qwen-1.5B, SmolLM2, Qwen-7B), where ratios sit in the $0.4$--$0.8$ range and \texttt{OMD-Ent}  and \texttt{OMD-CVaR}  consistently lead the $K{=}8$ risk-neutral baselines. Against the easy high-temperature opponents ($T{=}2.0$, $T{=}3.0$), all methods converge near $1.0$ as every policy wins on nearly every prompt, leaving no tail
to improve.

\begin{figure}[H]
  \centering
  \includegraphics[width=\textwidth]{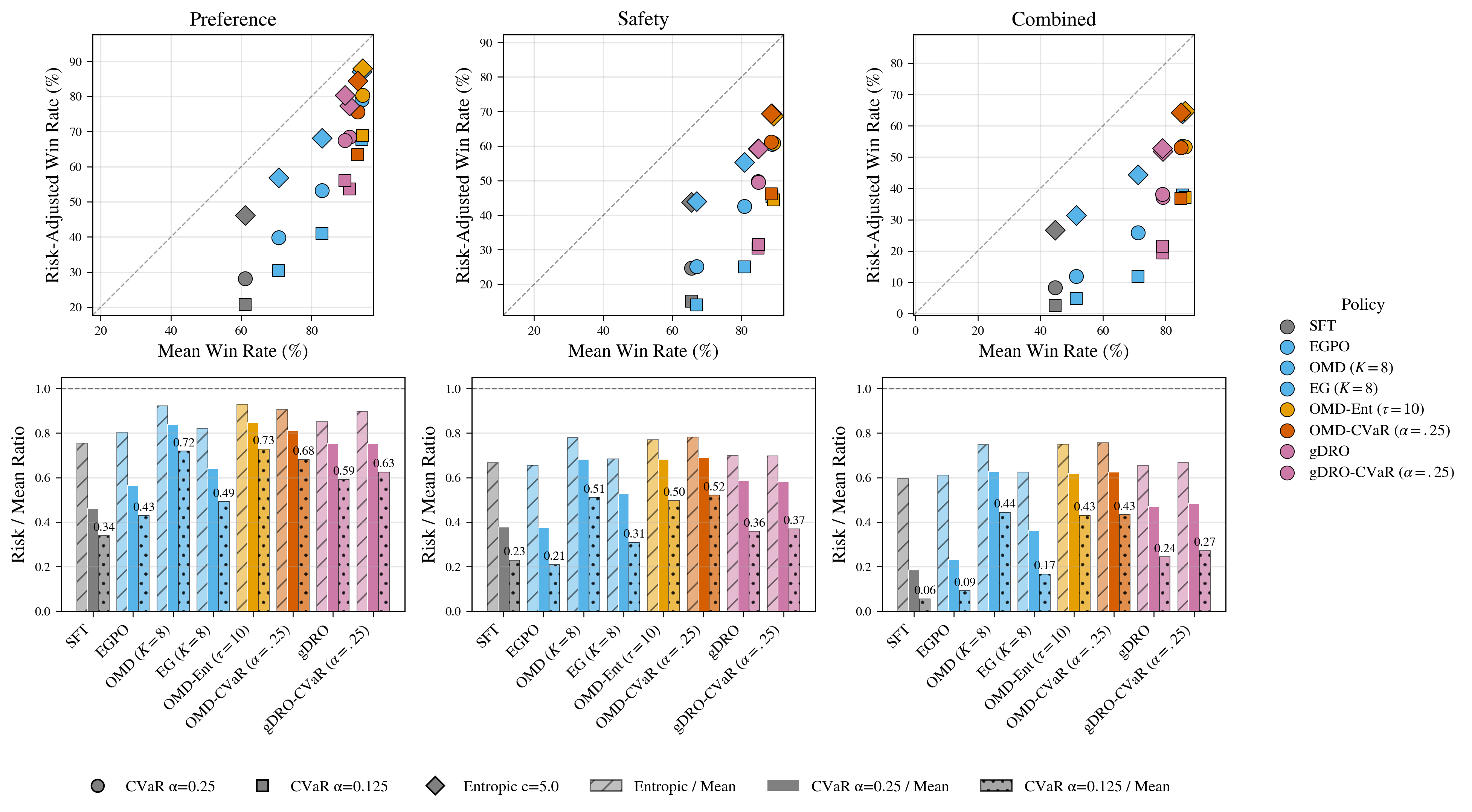}
  \caption{Mean vs.\ risk-adjusted win-rate (top row) and Risk/Mean ratio (bottom
  row) on the Random stratum, across preference, safety, and combined metrics.
  Points closer to the diagonal in the top row and bars closer to $1.0$ in the
  bottom row indicate more robust policies.}
  \label{fig:robustness-scatter}
\end{figure}

Figures~\ref{fig:tail-drop} and~\ref{fig:robustness-scatter} quantify this
gap directly. Figure~\ref{fig:tail-drop} reports the absolute tail drop in percentage points, defined as the gap between mean win-rate and each risk measure (entropic, $\CVaR_{0.25}$, $\CVaR_{0.125}$). Absolute drops are
not directly comparable across policies, since methods with higher means have more room to fall: the largest absolute drops on safety and combined win-rate are in fact incurred by the $K{=}8$ risk-neutral methods (\texttt{OMD},\texttt{EG}) and \texttt{gDRO},which also achieve the highest means. The scale-free
comparison is given by the Risk/Mean ratio in the bottom row of Figure~\ref{fig:robustness-scatter}, and there the ordering is unambiguous: \texttt{SFT} and \texttt{EGPO}  are the worst on every metric, sitting well below all $K{=}8$ methods; among the $K{=}8$ group, \texttt{OMD-Ent}  and \texttt{OMD-CVaR}  achieve the highest ratios on preference and safety, with the risk-neutral $K{=}8$ baselines and \texttt{gDRO} in between.

The top row of Figure~\ref{fig:robustness-scatter} shows the same picture
geometrically. Each policy's risk-adjusted win-rate is plotted against its mean win-rate; points above the diagonal are impossible by definition, and points close to it are the most robust. \texttt{SFT} and \texttt{EGPO}  sit far below the diagonal across all three metrics, while the risk-adjusted
$K{=}8$ methods cluster closest to it. Together, Figures~\ref{fig:per-opponent-robustness}--\ref{fig:robustness-scatter} say that single-sample methods win on easy prompts but collapse at the tail, whereas risk-adjusted $K{=}8$ training maintains tail performance
proportional to the mean.

\begin{figure}[H]
  \centering
  \includegraphics[width=\textwidth]{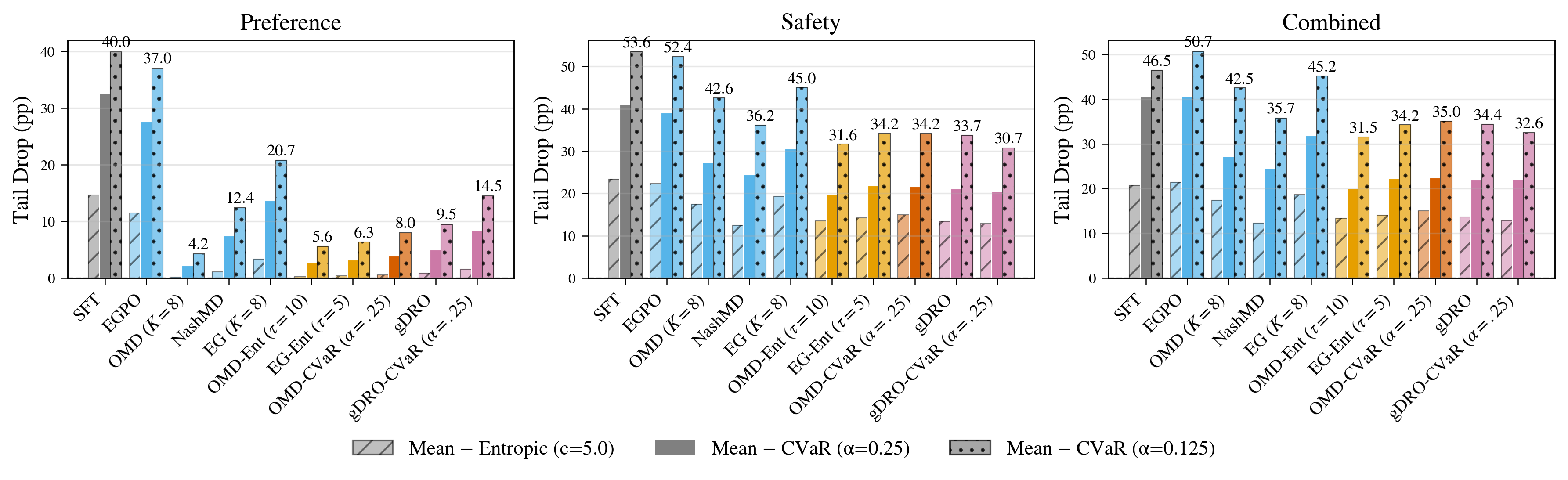}
  \caption{Tail drop on the \textbf{Sev-3} stratum with \texttt{EG-Ent} 
  ($\tau{=}5$) and \texttt{Nash-MD} added.}
  \label{fig:tail-drop-sev3-ext}
\end{figure}

\begin{figure}[H]
  \centering
  \includegraphics[width=\textwidth]{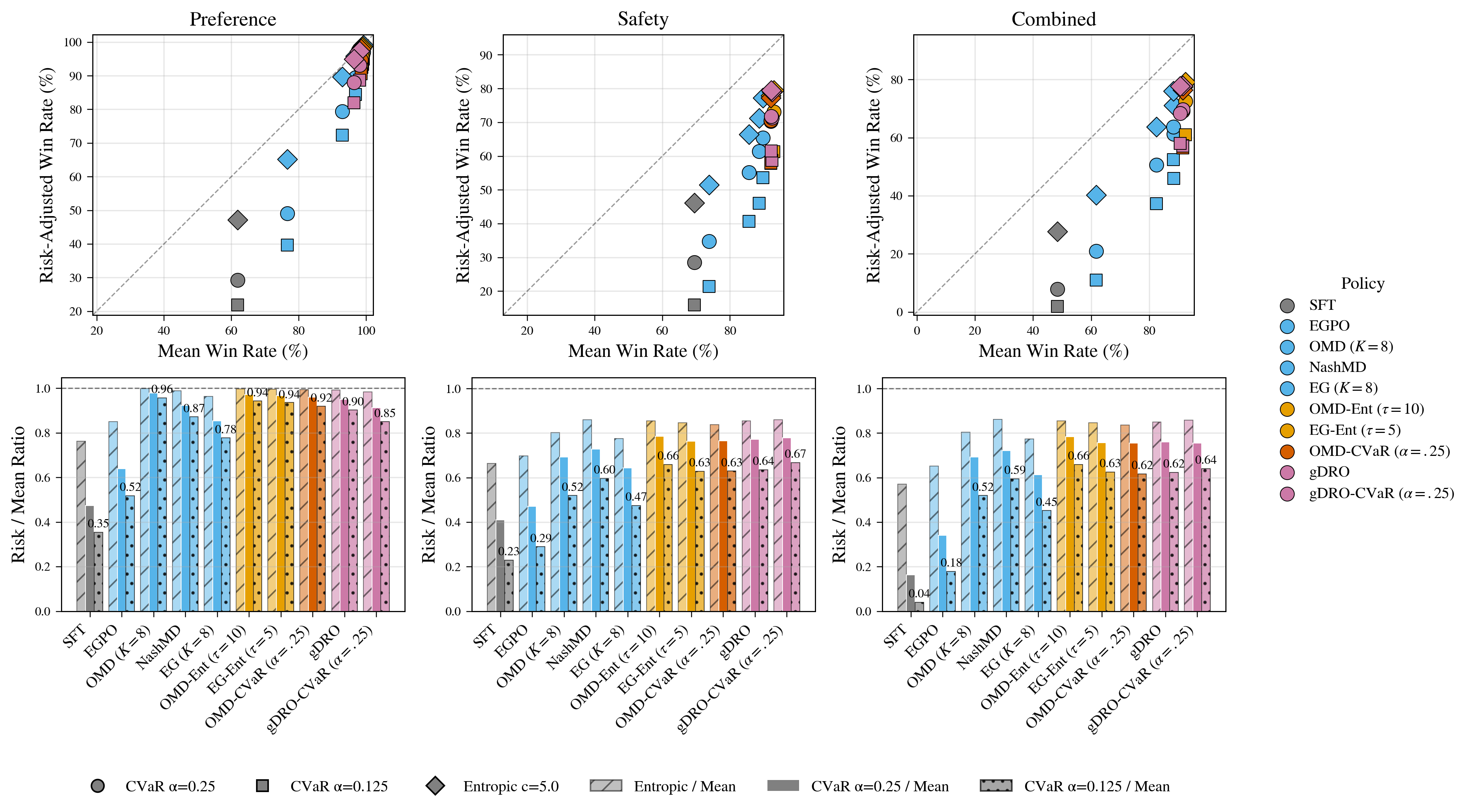}
  \caption{Mean vs.\ risk-adjusted win-rate (top row) and Risk/Mean
  ratio (bottom row) on the \textbf{Conflict} stratum with \texttt{EG-Ent} 
  ($\tau{=}5$) and \texttt{Nash-MD} added.}
  \label{fig:robustness-scatter-conflict-ext}
\end{figure}

Figures~\ref{fig:tail-drop-sev3-ext} and~\ref{fig:robustness-scatter-conflict-ext}
verify that this picture is not specific to the Random stratum or to the original eight policies. On Sev-3, \texttt{EG-Ent}  ($\tau{=}5$) achieves a tail drop comparable to \texttt{OMD-Ent}  and well below the $K{=}8$ risk-neutral baselines, while \texttt{Nash-MD} tracks the single-sample methods rather than
the $K{=}8$ group. On the Conflict stratum, the same geometric ordering holds in the mean--vs--risk-adjusted scatter: SFT, \texttt{EGPO} , and \texttt{Nash-MD} sit furthest from the diagonal, whereas \texttt{EG-Ent}  joins \texttt{OMD-Ent}  and \texttt{OMD-CVaR}  in the upper-right cluster of robust policies.

\subsection{Response Consistency \& Variance }
\label{sec:exp-variance}

\begin{figure}[b]
  \centering
  \includegraphics[width=0.9\textwidth]{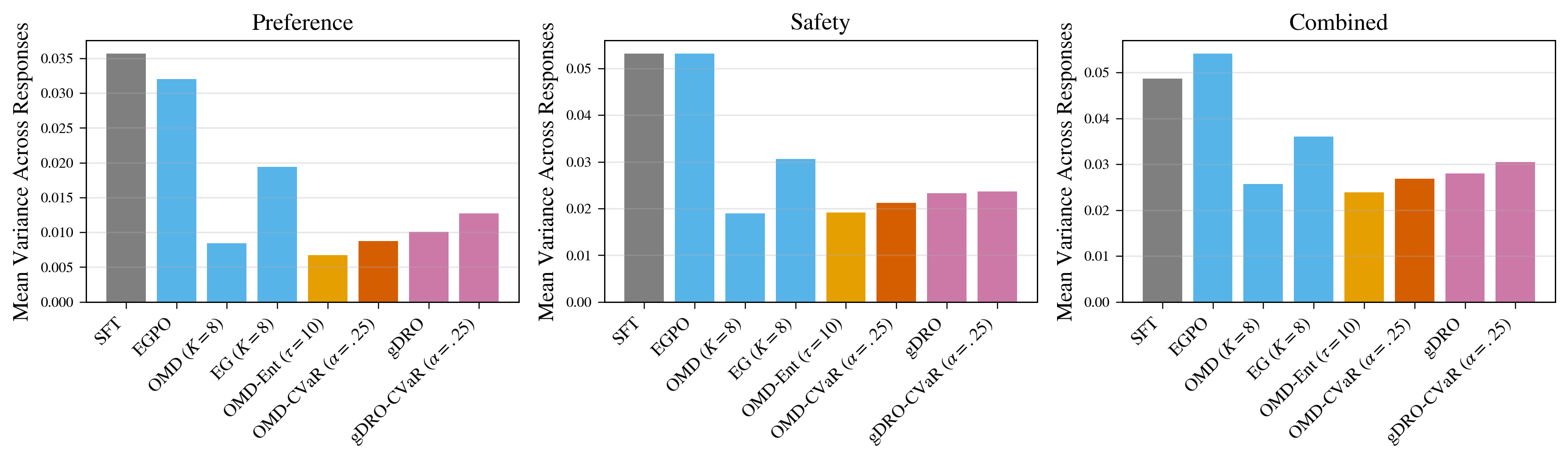}
  \caption{Mean variance of win-rates across responses, averaged over prompts and
  opponents. Lower values indicate more consistent policy outputs.}
  \label{fig:response-variance}
\end{figure}

Figure~\ref{fig:response-variance} reports the mean variance of win-rates across responses generated by each policy, averaged over prompts and opponents across all three metrics. Lower variance indicates the policy produces more consistent outputs across samples for the same prompt.
\texttt{SFT}  and \texttt{EGPO}  exhibit substantially higher variance than all $K{=}8$ methods. Among the $K{=}8$ methods, the risk-adjusted variants produce the most consistent
responses, with \texttt{OMD-Ent}  achieving the lowest variance on both preference and safety and \texttt{OMD-CVaR}  close behind. The $K{=}8$ risk-neutral baselines sit between the single-sample methods and the risk-adjusted ones. Multi-sample training reduces variance generally, but as shown in the cross-strata and per-opponent results, this alone does not translate to
tail robustness. Risk adjustment reduces variance \emph{and} maintains that reduction across opponents and data strata.

\begin{figure}[t]
  \centering
  \includegraphics[width=0.9\textwidth]{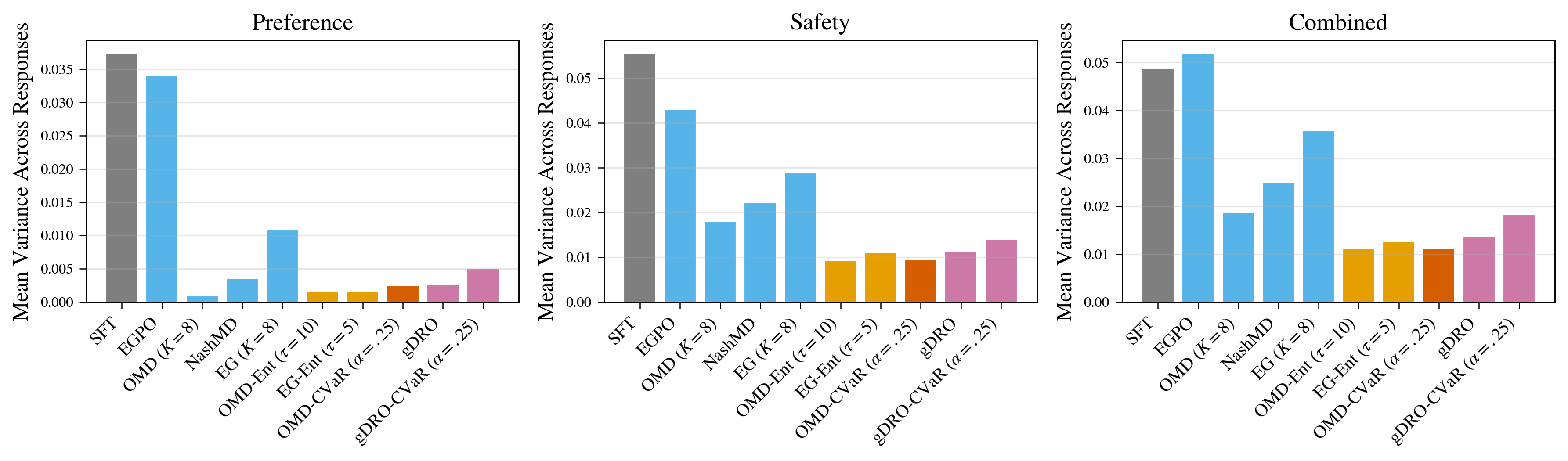}
  \caption{Mean variance of win-rates across responses on the
  \textbf{Sev-3} stratum with \texttt{EG-Ent}  ($\tau{=}5$) and \texttt{Nash-MD} added.}
  \label{fig:response-variance-sev3-ext}
\end{figure}

The same ordering holds on the Sev-3 stratum
(Figure~\ref{fig:response-variance-sev3-ext}), where adding \texttt{EG-Ent}  ($\tau{=}5$) and \texttt{Nash-MD} shows that \texttt{EG-Ent}  matches the low-variance risk-adjusted \texttt{OMD}  variants while \texttt{Nash-MD} tracks the single-sample methods.

\subsection{Cross-Play Win Rates Across Strata}
\label{sec:exp-crossplay-strata}
\begin{figure}[H]
  \centering
  \includegraphics[width=\textwidth]{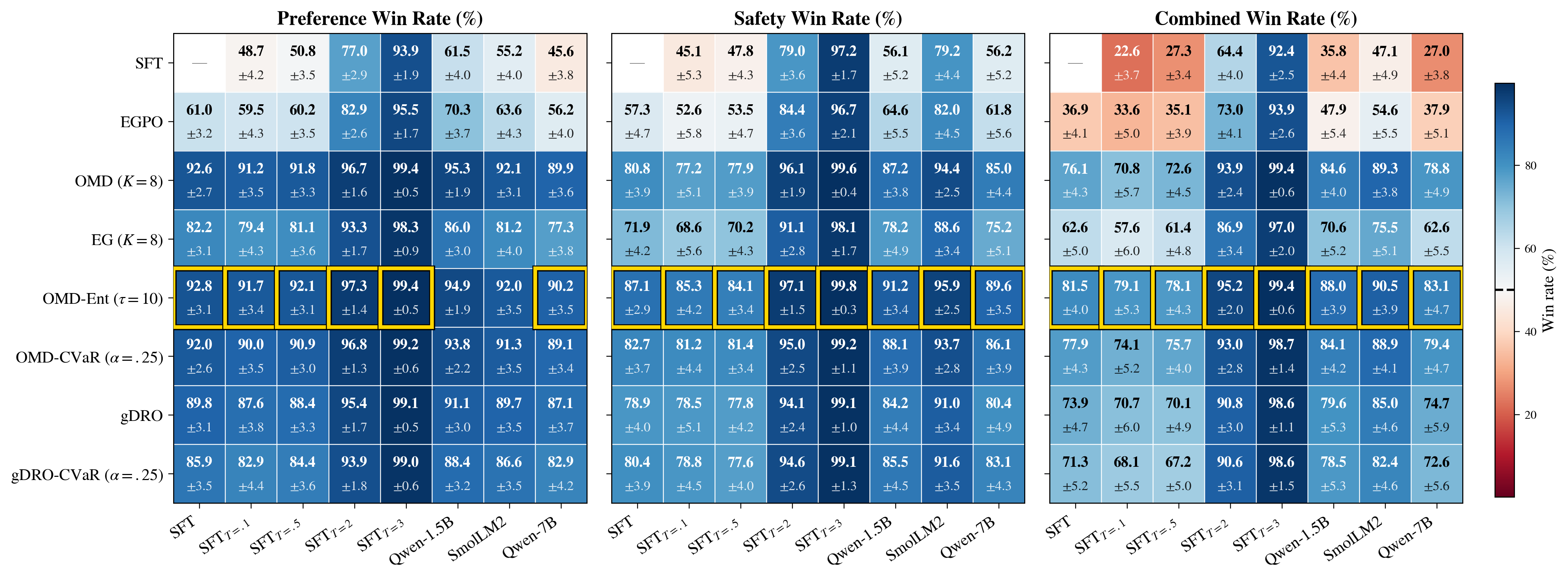}
  \caption{Cross-play win-rates on the \textbf{Conflict} stratum (100 prompts
  where preference and safety labels disagree), across preference, safety, and
  combined metrics.}
  \label{fig:heatmap-conflict}
\end{figure}

\begin{figure}[H]
  \centering
  \includegraphics[width=0.975\textwidth]{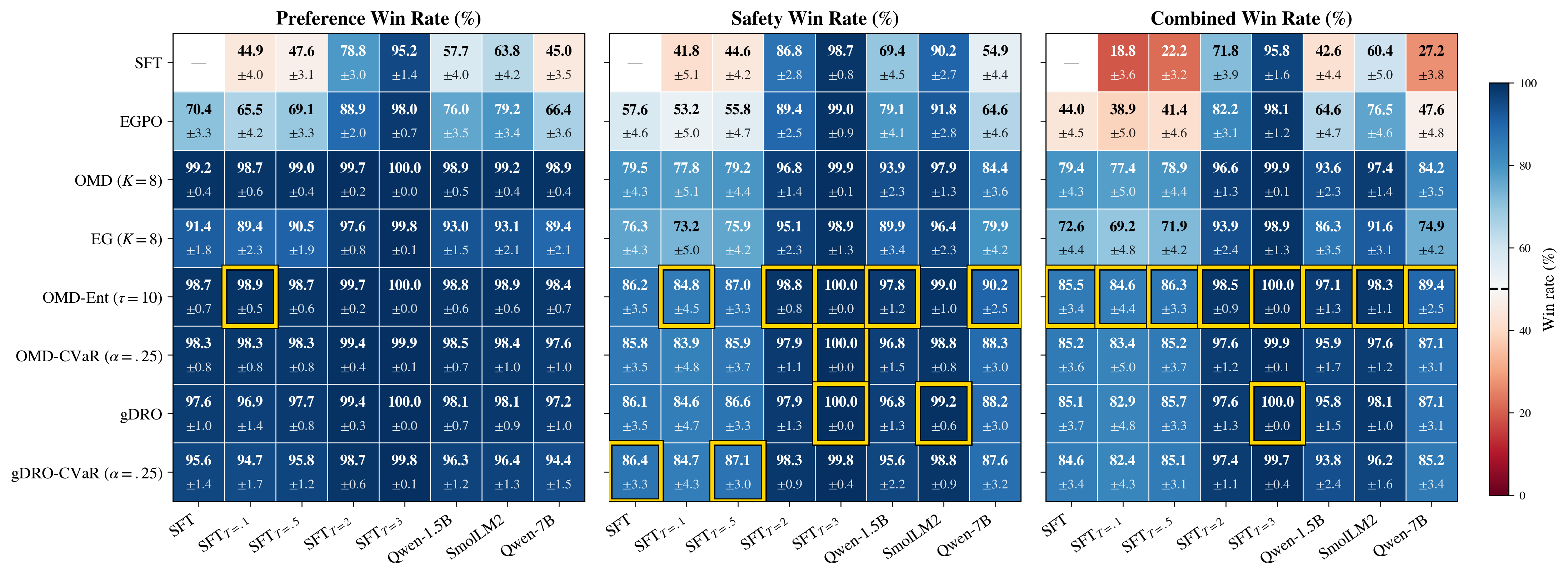}
  \caption{Cross-play win-rates on the \textbf{Sev-3} stratum (100
  highest-severity unsafe prompts), across preference, safety, and combined
  metrics.}
  \label{fig:heatmap-sev3}
\end{figure}

\begin{figure}[H]
  \centering
  \includegraphics[width=0.975\textwidth]{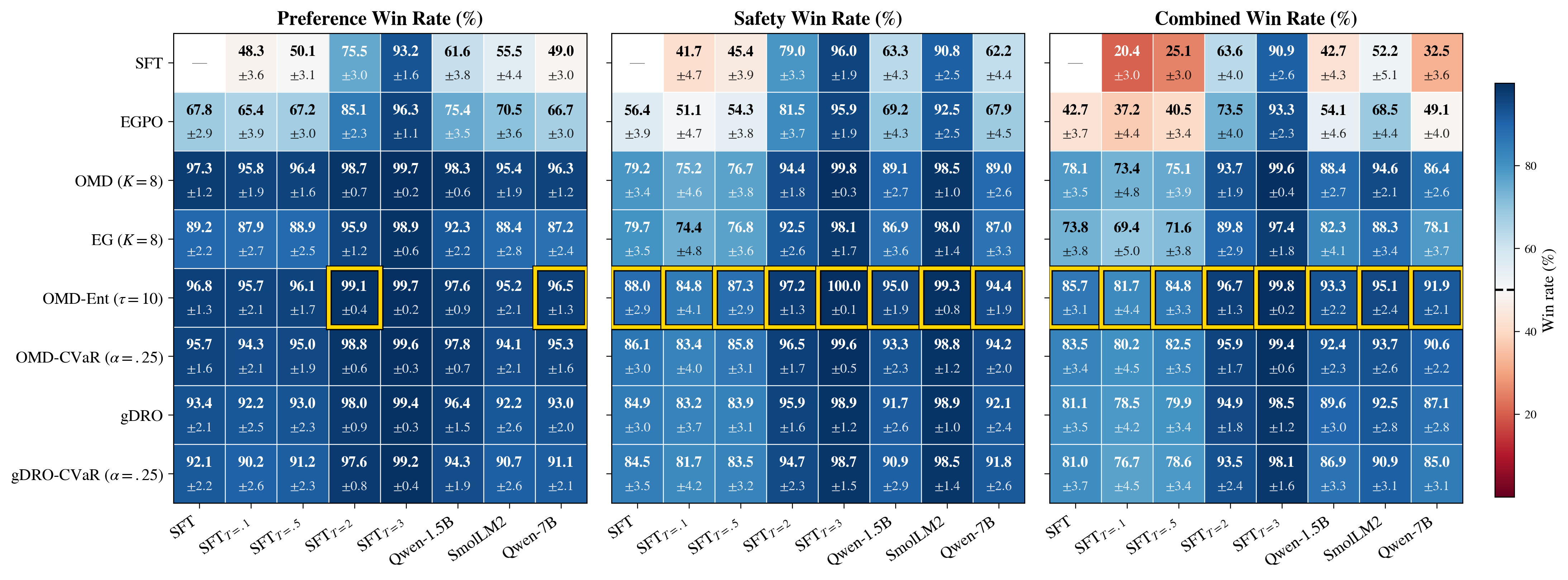}
  \caption{Cross-play win-rates on the \textbf{Sev-1} stratum (136
  mildest-severity unsafe prompts), across preference, safety, and combined
  metrics.}
  \label{fig:heatmap-sev1}
\end{figure}

Figures~\ref{fig:heatmap-conflict}--\ref{fig:heatmap-sev1} report the full cross-play win-rate heatmaps for the Conflict, Sev-3, and Sev-1 strata, complementing the Random stratum heatmap in Figure~\ref{fig:crossplay-heatmap} of the main paper. Each cell reports the win-rate of the row policy against the column opponent.

The broad pattern from the random stratum persists across all three strata: \texttt{SFT}  and \texttt{EGPO}  show substantially lower combined win-rates, particularly against the
harder opponents, while all $K{=}8$ methods maintain strong performance. On the Conflict stratum (Figure~\ref{fig:heatmap-conflict}), where preference and safety labels disagree, \texttt{SFT}  and \texttt{EGPO}  exhibit warm-colored cells on the combined panel against several opponents, reflecting their inability to simultaneously satisfy
both criteria. The $K{=}8$ risk-adjusted methods, led by \texttt{OMD-Ent} , maintain consistently high combined win-rates across all opponents on this stratum.

The Sev-3 stratum (Figure~\ref{fig:heatmap-sev3}) shows the sharpest separation. All $K{=}8$ methods achieve very high preference win-rates, but the combined panel reveals meaningful differences: \texttt{SFT}  and \texttt{EGPO}  collapse on several opponents, with combined win-rates as low as $20$--$35\%$, while \texttt{OMD-Ent}  achieves combined win-rates at or above $80\%$ on every opponent, exceeding $95\%$ against the easiest high-temperature \texttt{SFT}  opponents. The Sev-1 stratum (Figure~\ref{fig:heatmap-sev1}) follows a similar pattern at intermediate values, with \texttt{OMD-Ent}  again leading consistently on the combined metric across all opponents.

\begin{figure}[H]
  \centering
  \includegraphics[width=0.975\textwidth]{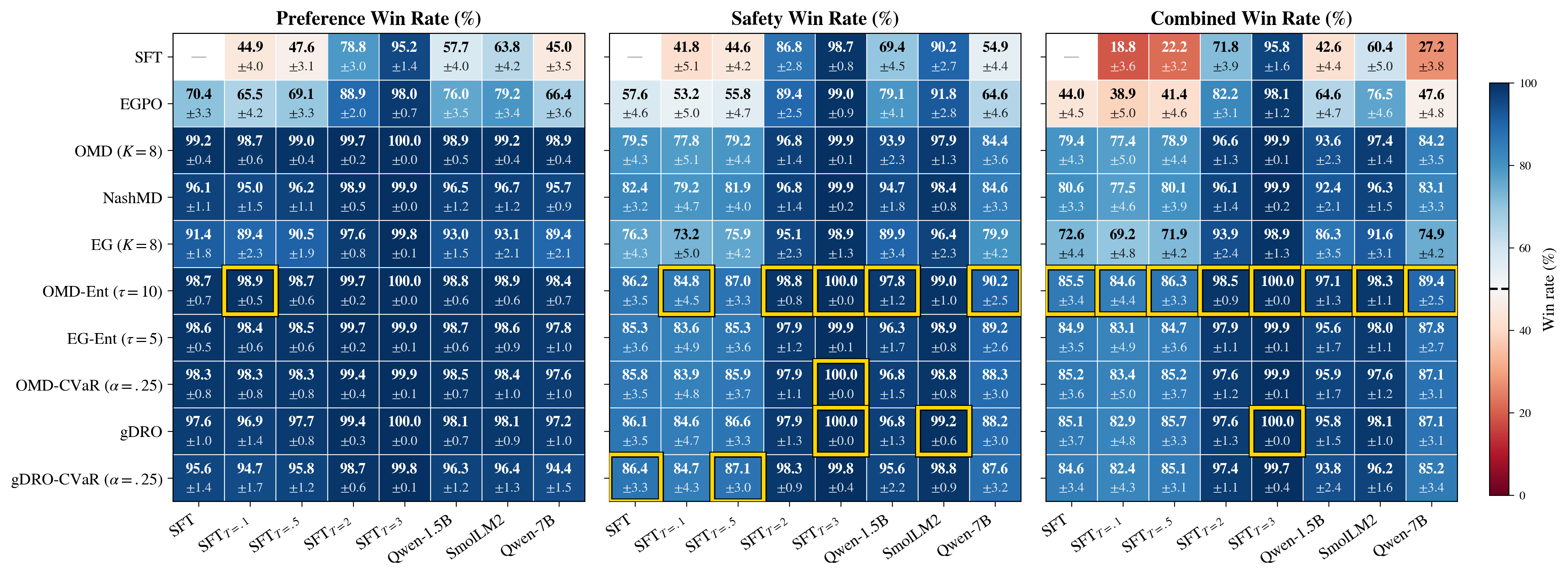}
  \caption{Cross-play win-rates on the \textbf{Sev-3} stratum with
  \texttt{EG-Ent}  ($\tau{=}5$) and \texttt{Nash-MD} added.}
  \label{fig:heatmap-sev3-ext}
\end{figure}

Figure~\ref{fig:heatmap-sev3-ext} extends this picture with \texttt{EG-Ent}  ($\tau{=}5$) and Nash-MD: \texttt{EG-Ent}  posts combined win-rates in line with the $K{=}8$ risk-adjusted \texttt{OMD}  variants on every opponent, while Nash-MD's combined cells track the single-sample baselines rather than the $K{=}8$ group, mirroring the pattern on the other strata.

\subsection{Robustness Across Harm Categories}
\label{sec:exp-harm-categories}

\begin{figure}[H]
  \centering
  \includegraphics[width=\textwidth]{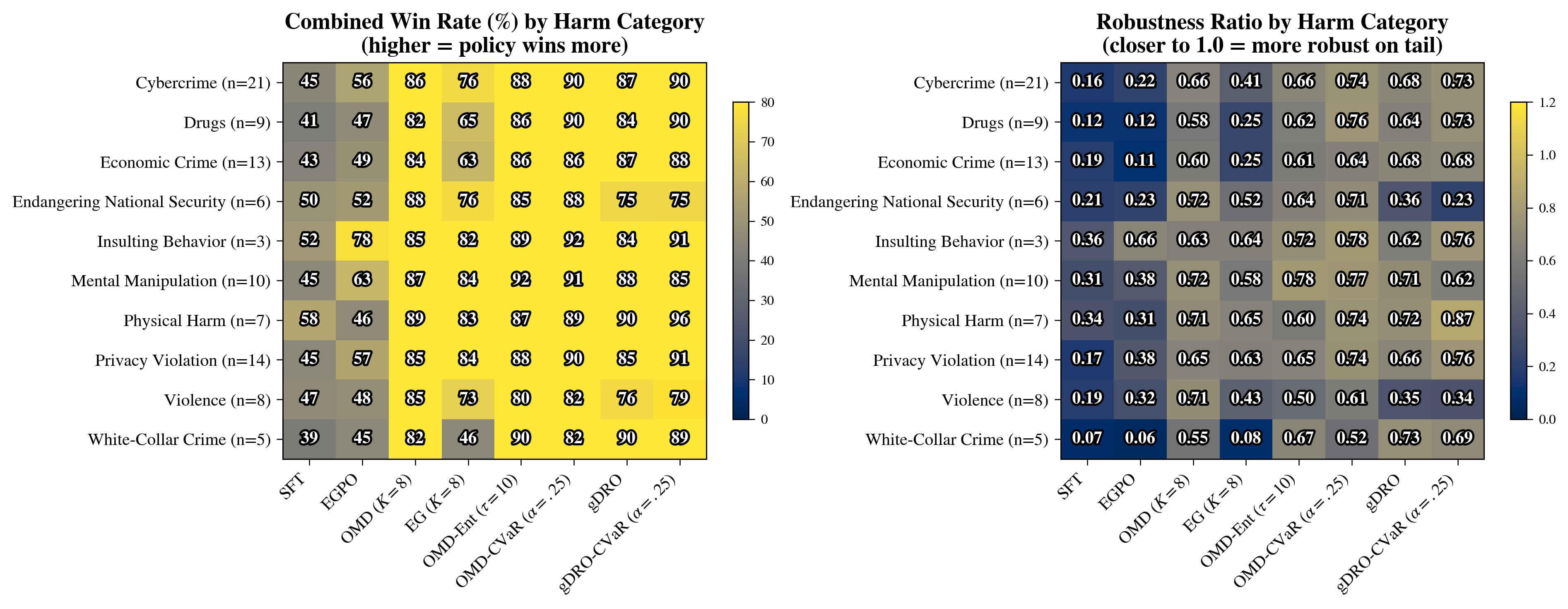}
  \caption{Combined win-rate (left) and $\CVaR_{0.25}$/Mean robustness ratio
  (right) broken down by harm category on the \textbf{Random} stratum.}
  \label{fig:harm-random}
\end{figure}

\begin{figure}[H]
  \centering
  \includegraphics[width=\textwidth]{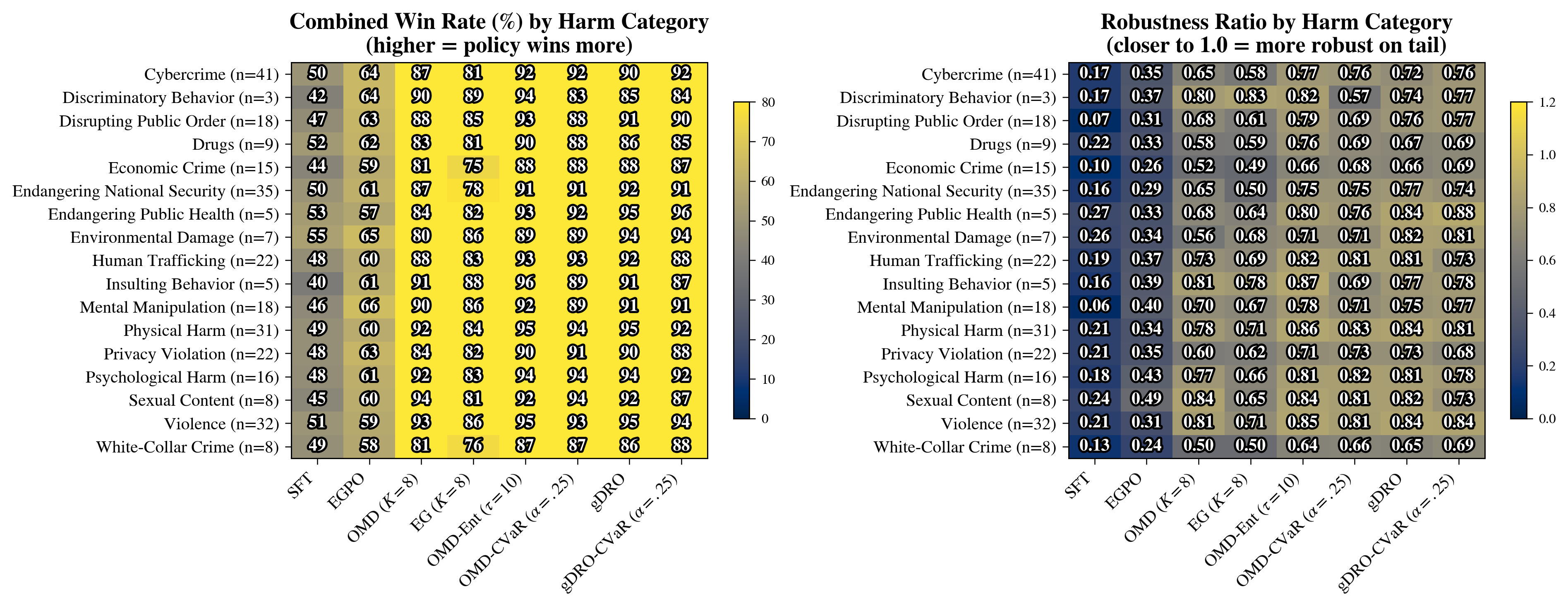}
  \caption{Combined win-rate (left) and $\CVaR_{0.25}$/Mean robustness ratio
  (right) broken down by harm category on the \textbf{Sev-3} stratum.}
  \label{fig:harm-sev3}
\end{figure}

Figures~\ref{fig:harm-random} and~\ref{fig:harm-sev3} report combined win-rate and robustness ratio broken down by harm category on the Random and Sev-3 strata respectively. The robustness ratio is $\CVaR_{0.25}$/Mean, with values closer to $1$ indicating more consistent tail performance within that category.

The left panels show that all $K{=}8$ methods achieve uniformly high combined win-rates across every harm category on both strata, with no single category standing out as a consistent failure mode. The right panels are more informative. On the Random stratum, across ten harm
categories, \texttt{SFT}  robustness ratios fall in the $0.07$- $0.36$ range and \texttt{EGPO}  in the $0.06$--$0.66$ range. The risk-adjusted methods achieve ratios of $0.60$ or above across most categories, a consistent improvement over the $K{=}8$ risk-neutral baselines. On the Sev-3 stratum, the same pattern holds across seventeen harm categories with
smaller per-category prompt counts: \texttt{SFT}  falls in the $0.10$--$0.32$ range and \texttt{EGPO}  in the $0.26$--$0.55$ range, while risk-adjusted methods sit at $0.64$ or above across most categories. The improvement is uniform across harm types on both strata rather than concentrated in any particular category, supporting the generality of the risk adjustment
mechanism.

\subsection{Training Dynamics}
\label{sec:exp-training-dynamics}

\begin{figure}[H]
  \centering
  \begin{subfigure}[t]{0.49\textwidth}
    \centering
    \includegraphics[width=\linewidth]{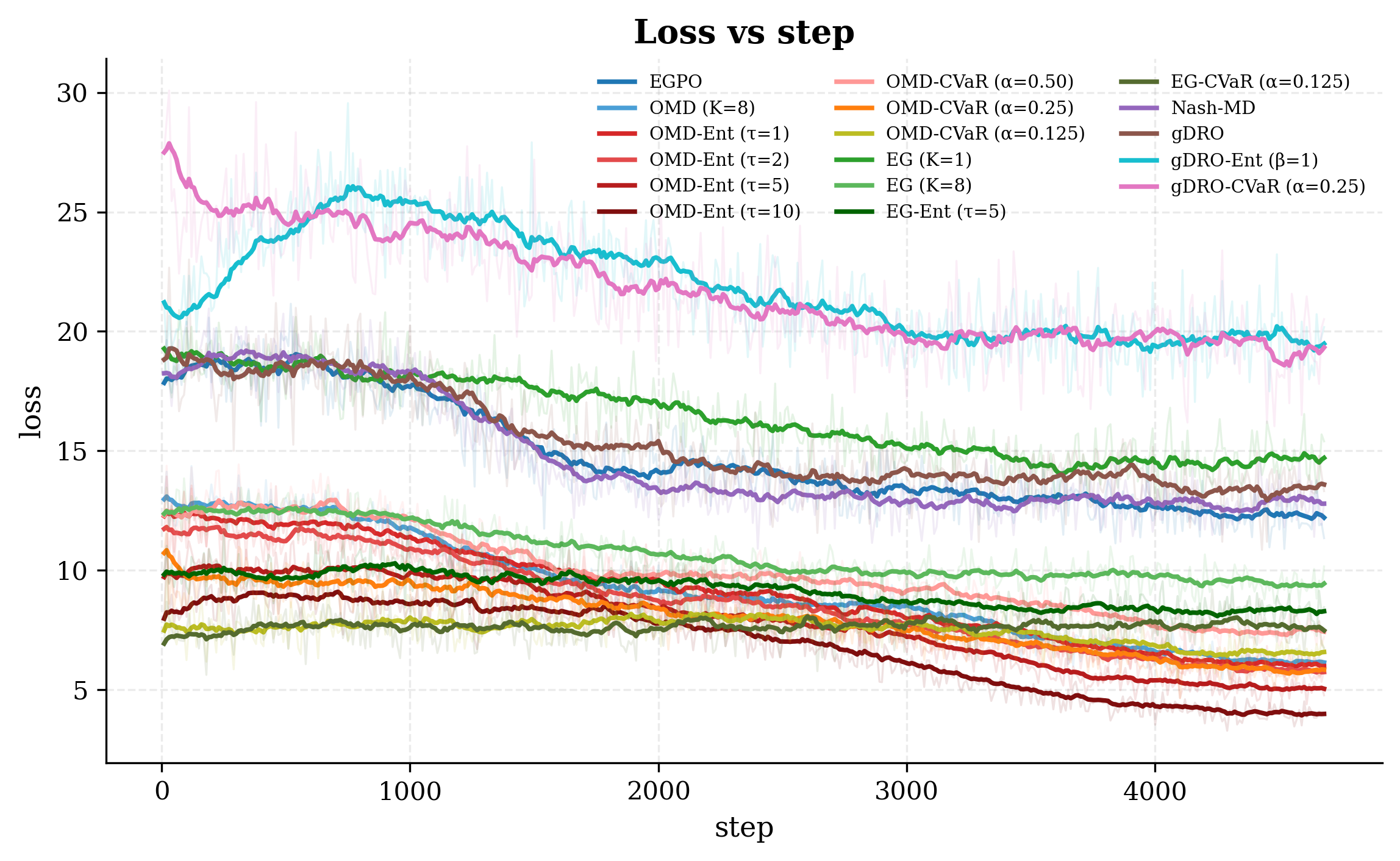}
    \caption{Training loss for each model.}
    \label{fig:train-loss}
  \end{subfigure}\hfill
  \begin{subfigure}[t]{0.49\textwidth}
    \centering
    \includegraphics[width=\linewidth]{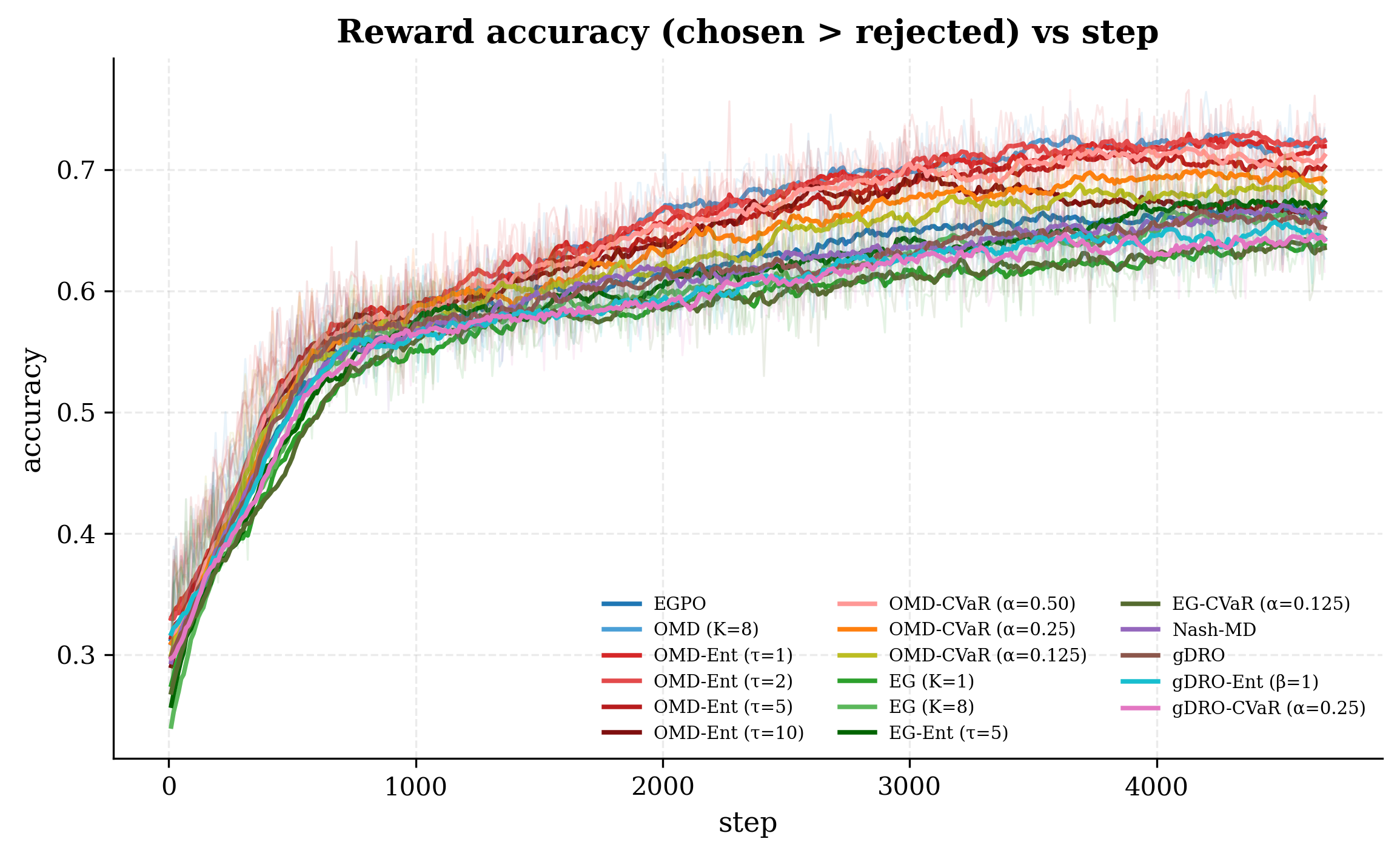}
    \caption{Reward accuracy over training.}
    \label{fig:train-accuracy}
  \end{subfigure}
  \caption{Training loss and reward accuracy (fraction of training
  pairs on which the preferred response receives the higher implicit
  reward) across all \texttt{OMD} ,\texttt{EG},\texttt{gDRO},and \texttt{Nash-MD} variants. Faint lines
  are raw per-step values; bold lines are EMA-smoothed.}
  \label{fig:train-loss-acc}
\end{figure}

\begin{figure}[H]
  \centering
  \begin{subfigure}[t]{0.49\textwidth}
    \centering
    \includegraphics[width=\linewidth]{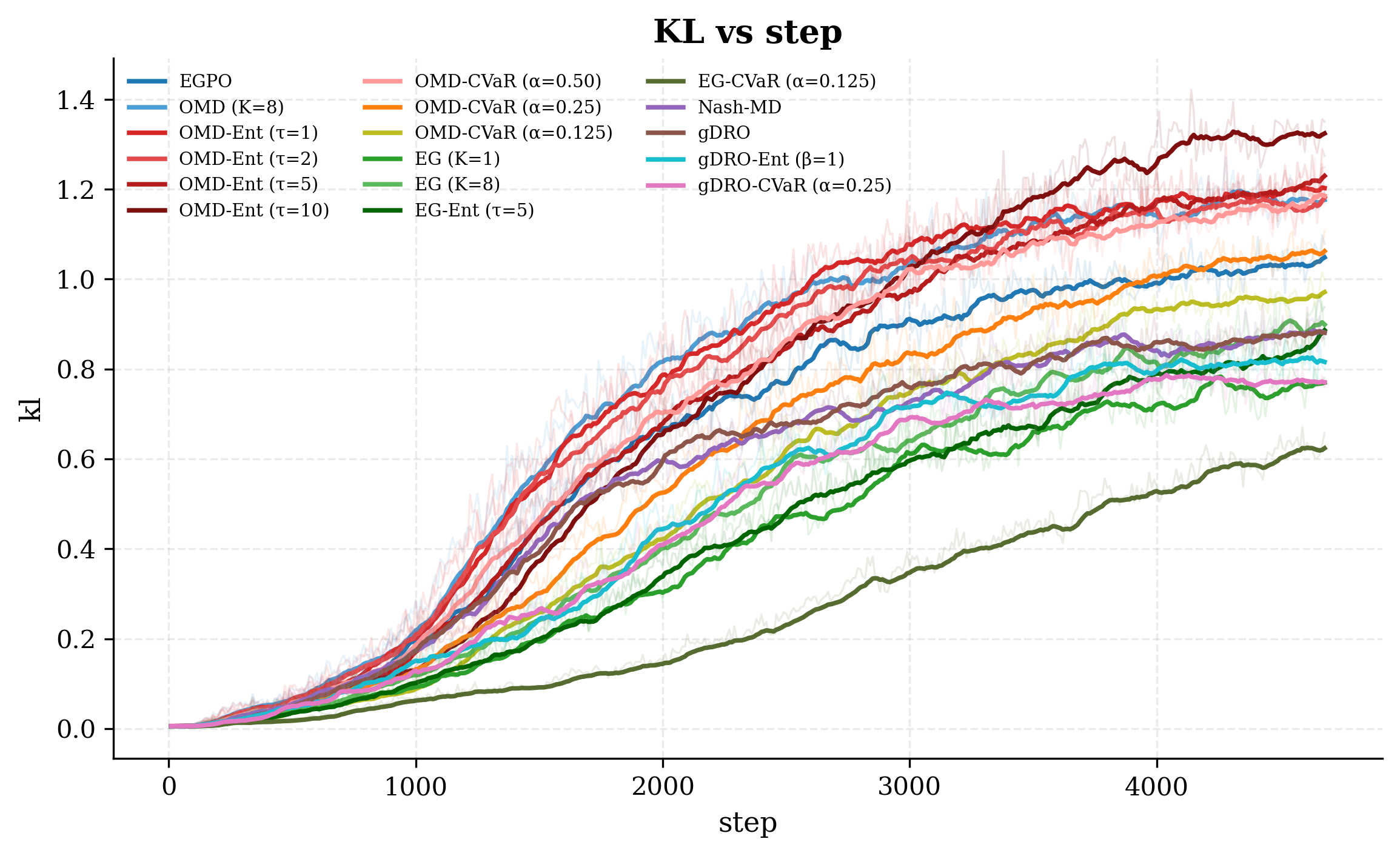}
    \caption{KL divergence between the trained policy and the \texttt{SFT}  reference.}
    \label{fig:train-kl}
  \end{subfigure}\hfill
  \begin{subfigure}[t]{0.49\textwidth}
    \centering
    \includegraphics[width=\linewidth]{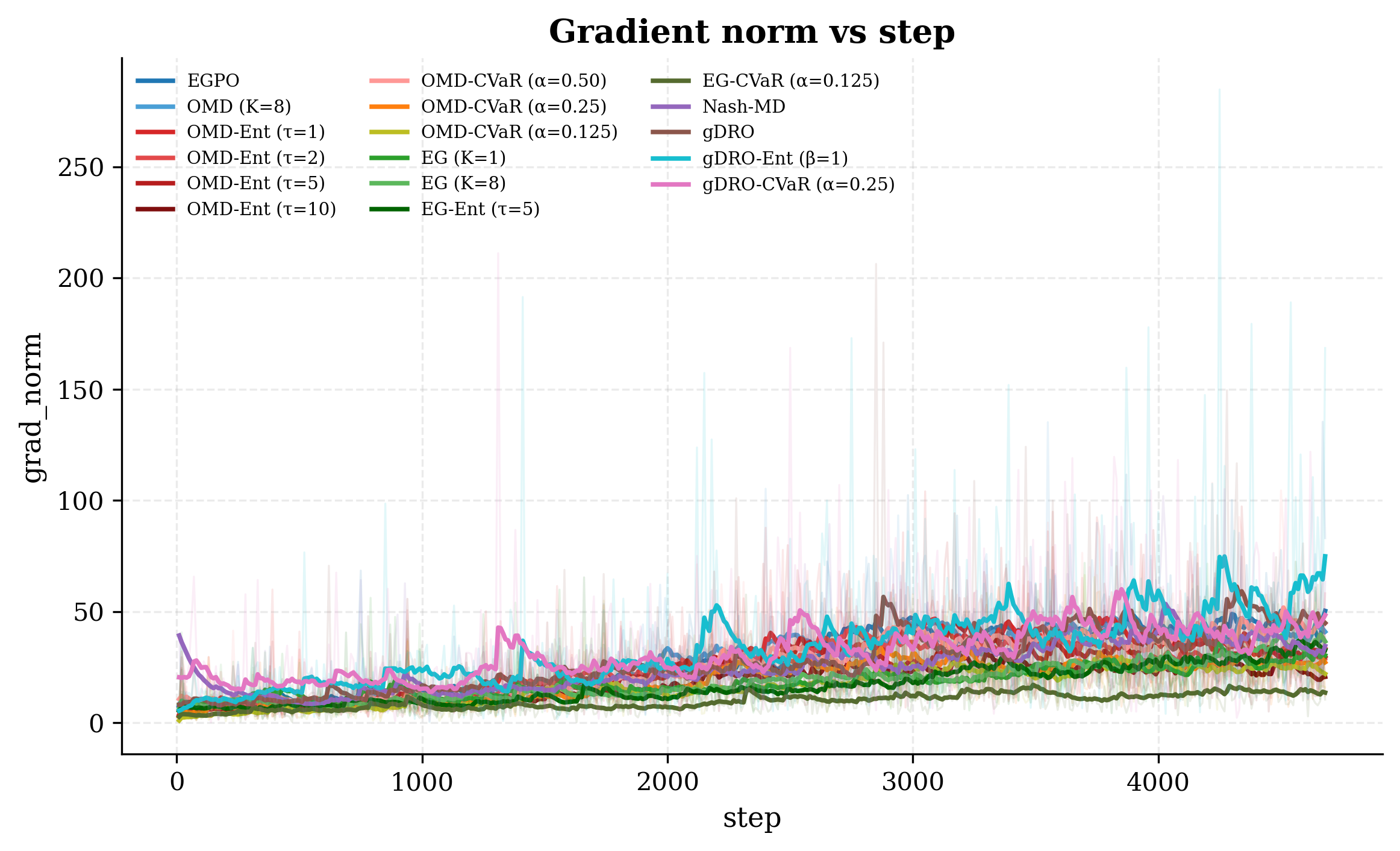}
    \caption{Gradient norm over training.}
    \label{fig:train-grad-norm}
  \end{subfigure}
  \caption{KL drift to the \texttt{SFT} reference and gradient norm over training.}
  \label{fig:train-kl-grad}
\end{figure}

Figures~\ref{fig:train-loss-acc} and~\ref{fig:train-kl-grad} report training-time diagnostics --- loss, reward accuracy, KL to the \texttt{SFT}  reference, and gradient norm --- across the full set of policies we trained, including parameter sweeps and additional families that are not reported in the cross-play evaluation: the entropic temperature sweep $\tau \in \{1,2,5,10\}$ and CVaR sweep $\alpha \in \{0.125,0.25,0.50\}$ for \texttt{OMD} ,the analogous \texttt{EG} risk variants (\texttt{EG-Ent} , \texttt{EG-CVaR}), \texttt{Nash-MD}, and the \texttt{gDRO} family. All curves are clipped to the canonical $10$-epoch horizon ($4680$ steps); faint lines are raw per-step values, bold lines are EMA-smoothed.

Loss values are not directly comparable across families because each family minimizes a different objective (the IPO loss for \texttt{OMD} ,the extragradient loss for \texttt{EG},and the group-reweighted variants for gDRO), but within a family the loss tracks learning progress, and within \texttt{OMD}  the entropic temperature $\tau$ orders the runs as expected, with $\tau{=}10$ producing the steepest descent. KL drift rises steadily and is bounded for every method, with risk-adjusted \texttt{OMD}  variants drifting slightly further than their risk-neutral counterparts and \texttt{EG} drifting less than \texttt{OMD}  at matched risk parameters. Gradient norms remain in the $\sim$$10$--$50$ band throughout, with no evidence of explosion or collapse, and reward accuracy plateaus by $\sim$$2000$ steps with \texttt{OMD-Ent}  variants edging the risk-neutral \texttt{OMD}  ($K{=}8$) baseline. Together these diagnostics say that risk adjustment, opponent-sample aggregation, and group
reweighting all train stably under our shared configuration: the evaluation differences in Sections~\ref{sec:exp-cdfs}- \ref{sec:exp-harm-categories} arise from what the methods optimize, not from differences in how well they optimize it.

\subsection{Risk-Sensitive IPO with Uncertainty over Safety Categories}
\label{sec:group_weighting}
In this appendix section, we discuss the empirical implementation details for the group weighted RSPG. 
Let
\begin{equation}
Z(\pi,\pi';\xi) \;=\; \mc{P}^{\xi}(y \succ y' \mid x) 
\;-\; \tau \log \frac{\pi(y\mid x)}{\mu(y\mid x)} 
\;+\; \tau \log \frac{\pi'(y'\mid x)}{\mu(y'\mid x)},
\label{eq:Z}
\end{equation}
where $\xi = (x,a,e,\omega,y,y')$ collects prompt, annotator, environment/subgroup, latent noise, and sampled responses, and $\mu$ is the reference policy.

The nominal (risk-neutral) IPO population objective is
\begin{equation}
\mathcal{L}(\pi,\pi') \;=\; \mb{E}_\xi\left[Z(\pi,\pi';\xi)\right].
\label{eq:L-nominal}
\end{equation}
Risk-sensitivity replaces the expectation with the entropic risk functional of parameter $\beta$: 
\begin{equation}
\Risk_\beta[Z] \;=\; \frac{1}{\beta}\log \mb{E}_\xi\left[e^{\beta Z}\right]
\;=\; \sup_{q \ll p_\xi}\left\{\mb{E}_q[Z] - \frac{1}{\beta}\KL(q \,\|\, p_\xi)\right\}.
\label{eq:variational}
\end{equation}
The variational form in~\eqref{eq:variational} makes the robustness interpretation explicit: the learner behaves as if the distribution over $\xi$ were chosen by an adversary within a KL-ball around the nominal $p_\xi$, with ball radius controlled by $1/\beta$.

Different choices of which component of $\xi$ the adversary is allowed to reweight correspond to different robustness claims. We focus on risk taken over the safety category~$e$.
\subsubsection{What "risk over group uncertainty" actually means here}

The variational identity
\[\Risk_\beta[Z] \;=\; \frac{1}{\beta}\log \mb{E}_\xi\left[e^{\beta Z}\right]
\;=\; \sup_{q \ll p_\xi}\left\{\mb{E}_q[Z] - \frac{1}{\beta}\KL(q \,\|\, p_\xi)\right\}.\]
says that entropic risk is
 DRO over a KL ball around the nominal. When you restrict the adversary to reweight only $q(e)$, the KL collapses to $\KL(q(e)\| p(e))$ and the ambiguity set is 
 \[\mc{Q}_{\eta}=\{q(e)\mid \KL(q(e)\| p(e))\leq \eta\}\]
with $\eta$ implicit in $\beta$ (Lagrangian form). 
So "risk with respect to group uncertainty" means: the training-time marginal over safety categories $p(e)$
 is not trusted; deployment might see $q(e) \ne p(e)$ and the learner prepares for the worst such $
q$ within a KL ball.
Concretely for PKU-SafeRLHF with  \[e \in \{\text{safe}, \text{unsafe-low}, \text{unsafe-med}, \text{unsafe-high}\}\]
training data has some mix, say $p = (0.6, 0.2, 0.15, 0.05)$. At deployment you might get $q = (0.3, 0.3, 0.2, 0.2)$---i.e., more unsafe-high prompts than training. Risk-sensitive training asks: among plausible reweightings, which one makes me look worst, and can I minimize my loss against that?

Importantly, risk sits on the \emph{outer} expectation over 
$e$, not inside per-sample preferences. So this is risk over \emph{aleatoric category composition}, not risk over, e.g., annotator noise or preference-model epistemic uncertainty. Those would be different choices of which coordinate of $\xi$
 the adversary reweights.

\subsubsection{Risk over the Safety Category $\xi=e$}
Factor the nominal distribution as
\begin{equation}
p(x,e,y,y') \;=\; p(e)\, p(x \mid e)\, \pi(y \mid x)\, \pi'(y' \mid x),
\end{equation}
and restrict the adversary in~\eqref{eq:variational} to reweight only the marginal over $e$:
\begin{equation}
q(x,e,y,y') \;=\; q(e)\, p(x \mid e)\, \pi(y \mid x)\, \pi'(y' \mid x).
\end{equation}
Under this restriction, the KL divergence between joints collapses to a KL between marginals,
\begin{equation}
\KL(q \,\|\, p) \;=\; \KL\big(q(e) \,\|\, p(e)\big).
\end{equation}
This encodes the statement: \emph{the category frequencies may be mis-specified at deployment, but within a category the prompt distribution, the policies, and the preference process behave as nominal}.


Define the category-conditional mean advantage
\begin{equation}
Z_e(\pi,\pi') \;:=\; \E\left[Z(\pi,\pi';\xi) \,\big|\, e\right]
\;=\; \E_{x\mid e}\,\E_{\stackrel{y\sim\pi}{\tiny y'\sim\pi'}}\left[P(y \succ y' \mid x) - \tau\log\frac{\pi(y\mid x)}{\mu(y\mid x)} + \tau\log\frac{\pi'(y'\mid x)}{\mu(y'\mid x)}\right]
\label{eq:Ze}
\end{equation}
where here $\mu$ is the reference policy. 
The variational problem over $q(e)$ then admits the closed form (Donsker--Varadhan for the discrete variable $e$):
\begin{equation}
\mathcal{L}_\beta^{e}(\pi,\pi') \;=\; \frac{1}{\beta}\log \sum_{e} p(e)\, \exp\!\big(\beta\, Z_e(\pi,\pi')\big),
\label{eq:L-risk-e}
\end{equation}
with worst-case category distribution
\begin{equation}
q^\star(e) \;\propto\; p(e)\, \exp\!\big(\beta\, Z_e(\pi,\pi')\big).
\end{equation}
As $\beta \to 0$, $\mathcal{L}_\beta^{e} \to \E_e[Z_e]$ (nominal). As $\beta \to \infty$, $\mathcal{L}_\beta^{e} \to \max_e Z_e$ (worst category).

\subsubsection{Sign convention: robust vs.\ optimistic}

The sign of $\beta$ determines whether the adversary hurts or helps the learner. Suppose $Z$ is player $1$'s payoff advantage (higher is better for player $1$) and player $1$ seeks \emph{robustness} across safety categories, i.e., wants to perform well in the worst category. Then player $1$ should maximize
\begin{equation}
\widetilde{\mathcal{L}}_\beta^{e}(\pi,\pi') \;=\; -\frac{1}{\beta}\log \sum_{e} p(e)\, \exp\!\big(-\beta\, Z_e(\pi,\pi')\big), \qquad \beta > 0,
\label{eq:L-robust}
\end{equation}
which is a soft-min over categories. Gradient mass concentrates on the categories where player $1$ is doing \emph{worst}. This is the direction to use for DRO-style safety-category robustness; using~\eqref{eq:L-risk-e} with $+\beta$ would instead produce optimism over categories.

\subsubsection{Sample-Level Implementation}

Let $\{(x_i, e_i, y_i, y_i')\}_{i=1}^B$ be a batch, and let $\mathcal{B}_e = \{i : e_i = e\}$. Compute per-category empirical advantages
\begin{equation}
\hat Z_e \;=\; \frac{1}{|\mathcal{B}_e|}\sum_{i \in \mathcal{B}_e} Z(\pi,\pi';\xi_i),
\end{equation}
and form the robust objective
\begin{equation}
\hat{\mathcal{L}}_\beta^{e} \;=\; -\frac{1}{\beta}\log\left(\sum_{e} \hat p(e)\, \exp(-\beta\, \hat Z_e)\right),
\label{eq:L-empirical}
\end{equation}
where $\hat p(e)$ is the nominal or empirical category frequency. The gradient is a category-reweighted IPO gradient,
\begin{equation}
\nabla \hat{\mathcal{L}}_\beta^{e} \;=\; \sum_{e} w_e(\beta)\, \nabla \hat Z_e, \qquad 
w_e(\beta) \;=\; \frac{\hat p(e)\, e^{-\beta \hat Z_e}}{\sum_{e'} \hat p(e')\, e^{-\beta \hat Z_{e'}}},
\label{eq:grad}
\end{equation}
i.e., the usual IPO/NLHF gradient with per-category softmax-worst-case weights $w_e$. This is a one-line modification of the IPO training loop: group sample losses by $e$, compute $w_e$, and sum.

\paragraph{Connection to group DRO.} Equation~\eqref{eq:L-empirical} is exactly the group-DRO objective of \citet{sagawa2020investigation} with temperature $\beta$, and the variational form~\eqref{eq:variational} is the KL-ball DRO justification \cite{hashimoto2018fairness}.

\end{document}